\documentclass[twoside,12pt]{article}
\pdfoutput=1

\usepackage[utf8]{inputenc} 
\usepackage{morefloats}
\usepackage{lineno}
\usepackage{graphicx}
\usepackage{booktabs}
\usepackage{threeparttable}
\usepackage{afterpage}
\usepackage{cite}
\usepackage{amssymb}
\usepackage{amsmath}
\usepackage{dsfont}
\usepackage{multirow}
\usepackage{url}
\usepackage{color}
\usepackage{verbatim}
\usepackage{rotating}
\usepackage{subfig}
\usepackage{colortbl}
\usepackage[]{xcolor}
\usepackage{siunitx} 
\usepackage{url}
\usepackage[colorlinks, citecolor=blue,anchorcolor=blue,menucolor=blue, 
linkcolor=blue,filecolor=blue,runcolor=blue,urlcolor=blue,frenchlinks=blue]{hyperref}

\newcommand{\be}{\begin{equation}}
\newcommand{\ee}{\end{equation}}
\newcommand{\bea}{\begin{eqnarray}}
\newcommand{\eea}{\end{eqnarray}}
\newcommand{\bi}{\begin{itemize}}
\newcommand{\ei}{\end{itemize}}
\newcommand{\ben}{\begin{enumerate}}
\newcommand{\een}{\end{enumerate}}
\newcommand{\la}{\left\langle}
\newcommand{\ra}{\right\rangle}
\newcommand{\lc}{\left[}
\newcommand{\rc}{\right]}
\newcommand{\lp}{\left(}
\newcommand{\rp}{\right)}

\def\frac#1#2{{{#1}\over {#2}}}
\def\gsim{\mathrel{\rlap{\lower4pt\hbox{\hskip1pt$\sim$}}
    \raise1pt\hbox{$>$}}}         
\def\lsim{\mathrel{\rlap{\lower4pt\hbox{\hskip1pt$\sim$}}
    \raise1pt\hbox{$<$}}}         
\newcommand{\mrexp}{\mathrm{exp}}

\newcommand{\art}{\mathrm{art}} 
\newcommand{\rep}{\mathrm{rep}}

\newcommand{\draft}[1]{}
\def\beq{\begin{equation}}  
\def\eeq{\end{equation}}

\def \n0{N_j^{(0)}}

\def\lapprox{\lower .7ex\hbox{$\;\stackrel{\textstyle <}{\sim}\;$}}
\def\gapprox{\lower .7ex\hbox{$\;\stackrel{\textstyle >}{\sim}\;$}}

\def\bstar{{\color{blue}$\bigstar$}}
\def\bcirc{\raisebox{-1pt}{\scalebox{1.5}{\color{blue}$\circ$}}}
\def\rsquare{{\raisebox{-1pt}{\color{red}$\blacksquare$}}}

\numberwithin{equation}{section}
\numberwithin{figure}{section}
\numberwithin{table}{section}

\makeatletter
\g@addto@macro\bfseries{\boldmath}
\makeatother

\begin{document}

\vspace{.3cm}

\begin{center}
{\Large \bf Parton distributions and lattice QCD calculations:
\\[0.2cm] a community white paper}
\vspace{.4cm}

{\small 
  Huey-Wen~Lin$^{1,2}$,
  Emanuele~R.~Nocera$^{3,4}$,
  Fred~Olness$^5$,
  Kostas~Orginos$^{6,7}$,
  Juan~Rojo$^{8,9}$ (editors),
  Alberto~Accardi$^{7,10}$, 
  Constantia~Alexandrou$^{11,12}$, 
  Alessandro~Bacchetta$^{13}$, 
  Giuseppe~Bozzi$^{13}$, 
  Jiunn-Wei~Chen$^{14}$,
  Sara~Collins$^{15}$, 	
  Amanda~Cooper-Sarkar$^{16}$,
  Martha~Constantinou$^{17}$, 
  Luigi~Del~Debbio$^{4}$, 
  Michael~Engelhardt$^{18}$, 
  Jeremy~Green$^{19}$, 
  Rajan~Gupta$^{20}$, 
  Lucian~A.~Harland-Lang$^{3,21}$, 
  Tomomi~Ishikawa$^{22}$, 
  Aleksander~Kusina$^{24}$, 
  Keh-Fei~Liu$^{25}$, 	
  Simonetta~Liuti$^{26,27}$, 		
  Christopher~Monahan$^{28}$, 		
  Pavel~Nadolsky$^{5}$,
  Jian-Wei Qiu$^{7}$,
  Ingo~Schienbein$^{23}$, 	
  Gerrit~Schierholz$^{29}$,
  Robert~S.~Thorne$^{21}$,
  Werner~Vogelsang$^{30}$,\\
  Hartmut Wittig$^{31}$, 
  C.-P. Yuan$^{1}$, and
  James Zanotti$^{32}$
}

\vspace{.2cm}
{\it \footnotesize
~$^{1}$Department of Physics and Astronomy, 
Michigan State University, East Lansing, MI 48824, USA\\
~$^{2}$Department of Computational Mathematics, Science and Engineering,\\
Michigan State University, East Lansing, MI 48824, USA\\
~$^{3}$Rudolf Peierls Centre for Theoretical Physics, 1 Keble Road,\\ 
University of Oxford, OX1 3NP Oxford, United Kingdom\\
~$^{4}$Higgs Centre for Theoretical Physics, School of Physics and Astronomy,\\ 
University of Edinburgh, EH9 3FD, UK\\
~$^{5}$Department of Physics, 
Southern Methodist University, Dallas, TX 75275, USA\\
~$^{6}$Physics Department, 
College of William and Mary, Williamsburg, VA 23187, USA\\
~$^{7}$Thomas Jefferson National Accelerator Facility, 
Newport News, VA 23606, USA\\
~$^{8}$Department of Physics and Astronomy, VU University Amsterdam,\\
De Boelelaan 1081, NL-1081, HV Amsterdam, The Netherlands\\
~$^{9}$Nikhef, Science Park 105, NL-1098 XG Amsterdam, The Netherlands\\
~$^{10}$Hampton University, Hampton, VA 23668, USA\\
~$^{11}$Department of Physics, 
University of Cyprus, P.O. Box 20537, 1678 Nicosia, Cyprus\\
~$^{12}$Computation-based Science and Technology Research Research,\\
The Cyprus Institute, 20 Kavafi Str., Nicosia 2121, Cyprus\\
~$^{13}$Dipartimento di Fisica, Universit\`{a} degli Studi di Pavia, and INFN, 
Sezione di Pavia, 27100 Pavia, Italy\\
~$^{14}$Department of Physics, Center for Theoretical Sciences,
and Leung Center for Cosmology and Particle Astrophysics,
National Taiwan University, Taipei, Taiwan 106\\
~$^{15}$Institute for Theoretical Physics, 
Universit\"at Regensburg, D-93040 Regensburg, Germany\\
~$^{16}$Department of Physics, University of Oxford,\\
Denys Wilkinson Building, 1 Keble Road, OX1 3RH Oxford, United Kingdom\\
~$^{17}$ Department of Physics, 
Temple University, Philadelphia, PA 19122, USA\\
~$^{18}$Department of Physics, 
New Mexico State University, Las Cruces, NM 88003-8001, USA\\
~$^{19}$NIC, Deutsches Elektronen-Synchrotron, 15738 Zeuthen, Germany\\
~$^{20}$Los Alamos National Laboratory, Theoretical Division, 
T-2, Los Alamos, NM 87545, USA\\
~$^{21}$ Department of Physics and Astronomy, 
University College London, WC1E 6BT, United Kingdom\\
~$^{22}$T.~D.~Lee Institute, 
Shanghai Jiao Tong University, Shanghai, 200240, P.~R.~China\\
~$^{23}$Laboratoire de Physique Subatomique et de Cosmologie, 
Universit\`e Grenoble-Alpes,\\ 
CNRS/IN2P3, 53 avenue des Martyrs,  38026 Grenoble, France. \\
~$^{24}$Institute of Nuclear Physics Polish Academy of Sciences, 
PL-31342 Krakow, Poland\\
~$^{25}$Department of Physics and Astronomy, University of Kentucky, 
Lexington, KY 40506, USA\\
~$^{26}$Physics Department, University of Virginia, 382 MCormick Road,
Charlottesville, VA 22904, USA\\
~$^{27}$Laboratori Nazionali di Frascati, INFN, Frascati, Italy\\
~$^{28}$Institute for Nuclear Theory, 
University of Washington, Seattle, WA 98195, USA\\
~$^{29}$Deutsches Elektronen-Synchrotron DESY, 22603 Hamburg, Germany\\
~$^{30}$Institute for Theoretical Physics, 
T\"ubingen University, D-72076 T\"ubingen, Germany\\
~$^{31}$PRISMA Cluster of Excellence and Institute for Nuclear Physics 
Johannes Gutenberg University of Mainz,\\ 
Johann-Joachim-Becher-Weg 45, 55128 Mainz, Germany\\
~$^{32}$CSSM, Department of Physics, 
The University of Adelaide, Adelaide, SA, Australia 5005\\
}
\vspace{0.3cm}
       {\it Preprint numbers:} \\
DESY 17-185,
IFJPAN-IV-2017-19,       
INT-PUB-17-042, 
JLAB-THY-17-2604,
MSUHEP-17-017, 
Nikhef-2017-047,
OUTP-17-15P,
SMU-HEP-17-08.

\clearpage

{\bf \large Abstract}

\end{center}

\vspace{-0.4cm}
In the framework of quantum chromodynamics (QCD), parton distribution 
functions (PDFs) quantify how the momentum and spin of a hadron are divided 
among its quark and gluon constituents.
Two main approaches exist to determine PDFs.
The first approach, based on QCD factorization theorems, realizes a QCD 
analysis of a suitable set of hard-scattering measurements, often using a 
variety of hadronic observables.
The second approach, based on first-principle operator definitions of PDFs,
uses lattice QCD to compute directly some PDF-related quantities, such 
as their moments.
Motivated by recent progress in both approaches, in this document we present
an overview of lattice-QCD and global-analysis techniques used to 
determine unpolarized and polarized proton PDFs and their moments. 
We provide benchmark numbers to validate present and future lattice-QCD 
calculations and we illustrate how they could be used to reduce
the PDF uncertainties in current unpolarized and polarized global analyses.
This document represents a first step towards establishing a common
language between the two communities, to foster dialogue and
to further improve our knowledge of PDFs.
\vspace{-0.5cm}

\tableofcontents

\section{Introduction and motivation}

The detailed understanding of the inner structure of nucleons is an 
active research field with phenomenological implications in 
high-energy, hadron, nuclear and astroparticle physics.
Within quantum chromodynamics (QCD), information on this structure ---
specifically on how the nucleon's momentum and spin are divided among quarks 
and gluons --- is encoded in parton distribution functions (PDFs).

There exist two main methods to determine PDFs.\footnote{In this paper, we do
not discuss nonperturbative, QCD-based models of nucleon structure. 
We refer the reader to~\cite{Ball:2016spl,Nocera:2014uea} and references 
therein for details on unpolarized and polarized PDFs respectively.}

The first method is the {\it global QCD analysis}~\cite{Perez:2012um,
DeRoeck:2011na,Alekhin:2011sk,Ball:2012wy,Forte:2013wc,Jimenez-Delgado:2013sma,
Rojo:2015acz,Butterworth:2015oua,Accardi:2016ndt,Gao:2017yyd}.
It is based on QCD factorization of physical observables, {\it i.e.}
the fact that a class of hard-scattering cross-sections can be expressed as a 
convolution between short-distance, perturbative, matrix 
elements and long-distance, nonperturbative, PDFs.
By combining a variety of available hard-scattering experimental data with 
state-of-the-art perturbative calculations, complete PDF sets, including 
the gluon and various combinations of quark flavors, are currently determined
for protons, in both the unpolarized~\cite{Ball:2017nwa,Harland-Lang:2014zoa,
Dulat:2015mca,Alekhin:2017kpj,Accardi:2016qay} and the
polarized~\cite{Nocera:2014gqa,deFlorian:2009vb,Sato:2016tuz,Hirai:2008aj} case.

Recent progress in global QCD analyses has been driven, on the one hand, 
by the increasing availability of a wealth of high-precision measurements from 
Jefferson Lab, HERA, RHIC, the Tevatron and the LHC and, on the other hand, 
by the advancement in perturbative calculations of QCD and 
electroweak (EW) higher-order corrections.
Parton distributions are now determined with unprecedented precision, 
in many cases at the few-percent level.
A paradigmatic illustration of this progress is provided by both the 
unpolarized and polarized gluon PDFs, which were affected by rather large 
uncertainties until recently, due to the limited experimental information 
available.
In the unpolarized case, the gluon PDF is now constrained quite accurately from 
small to large $x$ thanks to the inclusion of processes such as 
inclusive deep-inelastic scattering (DIS)~\cite{Abramowicz:2015mha}, 
$D$-meson production~\cite{Zenaiev:2015rfa,Gauld:2016kpd},
the transverse momentum of $Z$ bosons~\cite{Boughezal:2017nla},
inclusive jet production~\cite{Currie:2016bfm}, and top-quark pair
distributions~\cite{Czakon:2016olj,Guzzi:2014wia}.
In the polarized case, the gluon PDF is now constrained from double 
spin-asymmetries for high-$p_T$ jet and pion production in proton-proton 
collisions~\cite{deFlorian:2014yva,Nocera:2014gqa}, 
although only in the medium-to-large $x$ region.

The second method is {\it lattice QCD}~\cite{Olive:2016xmw,Gupta:1997nd}.
It is based on the direct computation of the QCD path integral in a 
discretized finite-volume Euclidean space-time, providing a suitable 
ultraviolet cut-off.
To connect with experimental measurements, extrapolations to the 
continuum and infinite-volume limits are necessary so that any  
cut-off dependence and finite-volume effects, respectively, are removed.
Lattice-QCD calculations require minimal external input: one needs only to 
set the hadronic scale $\Lambda_\text{QCD}$ and the values of the quark masses.
For calculations relevant to low-energy hadron structure, this means
setting the up, down and strange quark masses,
which is usually done using the pion and kaon masses as external inputs.
The overall hadronic scale can be set using well-determined baryon masses 
such as that of the $\Omega$ baryon.
A variety of QCD quantities can then be computed using lattice QCD, including
moments of PDFs or of certain quark flavor PDF combinations.

Early lattice-QCD attempts to determine the proton PDFs were limited by the 
available computational resources and various technical challenges, with most 
results restricted to the first few moments of nonsinglet PDFs at relatively 
large (unphysical) quark masses.
Overcoming these limitations, recent progress has been mostly
driven by advances in two main areas. 
First, by improved systematic control (physical pion mass, excited-state 
contamination, large volumes) for quantities such as the nucleon matrix 
elements corresponding to the low moments of PDFs.
Second, by the  development of novel strategies
for the computation of the first few 
moments~\cite{Constantinou:2014tga,Syritsyn:2014saa,Lin:2012ev},
the determination of more challenging quantities 
such as gluon and flavor-singlet matrix elements, and
for the direct calculation of the 
Bjorken-$x$ dependence of PDFs~\cite{Lin:2014zya,Alexandrou:2015rja,
Chen:2016utp,Alexandrou:2016jqi}.

These developments have pushed lattice-QCD calculations to the point where, 
for the first time, it is possible to provide information on the PDF shape
of specific flavor combinations, both for quarks and for antiquarks, 
and where meaningful comparisons with global fits can be made.
Indeed, one of the main motivations for these lattice-QCD efforts is to 
achieve a sufficient accuracy to constrain the PDFs obtained from global 
analyses.

Despite these developments in both the global QCD analysis and lattice-QCD 
methods, interplay between the two --- and communication between the 
respective communities of physicists --- have been rather limited so far.
This situation led some of us to organize the first workshop on
{\it Parton Distributions and Lattice Calculations in the LHC Era}
(PDFLattice2017), which took place in Balliol College, University of 
Oxford, in March 2017.\footnote{\url{http://www.physics.ox.ac.uk/confs/PDFlattice2017/index.asp}}
The main goal of this workshop was to establish a common ground 
and language for discussions between the two communities.
In addition, we aimed to carry out a first quantitative exploration of how PDF 
fits can be exploited to benchmark existing and future lattice calculations,
and of how lattice-QCD calculations could be used to improve global PDF fits.
In this context, some of the questions that were addressed during this workshop
included the following.
\begin{itemize}
\item What information from PDF fits is relevant to constrain, 
  test, or validate lattice calculations?

\item What PDF-related quantities are most compelling
  to compute in lattice QCD in terms of phenomenological relevance?

\item What accuracy do we need from lattice quantities 
  in order to have a significant impact on global PDF fits?

\item What information does lattice QCD provide on the
  shape (Bjorken-$x$ dependence) of the PDFs? Which specific
  PDF moments can be computed?
  
\item How do we consistently quantify the systematic uncertainties 
  in lattice-QCD calculations?

\item To what extent do available lattice results agree with the results of
  global PDF fits? Is there a tension between global PDF fits, PDF
  fits based on reduced datasets, and PDF calculations from the lattice?

\item What is the accuracy that can be expected from lattice-QCD
  calculations in the near and medium future? What will be their
  constraining power on PDFs?

\end{itemize}

This white paper summarizes the joint effort between the two communities to 
address some of these questions, and follows up on the very fruitful 
discussions and interactions that took place both during 
the workshop and in the subsequent months.
While this document does not represent the final word on this topic, it 
provides a solid starting point for further collaborative efforts, and 
should facilitate smooth interactions between the two communities in the future.

The outline of this white paper is the following.
In Sec.~\ref{sec:theoryoverview} we review the global QCD analysis and 
lattice-QCD methods for the determination of polarized and unpolarized PDFs.
In Sec.~\ref{sec:benchmarking} we present state-of-the-art benchmarks 
for selected PDF moments between the most recent lattice-QCD calculations and 
global QCD analyses.
In Sec.~\ref{sec:projections} we quantitatively assess the impact that
lattice calculations of PDF-related quantities could have on unpolarized
and polarized global analyses, assuming different scenarios for the 
uncertainties in the lattice-QCD calculations.
In Sec.~\ref{sec:outlook} we conclude
and discuss future interactions between
the global-analysis and lattice-QCD communities.
In Appendix~\ref{app:notation} we summarize the conventional notation
adopted in this document for the definition of the PDF moments; 
in Appendix~\ref{sec:LQCDtables} we compile bibliographical tables for
existing lattice-QCD calculations of PDF moments;
and in Appendix~\ref{app:Hmoms} we collect some
additional results of PDF moments from global QCD analyses.

\section{Theory overview}
\label{sec:theoryoverview}

In this section we summarize the theoretical background that underlies
lattice-QCD calculations of PDF-related quantities, on the one hand, and 
phenomenological fits of PDFs, on the other hand.
We first review the general framework in which unpolarized and 
polarized PDFs are defined, then we present the available lattice-QCD 
and global-fit approaches to determine them. 
The discussion is restricted to the information required to connect the
lattice-QCD and global-fit methods.
We have devoted particular attention to ensuring a unified and
consistent notation between the two.
An extended treatment of the subjects discussed in this section
can be found in dedicated reviews.
We refer the interested reader to Refs.~\cite{Olive:2016xmw,Gupta:1997nd}
for lattice QCD and to Refs.~\cite{Perez:2012um,DeRoeck:2011na,Alekhin:2011sk,
Ball:2012wy,Forte:2013wc,Jimenez-Delgado:2013sma,Rojo:2015acz,
Butterworth:2015oua,Accardi:2016ndt,Gao:2017yyd} for global fits.
Details on the framework underlying PDFs can be found in general 
textbooks~\cite{Ellis:1991qj,Leader:2001gr,Collins:2011zzd,
DeGrand:2006zz,Gattringer:2010zz}.

\subsection{Parton distribution functions}
\label{Sec:IntroPDFs}

Quantum chromodynamics is the non-abelian quantum field 
theory that describes the strong interaction.
It provides the theoretical foundation for the phenomenological ideas of 
quark model, color charge, and partons as hadron constituents.
The power of QCD to describe physics from the pion mass scale all the way up 
to the scale of high-energy colliders, such as the LHC, relies on the 
remarkable properties of asymptotic freedom~\cite{Gross:1973ju,Gross:1973id,
Gross:1974cs,Politzer:1974fr} and 
factorization~\cite{Collins:1987pm,Collins:1989gx}.

At high energies, or short distances, the QCD coupling is small 
and perturbation theory can accurately characterize the relevant scattering 
processes~\cite{Campbell:2006wx}.
At low energies, or larger distances, nonperturbative effects give rise to 
quark confinement and spontaneous chiral symmetry breaking~\cite{Gasser:1983yg}.
The connection between low- and high-energy dynamics is provided by QCD 
factorization theorems~\cite{Collins:1987pm,Collins:1989gx}: 
short-distance physics above the factorization scale $\mu$ is captured by 
partonic hard-scattering cross-sections calculated perturbatively as a 
power series expansion in the QCD coupling, while the 
long-distance physics below the factorization scale $\mu$ is described by 
nonperturbative quantities.
In a collinear, leading-twist factorization framework, these quantities are
universal ({\it i.e.} process-independent) PDFs.
Depending on the helicity state of the parent hadron, one usually 
distinguishes between helicity-averaged (unpolarized, henceforth)
and helicity-dependent (polarized, henceforth) PDFs.

Unpolarized PDFs are denoted as 
\begin{equation}
f(x,\mu^2)\equiv f^{\rightarrow}(x,\mu^2) + f^{\leftarrow}(x,\mu^2)\mbox{,}\qquad 
f=\{g,u,\bar{u},d,\bar{d},s,\bar{s},...\}
\,\mbox{,}
\label{eq:unpPDFs}
\end{equation}
where $x$ is the fraction
of the hadron longitudinal momentum carried by the parton,
and the sum over parton's helicities aligned along ($\rightarrow$) and 
opposite ($\leftarrow$) the parent's nucleon helicity is made explicit.
An additional index could be used to denote the hadronic species (proton,
neutron, pion, \dots).
However, we omit such a designation, as we only refer to the proton
in this paper.

At leading order (LO) in the QCD coupling series, unpolarized PDFs 
describe the probability distribution of a parton with a specified 
momentum fraction $x$.
The total momentum carried by each parton flavor is then given by 
the first moment of the corresponding PDF, for instance
\begin{align}
\int_{0}^{1}dx\ x\ \left[u(x,\mu^2)\right] 
= & {}  
\left\langle x\right\rangle _{u}(\mu^2)\,, \label{eq:umoment1}\\
\int_{0}^{1}dx\ x\ \left[u(x,\mu^2)+\bar{u}(x,\mu^2)\right] 
= & {} 
\left\langle x\right\rangle _{u^{+}}(\mu^2)\,. \label{eq:uplusmoment1}
\end{align}
Here, $\left\langle x\right\rangle _{u}$ is the momentum
carried by the up-quark, and $\left\langle x\right\rangle _{u^{+}}$ is
the momentum carried by the sum of up and anti-up quarks~\footnote{We always 
 refer to $q^+$ to indicate the sum of the quark and anti-quark PDFs of the 
 same flavor.},
see Appendix~\ref{app:notation} for our notational conventions.

Polarized PDFs describe the extent to which quarks and gluons 
with a given momentum fraction $x$ have their spins aligned with the spin 
direction of a fast moving nucleon in a helicity eigenstate. 
They are denoted as 
\begin{equation}
\Delta f(x,\mu^2) \equiv f^{\rightarrow}(x,\mu^2) - f^{\leftarrow}(x,\mu^2)
\mbox{,}\qquad f=\{g,u,\bar{u},d,\bar{d},s,\bar{s},...\}
\,\mbox{,}
\label{eq:polPDFs}
\end{equation}
where, as in Eq.~\eqref{eq:unpPDFs}, $x$ is the fractional 
momentum carried by the parton,
and the parton's spin alignment along ($\rightarrow$) or opposite 
($\leftarrow$) the polarization direction of its parent nucleon
is made explicit.

Much of the interest in polarized PDFs is related to the fact that 
their zeroth moments can be interpreted as the fractions of the proton's 
spin carried by the corresponding partons.
They are therefore the key to one of the most fundamental, 
but not yet satisfactorily answered questions in hadronic physics,
{\it i.e.}, how the spin of the proton is distributed among its constituents.
Specifically, the zeroth moments of the singlet and the gluon polarized PDFs,
\begin{align}
\Delta\Sigma(\mu^2)
& =
\sum_{q}^{N_f}\int_0^1 dx 
\left[\Delta q(x, \mu^2) + \Delta\bar{q}(x, \mu^2)\right]
\equiv
\sum_q^{N_f}\langle 1 \rangle_{\Delta q^+}(\mu^2)\,,
\label{eq:singletmom}
\\
\Delta G(\mu^2)
& =
\int_0^1 dx \Delta g(x,\mu^2)
\equiv
\langle 1 \rangle_{\Delta g}(\mu^2)
\,,
\label{eq:moments}
\end{align}
where $N_f$ is the number of active flavors,
directly contribute to the proton spin sum rule~\cite{Leader:2013jra}.

Beyond LO, PDFs are renormalization scheme-dependent 
quantities, typically worked out in the $\overline{\rm MS}$ 
scheme~\cite{tHooft:1973mfk,Weinberg:1951ss}.
When PDFs are convolved with the appropriate partonic hard-scattering 
cross-sections, computed in the same scheme, the corresponding physical 
observables are scheme-independent, up to subleading spurious 
terms in the perturbative expansion. 

Both unpolarized and polarized PDFs are accessible, theoretically and 
experimentally, through the forward Compton scattering amplitude
\begin{equation}
\label{eq:Compton}
T_{\mu\nu}(p,q,s) 
= 
\int {\rm d}^4\!z\, e^{iqz}  \langle p,s |T J_\mu(z) J_\nu(0)|p,s\rangle
\end{equation}
at large virtual photon momenta $q^2=-Q^2$. 
Here $T$ is the time-ordering operator, $J_\mu(z)$ and $J_\nu(0)$ are vector
currents at space-time points $z$ and $0$ respectively, and the 
external states are hadronic states with momentum $p$ and spin $s$.

The most general form of the Compton amplitude $T_{\mu\nu}(p,q)$ 
reads~\cite{Manohar:1992tz}
\begin{align}
T_{\mu\nu}(p,q,s) 
= {} & 
  \left(-g_{\mu\nu}+\frac{q_\mu q_\nu}{q^2}\right)\mathcal{F}_1(\omega,Q^2) 
+ \left(p_\mu-\frac{p\cdot q}{q^2}q_\mu\right) \left(p_\nu-\frac{p\cdot q}{q^2}q_\nu\right) \frac{1}{p\cdot q} \mathcal{F}_2(\omega,Q^2)
\nonumber\\ 
& {} \quad  
+ i\,\epsilon_{\mu\nu\lambda\sigma}q^\lambda s^\sigma \frac{1}{p\cdot q}\mathcal{G}_1(\omega,Q^2)
+ i\,\epsilon_{\mu\nu\lambda\sigma}q^\lambda \left(p\cdot q\, s^\sigma - s\cdot q\, p^\sigma\right) \frac{1}{(p\cdot q)^2}\mathcal{G}_2(\omega,Q^2)\,,
\label{eq:Comptampl}
\end{align}
where $\omega=2p\cdot q/q^2$ and $\mathcal{F}_1$, $\mathcal{F}_2$, 
$\mathcal{G}_1$ and $\mathcal{G}_2$ are the Compton amplitude structure 
functions.
They can be related to the electromagnetic structure functions
$F_1$, $F_2$, $g_1$ and $g_2$, used to parametrize the deep-inelastic 
scattering (DIS) hadronic tensor\footnote{A more
 general expression of the DIS hadronic tensor including electroweak currents
 can be worked out, see~\cite{Anselmino:1993tc,Anselmino:1992rn}.}
\begin{align}
W_{\mu\nu}(p,q,s)
= {} &
\frac{1}{4\pi}\int d^4z e^{iqz}\langle p,s |[J_\mu(z),J_\nu(0)]|p,s\rangle
\nonumber
\\
= {} &
\left(-g_{\mu\nu} +  \frac{q_\mu q_\nu}{q^2}\right) F_1(x,Q^2)
+\left( p_\mu - \frac{p\cdot q}{q^2}q_\mu \right)
 \left(p_\nu - \frac{p\cdot q}{q^2}q_\nu \right) \frac{1}{p\cdot q}
F_2(x, Q^2)
\nonumber
\\
& +i\,\epsilon_{\mu\nu\lambda\sigma}q^\lambda s^\sigma
\frac{1}{p\cdot q} g_1(x,Q^2)
+ i\,\epsilon_{\mu\nu\lambda\sigma}q^\lambda(p\cdot q\, s^\sigma - s\cdot q\, p^\sigma)
\frac{1}{(p\cdot q)^2}g_2(x,Q^2)\,,
\label{eq:hadtensor}
\end{align}
where $x=1/\omega$ is the Bjorken variable identified with the parton
fractional momentum at Born level; 
see~\cite{Anselmino:1992rn,Manohar:1992tz} for details.
Specifically, given the definitions in \eqref{eq:Comptampl} and 
\eqref{eq:hadtensor}, the optical theorem implies that twice the imaginary 
part of $T_{\mu\nu}$ is equal to $W_{\mu\nu}$ times $4\pi$.
Neglecting target mass corrections, one has
\begin{align}
\mathcal{F}_1(\omega,Q^2) 
= {} & 2 \omega^2 \int_0^1 dx\,  \frac{xF_1(x,Q^2)}{1-(\omega x)^2} 
= \sum_{n=2,4,\cdots}^\infty 2\omega^n \int_0^1 dx\, x^{n-1} F_1(x,Q^2) \,, \\
\mathcal{G}_1(\omega,Q^2) 
= {} & 2 \omega \int_0^1 dx\, \frac{g_1(x,Q^2)}{1-(\omega x)^2} 
= \sum_{n=1,3,\cdots}^\infty 2\omega^n \int_0^1 dx\, x^{n-1} g_1(x,Q^2)\,.
\end{align}

At a sufficiently high momentum transfer $Q^2$, power corrections can be 
neglected and QCD factorization allows one to write the structure functions 
$F_1(x,Q^2)$ and $g_1(x,Q^2)$ as a convolution between perturbatively-computable
hard-scattering cross-sections and nonperturbative parton distributions:
\begin{align}
F_1(x,Q^2) 
= {} 
& x\sum_f \int_x^1 \frac{{\rm d}z}{z}\,C_{1,f}\left(\frac{x}{z},\alpha_s(Q^2)\right)f(z,Q^2) \,, \label{eq:Fi}\\
g_1(x,Q^2) 
= {} 
& \sum_f \int_x^1\frac{dz}{z}\, \Delta C_{1,f}\left(\frac{x}{z},\alpha_s(Q^2)\right) \Delta f(z,Q^2) \,.
\label{pdf}
\end{align}
Here, the sums run over the number of active
flavors at the scale $Q^2$ (including the gluon), $C_{1,f}$ and 
$\Delta C_{1,f}$ are the perturbative partonic hard-scattering cross-sections,
$\alpha_s$ is the QCD strong coupling, and $f(x,Q^2)$ and $\Delta f(x,Q^2)$ 
are the unpolarized and polarized PDFs.

Parton distributions allow for a proper field-theoretic definition as matrix 
elements in a hadron state of bilocal operators that act to count the number 
of quarks and gluons carrying a fraction $x$ of the hadron's momentum.
The definitions are usually stated in the light-cone frame, where 
the hadron carries momentum $p$ with plus/minus components
$p^\pm=(p^0\pm p^3)/\sqrt{2}$, and transverse components equal to zero.
For example, in the case of unpolarized and polarized quark PDFs, one has
\begin{align}
q(x) & = \frac{1}{4\pi}
\int dy^-e^{-iy^-xp^+}\langle p|\bar{\psi}(0,y^-,\mathbf{0}_\perp)
\gamma^+\mathcal{G}\psi(0,0,\mathbf{0})|p\rangle\,,
\label{eq:LCdefunp}\\
\Delta q(x) & = \frac{1}{4\pi}
\int dy^-e^{-iy^-xp^+}\langle p, s|\bar{\psi}(0,y^-,\mathbf{0}_\perp)
\gamma^+\gamma^5\mathcal{G}\psi(0,0,\mathbf{0})|p, s\rangle\,,
\label{eq:LCdefpol}
\end{align}
where $\psi$ is the quark field and $\mathcal{G}$ is an appropriate gauge link
required to make Eqs.~\eqref{eq:LCdefunp}--\eqref{eq:LCdefpol} gauge invariant.
See Refs.~\cite{Collins:1981uw,Curci:1980uw,Baulieu:1979mr,Collins:1989gx} 
for the definition of $\mathcal{G}$ and for
explicit light-cone formul{\ae} of unpolarized and polarized gluon PDFs.

While PDFs cannot be calculated perturbatively, their dependence on the scale 
$\mu$ resulting from factorization can be.
This is done by means of the
DGLAP (Dokshitzer-Gribov-Lipatov-Altarelli-Parisi) 
evolution equations~\cite{Dokshitzer:1977sg,Gribov:1972ri,Altarelli:1977zs},
a set of integro-differential coupled equations of the form
\begin{equation}
  \label{eq:dglapunp}
\frac{\partial f^\prime(x,\mu^2)}{\partial \ln \mu^2}
=
\sum_{f=g,q,\bar{q}}\int_x^1 
\frac{{\rm d}z}{z}P_{f^\prime f}\left(\frac{x}{z},\alpha_s(\mu^2)\right)f(z,\mu^2)\, ,
\end{equation}
\begin{equation}
  \label{eq:dglappol}
\frac{\partial \Delta f^\prime(x,\mu^2)}{\partial \ln \mu^2}
=
\sum_{f=g,q,\bar{q}}\int_x^1 
\frac{{\rm d}z}{z}\Delta P_{f^\prime f}\left(\frac{x}{z},\alpha_s(\mu^2)\right)\Delta f(z,\mu^2)\, .
\end{equation}
In short, the logarithmic derivative of the PDF is determined by a convolution
of the PDFs with the unpolarized (polarized) DGLAP kernels $P_{f^\prime f}$
($\Delta P_{f^\prime f}$), which can be 
computed perturbatively in powers of $\alpha_{s}$.
The unpolarized splitting functions $P_{f^\prime f}$ are currently completely 
known up to NNLO~\cite{Moch:2004pa,Vogt:2004mw} in the $\overline{\rm MS}$ 
renormalization scheme.
Results for the unpolarized nonsinglet splitting functions have appeared 
recently at N$^3$LO~\cite{Davies:2016jie,Moch:2017uml}.
The polarized splitting functions $\Delta P_{f^\prime f}$ are currently known 
up to  NNLO~\cite{Moch:2014sna} in the $\overline{\rm MS}$ scheme.
The DGLAP evolution equations can be solved numerically using
either $x$-space or Mellin $N$-space techniques that are widely available 
in various public codes~\cite{Vogt:2004ns,Salam:2008qg,Botje:2010ay,
Bertone:2013vaa,Bertone:2015cwa}.
The typical level of agreement for the results of the PDF evolution 
has been demonstrated to be of 
$\mathcal{O}(10^{-5})$~\cite{Giele:2002hx,Dittmar:2005ed}.

\subsection{Lattice QCD}
\label{Sec:IntroLQCD}

The lattice-QCD method is based on regularizing QCD on a finite Euclidean 
lattice and is generally studied by numerical computation of QCD correlation 
functions in the path-integral formalism~\cite{DiPierro:2000nt,Lepage:1998dt,
Luscher:1998pe,Gupta:1997nd}, using methods adapted from statistical 
mechanics~\cite{Binder:2015klx,Newman:1999mng}.
To make contact with experimental data, the numerical results are extrapolated 
to the continuum and infinite-volume limits.
The past decade has seen significant progress in the development of efficient 
algorithms for the generation of ensembles of gauge field configurations and 
tools for extracting the relevant information from lattice-QCD
correlation functions.
In this respect, lattice-QCD calculations have reached a level where
they not only complement, but also guide current and forthcoming
experimental programs~\cite{Brodsky:2015aia,Aschenauer:2014twa}.

In this section, we discuss the sources of systematic uncertainties
that affect current lattice QCD calculations, and we present 
lattice-QCD methods to determine either the Mellin moments of PDFs
or the complete PDF $x$-dependence.

\subsubsection{Systematic uncertainties}
Lattice-QCD calculations must demonstrate control over all sources of
systematic uncertainty introduced by the discretization of QCD on the
lattice to make meaningful contact with experimental data.
These
include discretization effects that vanish in the continuum limit;
extrapolation from unphysically heavy pion masses; finite volume
effects; and renormalization of composite operators.
To take the continuum limit requires accurate determinations of the 
lattice spacing.
We briefly review these main sources of systematic uncertainty here; for a 
fuller account see, for example, Ref.~\cite{Aoki:2016frl}.

\begin{itemize}

\item {\bfseries Discretization effects and the continuum limit.} 
There is a fair degree of flexibility in discretizing the QCD action. 
This has led to a variety of formulations, which differ mainly in the choice of
the action for quarks.
In the continuum limit, which corresponds to taking
the lattice spacing $a$ to zero with all physical quantities fixed,
the simplest discretizations differ from continuum QCD at ${\mathcal
O}(a)$.
In practice, one cannot afford to perform numerical
simulations at arbitrarily small lattice spacings, because the cost of
computation increases with a large inverse power of the lattice
spacing, and ${\cal O}(a)$ effects can be significant even
with current lattice spacings ranging from $0.15 \,\mbox{fm}$ to
$0.05 \,\mbox{fm}$.
To accelerate the convergence to the continuum
limit, improved quark and gluon actions are widely used, which include
higher-dimension operators to reduce the discretization errors to
${\cal O}(a^2)$ or better.
Chiral fermions with automatic $\mathcal{O}(a)$ improvement and small 
$\mathcal{O}(a^2)$ discretization errors are also adopted to admit 
calculations on coarser lattice spacings~\cite{Creutz:2011hy,Vladikas:2011bp,
Chandrasekharan:2004cn}. 

\item {\bfseries Pion mass dependence.} 
The computational cost of the fermion contribution to the path
integral increases with a large inverse power of the bare quark mass
(or, equivalently, the pion mass).
Lattice-QCD calculations are therefore
often performed at unphysically heavy pion masses, although results calculated
directly with physical pion masses have become increasingly common, albeit with larger
errors.
To obtain results at the physical pion mass, lattice data are
generated at a sequence of pion masses and then extrapolated to the
physical pion mass.
To control the associated systematic
uncertainties, these extrapolations are guided by effective
theories.
In particular, the pion-mass dependence can be parametrized
using chiral perturbation theory ($\chi$PT)~\cite{Golterman:2009kw}, 
which accounts for the
Nambu-Goldstone nature of the lowest excitations that occur in the
presence of light quarks. 
%

\item {\bfseries Finite volume effects.} Numerical lattice-QCD 
calculations are necessarily restricted to a finite space-time
volume, {\it e.g.}, a hypercube of side $L$.
For most simple quantities, these effects decay exponentially
with the size of the lattice~\cite{Luscher:1985dn,Luscher:1986pf}, and 
therefore the easiest way to
minimize or eliminate finite volume effects is to choose the volume
sufficiently large in physical units.
Unfortunately, this can be
prohibitively expensive as one approaches the continuum limit, requiring the
number of lattice sites to grow as $L/a$ in all four directions. 
Finite volume $\chi$PT is the preferred
tool to develop systematic expansions that provide quantitative
information on finite-volume effects.
In general, finite volume
effects of hadrons are dominated by their interactions with pions,
which can travel around the (periodic) lattice many times.
Numerical evidence suggests that lattice sizes of $m_\pi L \geq 4$, where
$m_\pi$ is the pion mass, are generally sufficiently large that finite
volume effects are negligible for mesons, within the current precision 
of lattice-QCD calculations.
From the studies of the pseudoscalar and electromagnetic form factors of the 
nucleon, it is evident that larger physical volumes are needed for the 
baryons.

\item {\bfseries Excited state contamination.} 
At small Euclidean times, a lattice-QCD correlation function
is a sum over a tower of states that behave as $e^{-m_it}$, where $m_i$ is the 
energy of the state and $t$ is the Euclidean time. 
Thus, at large Euclidean times,
ground-state quantities can be extracted by fitting to the dominant 
exponential behavior.
Unfortunately, the signal-to-noise ratio is exponentially suppressed 
as $e^{-(E_N-3m_\pi/2)t}$, where $E_N$ is the nucleon energy~\cite{Lepage:1989hd}.
Thus, lattice-QCD results
are extracted from an intermediate region in which excited state contributions 
are either small or well-controlled and the signal-to-noise ratio is 
sufficiently large that the signal can be reliably extracted. 
This is a particular challenge for baryons and is one of the largest 
sources of systematic uncertainties for nucleon matrix elements.

\item {\bfseries Renormalization.} The matrix elements extracted from a 
lattice-QCD calculation at a given lattice spacing are bare matrix elements,
rendered finite by the presence of the lattice spacing, which serves
as a gauge-invariant UV regulator. 
To take the continuum limit, {\it i.e.}, remove the regulator, one must 
renormalize the corresponding operators and fields and match them to some 
common scheme and scale used by phenomenologists. 
Although renormalization is traditionally
discussed in the framework of perturbation theory, at hadronic energy
scales the renormalization constants should be computed
nonperturbatively to avoid uncontrolled uncertainties due to 
truncated perturbative results.
To compare with phenomenology, which uses the $\overline{\rm MS}$ scheme, 
a conversion factor from the nonperturbative scheme must be computed 
perturbatively. 
This requires a renormalization condition that can be implemented on the 
lattice and in continuum perturbation theory. 
In QCD with only light quarks it is technically advantageous to employ 
so-called mass-independent renormalization schemes. 
A common choice is the regularization-independent/momentum (RI/MOM) 
scheme~\cite{Martinelli:1994ty}.

In addition, on a hypercubic lattice, the orthogonal group $O(4)$ of
continuum Euclidean space-time is reduced to the hypercubic group
$H(4)$.
Thus, operators are classified according to irreducible
representations of $H(4)$~\cite{Gockeler:1996mu}.
Different irreducible representations belonging to the same $O(4)$ multiplet
will, in general, give different answers at finite lattice spacing, an effect 
that can be reduced by improving the operators~\cite{Gockeler:2004wp}.
Conversely, operators that lie in different irreducible representations of 
$O(4)$, but the same irreducible representations of $H(4)$, will mix at finite 
lattice spacing but not in the continuum. 
When these operators have lower mass dimensions,
the mixing coefficients scale with the inverse lattice spacing to some
power, and diverge in the continuum limit.
This power-divergent mixing
must be removed nonperturbatively, and is a particular challenge for
lattice calculations of the Mellin moments of PDFs (see
Sec.~\ref{Sec:MomentsLQCD}).


\item {\bfseries Lattice-spacing determination.} 
Numerical lattice-QCD calculations naturally determine all dimensionful 
quantities in units of the lattice spacing. 
Thus, extracting physical values requires the determination of the lattice 
scale. 
This is achieved by matching a quantity with mass dimension to its experimental 
value or through a well-defined theoretical procedure, that is referred to as
{\it scale-setting}. 
Popular reference scales include light decay constants, hadron masses, 
scales defined in terms of the heavy quark potential or, most recently, 
the length scales $\sqrt{t_0}$~\cite{Luscher:2010iy} and 
$w_0$~\cite{Borsanyi:2012zs} defined via the Wilson  gradient 
flow~\cite{Luscher:2010iy}. 
These scales can be computed cheaply and can be used to 
match scales between different gauge ensembles very accurately.  
However, a hadron mass or a decay constant --- which are known accurately 
from experiment and can be computed precisely in lattice-QCD --- 
have to be used for absolute scale setting. 
A popular hadronic mass for this purpose is the mass of the triply strange 
$\Omega$ baryon~\cite{Durr:2008zz} or the 2S-1S splitting in the Upsilon 
spectrum~\cite{Kendall:2008zz}.

\end{itemize}

These sources of systematic uncertainty all need to be under control
when confronting experimental data with lattice results, or vice
versa.
For a coherent assessment of the present state of lattice-QCD
calculations of various quantities, the degree to which each
systematic has been controlled in a given calculation is an important
consideration.
In Sec.~\ref{subsubsec:BClQCD}, we characterize the
quality of the lattice calculations, based on criteria inspired by the
FLAG analysis of flavor physics on the lattice~\cite{Aoki:2016frl}.

\subsubsection{Mellin moments of PDFs from lattice QCD}
\label{Sec:MomentsLQCD}

Parton distributions cannot be directly determined in Euclidean lattice QCD, 
because their field-theoretic definition involves fields at light-like 
separations.
Instead, the traditional approach for lattice-QCD calculations has been to 
determine the matrix elements of local twist-two operators, where twist is the 
dimension minus the spin, that can be related to the Mellin moments of PDFs.
In principle, given a sufficient number of Mellin moments, PDFs can be 
reconstructed from the inverse Mellin transform. 
In practice, however, the calculation is limited to the lowest three moments, 
because power-divergent mixing occurs between twist-two operators on the 
lattice.
Three moments are insufficient to fully reconstruct the momentum dependence of 
the PDFs without significant model dependence~\cite{Detmold:2003rq}.
The lowest three moments do provide, however, useful information, both as 
benchmarks of lattice-QCD calculations and as constraints in global extractions 
of PDFs. 
Here we briefly review the determination of Mellin moments of PDFs from lattice 
QCD. 
We emphasize that the order of each moment is counted from zero ({\it i.e.},
the lowest moment is the zeroth moment), see Appendix~\ref{app:notation} 
for explicit definitions.

Using the operator product expansion (OPE)~\cite{Zimmermann:1972tv}, the Mellin 
moments of the structure functions and the corresponding PDFs can be expressed, 
up to higher-twist effects, in terms of matrix elements of local operators:
\begin{align}
\!\!\!2 \int_0^1 dx\, x^{n-1} F_1(x,Q^2) &= \sum_a C_{1,a}^n(\mu^2)\, v_a^n(\mu^2)|_{\mu^2=Q^2} = \sum_a C_{1,a}^n(Q^2)\, \int_0^1 dx\, x^{n-1} f_a(x,Q^2)\,,\\
4 \int_0^1 dx\, x^n g_1(x,Q^2) &= \sum_a \Delta C_{1,a}^n(\mu^2)\, a_a^n(\mu^2)|_{\mu^2=Q^2} = \sum_a \Delta C_{1,a}^n(Q^2)\, \int_0^1 dx\, x^n\, 2 \Delta f_a(x,Q^2)\,,
\end{align}
where $v_i^n(\mu^2)$ and $a_i^n(\mu^2)$ are reduced matrix elements of the appropriate twist-two operators~\cite{Gockeler:1995wg},
\begin{align}
\frac{1}{2} \sum_s \langle p,s|\mathcal{O}^i_{\{\mu_1,\cdots,\mu_n\}}|p,s\rangle = {} & 2 v_i^n\, [p_{\mu_1}\cdots p_{\mu_n} - {\rm traces}] \,, \label{eq:twist2me}\\
\langle p,s|\mathcal{O}^{5\,i}_{\{\sigma \mu_1,\cdots,\mu_n\}}|p,s\rangle = {} & \frac{1}{n+1} a_i^n\, [s_\sigma p_{\mu_1}\cdots p_{\mu_n} - {\rm traces}]\,,
\end{align}
and $C_{1,i}^n(\mu^2)$ and $\Delta C_{1,i}^n(\mu^2)$ are the Mellin moments of the corresponding Wilson coefficients
\begin{equation}
C_{1,i}^n(\mu^2) = \int_0^1 dy\, y^{n-1} C_{1,i}(y,\mu^2)\,, \quad
\Delta C_{1,i}^n(\mu^2) = \int_0^1 dy\, y^n \Delta C_{1,i}(y,\mu^2)\,.
\end{equation}
The trace terms include operators with at least one factor of the metric 
tensor $g^{\mu_i \mu_j}$ multiplied by operators of dimension $(n+2)$ with 
$n-2$ Lorentz indices. 
The operators relevant for the lowest two moments are listed in 
Table~\ref{Tab:twist2}. 
The operator $\mathcal{O}^q_{\mu_1\mu_2}$ decomposes into two different 
representations of $H(4)$~\cite{Gockeler:1996mu}, each with different 
lattice artifacts and renormalization factors. 
In the continuum limit, however, both operators should lead to the same result. 
In contrast, the operator $\mathcal{O}^q_{\mu_1\mu_2\mu_3}$ splits into several 
representations of $O(4)$ transforming identically under $H(4)$ and causing 
the corresponding operators to mix under renormalization on the lattice.

\begin{table}
\renewcommand{\arraystretch}{1.6} 
\centering
\begin{tabular}{@{}ccc@{}}
\toprule
Matrix element & Operator & PDF moment \\ 
\midrule
$v_q^2$\,, $v_{\bar{q}}^2$  & 
$\displaystyle \left({\rm i}/2\right) \bar{q}(x)\gamma_{\mu_1} \overleftrightarrow{D}_{\mu_2} q(x)$ & 
$\langle x \rangle_{q^+}$\\
$v_q^3$\,, $v_{\bar{q}}^3$  & $\displaystyle \left({\rm i}/2\right)^2 \bar{q}(x)\gamma_{\mu_1} \overleftrightarrow{D}_{\mu_2} \overleftrightarrow{D}_{\mu_3} q(x)$ & $\langle x^2 \rangle_{q^-}$\\
$a_q^0$ & $\displaystyle \bar{q}(x)\gamma_{\sigma} \gamma_5 q(x)$ & 
$2\, \langle 1 \rangle_{\Delta q^+}$ \\
$a_q^1$ & $\displaystyle \left({\rm i}/2\right) \bar{q}(x)\gamma_{\sigma} \gamma_5 \overleftrightarrow{D}_{\mu_1} q(x)$ & $2\, \langle x \rangle_{\Delta q^-}$ \\
$v_g^2$ & $\displaystyle - {\rm Tr}\, F_{\mu_1\alpha}F_{\mu_2\alpha}$ & $\langle x \rangle_g$ \\
\bottomrule
\end{tabular}
\caption{\label{Tab:twist2}
\small List of operators relevant for the computation of the lowest two 
Mellin moments of polarized and unpolarized PDFs.
Here we indicate, for each operator, the corresponding matrix element and
the specific PDF moment that can be evaluated (see
Appendix~\ref{app:notation} for the notation used).
}
\end{table}

\paragraph*{Higher-twist contributions.}
The discussion so far has focused on the limit in which higher-twist 
contributions, suppressed by powers of the momentum transfer, have been ignored.
In fact, higher-twist contributions to the lowest moment of the structure 
function $F_1(x,Q^2)$ are found to be of 
${\cal O}(1\mbox{ GeV}^2/Q^2)$~\cite{Blumlein:2008kz}.
For lattice QCD, typically $Q^2 \simeq 1/a^2$, and at present lattice spacings 
this corresponds to $Q^2 = {\cal O}(10\mbox{ GeV}^2)$ or a higher-twist 
contribution of 5--$10\%$. 
With contributions of higher-twist included, the OPE for $F_1$ reads
\begin{equation}
2 \int_0^1 dx\, x F_1^q(x,Q^2) = C_{1,q}^2(\mu^2)\, v_q^2(\mu^2)|_{\mu^2=Q^2} + \frac{\bar{C}_{1,q}^2(\mu^2)}{Q^2}\, \bar{v}_q^2(\mu^2)|_{\mu^2=Q^2} + \cdots \,,
\label{tex}
\end{equation}
where $\bar{C}_{1,q}^2$ and $\bar{v}_q^2(\mu^2)$ are the Wilson coefficient and 
reduced matrix element of a generic twist-four operator. 
Both twist-two and four contributions mix under renormalization, to the extent 
that the perturbative series for the Wilson coefficients $C_{1,q}^2(\mu^2)$ 
diverges due to the presence of infrared (IR) renormalon singularities.
This ambiguity is canceled by that in the twist-four matrix element 
$\bar{v}_q^2(\mu^2)$ that arises as a result of an ultraviolet (UV) 
renormalon singularity~\cite{Martinelli:1996pk}. 
If mixing effects are ignored, the uncertainties will be, at least, comparable 
to the power corrections themselves.
Power corrections can be assessed most efficiently, and the twist expansion 
tested, by a direct lattice-QCD evaluation of the Compton amplitude, which we 
discuss in Sec.~\ref{Sec:InversionMethod}.

\paragraph*{Beyond the first three moments.}
Moving beyond the lowest three moments requires overcoming the challenge of 
power-divergent mixing for lattice-QCD twist-two operators.
One novel approach to this problem~\cite{Davoudi:2012ya} builds upon the 
physical intuition that as long as the scale associated with the operator 
(for the twist-two operators, this is the renormalization scale $\mu$) is taken 
to be much larger than the hadronic scale but much smaller than the inverse 
lattice spacing, no singularity necessarily arises as one takes the continuum 
limit.
The operator can still probe the correct hadron structure at the scale $\mu$, 
but should be insensitive to the details of the discretization of the operator 
at shorter distances.
A simple way to incorporate an intrinsic {\it smearing} scale for an operator 
is to sum over bilinears of quark fields that are displaced over many lattice 
sites in a small (compared to the scale $1/\mu$) region of Euclidean space-time 
(an alternative approach appears in Ref.~\cite{Monahan:2015lha}).

To ensure that the correct $SO(4)$ transformation properties of the matrix 
elements are recovered in the continuum limit, one must project the sum using 
hyper-spherical harmonics.
The properties of these operators, such as their mixing patterns and scaling 
properties, are discussed in detail in Ref.~\cite{Davoudi:2012ya}.
In particular, while the classical mixing with lower and higher spin operators 
are both suppressed by $\sim a^2$ for spatially improved operators, the mixing 
at one-loop in lattice perturbation theory is suppressed 
by ${\cal O}(\alpha_s a)$ or ${\cal O}(\alpha_s a^2)$. 
The suppression depends on the lattice action used, provided that the gauge 
action adopted to construct the gauge-invariant bilinears is tadpole-improved 
and smeared over a region whose physical size is held fixed as the continuum 
limit is taken. 
In principle, this allows higher moments of PDFs to be obtained from lattice 
QCD, without power divergences. Numerical investigations of this approach,
which requires gauge configurations with very fine lattice spacings, are 
underway.
Other approaches that avoid power-divergent mixing have also been suggested, 
including coupling fictitious heavy quarks to light-quark 
currents~\cite{Detmold:2005gg}, and calculating current correlators in 
position space~\cite{Braun:2007wv}. 
The practical application of these ideas is yet to be studied nonperturbatively.

\subsubsection{The $x$-dependence of PDFs from lattice QCD}
\label{sec:xdependence}

While the lowest three moments of PDFs can provide important benchmarks for 
lattice-QCD calculations of nucleon structure, and useful constraints in global 
extractions of PDFs, they are not in themselves sufficient to determine the 
$x$-dependence of PDFs.
In the following section we summarize recent approaches to determining the 
$x$-dependence of PDFs directly from lattice QCD.

\paragraph*{Hadronic tensor.} 
In principle, PDFs can be determined from hadronic tensors provided the 
higher-twist contributions, which have different $Q^2$ dependence than the 
leading-twist, can be subtracted. 
Calculating the hadronic tensor in the Euclidean path-integral approach
has the advantage that no renormalization is required if conserved vector      
currents are used in the current-current correlation and only finite 
renormalizations are needed for the local currents.
Furthermore, since the structure functions are frame-independent, they 
can be calculated in any momentum frame of the nucleon. 
One can choose the nucleon momenta and momentum transfers judiciously 
to have a desirable coverage of $x$ for a given $Q^2$. 
However, the inverse Laplace transform that is needed to convert the hadronic tensor from Euclidean space to Minkowski space can be a 
challenge~\cite{Liu:1993cv,Liu:1999ak}. 
Three numerical approaches, the Backus-Gilbert method~\cite{Hansen:2017mnd}, 
improved maximum entropy, and fitting with model spectral functions, 
are suggested to tackle this inverse Laplace-transform 
problem~\cite{Liu:2016djw}. 
In Ref.~\cite{Liu:1993cv} sea partons are separated into {\it connected sea} 
and {\it disconnected sea} contributions, based on the distinct topologies of 
the diagrams in a lattice computation. 
This distinction can help identify the impact on PDF uncertainties of 
improving the uncertainties associated with disconnected diagrams determined 
using lattice-QCD.
The extended evolution equations to accommodate both the connected sea 
(CS) and disconnected sea (DS) partons are derived in Ref.~\cite{Liu:2017lpe}.

\paragraph*{The inversion method.} 
\label{Sec:InversionMethod}

The Compton amplitude $T_{\mu\nu}(p,q)$, Eq.~\eqref{eq:Compton}, can be
directly obtained in lattice QCD, including disconnected contributions,  
by a simple extension~\cite{Chambers:2017dov} of existing implementations of 
the Feynman-Hellmann technique to lattice QCD~\cite{Horsley:2012pz,
Chambers:2014qaa,Chambers:2015bka}.
Provided one works at sufficiently large $Q^2$, the Compton amplitude will be 
dominated by twist-two contributions.
Varying $Q^2$ allows one to test the twist expansion and, in particular, 
isolate twist-four contributions. Moreover, one can distinguish between 
contributions from up, down and strange quarks, connected and disconnected, 
by appropriate insertions of the electromagnetic current.

To compute the Compton amplitude from the Feynman-Hellmann relation, a 
perturbation to the QCD Lagrangian is introduced, for example,
\begin{equation}
\mathcal{L}(x) 
\rightarrow 
\mathcal{L}(x) + \lambda \mathcal{J}_3(x)\,, 
\quad 
\mathcal{J}_3(x)
=
Z_V\cos(\vec{q} \cdot \vec{x})\; 
e_q \,\bar{q}(x)\gamma_3 q(x) 
\label{in}
\end{equation}
where $q$ is the quark field to which the photon is attached, and $e_q$ its 
electric charge. 
For simplicity, we consider the local vector current only, so that the 
renormalization factor $Z_V$ is known and no further renormalization is needed. 
Taking the second derivative of the nucleon two-point function 
\begin{equation}
\langle N(\vec{p},t) \bar{N}(\vec{p},0)\rangle_\lambda 
\simeq 
C_\lambda\, e^{-E_\lambda(p,q)\,t}
\end{equation}
with respect to $\lambda$ on both sides, gives
\begin{equation}
-2 E_\lambda(p,q)\, 
\frac{\partial^2}{\partial\lambda^2}  E_\lambda(p,q)\,\big|_{\lambda=0} 
= 
T_{33}(p,q) \,.
\end{equation}
For $p_3=q_3=q_4=0$ this leaves us with
\begin{equation}
T_{33}(p,q) 
= 
4 \omega^2 \int_0^1 dx\,  \frac{xF_1(x,Q^2)}{1-(\omega x)^2} \,.
\label{ff}
\end{equation}
Extracting the polarized structure functions requires insertions of two 
different currents with $\mu\neq \nu$. 
The idea is then to solve Eq.~\eqref{ff} for $F_1(x,Q^2)$ numerically.
In Refs.~\cite{Ji:2001wha,Chambers:2017dov} it was shown that the unpolarized 
structure function $F_1(x,Q^2)$ can be computed from a lattice calculation 
of the Compton amplitude, devoid of any renormalization and mixing issues. 
With the same method, PDFs can be computed directly without the need to go 
through the structure functions, provided $Q^2$ is sufficiently large that 
power corrections can be neglected. 

\paragraph*{Quasi-PDFs.}
Quasi-PDFs provide an alternative approach to determining the $x$-dependence 
of PDFs directly from lattice QCD~\cite{Ji:2013dva,Ji:2014gla}. 
In the following discussion, we focus on the flavor-nonsinglet quasi-PDF, 
for which we can ignore mixing with the gluon quasi-PDF. 
The unpolarized quark quasi-PDF is defined as the momentum-dependent
nonlocal forward matrix element
\begin{align}\label{eq:qPDF}
\widetilde{q}(x,\Lambda,p_z)  
= {} &  \int \frac{dz}{2\pi} e^{-i x z p_z} p_z h(z,p_z), \nonumber \\
h(z,p_z) 
= {} &
\frac{1}{4 p_{\alpha}}\sum_{s=1}^2\left\langle p,s\right\vert \bar{\psi}(z)\gamma_\alpha e^{ig\int_0^z
A_z(z^\prime) dz^\prime} \psi(0) \left\vert p,s\right\rangle,
\end{align}
where $\Lambda$ is an UV cut-off scale, such as the inverse lattice spacing 
$1/a$. 
The Lorentz index $\alpha$ of the matrix $\gamma_\alpha$ is generally chosen 
to be spatial, $\alpha = z$, but the alternative choice $\alpha = 4$ is also 
possible and removes part of the leading order twist-4 
contamination~\cite{Xiong:2013bka,Radyushkin:2016hsy}. 
Because $p$ is finite, the momentum fraction $x$ can be larger than unity.

The quasi-PDF is defined for nucleon states at finite momentum and must be 
related to the corresponding light-front PDF\footnote{In this context the term 
 light-front PDF is used to distinguish ordinary PDFs, 
 Eqs.~\eqref{eq:unpPDFs}--\eqref{eq:polPDFs} from quasi-PDFs, 
 Eq.~\eqref{eq:qPDF}.}, 
for which the nucleon momentum is taken to infinity.
In the  large-momentum  effective field theory (LaMET) approach, the
quasi-PDF $\widetilde{q}(x,\Lambda,p_z)$ can be related to the $p_z$-independent
light-front PDF $q(x,Q^2)$ through~\cite{Ji:2013dva,Ji:2014gla}
\begin{equation} \label{eq:qPDFmatching}
\widetilde{q}(x,\Lambda ,p_z) = 
  \int_{-1}^1 \frac{dy}{\left\vert y\right\vert} 
    Z\left( \frac{x}{y}, \frac{\mu}{p_z}, \frac{\Lambda}{p_z}\right)_{\mu^2 = Q^2} q(y,Q^2) +
  \mathcal{O}\left( \frac{\Lambda_\text{QCD}^2}{p_z^2},\frac{M^2}{p_z^2}\right), 
\end{equation}
where $\mu$ is the renormalization scale,
$Z$ is a matching kernel and $M$ is the nucleon mass.
Here the $\mathcal{O}\left(M^2/p_z^2\right)$ terms are target-mass corrections 
and the $\mathcal{O}\left(\Lambda_\text{QCD}^2/p_z^2\right)$ terms are 
higher-twist effects, both of which are suppressed at large nucleon momentum. 
%
%
An alternative, but related, construction is proposed in 
Refs.~\cite{Radyushkin:2016hsy,Radyushkin:2017cyf} and explored in 
Ref.~\cite{Orginos:2017kos}.

Preliminary results from lattice calculations of quasi-PDFs have been 
encouraging~\cite{Lin:2014zya,Alexandrou:2015rja,Chen:2016utp,
Alexandrou:2016jqi}. 
However, there are a number of remaining challenges that must be overcome for 
an {\it ab initio} determination of the $x$-dependence of PDFs directly from 
lattice QCD that incorporates complete control over systematic uncertainties. 
Lattice calculations of quasi-PDFs are subject to the same sources of 
systematic uncertainty that affect all lattice calculations, see 
Sec.~\ref{Sec:IntroLQCD}. 
Here we focus on systematic uncertainties that are more specific to quasi-PDFs.
These are uncertainties associated with the finite nucleon momentum of the 
lattice calculations and to the renormalization of quasi-PDFs.

\begin{itemize}

\item Preliminary nonperturbative studies of the quasi-PDF used nucleon 
momenta in the range $p_z = 2\pi/L$ to $10\pi/L$, where $L$ is the physical 
extent of the lattice, corresponding to $p_z = 0.5$ to 
$2.5$~GeV~\cite{Lin:2014zya,Alexandrou:2015rja,Chen:2016utp,Alexandrou:2016jqi}.
At such low momenta, higher-twist and target mass corrections are likely to be 
considerable.

Target mass corrections can be removed to all orders~\cite{Chen:2016utp}, and 
twist-4 contributions can be removed in 
principle~\cite{Chen:2016utp,Radyushkin:2016hsy}, leaving higher-twist 
contamination. 
To reduce these remaining effects starting at $O(\Lambda_{\rm QCD}^2/p_z^2)$, 
the authors of Refs.~\cite{Lin:2014zya,Chen:2016utp} extrapolated to infinite 
nucleon momentum using the fit ansatz $a + b/p_z^2$ for each value of $x$. 
Although the effects of finite nucleon momentum can be mitigated, a quark-model 
study asserts that reducing systematic uncertainties to less than 20\% at 
moderate values of $x$ requires significantly larger values of nucleon 
momentum~\cite{Gamberg:2014zwa}, and at larger values of $x$ 
(roughly $x\simeq 1$) requires nucleon momentum as large as $p_z > 4$~GeV.

The size of the nucleon momentum is currently limited by the decreasing 
signal-to-noise ratio at large momenta, which requires very high statistics 
to extract a signal. 
New approaches to high-momentum nucleons are being investigated, with the most 
promising an approach that employs momentum smearing~\cite{Bali:2016lva}. 
This method has been applied to quasi-PDFs in 
Refs.~\cite{Alexandrou:2016jqi,Green:2017xeu}, demonstrating a large 
improvement in the signal-to-noise ratio by reaching momenta of about $2.5$~GeV.

\item The leading-twist quasi-PDFs and light-front PDFs are connected through 
the matching (or {\it factorization}) relation, Eq.~\eqref{eq:qPDFmatching}. 
Provided the quasi and light-front PDFs share the same IR behavior, the 
matching kernel can be determined in perturbation theory~\cite{Xiong:2013bka}. 
The one-loop matching kernel including gluon channel has been recently 
reported~\cite{Wang:2017qyg}.
The factorization of the IR structure of quasi-PDFs into light-front PDFs and an IR-safe matching kernel was claimed to hold to all orders in Refs.~\cite{Ma:2014jla,Ma:2014jga,Ma:2017pxb}.
More specifically, Refs.~\cite{Ma:2014jla,Ma:2014jga} claim that the 
factorization holds to all orders provided that UV divergences 
are properly renormalized.
However, Ref.~\cite{Li:2016amo} asserted that there might be subtleties beyond 
leading order in perturbation theory. 
A distinct, but similar, issue is the IR structure of extended operators in 
Euclidean and Minkowski space-time. 
There are again subtleties in perturbation theory~\cite{Carlson:2017gpk}, 
but arguments based on general field-theoretic grounds demonstrate that the 
quasi-PDF extracted from an Euclidean correlation function is exactly the 
same matrix element as that determined from the LSZ reduction formula in 
Minkowski space-time~\cite{Briceno:2017cpo}.

In contrast to the IR structure, the UV structure of the quasi-PDF is quite 
different from the UV structure of the light-front PDF: the former has both 
linear and logarithmic divergences, while the latter contains only logarithmic 
divergences. 
Although there are no power-divergences in dimensional regularization, 
quasi-PDFs determined on the lattice are regulated by the inverse lattice 
spacing. 
In the continuum limit (for which $a\to 0$, with all physical quantities held 
fixed) there is a divergence, associated with the length of the Wilson line $z$, 
that scales as $z/a$. This divergence must be removed nonperturbatively.

For a general nonlocal bilinear operator with Lorentz structure $\Gamma$, 
the renormalized operator $O_{\Gamma}^{\rm (ren)}(z,\mu)$ is related to its bare 
operator $O^{(0)}_{\Gamma}(z)$ by~\cite{Dotsenko:1979wb,Arefeva:1980zd, 
Craigie:1980qs,Stefanis:1983ke,Dorn:1986dt}
\begin{equation}\label{eq:renorm_non-local}
O_{\Gamma}^{\rm (ren)}(z,\mu)=e^{\delta m(\mu)|z|}Z_{\psi, z}(\mu,z)O^{(0)}_{\Gamma}(z),
\end{equation}
where $\delta m$ is the mass renormalization of a test particle moving along 
the Wilson line of length $z$ and $Z_{\psi, z}(\mu,z)$ removes the remaining 
logarithmic divergences associated with the Wilson line endpoints 
(the quark fields). 
This result holds to all orders in perturbation theory: the exponentiated 
counterterm $\delta m(\mu)$ completely removes the linear divergence and the 
quasi-PDF can be renormalized 
multiplicatively~\cite{Ji:2017oey,Ishikawa:2017faj}. 
The exponentiated counterterm can be determined using a static heavy quark 
potential, which shares the same power-law divergence as the nonlocal quark 
bilinear~\cite{Musch:2010ka,Ishikawa:2016znu,Chen:2016fxx,Green:2017xeu}. 
An alternative approach for controlling the power divergence has been proposed 
in Ref.~\cite{Monahan:2016bvm}.

Once the linear divergence has been removed nonperturbatively, lattice 
perturbation theory can be used to renormalize the remaining logarithmic
divergences in the quasi-PDF~\cite{Ishikawa:2016znu,Chen:2016fxx,
Carlson:2017gpk,Xiong:2017jtn}. 
A delicate point regarding the renormalization is the mixing among certain 
subsets of these nonlocal operators. 
Such a mixing has been identified at 
one-loop in perturbation theory in Ref.~\cite{Constantinou:2017sej} 
for a variety of fermion/gluon actions or nonperturbatively based on 
symmetries~\cite{Chen:2017mzz,Chen:2017mie}. 
The mixing coefficients are necessary to disentangle the individual matrix 
elements for each quasi-PDF from lattice calculation data. 
Of particular interest is the case of the unpolarized quasi-PDF, which mixes 
with the scalar quasi-PDF if the Lorentz index of Eq.~\eqref{eq:qPDF} is in the 
same direction as the Wilson line. 
In contrast, the polarized and transversity PDFs with a Lorentz index in the 
Wilson line direction do not exhibit any mixing (to one-loop in 
perturbation theory). 

In addition, nonperturbative schemes, such as the 
RI/MOM scheme~\cite{Martinelli:1994ty}, 
can be used to renormalize matrix elements determined on the lattice. 
Nonperturbative schemes avoid the use of lattice perturbation theory at 
low energy scales (usually chosen to be $\mu = \pi/a$), although perturbative 
matching between renormalization schemes is still necessary for PDFs expressed 
in the $\overline{\rm MS}$ scheme. 
Combining a nonperturbative renormalization scheme with a step-scaling 
procedure~\cite{Luscher:1991wu} significantly reduces perturbative truncation 
uncertainties by providing a nonperturbative method for reaching high energy 
scales.
Nonperturbative renormalization methods for quasi-PDFs have recently been 
constructed and applied in 
Refs.~\cite{Alexandrou:2017huk,Chen:2017mzz,Green:2017xeu}.

These nonperturbative procedures also remove the mixing between the unpolarized 
quasi-PDF and the twist-3 scalar operator, which occurs for lattice 
regularization that break chiral symmetry, through the construction 
of a $2\times2$ mixing matrix. 
The mixing coefficients do not contain any divergences. 
Further details can be found in Refs.~\cite{Alexandrou:2017huk,Chen:2017mzz}. 

It was recently observed that potential problems with the power divergent 
mixing patterns of DIS operators may arise when lattice regularization
is used~\cite{Rossi:2017muf}.
Further investigations into this issue would be interesting.

\end{itemize}

Lattice calculations of the $x$-dependence of PDFs have not matured 
up to the point to control all these sources of systematic uncertainty.       
Recent progress, however, has led to preliminary results that are encouraging. 
Here we highlight these results for the $x$-dependence
of the unpolarized and polarized PDFs extracted from lattice QCD. 

\begin{figure}[!t]
\centering
\includegraphics[scale=0.22,angle=270]{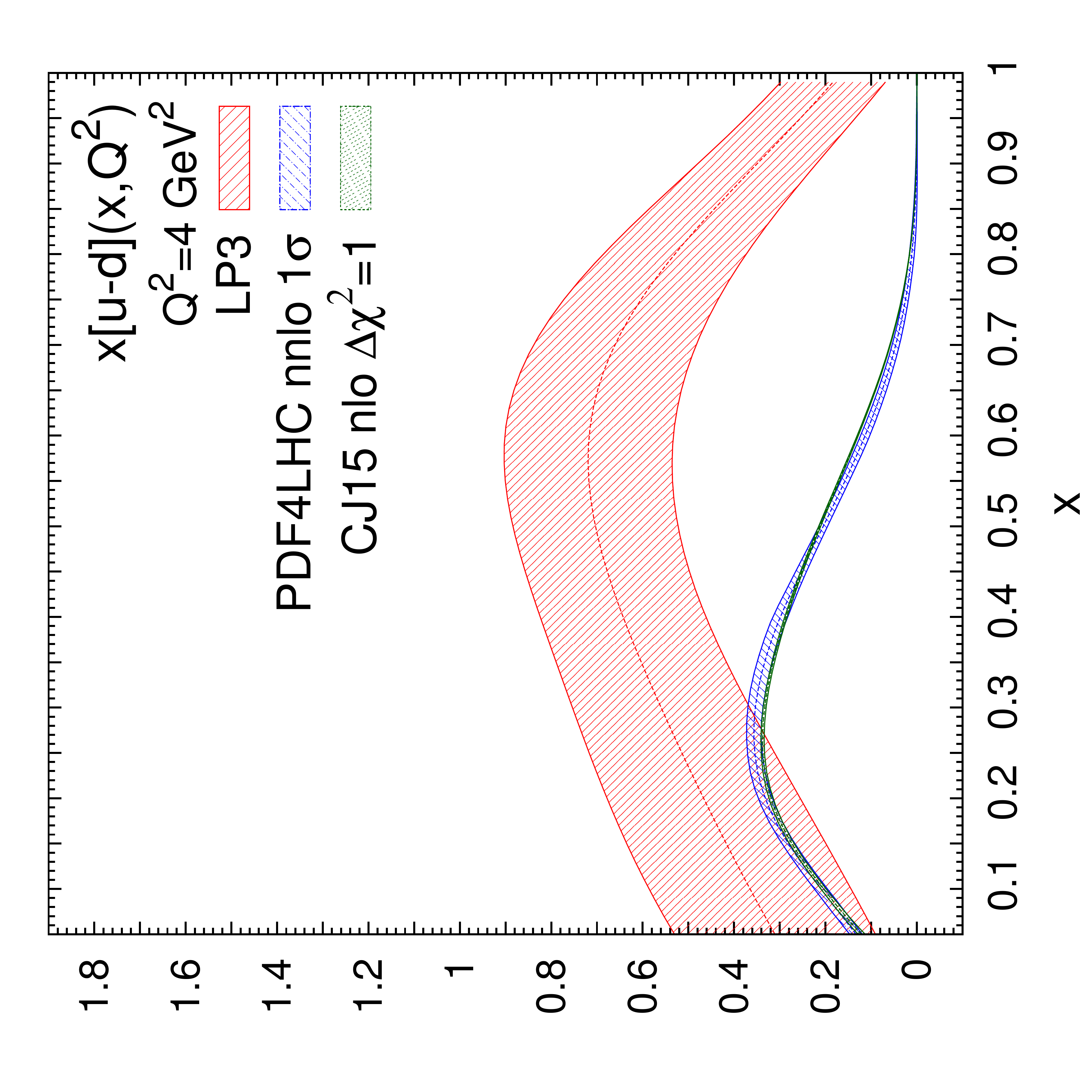}
\includegraphics[scale=0.22,angle=270]{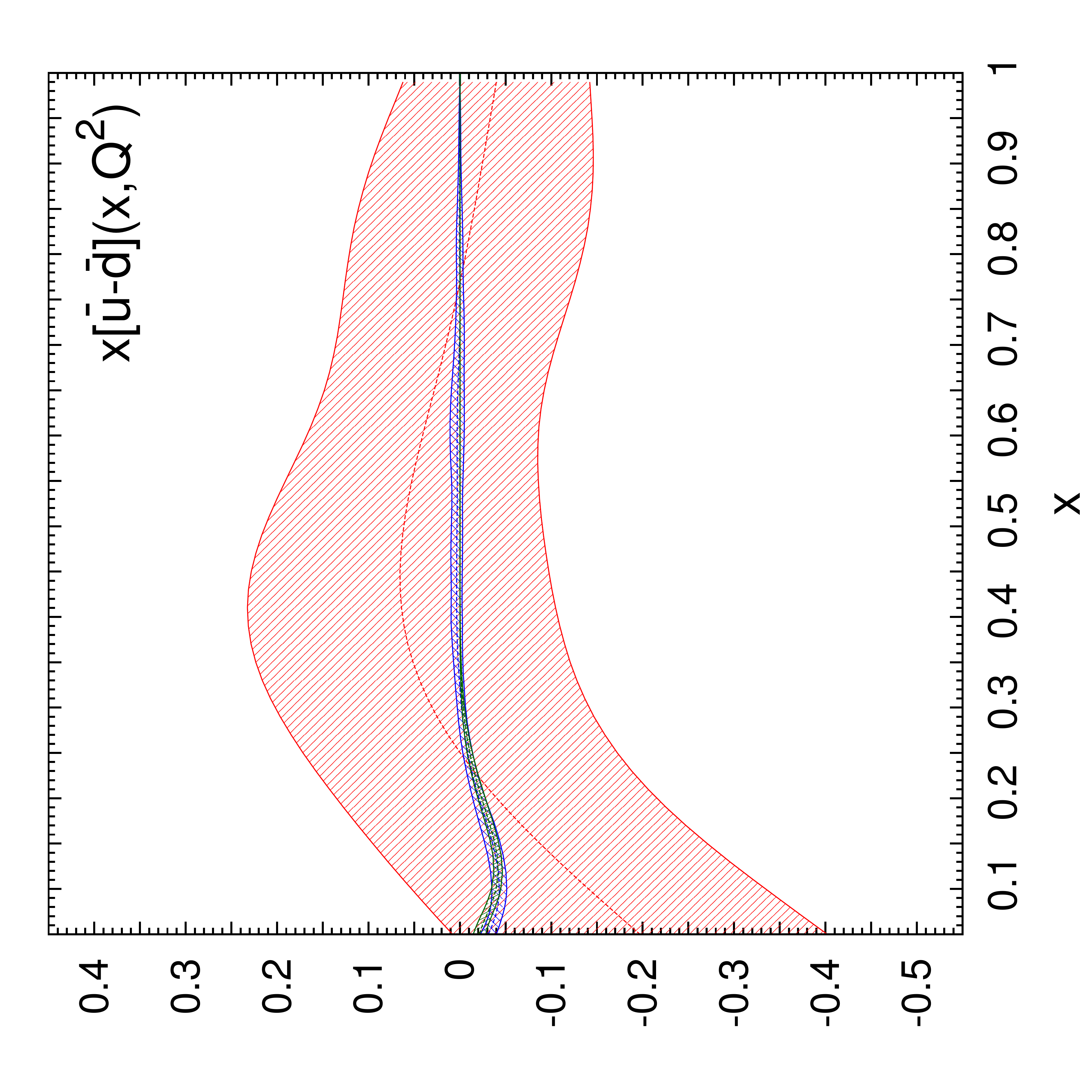}\\
\includegraphics[scale=0.22,angle=270]{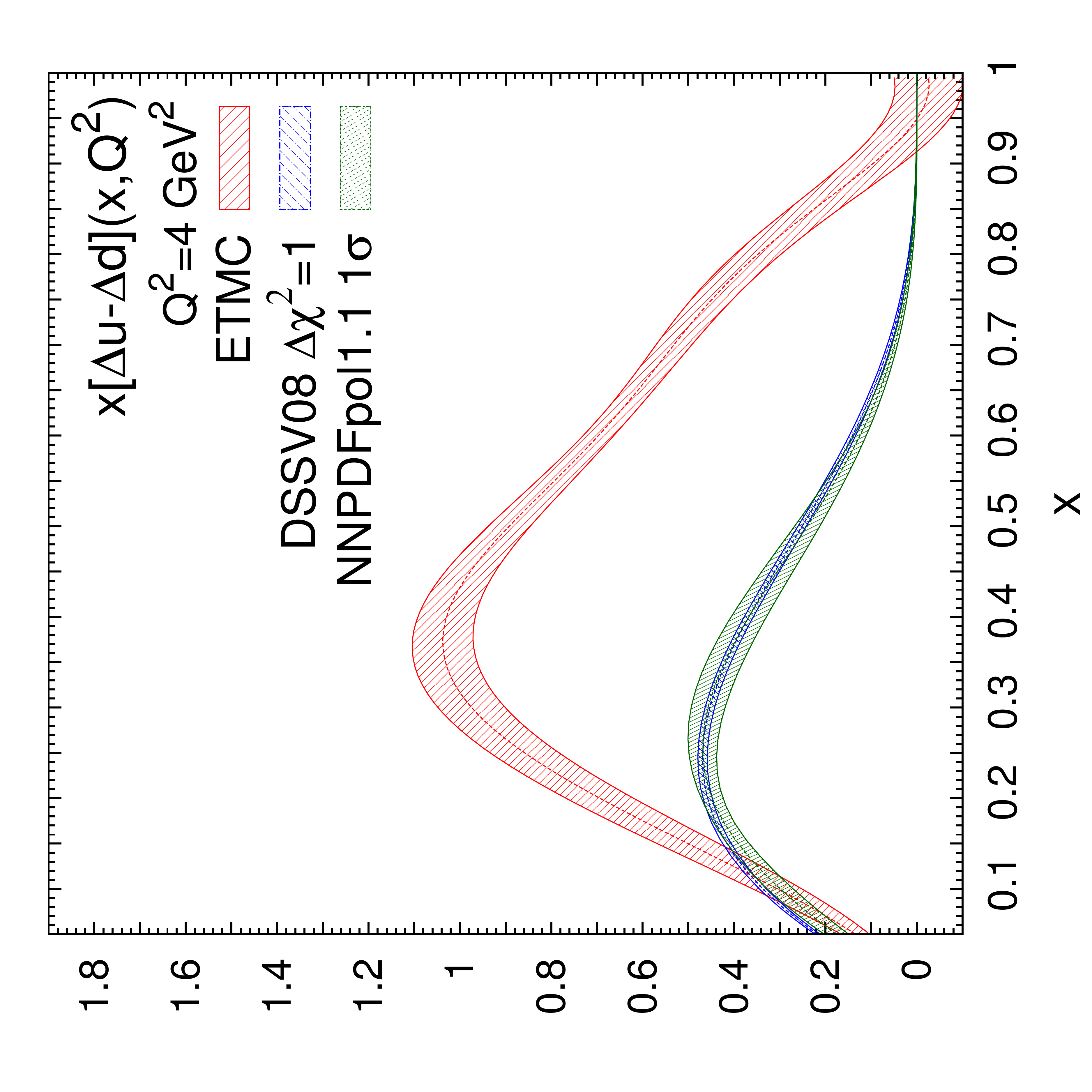}
\includegraphics[scale=0.22,angle=270]{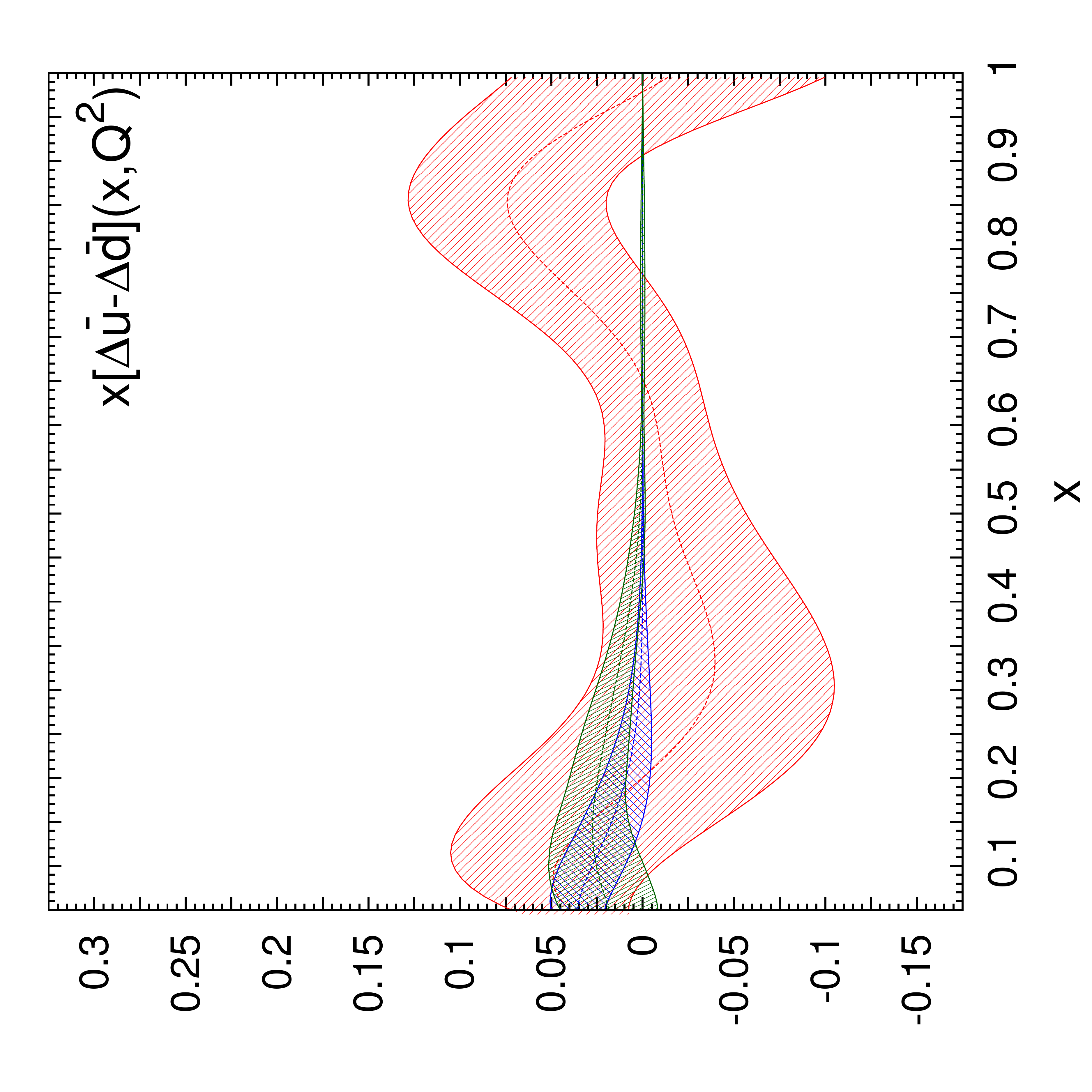}\\
\caption{\small LP3's renormalized unpolarized isovector quark (top left) and 
  antiquark (top right) PDF combinations at physical pion mass with the renormalization scale 
  $\mu=2$~GeV~\cite{Lin:2017ani}. 
  ETMC's renormalized polarized isovector quark (bottom left) and antiquark
  (bottom right) PDF combinations at pion mass of 
  375~MeV~\cite{Alexandrou:2017huk}.
  Note that only statistical errors are shown here; the systematics are yet to be addressed. The small-$x$ region ($x< 0.2$) can suffer larger systematics than the rest of the distribution due to the limited nucleon boost momentum.
  } 
\label{fig:qPDF-demo}
\end{figure}

Fig.~\ref{fig:qPDF-demo} shows example results for the renormalized unpolarized 
PDFs from Ref.~\cite{Chen:2017mzz} and polarized PDF from 
Ref.~\cite{Alexandrou:2017huk}.
In both cases, a nonperturbative renormalization procedure is applied to the 
bare matrix elements that appeared in earlier work~\cite{Lin:2014zya,
Alexandrou:2015rja,Chen:2016utp,Alexandrou:2016jqi,Alexandrou:2016eyt}.
For the unpolarized PDF, the calculation is carried out at a pion mass of 
310~MeV, includes one-loop matching and target mass corrections at the 
renormalization scale $\mu^2=4$~GeV$^2$, and the leading higher-twist 
$O(\Lambda_\text{QCD}^2/p_z^2)$ contributions have been 
removed~\cite{Chen:2016utp}. 
Multiple source-sink separations are used to take into account the effects of 
excited-state contamination, which become more important at large momentum. 
Mixing under renormalization has been estimated to be a small effect but is not 
yet computed explicitly. 
More recent work at the physical pion mass~\cite{Lin:2017ani} uses a different 
operator to avoid mixing effects. 
The polarized PDF has the advantage that is free from mixing, and is computed in
Ref.~\cite{Alexandrou:2017huk}  with fully renormalized matrix element, 
at a pion mass of 375 MeV. 
The matching to $\overline{\rm MS}$ at $\mu^2=4$~GeV$^2$ does not include any 
linearly divergent term, as the matrix element in 
coordinate space is renormalized.
Note that in both cases, the antiquark asymmetry is compatible with zero 
within current uncertainties, contrary to earlier unrenormalized 
results~\cite{Lin:2014zya,Alexandrou:2015rja,Chen:2016utp,Alexandrou:2016eyt}.
This is mainly due to the rapid increase of the renormalization factor with 
Wilson-line length, which amplifies the finite-volume effect from truncating 
long-range correlations. 
Ref.~\cite{Lin:2017ani} showed that this truncation causes unphysical 
oscillations in the sea-flavor asymmetry and proposed that the oscillations 
can be removed by either imposing a filter to reduce the weighting of 
long-range correlations or by taking the derivative of the matrix element in 
coordinate space. 
The effectiveness of both these two methods is 
demonstrated in Refs.~\cite{Alexandrou:2017dzj,Lin:2017ani}. 

\paragraph*{Pseudo-PDFs.} 
The general dependence of the  matrix element $h(z,p_z)$ of Eq.~\eqref{eq:qPDF} 
on the hadron momentum $p$ and the displacement of the quark and antiquark 
fields $z$ can be expressed as a function of the Lorentz invariants 
$\nu=z\cdot p$ (Ioffe time~\cite{Ioffe:1969kf,Braun:1994jq}) 
and $z^2$, where $z$ and $p$ are general 4-vectors.  
We can thus introduce
\begin{equation}
\overline{h}(\nu,z^2) \equiv h(z,p_z)\,.
\end{equation}

The pseudo-PDF is then defined by the Fourier transform
\begin{equation}
{\mathcal P}(x,z^2)=\int \frac{d\nu}{2\pi} e^{-ix\nu} \overline{h}(\nu,z^2),
\end{equation}
which has support only in the physical range 
$x=[-1,1]$ \cite{Radyushkin:2016hsy,Radyushkin:2017cyf}. 
As discussed in Refs.~\cite{Radyushkin:2016hsy,Radyushkin:2017cyf}, 
the pseudo-PDF is directly related to both the PDFs and the 
transverse-momentum-dependent PDFs (TMDs).
In Ref.~\cite{Radyushkin:2017cyf}, using the temporal gamma matrix in the 
matrix element, an approximate factorization of the primordial TMD 
${\cal F}(x,k_\perp^2)$ as 
\begin{equation}
{\cal F}(x,k_\perp^2)\approx K(k_\perp^2) q(x)
\label{eq:primordial}
\end{equation}
was conjectured. 
Here $k_\perp$ is the transverse momentum of the quark in the hadron and $q(x)$ 
is the PDF. 
This conjecture implies that the ratio
\begin{equation}
{\mathcal M}(\nu,z^2) =\frac{\overline h(\nu,z^2)}{\overline h(0,z^2)}
\label{eq:RatioPseudo}
\end{equation}
is directly related to the PDFs as 
\begin{equation}
{\mathcal M}(\nu,z^2) =Q(\nu,\mu^2) + {\cal O}(z^2)\,,
\label{eq:IoffePDF}
\end{equation}
with $\mu^2=1/z^2$.
Here $Q(\nu,\mu^2)$ is the Ioffe time PDF~\cite{Ioffe:1969kf,Braun:1994jq}, 
which is the Fourier transform of the PDFs,
\begin{equation}
{q}(x,\mu^2)=\int \frac{d\nu}{2\pi} e^{-ix\nu} Q(\nu,\mu^2).
\end{equation}
The ratio in Eq.~\eqref{eq:RatioPseudo} has a well-defined continuum 
limit and requires no renormalization. 
The polynomial corrections in Eq.~\eqref{eq:IoffePDF} are due to violations of 
the factorization conjecture, while the PDF ${q}(x,\mu^2)$ is the PDF in a 
particular scheme defined at scale $\mu^2=1/z^2$. 
Matching to $\overline{\rm MS}$ 
can be performed in perturbation theory following standard methodology. 
One loop results can be found in Refs.~\cite{Ji:2017rah,Radyushkin:2017lvu}.
A preliminary study  in quenched QCD was presented in 
Refs.~\cite{Orginos:2017kos,Karpie:2017bzm}, 
where it was shown that indeed the conjectured factorization is observed and 
the residual corrections are small. The same conclusion had also been reached 
in unquenched studies of TMDs in~\cite{Musch:2010ka}.
Furthermore, evidence of the expected perturbative evolution of the Ioffe time 
PDFs was also observed. 
This methodology  is currently under study and results from realistic 
calculations are soon to be expected.
It should be noted that the basic function $\overline h(\nu,z^2)$ used here or 
in the computation of quasi-PDFs is related to both the longitudinal and 
transverse structure of the hadron. 
This relationship is discussed in detail 
in~\cite{Musch:2010ka,Radyushkin:2017cyf,Broniowski:2017gfp}.

Finally, both the quasi-PDF and the Ioffe-time PDF approaches are faced with 
the technical problem of inferring the PDF from a Fourier transform where data 
are known only in  a limited range of $\nu$ or $z$. 
This introduces systematic errors that require careful study. 
Such effects have been discussed 
in~\cite{Chen:2017lnm,Lin:2017ani,Broniowski:2017gfp}. 
In particular, because $x$ is the Fourier dual of $\nu$, accessing a limited 
range of $\nu$ (or $z$) has the effect of introducing uncontrolled systematic 
errors at small $x$ (roughly $x\lesssim 0.15 $ for existing lattice 
calculations).
These systematic errors  can be controlled using increasingly large values of 
$\nu$, although this requires an increased computational power.  
Therefore, improved computational methods are required to reliably 
extract PDFs at small $x$.

\paragraph*{Lattice cross-sections.}  
Similarly to a global QCD analysis of high energy scattering data,
PDFs can also be extracted from analyzing {\it data} generated by lattice-QCD 
calculation of good {\it lattice cross-sections}~\cite{Ma:2014jla,Ma:2014jga}. 
A lattice cross-section is defined as a single-hadron matrix element of a 
time-ordered, renormalized, nonlocal operator ${\cal O}_n(z)$: ${\sigma}_{n}(\nu,z^2,p^2)=\langle p| {T}\{{\cal O}_n({z})\}|p\rangle$ with four-vectors 
$p$, $z$ and $\nu$ defined above, and renormalization scale suppressed. 
The four-vectors $p$ and $z$ effectively define the {\it collision} kinematics, 
and the choice of ${\cal O}_n$ determines the dynamical features of the lattice 
cross-section. 
A good lattice cross-section should have the following three key properties: 
(1) it is calculable in lattice-QCD with an Euclidean time, (2) it has a 
well-defined continuum limit as the lattice spacing $a\to 0$, and (3) 
it has the same and factorizable logarithmic collinear (CO) divergences as 
that of PDFs, which connects the good lattice cross-sections to PDFs, 
just like how high energy hadronic cross-sections are related to PDFs 
in terms of QCD factorization.  

A class of good lattice cross-sections can be constructed in terms of a 
correlation of two {\it renormalizable} currents, 
${\cal O}_{j_1j_2}(z)\equiv z^{d_{j_1}+d_{j_2}-2} Z_{j_1} Z_{j_2}\, j_1(z) j_2(0)$, 
with dimension $d_j$ and renormalization constant $Z_j$ of the current $j$.
There could be many choices for the current, such as a vector quark current, 
$j_q^V(z) = \overline{\psi}_q(z)\gamma\cdot{z}\, {\psi}_{q}(z)$, or a tensor 
gluonic current, $j_g^{\mu\nu}(z)\propto F^{\mu\rho}(z){F_{\rho}}^\nu(z)$~\cite{Ma:2017pxb}.  
Different combinations of the two currents could help enhance the lattice 
cross-sections' flavor dependence.  
If $z^2$ is sufficiently small, the lattice cross-section constructed from two 
renormalizable currents can be factorized into PDFs~\cite{Ma:2017pxb},
\begin{equation}
\label{eq:fac}
{\sigma}_{n}(\nu,z^2,p^2)=\sum_{a}\int_{-1}^1 \frac{dx}{x}\, f_{a}(x,\mu^2) 
K_{{n}}^{a}(x\nu,z^2,x^2p^2,\mu^2) +O(z^2\Lambda_{\rm QCD}^2)\, ,
\end{equation}
where $\mu$ is the factorization scale, $K_n^{a}$ are perturbatively calculable 
hard coefficients, and $f_{a}$ is the PDF of flavor $a=q,g$ with anti-quark 
PDFs expressed from quark PDFs using the equivalence
$f_{\bar{a}}(x,\mu^2)=-f_{{a}}(-x,\mu^2)$.  
PDFs could be extracted from global fits of lattice-QCD generated data for 
various lattice cross-sections $\sigma_{n}(\nu,z^2,p^2)$ with corresponding 
perturbatively calculated coefficients $K_n^{a}$ in Eq.~(\ref{eq:fac}).

The quasi-PDFs and pseudo-PDFs introduced above could be derived by choosing 
\begin{equation}
{\cal O}_{q}(z)=Z_q(z^2)\overline{\psi}_q(z)\gamma\cdot {z}\, 
\Phi(z,0){\psi}_q(0)\,,
\end{equation}
with the renormalization constant $Z_q(z^2)$ and the path ordered gauge link 
$\Phi(z,0)={\cal P}e^{-ig\int_0^{1} z\cdot A(\lambda z)\,d\lambda}$~\cite{Ma:2017pxb}.  
With $K^{q(0)}_{q}(x \nu,z^2,0,\mu)= 2 x \nu  e^{i x \nu}$, one finds
\begin{equation}
\label{eq:lcsQuasi}
\int \frac{d \nu}{\nu}\, \frac{e^{-i x \nu}}{4\pi} \sigma_{q}(\nu,z^2,p^2)
\approx f_{q}(x,\mu)\, ,
\end{equation}
modulo $\mathcal{O}(\alpha_s)$ and higher twist corrections.  
By choosing $z_0=0$ and both $\vec{p}$ and $\vec{z}$ along the ``3"-direction, 
one finds that $\nu=-z_3\, p_3$ and the left hand side of 
Eq.~(\ref{eq:lcsQuasi}) is the quasi-quark distribution introduced in 
Ref.~\cite{Ji:2013dva} if the integral is performed by fixing $p_3$, 
while it is effectively the pseudo-quark distribution used in 
Ref.~\cite{Orginos:2017kos} if the integral is performed by fixing $z_3$. 
That is, these two approaches for extracting PDFs are equivalent if matching 
coefficients are calculated at the lowest order in $\alpha_s$ neglecting all 
power corrections, but different if contributions from either higher order in 
$\alpha_s$ or higher powers in $z^2$ need to be considered.
Furthermore, Eq.~(\ref{eq:lcsQuasi}) indicates that the quasi-PDFs and 
pseudo-PDFs are two special cases of good lattice cross-sections. 

\subsection{Global PDF fits}
\label{Sec:IntroGlobalFits}

Global PDF fits realize a QCD analysis of hard-scattering measurements,
often using a variety of hadronic observables.
Parton distributions are parametrized at an initial energy scale, 
evolved up to the scale of the data via DGLAP 
equations~\eqref{eq:dglapunp}--\eqref{eq:dglappol}, and used to build up the 
theoretical predictions for the relevant observables.
In the corresponding factorization formul\ae, the factorization scale, $\mu$,
is usually set equal to the characteristic scale of the process, $Q$.
The best-fit parameters of PDFs are then determined by minimization of a 
proper figure of merit, such as the log-likelihood $\chi^2$.
In this section, we present the general global PDF fitting framework.
We discuss how PDFs are determined from hard-scattering observables,
paying attention to the assessment of PDF uncertainties.
We highlight the theory and the data used to fit both unpolarized and 
polarized PDFs and present a brief review of their state-of-the-art 
determination.

\subsubsection{General framework}
\label{sec:genframework}

\paragraph*{Fitting PDFs from hard-scattering data.} 
Parton distributions appear in the factorization formul{\ae} of a class of 
sufficiently inclusive processes, among which are deep-inelastic scattering 
(DIS) and proton-(anti)proton collisions.
The factorization formul{\ae} for the unpolarized and polarized structure 
functions $F_1$ and $g_1$ were introduced in Eqs.~\eqref{eq:Fi}--\eqref{pdf}.
For the hadroproduction of a generic final-state $X$ in unpolarized 
proton-proton ($pp$) collisions, the corresponding factorized expression reads
\begin{align}
\sigma_{pp\to X}(s,\mu^2_F,\mu^2_R)=&\sum_{a,b}\int {\rm d}x_1 {\rm d}x_2\, 
f_a(x_1,\mu_F^2)f_b(x_2,\mu_F^2)\,
\hat{\sigma}_{ab\to X}(x_1,x_2,s;\mu^2_{F},\mu^2_R)\;,
\label{eq:LHCxsecunp}
\end{align}
where the unpolarized hard cross-section 
$\hat{\sigma}_{ab\to X}$ can be calculated 
perturbatively as an expansion in the QCD and electroweak (EW) 
running couplings.
The specific values of the momentum fractions
$x_i$ can be related to the kinematics of the final state.
For example, for the production of a heavy final state, 
it can be shown that, at LO,
\be
x_{1,2}=\frac{M_X}{\sqrt{s}}e^{\pm y_X} \, ,
\ee
where $M_X$ and $y_X$ are the invariant mass and rapidity of the produced 
system and $\sqrt{s}$ is the center-of-mass energy.
The factorization and renormalization scales, $\mu_F$ and $\mu_R$, are 
usually taken equal to the hard scale of the process, $\mu_F=\mu_R=\mu=Q$.
Factorization formul{\ae} analogous to Eq.~\eqref{eq:LHCxsecunp} can be
written in the polarized case for $pp$ collisions where only one or
both proton beams are longitudinally polarized, see {\it e.g.}
Refs.~\cite{Stratmann:2001pb,Nadolsky:2003fz}.

When one performs a global fit, the DGLAP evolution equations of the PDFs, 
Eqs.~\eqref{eq:dglapunp}--\eqref{eq:dglappol}, derive the PDFs at any scale
relevant to comparisons with the data from PDF parametrizations at an
arbitrary input scale, typically $Q_0\sim 1$~GeV.
The contribution of heavy quark flavors to any process is 
power-suppressed at scales which are below the threshold for their 
production~\cite{Collins:1978wz}. 
Therefore, whereas in principle the QCD Lagrangian contains six quark flavors, 
in practice only a smaller number of active flavors $N_f$ are included in 
loops, and thus in particular in the solution of evolution equations. 
When expressing predictions for processes at various disparate scales in terms 
of a single set of PDFs it is thus necessary to use a so-called variable-flavor 
number (VFN) scheme, whereby different numbers of active flavors are adopted 
at different scales (with up to $N_f=5$ active flavors in most of PDF sets). 
Use of a fixed-flavor number (FFN)
scheme only allows comparison with the data in a restricted range of scales.

The input PDF parametrization is usually chosen as
\begin{equation}
\label{eq:pdffunc}
f(x,Q_0,\{a_i\})= x^{a_1}(1-x)^{a_2}\:C(x,\{a_j\})\, ,
\end{equation}
where the parameters $\{a_i\}$ determine the PDF shape
and are different for each PDF flavor combination probed by the data.
The $(1-x)^{a_2}$ term, with $a_{2}>0$, ensures that the PDFs vanish in the 
elastic limit $x\to 1$. 
Specific values of the exponent $a_2$ are predicted by counting 
rules~\cite{Brodsky:1973kr}, although they are not always clearly
supported by phenomenological fits~\cite{Ball:2016spl,Nocera:2014uea}, 
and are not always used.
The $x^{a_1}$ term governs the low-$x$ PDF behavior. 
It is expected from considerations based on Regge theory, 
which also provides the values of the exponents $a_1$.
However, as for $a_2$, the value of $a_1$ is left free in the global fits.
The interpolating function $C(x)$ in Eq.~\eqref{eq:pdffunc}
affects the behavior of the PDFs away from the $x\to 0$ and $x\to 1$
extrapolation regions.
This is assumed to be a smoothly varying function of $x$, for which a variety 
of parametrizations can be made.

The simplest ansatz, which has been very widely used, is to take a basic 
polynomial form in $x$ (or $\sqrt{x}$), such as
\begin{equation}\label{eq:lpower}
C(x)=1+a_2\sqrt{x}+a_3 x+...\;.
\end{equation}
Functional forms of this type are, for example, taken in CJ, HERAPDF and 
earlier MMHT and CT sets (see below for the references to each set). 
More recently, the CT and MMHT collaborations expand 
in terms of a basis of  Bernstein and Chebyshev polynomials, respectively.
While formally equivalent to the simple polynomial expansion
Eq.~\eqref{eq:lpower}, these are much more convenient for fitting as the 
number of free parameters $n_{\rm par}$ is large.
In the latest CT and MMHT sets, there are between 20 and 40 free parameters in 
total, though some of these are kept fixed when evaluating the
Hessian PDF uncertainties to reduce redundancy between the parameters.
Furthermore, the use of orthogonal polynomials, like Chebyshev 
polynomials, allows one to decouple the parameters in $C(x)$ and to uniformly
sample its possible functional shapes.

An alternative approach to the PDF parametrization Eq.~\eqref{eq:lpower}
is adopted by the NNPDF collaboration. 
Here, the interpolating function $C(x)$ is modeled with 
a multi-layer feed-forward neural network (NN).
In practice, this allows for a greatly increased number of free parameters, 
typically an order of magnitude higher than in the sets of other groups.
The form of Eq.~\eqref{eq:pdffunc} is still assumed, but
now $C(x)={\rm NN}(x)$, where ${\rm NN}(x)$ is a neural network.
The $x^{a_1}(1-x)^{a_2}$ term that multiplies ${\rm NN}(x)$ represents
a preprocessing factor that speeds up the minimization procedure
and that is determined via an iterative procedure.
Because of its parametric redundancy, the neural network parametrization
can be overtrained and learn the statistical noise of the data.
In order to avoid such a drawback, the data are split into validation and 
training sets, then the best-fit is determined by
cross-validation~\cite{Forte:2002fg,DelDebbio:2004xtd}.
A similar technique is used also in the JAM 
fits~\cite{Sato:2016tuz,Ethier:2017zbq}.

In most of PDF sets currently in use, the PDFs for charm and heavier quarks
are not parametrized as in Eq.~\eqref{eq:pdffunc}, but rather they are
generated by perturbative emission of gluons and light quarks.
In the vicinity of the threshold for heavy-quark production, the quark mass
cannot be neglected.
It is thus necessary to explicitly include terms suppressed by
powers of the heavy-quark mass in the coefficient functions, while 
subtracting the logarithmically enhanced, unsuppressed terms that are 
already generated by solving the evolution equations in order 
to avoid double counting.
Various schemes exist so far to do so, see {\it e.g.}
Refs.~\cite{Forte:2013wc,Gao:2017yyd} for an extensive summary.
The possibility of parametrizing and fitting the charm PDF
on the same footing as light quark PDFs has been also explored, see 
{\it e.g.}~\cite{Brodsky:2015fna,Ball:2016neh,Hou:2017khm}
and references therein.

Once the PDF parametrization is chosen, and the theoretical details 
of the analysis are defined (such as the perturbative order, the
treatment of heavy quarks, etc.), the best-fit PDF parameters
and their uncertainty should be determined via a fitting methodology
that minimizes a suitable statistical estimator, typically the $\chi^2$.
There exist different alternative definitions of the $\chi^2$
to be used in the global 
fits~\cite{Ball:2012wy,Gao:2013xoa,Alekhin:2017kpj,Abramowicz:2015mha}. 
For instance one frequently used definition is
\begin{equation}
\chi^2 
= 
\sum_{i,j}^{N_{\rm dat}} (T_i(\{a_k\}) - D_i) 
({\rm cov^{-1}})_{ij} (T_j(\{a_k\}) - D_j)\,,
\label{eq:chi2}
\end{equation}
where $N_{\rm dat}$ is the number of data points of a given experiment,
$T_i$ and $D_i$ are the corresponding theoretical predictions
and experimental data, and $({\rm cov^{-1}})_{ij}$ is the inverse of the 
experimental covariance matrix.
The theoretical predictions $T_i(\{a_k\})$ depend on the input set of 
parameters $\{a_k\}$ via the PDF parametrization, see Eq.~\eqref{eq:pdffunc}.
Therefore, Eq.~\eqref{eq:chi2} assesses the agreement between theory and data.

The covariance matrix $({\rm cov})_{ij}$ accounts for the various sources of 
experimental systematic uncertainties and also allows for several
different definitions.
One example is the so-called $t_{0}$ prescription~\cite{Ball:2009qv}, 
where a fixed theory prediction $T_{i}^{(0)}$ is used to define the  
contribution to the $\chi^2$ from the multiplicative systematic uncertainties, 
namely
\be
\label{eq:covmat_t00}
({\rm cov})_{ij}=
\delta_{ij} \sigma_{\rm stat}^2 + 
\sum_{\alpha=1}^{N_c}\sigma^{(c)}_{i,\alpha}\sigma^{(c)}_{j,\alpha}D_{i} D_{j}
+ \sum_{\beta=1}^{N_{\cal L}} \sigma_{i,\beta}^{({\cal L})}\sigma_{j,\beta}^{({\cal L})}
T^{(0)}_{i} T^{(0)}_{j}\, .
\ee
Here $\sigma_{\rm stat}$ is the uncorrelated uncertainty,
and $\sigma^{(c)}_{i,\alpha}$ ($\sigma^{(\cal L)}_{i,\beta}$) are the various sources 
of additive (multiplicative) systematic uncertainties.
The $t_0$ prescription is needed to avoid the D'Agostini 
bias~\cite{DAgostini:2003syq,DAgostini:1993arp}, a downwards
bias of the statistical estimators for the central value and the uncertainty
of the theoretical predictions due to the rescaling induced by  multiplicative 
uncertainties.
See~\cite{Ball:2009qv,Ball:2012wy} and references therein for details and the 
alternative {\it penalty-trick} prescription, and~\cite{Gao:2013xoa}
for the alternative {\it extended}-$t$ prescription.

\paragraph{PDF uncertainties.}
Determining the best-fit values of the PDF parameters is not enough: one also 
needs to estimate the associated PDF uncertainties, possibly separated into 
the various sources of experimental, methodological and theoretical 
uncertainties.
In this respect, there are two main methods to determine PDF uncertainties, the 
{\it Hessian} and the {\it Monte Carlo} (MC) methods.\footnote{The Lagrange 
multiplier method~\cite{Stump:2001gu} is also frequently used for dedicated 
studies of PDF uncertainties.}

The Hessian method~\cite{Pumplin:2001ct} is based on the parabolic
expansion of the $\chi^2$ in the vicinity of its minimum 
\be
\label{eq:hessianexpansion}
\Delta\chi^2 \equiv \chi^2- \chi^2_{\rm min}
=\sum_{i=1,j}^{n_{\rm par}}H_{ij}\lp a_i-a_i^0\rp
\lp a_j-a_j^0\rp \, ,
\ee
where the $n_{\rm par}$ PDF fit parameters are denoted by 
$\{a_1,\ldots,a_{n_{\rm par}}\}$, the best-fit values that minimize the
$\chi^2$ are indicated by $\{a_1^0,\ldots,a^0_{n_{\rm par}}\}$,
and the Hessian matrix is defined as
\be
H_{ij}\equiv \frac{1}{2} \frac{\partial^2\chi^2}{\partial a_i
\partial a_j}\Bigg|_{\{\vec{a}\}=
\{\vec{a}^0\}}\, .
\ee
By diagonalizing this Hessian matrix, it becomes possible
to represent PDF uncertainties in terms of orthogonal eigenvectors
within a fixed tolerance $T=\sqrt{\Delta\chi^2}$.
These eigenvectors can then be used to estimate the PDF uncertainty for 
arbitrary cross-sections, using the master formula of Hessian PDF sets for 
the uncertainty of the cross-section $\mathcal{F}$, such 
as~\cite{Pumplin:2002vw}
\be
\label{eq:hessianmaster2}
\sigma_{\mathcal{F}}=\frac{1}{2}\lp \sum_{i}^{n_{\rm par}}
\lc \mathcal{F}(S_i^+)-\mathcal{F}(S_i^-) \rc^2 \rp^{1/2} \, ,
\ee
where $S_i^{\pm}$ correspond to the $i$-th eigenvector
associated with positive and negative variations with respect
to the best fit value.

The Monte Carlo method~\cite{Giele:1998gw,Giele:2001mr,Forte:2002fg,
DelDebbio:2004xtd} is based 
on constructing a representation of the probability distribution of the 
experimental data in terms of a large number $N_{\rm rep}$ of {replicas},  
which encode all the information on central values, variances and 
correlations provided by the experiments.
Specifically, given an experimental measurement of a hard-scattering
observable $F_{I}^{\rm (exp)}$, with total uncorrelated uncertainty 
$\sigma_{I}^{\rm (stat)}$, $N_{\rm sys}$ fully correlated systematic uncertainties 
$\sigma^{\rm (corr)}_{I,c}$ and $N_a$ ($N_r$) absolute (relative) normalization 
uncertainties $\sigma^{\rm (norm)}_{I,n}$, the Monte Carlo replicas are 
constructed using the expression
\be
\label{eq:replicas}
F_{I}^{(\art)(k)}
=
S_{I,N}^{(k)} F_{I}^{\rm (\mrexp)}\lp 1
+
\sum_{c=1}^{N_{\rm sys}}r_{I,c}^{(k)}\sigma^{\rm (corr)}_{I,c}
+
r_{I}^{(k)}\sigma_{I}^{\rm (stat)}\rp
\ , \quad k=1,\ldots,N_{\rep} \ ,
\ee
where $S_{I,N}^{(k)}$ is a normalization prefactor.
The variables $r_{I,c}^{(k)},r_{I}^{(k)},r_{p,n}^{(k)}$ are
univariate Gaussian random numbers.
For each individual replica, the random fluctuations associated with a given 
fully-correlated systematic uncertainty will be the same
for all data points, $r^{(k)}_{I,c}=r^{(k)}_{I',c}$.

Parton distribution fits are then performed separately on each of the 
Monte Carlo replicas.
The resulting ensemble of PDFs samples the probability density in the space
of PDFs.
The expectation function of a generic observable $ \mathcal{F} [ \{  f \}]$,
depending on the fitted set of PDFs $\{f\}$,
is evaluated as an average over the replica sample,
\be
\label{masterave}
\la \mathcal{F} [ \{  f \}] \ra
= \frac{1}{N_{\rm rep}} \sum_{k=1}^{N_{\rm rep}}
\mathcal{F} [ \{  f^{(k)} \}] \, .
\ee
The corresponding uncertainty is determined as the variance of the
Monte Carlo sample,
\be
\sigma_{\mathcal{F}} =
\left( \frac{1}{N_{\rm rep}-1}
\sum_{k=1}^{N_{\rm rep}}   
\lp \mathcal{F} [ \{  f^{(k)} \}] 
-   \la \mathcal{F} [ \{  f \}] \ra\rp^2 
 \right)^{1/2}.
\label{mastersig}
\ee
Likewise, other properties of the underlying PDF probability distribution, 
such as skewness and kurtosis, could be readily computed.

Given a PDF set in the Hessian representation, it is possible to construct
the corresponding Monte Carlo representation~\cite{Watt:2012tq,Hou:2016sho}
and vice-versa~\cite{Gao:2013bia,Carrazza:2015aoa}.

So far, we discussed PDF uncertainties following from propagation of the
uncertainty of the experimental data that underlie the PDF determination.
Procedural uncertainties, associated with the methodology used to 
determine PDFs from data, can also be accounted for in the MC or Hessian
approaches, or reduced to a negligible size, as in the NNPDF approach.
There are however additional sources of uncertainty, mostly theoretical, 
that are not accounted for, either in the Hessian or in the MC methods.
These are extensively discussed 
in Refs.~\cite{Forte:2013wc,Butterworth:2015oua,Accardi:2016ndt,Gao:2017yyd} 
and briefly summarized as follows.

\begin{itemize}

\item The uncertainty due to finite uncertainties 
associated with the input values of the physical parameters used in the global 
fit, such as $\alpha_s(m_Z)$ and the charm mass $m_c(m_c)$, is 
evaluated by repeating the fits for different values of the physical 
parameters and then by suitably combining the results.

\item The missing higher order uncertainty (MHOU), due to the truncation
of the perturbative expansion, is usually inferred by comparing NLO to NNLO 
unpolarized PDFs and LO to NLO polarized PDFs.
While this is expected to be small for NNLO fits, currently its size is unknown.

\item The uncertainty due to different choices in the treatment of heavy quarks
was studied in Refs.~\cite{Binoth:2010nha,Thorne:2012az}, for unpolarized PDFs, 
by looking at their impact.
It was found that differences may not be entirely negligible at NLO in 
the vicinity of the quark threshold, though they rapidly decrease at 
NNLO~\cite{Binoth:2010nha}.

\item Uncertainties associated with missing higher-twist (power-suppressed) 
corrections (if they are not included in the factorized description of 
fit observables) are kept under control by removing data, below some low 
cut-off scale, that may be affected by them. 
Their impact can be studied by varying this 
cut-off~\cite{Martin:2003sk,Accardi:2009br}, or
by looking at the stability of the fit with and without inclusion of higher 
twist terms~\cite{Sato:2016tuz,Accardi:2016qay,Alekhin:2017kpj}.

\item Uncertainties associated with nuclear corrections, whenever they are
not included, affect some DIS data in which targets are deuterons or heavier 
nuclei, rather than just protons.
They have been studied by including such corrections according to various 
models~\cite{Martin:2009iq,Ball:2009mk,Sato:2016tuz,Accardi:2016qay}, 
or by attempting to fit the corrections 
directly~\cite{Martin:2009iq,Martin:2012da}.

\item Extrapolation uncertainties in the region not covered 
by experimental data are particularly delicate as far as full moments of PDFs
are concerned.
They are difficult to quantify, especially in the polarized case at small $x$
due to the lack of data.
The impact of extrapolation uncertainties in the unpolarized case at large $x$
has been studied in~\cite{Accardi:2011fa,Accardi:2016ndt}.

\end{itemize}
At present, the only way of dealing with such uncertainties is to make sure 
that they are small enough (in comparison to the data uncertainty)
in each PDF set.
Therefore, in the remainder of this paper, 
we will assume that they can be neglected.
We will point out to the reader how global-fit results can be affected 
by underestimation of these uncertainties in 
Secs.~\ref{subsubsec:GPDFfits}--\ref{subsec:BN}.

\subsubsection{Unpolarized PDFs}
\label{sec:unpPDFs}

\paragraph*{Theoretical features.}
While the general $x$ dependence of the PDFs is determined by
nonperturbative QCD dynamics, there are still a number
of theoretical constraints that any PDF set should satisfy. 
These should be imposed during the fit procedure.

First, since the proton has the quantum numbers of two up quarks and one 
down quark, the following quark number sum rules, given in terms of zeroth
moments, must be satisfied: 
\begin{eqnarray}
\int_{0}^{1}dx\ \left[u(x,\mu^2)-\bar{u}(x,\mu^2)\right] 
& =\left\langle 1\right\rangle _{u^{-}}= & 2 \, ,\nonumber \\
\int_{0}^{1}dx\ \left[d(x,\mu^2)-\bar{d}(x,\mu^2)\right] 
& =\left\langle 1\right\rangle _{d^{-}}= & 1 \, ,
\label{eq:valencesumrules}\\
\int_{0}^{1}dx\ \left[s(x,\mu^2)-\bar{s}(x,\mu^2)\right] 
& =\left\langle 1\right\rangle _{s^{-}}= & 0 \, .\nonumber
\end{eqnarray}
Similar constraints hold for heavy quarks: 
$\left\langle 1\right\rangle _{c^{-}}=\left\langle 1\right\rangle _{b^{-}}
=\left\langle 1\right\rangle _{t^{-}}=0$.
The valence sum rules, Eqs.~\eqref{eq:valencesumrules}, should be satisfied at 
any scale $\mu$. 
Indeed it can be shown that if they hold at the input parametrization scale 
$\mu=Q_0$, they are subsequently respected by DGLAP evolution.
Therefore, for these distributions we must have $a_1>-1$ in
Eq.~\eqref{eq:pdffunc}, otherwise Eqs.\eqref{eq:valencesumrules} 
would be ill-defined.

Second, PDFs should satisfy the conservation of energy-momentum derived from
the QCD Lagrangian.
In other words, the proton's total momentum should be equal 
to the sum of the momentum carried by all its constituents
(the so-called momentum sum rule):
\begin{equation}
\label{eq:mom}
1 
= 
\left\langle x\right\rangle _{g}
+
\left\langle x\right\rangle _{u^{+}}
+
\left\langle x\right\rangle _{d^{+}}
+
\left\langle x\right\rangle _{s^{+}}
+
\left\langle x\right\rangle _{c^{+}}
+
\left\langle x\right\rangle _{b^{+}}
+
\left\langle x\right\rangle _{t^{+}}+\ldots\,,
\end{equation}
where the ellipsis represents any other partonic components (such
as a photon). 
The first moments, $\left\langle x\right\rangle _{f}$, are defined in analogy 
to Eqs.~\eqref{eq:umoment1}--\eqref{eq:uplusmoment1}. 
In order to avoid a divergent contribution, we must have $a_1>-2$ in 
Eq.~\eqref{eq:pdffunc} for the non-valence distributions.
Typically it turns out that $-2<a_1<-1$ for such distributions, hence 
the number of soft partons grows very quickly at small $x$, although the 
momentum fraction carried by them is well-defined and finite.
As in the case of the valence sum rules, the momentum
sum rule is preserved by the DGLAP evolution equations.

Theoretical calculations of DIS and hadronic cross-sections at the highest 
perturbative order available should be used.
Currently, this implies using NNLO for the QCD corrections and NLO
for the EW and photon-induced effects~\cite{Manohar:2016nzj,Manohar:2017eqh}.
Thanks to recent progress in higher-order calculations, these results
are available for most of the processes entering the global
PDF fits~\cite{Currie:2016bfm,Campbell:2016lzl,Czakon:2016dgf,
Boughezal:2017nla,Li:2012wna}, including differential distributions with 
colored particles in the final state.

These calculations should be provided in
a format such that the evaluation of the hadronic
cross-sections, Eq.~\eqref{eq:LHCxsecunp}, is not too burdensome
from a computational point of view.
To bypass the limitations of the lengthy (N)NLO
computations, a number of fast interfaces have
been developed that allow for the efficient calculation
of NLO (and NNLO) fully differential hadronic cross-sections,
among which {\tt APPLgrid}~\cite{Carli:2010rw},
{\tt FastNLO}~\cite{Wobisch:2011ij} and {\tt aMCfast}~\cite{Bertone:2014zva}.

\paragraph*{Experimental data.}
A broad set of input hard-scattering cross-sections from DIS and
proton-(anti)proton collisions, providing information on the PDFs over a wide 
range of $x$ and for different flavor combinations, is used in modern PDF fits.
Inclusive DIS measurements have been realized with electron, muon and neutrino
(and the corresponding antiparticles) off protons, deuterons and
heavy nuclear targets. 
While traditional PDF fits were based mostly on DIS structure functions, 
and Drell-Yan and inclusive jet cross-sections, in recent years many other 
processes have proved important for constraining PDFs, among which
top-quark pair production~\cite{Czakon:2016olj}, the $p_T$ distribution of $Z$ 
bosons~\cite{Boughezal:2017nla} and $D$ meson production in 
the forward region~\cite{Gauld:2016kpd}.

\begin{figure}[!t]
\centering
\includegraphics[scale=0.60]{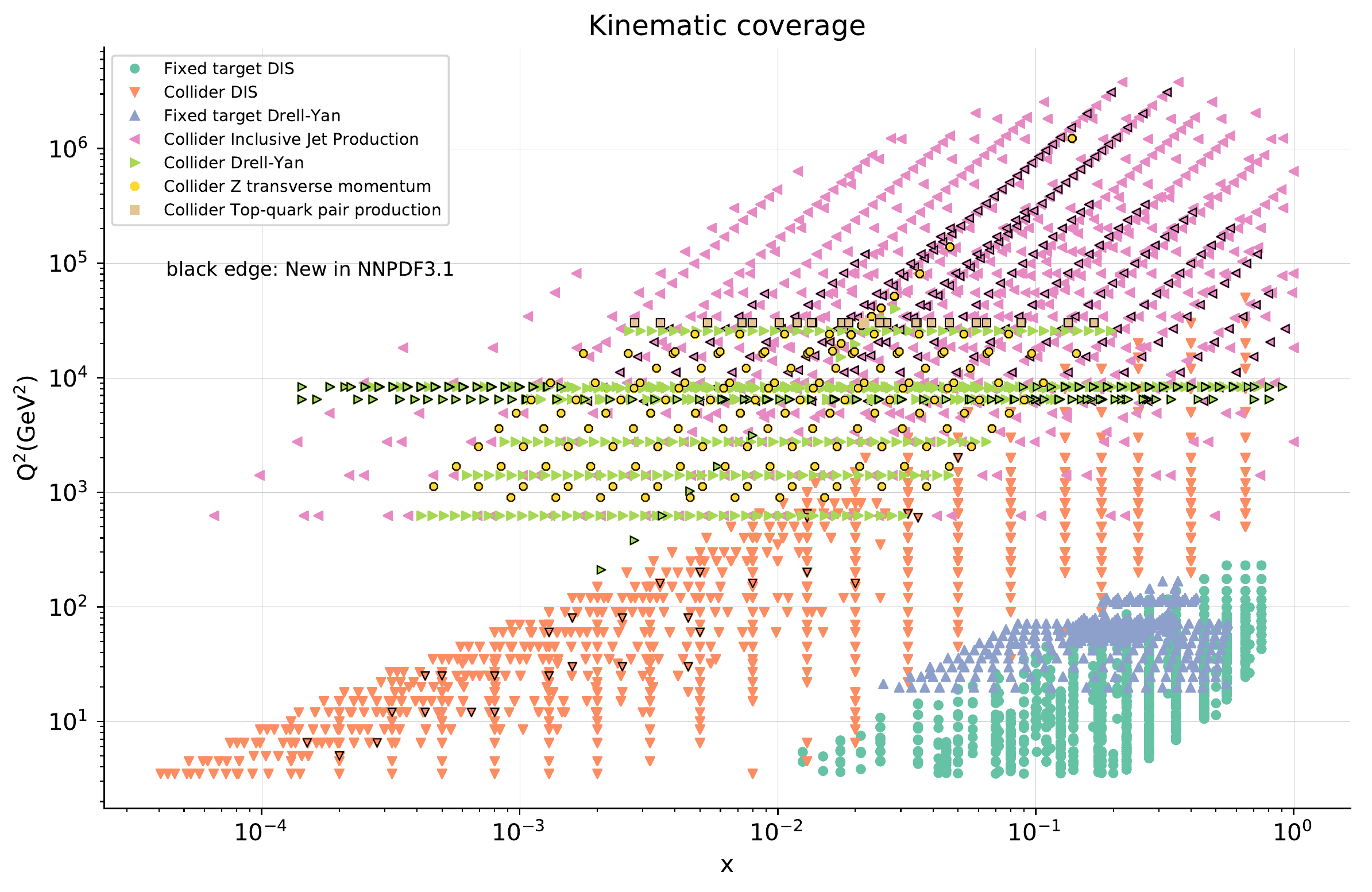}
\caption{\small Representative kinematic coverage in the $(x,Q^2)$ plane
 of the DIS and proton-(anti)proton hard-scattering measurements that are
 used as input in a typical fit of unpolarized PDFs, 
 NNPDF3.1~\cite{Ball:2017nwa}.
 Different datasets have been clustered together into families of
 related processes.
 For hadronic cross-sections, leading order kinematics is assumed to map
 each experimental bin to a pair of points in the $(x,Q^2)$ plane.
 Additional precise data from SLAC and Jefferson Lab exist also in the 
 bottom right corner of the $(x,Q^2)$ plane, although they were excluded from 
 the NNPDF3.1 fit by the cut on the invariant mass of the final 
 state $W^2<12.5$~GeV$^2$ adopted there.}
\label{fig:kinplot-report} 
\end{figure}

In Fig.~\ref{fig:kinplot-report} we show the representative kinematic coverage 
in the $(x,Q^2)$ plane of the DIS and proton-(anti)proton hard-scattering 
measurements that are used as input in a typical global fit of unpolarized 
PDFs, in this case NNPDF3.1~\cite{Ball:2017nwa}.
In order to facilitate visualization, different datasets have been clustered 
together into families of related processes.
For hadronic cross-sections, LO kinematics is assumed to map
each experimental bin into a pair of points in the $(x,Q^2)$ plane.
The fact that similar regions in the $(x,Q^2)$ plane are covered by
different processes is essential to achieve quark
flavor separation and to constrain the gluon PDF.

Abundant precise data from SLAC and Jefferson Lab exist also in the 
bottom right corner of the $(x,Q^2)$ plane, where however power corrections 
need to be accounted for 
in QCD fits~\cite{Alekhin:2017kpj,Owens:2012bv,Accardi:2016qay}.
They are not shown in Fig.~\ref{fig:kinplot-report} because they are excluded 
from the NNDPF3.1 fit by the kinematic cut on the invariant mass of the final
state $W^2<12.5$~GeV$^2$ adopted there.

\paragraph*{State-of-the-art global PDF fits.}
Various collaborations provide regular updates of their global unpolarized
PDF fits.
The latest fits from the three main global fitting collaborations
are CT14~\cite{Dulat:2015mca}, MMHT14~\cite{Harland-Lang:2014zoa} and
NNPDF3.1~\cite{Ball:2017nwa}.
These fits are performed up to NNLO in the strong coupling (with central value
$\alpha_s(m_Z)=0.118$), and include data from the HERA $e^{\pm} p$ collider, 
fixed (nuclear and proton) target experiments, the Tevatron $p\overline{p}$ 
collider and the LHC. 
The ABMP16~\cite{Alekhin:2017kpj} set fits to a similar global data set
(although excluding jet production) but differs in its treatment of errors,
 heavy flavors and the low-$Q^2$ and large-$x$ regions.
The HERAPDF2.0~\cite{Abramowicz:2015mha} set fits to the final combined HERA 
Run I + II data set only, with the aim of determining the PDFs from a 
completely consistent DIS data sample; in $x$ regions that are less constrained 
by HERA data, the uncertainties can be quite large.
The CJ15~\cite{Accardi:2016qay} NLO set focuses on constraining the PDFs at 
higher $x$ by lowering $Q^2$ and $W^2$ cuts in DIS.
This greatly increases the amount of available data, but requires additional 
modeling of power-like ${\cal O}(1/Q^2)$ corrections.

The features of each PDF set have been discussed in detail in 
Refs.~\cite{Butterworth:2015oua,Accardi:2016ndt}, including the 
dataset, the fitting methodology, the theoretical details of the 
corresponding QCD analyses, and, most importantly, the uncertainties
coming from each of these aspects.

\begin{figure}[!t]
\centering
 \includegraphics[scale=0.37]{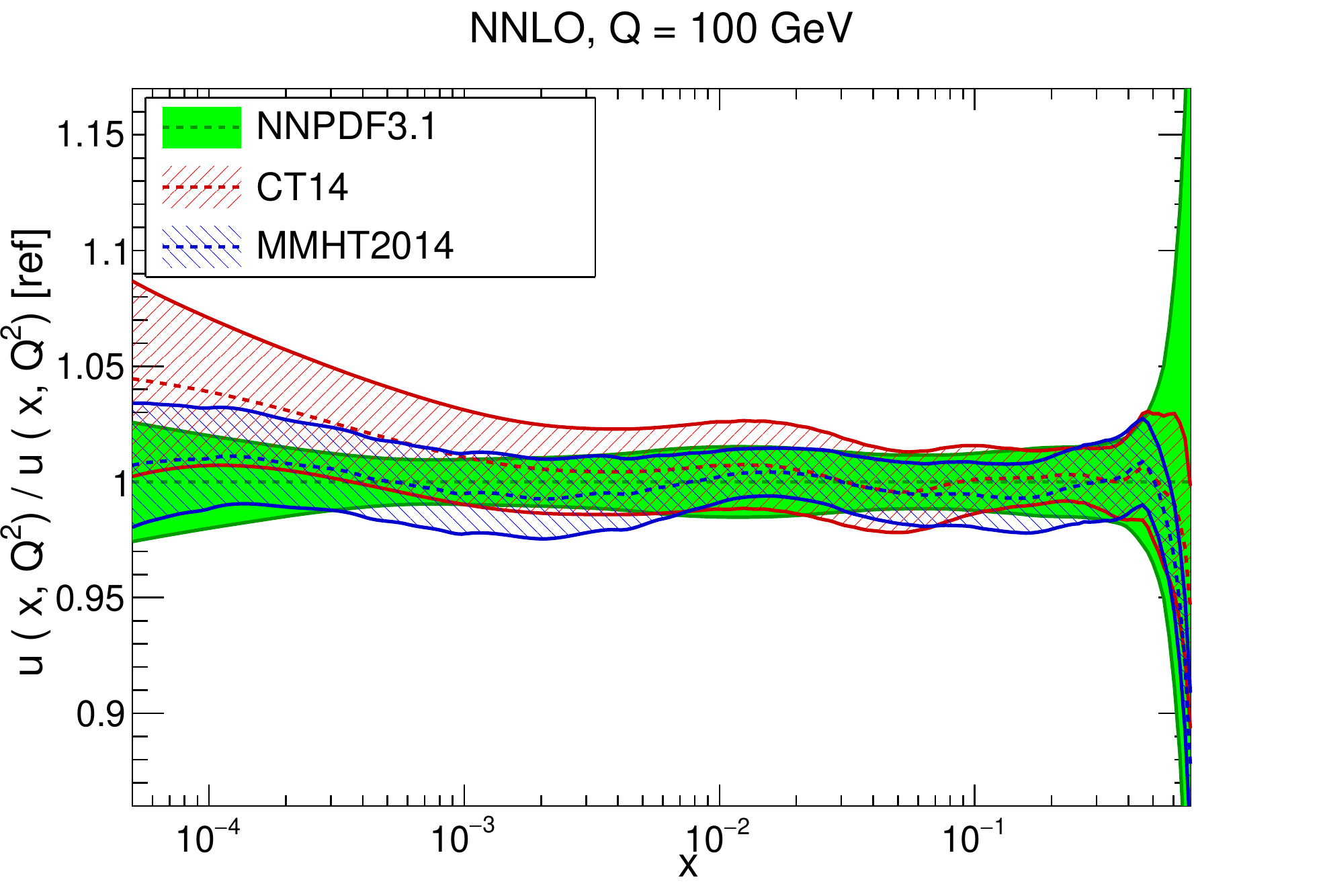}
 \includegraphics[scale=0.37]{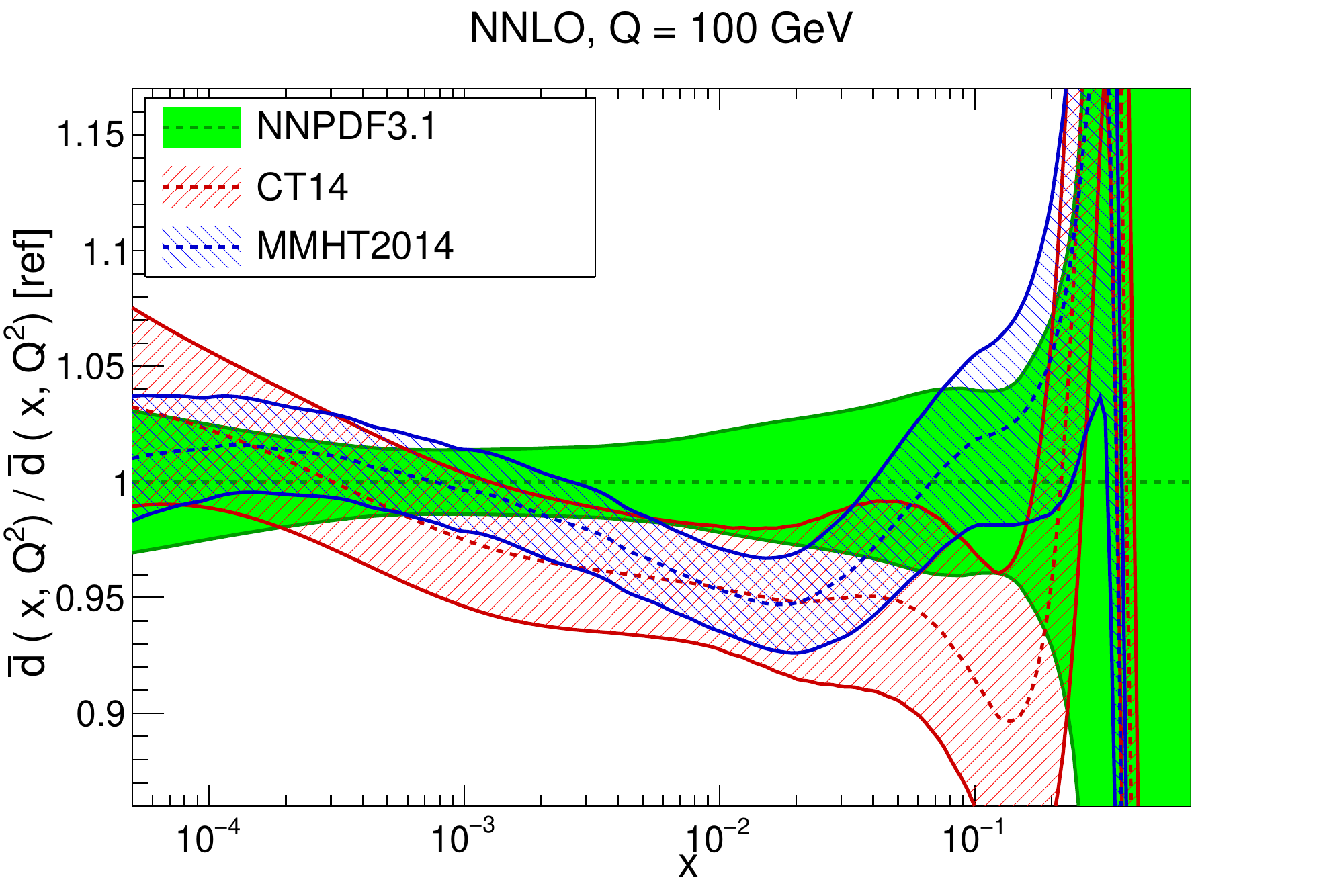}\\
 \includegraphics[scale=0.37]{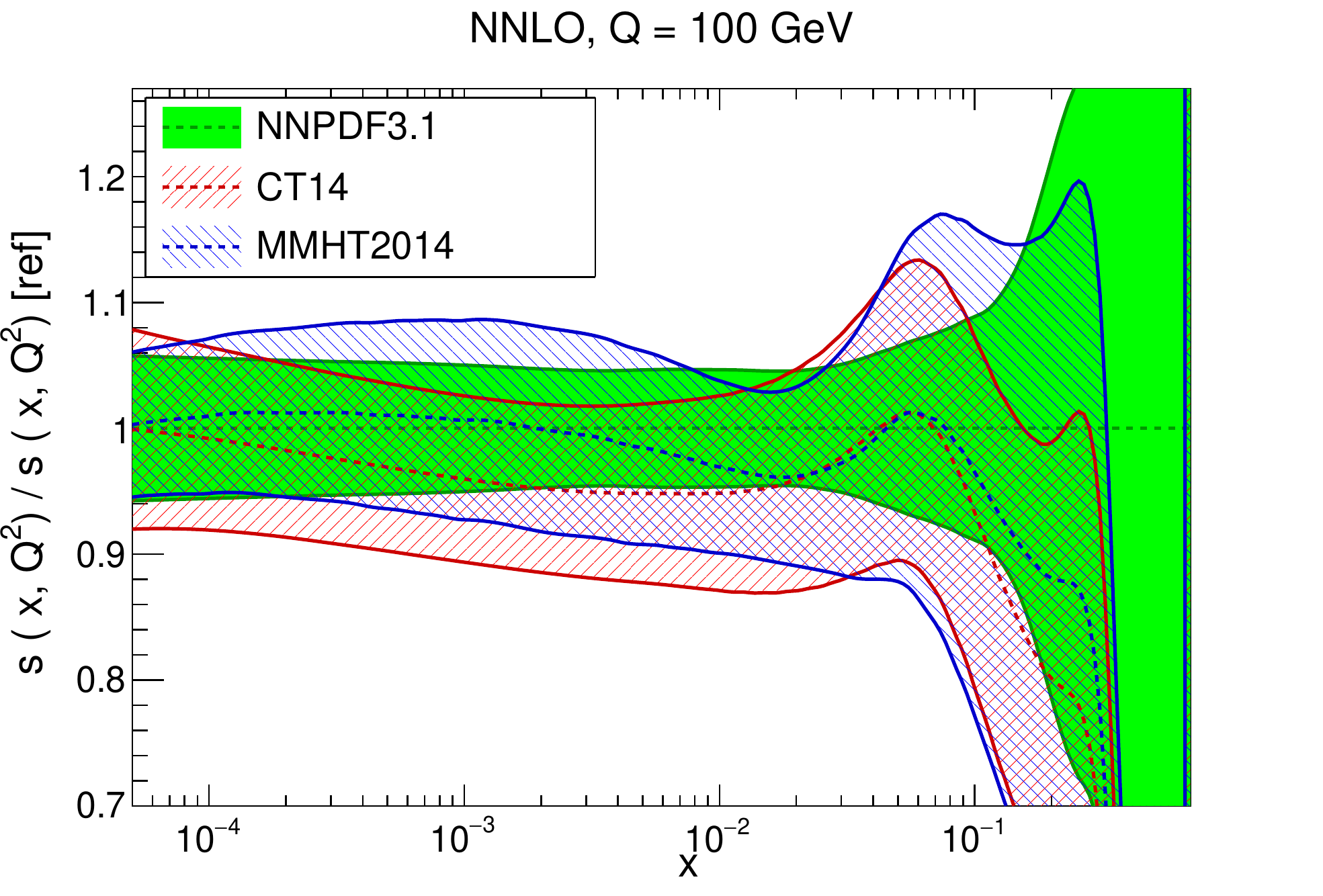}
 \includegraphics[scale=0.37]{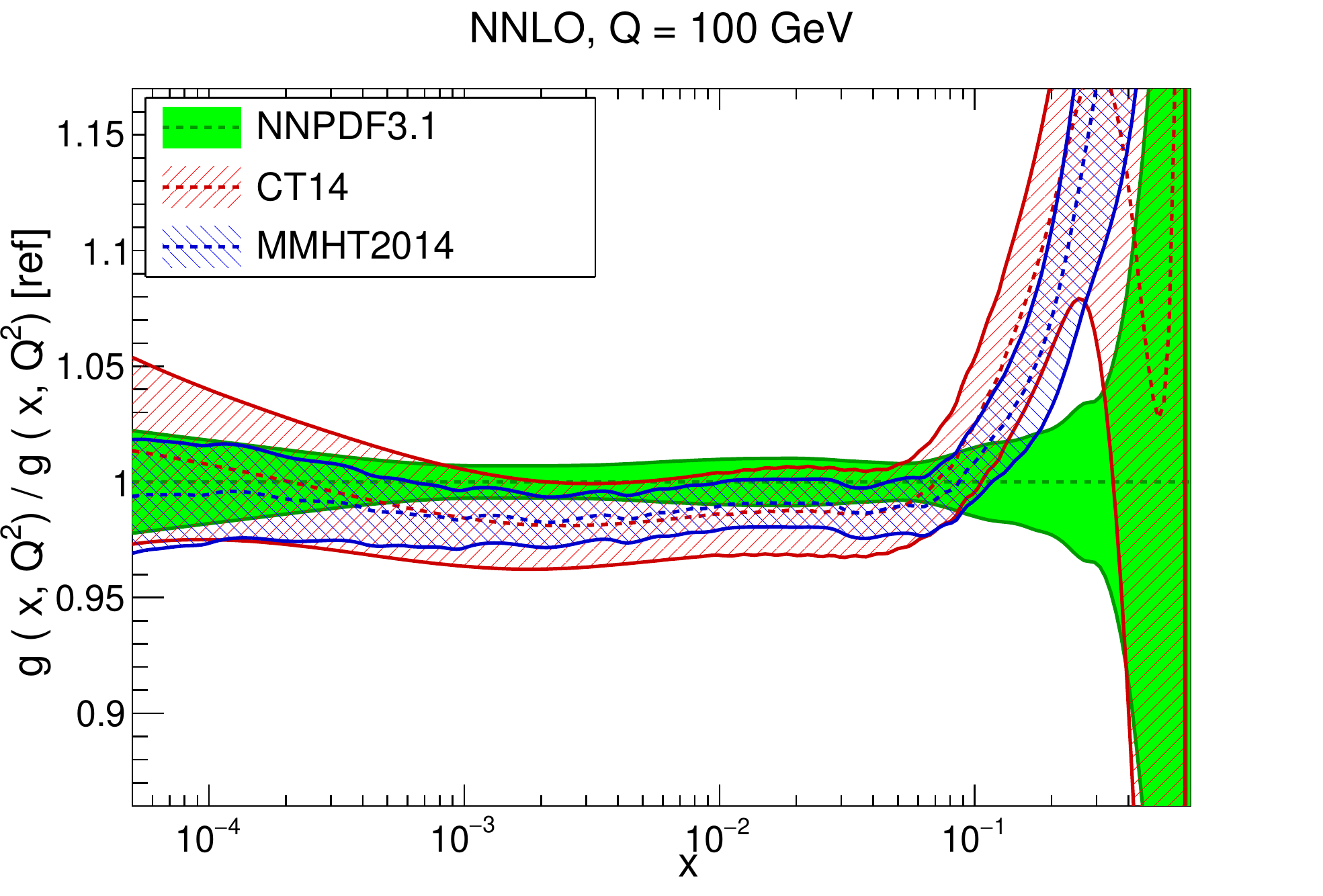}\\
 \caption{\small Comparison between the CT14, MMHT2014
  and NNPDF3.1 NNLO PDF sets at $Q=100$~GeV, normalized
  to the central value of the latter.
  From top to bottom and from left to right we show the
  $u$, $\bar{d}$ and $s$ quark PDFs as well as the gluon.
  The error bands indicate the 1-$\sigma$ PDF uncertainties
  associated with each set.
  These PDF comparison plots have been produced using the
  {\tt APFEL-Web} online plotting interface~\cite{Carrazza:2014gfa}.
    \label{fig:globalfits}
  }
\end{figure}

In Fig.~\ref{fig:globalfits}
we present a snapshot of the current understanding
of the proton structure in the global PDF fitting framework.
We compare the CT14, MMHT2014
and NNPDF3.1 NNLO PDF sets at $Q=100$~GeV, normalized
to the central value of the last.
From top to bottom and from left to right we show the
$u$, $\bar{d}$ and $s$ quark PDFs and the gluon PDF.
The error bands indicate the 68\%-confidence level (CL) PDF uncertainties
associated with each set, computed with the corresponding master formula.
We observe that differences for the up quark PDF
are small, at the few percent level, but greater differences
are observed for the sea quarks, in particular
in the medium and large-$x$ region.
For the gluon there is reasonable agreement except in the large-$x$ region, 
where NNPDF3.1 is softer than CT14 and MMHT14.
Any other comparison plots between PDFs can be straightforwardly
obtained using the {\tt APFEL-Web} online plotting 
interface~\cite{Carrazza:2014gfa}.

In addition to these latest versions of the global PDF fits,
there has recently been a significant development of techniques aiming
to construct combined PDF sets that are based on
a small number of Hessian eigenvectors or MC replicas and thus
are more efficient to use in lengthy higher-order
computations or Monte Carlo simulations.
In particular, the PDF4LHC15 PDF sets are based on the
combination of the CT14, MMHT14 and NNPDF3.1 NNLO PDF sets,
subsequently reduced to a small number of eigenvectors
(replicas) using the META-PDF~\cite{Gao:2013bia}
and MC2H~\cite{Carrazza:2015aoa}
(CMC~\cite{Carrazza:2015hva}) compression algorithms.
In this respect, Specialized Minimal PDF sets~\cite{Carrazza:2016htc}
(SM-PDFs) have also
been advocated, which
are tailored to specific physical processes and are based
on a minimal number of Hessian eigenvectors.
%

\begin{figure}[!t]
\centering
  \includegraphics[scale=0.8]{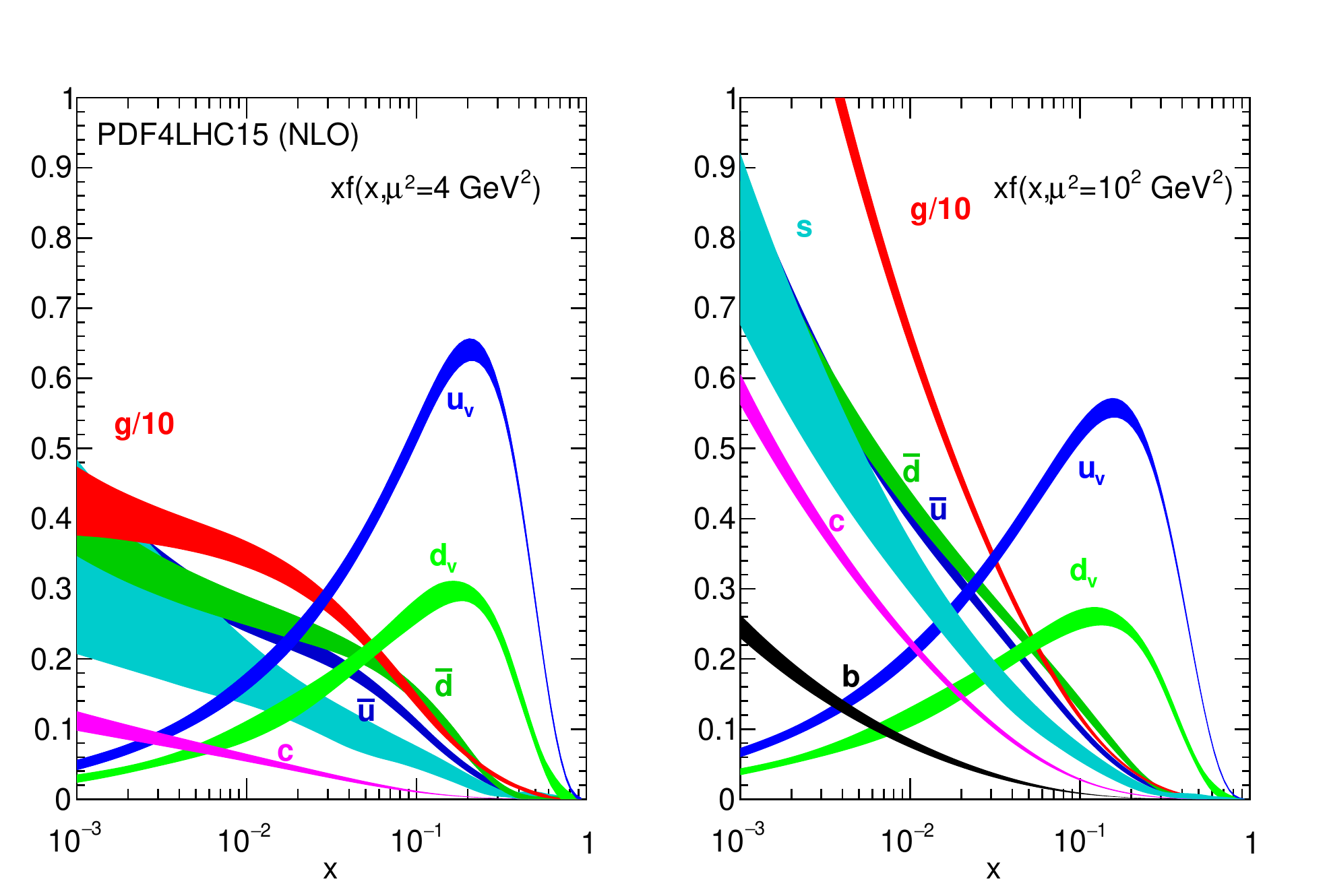}\\
  \caption{\small The PDF4LHC15 NLO PDFs at a low scale
    $\mu^2=Q^2=4~{\rm GeV}^2$ (left plot) and at 
    $\mu^2=Q^2=10^2~{\rm GeV}^2$ (right plot) as a function of $x$.
    We show the $u_v$ and $d_v$ valence combinations, the $\bar{u}$,
    $\bar{d}$, $s$ and $c$ sea quark PDFs, and the gluon (note that
    the latter is divided by a factor 10).
    \label{fig:nnlopdfs}
  }
\end{figure}

The PDF4LHC15 NLO set~\cite{Butterworth:2015oua} is displayed in 
Fig.~\ref{fig:nnlopdfs} at $\mu^2=Q^2=4~{\rm GeV}^2$ and at
$\mu^2=Q^2=10^2~{\rm GeV}^2$.
Specifically, we show the $u_v=u-\bar{u}$ and $d_v=d-\bar{d}$ valence 
combinations, the $\bar{u}$, $\bar{d}$, $s$ and $c$ sea quark PDFs, 
and the gluon (divided by a factor 10).
The evolution between $Q^2=4$~GeV$^2$ and $Q^2=10^2$~GeV$^2$ is completely
determined by the solution of the DGLAP evolution equations.
The shape of the $u_v~(u^{-})$ and $d_v~(d^{-})$ valence quark combinations
reflects the constraints from the valence sum rules 
Eq.~\eqref{eq:valencesumrules}.
At small $x$, there is a rapid growth of the gluon and the sea quark PDFs, 
implying that the higher the collision center-of-mass energy $\sqrt{s}$, 
the more important gluon- and sea-quark-initiated processes become.
The bands in Fig.~\ref{fig:nnlopdfs} represent the 68\% CL PDF uncertainties.

We strongly encourage the community to use the most recent versions
of global PDF fits when comparing with existing or new
lattice QCD calculations.
Comparing with deprecated sets, based on obsolete methodology
and in many cases experimental data that have been
superseded, should always be avoided.

\subsubsection{Polarized PDFs}
\label{sec:polPDFs}

\paragraph*{Theoretical features.}

The dependence on the momentum fraction $x$, fixed by nonperturbative QCD 
dynamics, should satisfy some theoretical constraints.
First, PDFs must lead to positive cross-sections.
At leading order (LO), this implies that polarized 
PDFs are bounded by their unpolarized counterparts\footnote{Beyond LO, more 
complicated relations hold~\cite{Altarelli:1998gn}; however they have little
effect on PDFs.}, $|\Delta f(x,\mu^2)|\leq f(x,\mu^2)$~\cite{Altarelli:1998gn}.
Second, PDFs must be integrable: this corresponds to the assumption 
that the nucleon matrix element of the axial current for each flavor is finite.
Third, SU(2) and SU(3) flavor symmetry, if assumed to be exact, imply that 
the zeroth moments of the nonsinglet $\mathcal{C}$-even PDF combinations,
$\Delta T_3=\Delta u^+ -\Delta d^+$ and 
$\Delta T_8 = \Delta u^+ +\Delta d^+ -2\Delta s^+$ 
(where $\Delta q^+=\Delta q+\Delta\bar{q}$, $q=u,d,s$), are respectively
related to the baryon octet $\beta$-decay constants, whose 
measured values are~\cite{Olive:2016xmw}
\begin{align}
 g_A  = a_3
 & =
 \int_0^1 dx \Delta T_3 (x,\mu^2)
 = \langle 1\rangle_{\Delta u^+} - \langle 1\rangle_{\Delta d^+}  = 1.2723 \pm 0.0023\,,
 \label{eq:a3}
 \\
 a_8
 & =
 \int_0^1 dx \Delta T_8 (x,\mu^2)
 = \langle 1 \rangle_{\Delta u^+} + \langle 1 \rangle_{\Delta d^+} -2\,\langle 1 \rangle_{\Delta s^+} 
 =0.585  \pm 0.025
 \,.
\label{eq:decayconst}
\end{align}
Fairly significant violations of SU(3) symmetry are advocated
in the literature (see {\it e.g.} Ref.~\cite{Cabibbo:2003cu} for a review). 
In this case, an uncertainty on the octet axial charge, which could be as 
large as 30\% of the experimental value of $a_8$ in Eq.~\eqref{eq:decayconst}, 
see Ref.~\cite{FloresMendieta:1998ii}. 

\paragraph*{Experimental data.}
The bulk of the experimental information on polarized PDFs comes from 
neutral-current (photon exchange) inclusive and semi-inclusive deep-inelastic
scattering (DIS and SIDIS) with charged lepton beams and nuclear targets. 
As photon scattering does not distinguish quarks and antiquarks, inclusive DIS 
data constrain only the total quark combinations $\Delta q^+$, 
while SIDIS data with identified pions or kaons in the final state 
constrain individual quark and antiquark flavors. 
In principle, both DIS and SIDIS are also sensitive to the gluon 
distribution $\Delta g$, as it directly enters the factorized expressions of
the corresponding structure functions beyond LO, and indirectly via DGLAP 
evolution.
In practice, the constraining power of DIS and SIDIS data on $\Delta g$ is 
rather weak because the $Q^2$ range covered by the data is limited,
especially if one restricts to the kinematic region not affected by
power-suppressed corrections and very precise data from JLab are therefore
excluded. 

Note that, in the case of SIDIS, a reliable knowledge of fragmentation 
functions (FFs) is required in the factorized expressions of the 
corresponding observables. 
Since FFs are nonperturbative objects on the same footing as PDFs, they are 
an additional source of uncertainty in PDF determinations, if not a bias.
A significant experimental and theoretical effort has been
invested in improving the independent determination of 
FFs~\cite{deFlorian:2014xna,deFlorian:2017lwf,
Hirai:2016loo,Sato:2016wqj,Bertone:2017tyb} and most recently in simultaneously 
fitting both PDFs and FFs~\cite{Ethier:2017zbq,Borsa:2017vwy}.

Besides DIS and SIDIS fixed-target data, a significant amount of data from
longitudinally polarized proton-proton collisions at the Relativistic 
Heavy Ion Collider (RHIC) has become available recently (see {\it e.g.} 
Ref.~\cite{Aschenauer:2015eha} for an overview), although in a limited range 
of momentum fractions, $0.05\lesssim x \lesssim 0.4$.
On the one hand, longitudinal (parity-violating) single-spin and 
(parity-conserving) double-spin asymmetries for $W^\pm$ boson production are 
sensitive to the flavor decomposition of polarized quark and antiquark 
distributions, because of the chiral nature of the weak 
interaction~\cite{Bourrely:1993dd}. 
On the other hand, double-spin asymmetries for jet, di-jet and $\pi^0$ 
production are directly sensitive to the gluon polarization in 
the proton, because of the dominance of gluon-gluon and quark-gluon initiated 
subprocesses in the kinematic range accessed by RHIC~\cite{Bourrely:1990pz}.

\begin{figure}[!t]
\centering
\includegraphics[width=0.9\textwidth]{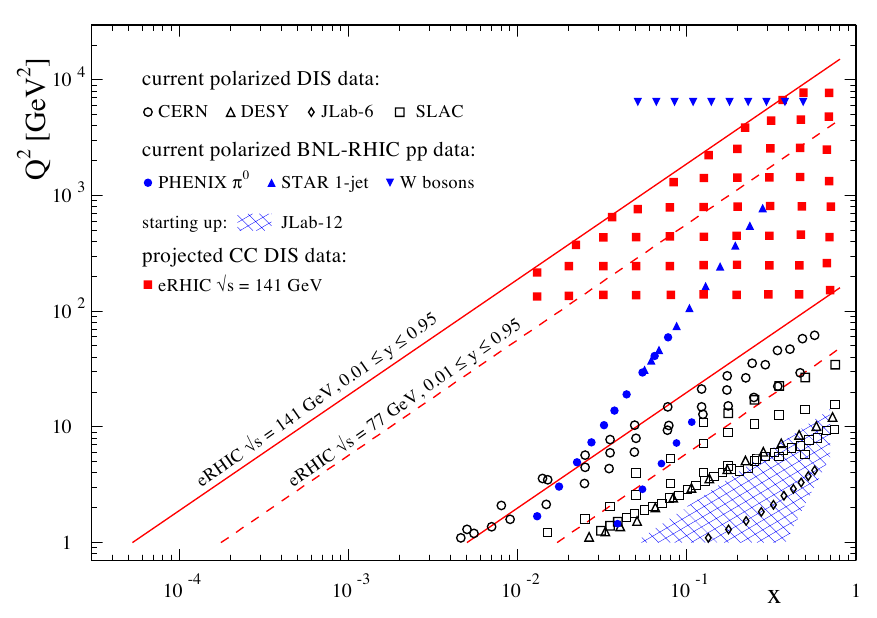}\\
\caption{\small Representative kinematic coverage, in the $(x,Q^2)$ plane,
of the (neutral current) DIS, SIDIS and proton-proton hard-scattering 
measurements that are used as input in a global polarized PDF fit.
The extended kinematic coverage achieved by 
JLab-12~\cite{Dudek:2012vr} and by an EIC~\cite{Accardi:2012qut}
(including projected charged-current (CC) DIS data and denoted as eRHIC) 
is also shown.
Figure taken from Ref.~\cite{Aschenauer:2014cki}.}
\label{fig:kinEIC}
\end{figure}

The kinematic coverage of the data that can be used to constrain polarized 
PDFs is displayed in Fig.~\ref{fig:kinEIC}.
A comparison with Fig.~\ref{fig:kinplot-report} makes it apparent that the
quantity of data points, their kinematic coverage and the variety of 
available hard-scattering processes are presently much more limited in the 
polarized case than in the unpolarized case.
Therefore, polarized PDFs can currently be determined with much less 
precision than their unpolarized counterparts and only over an $x$-range limited
to $x\gtrsim 0.005$.
The kinematic coverage is expected to be significantly extended in the future,
with DIS and SIDIS data from JLab-12~\cite{Dudek:2012vr} and a polarized 
high-energy Electron-Ion Collider (EIC)~\cite{Accardi:2012qut}.
Such an extended kinematic coverage is also displayed in Fig.~\ref{fig:kinEIC},
where it is denoted as eRHIC.

\begin{figure}[!t]
\centering
\includegraphics[scale=0.8]{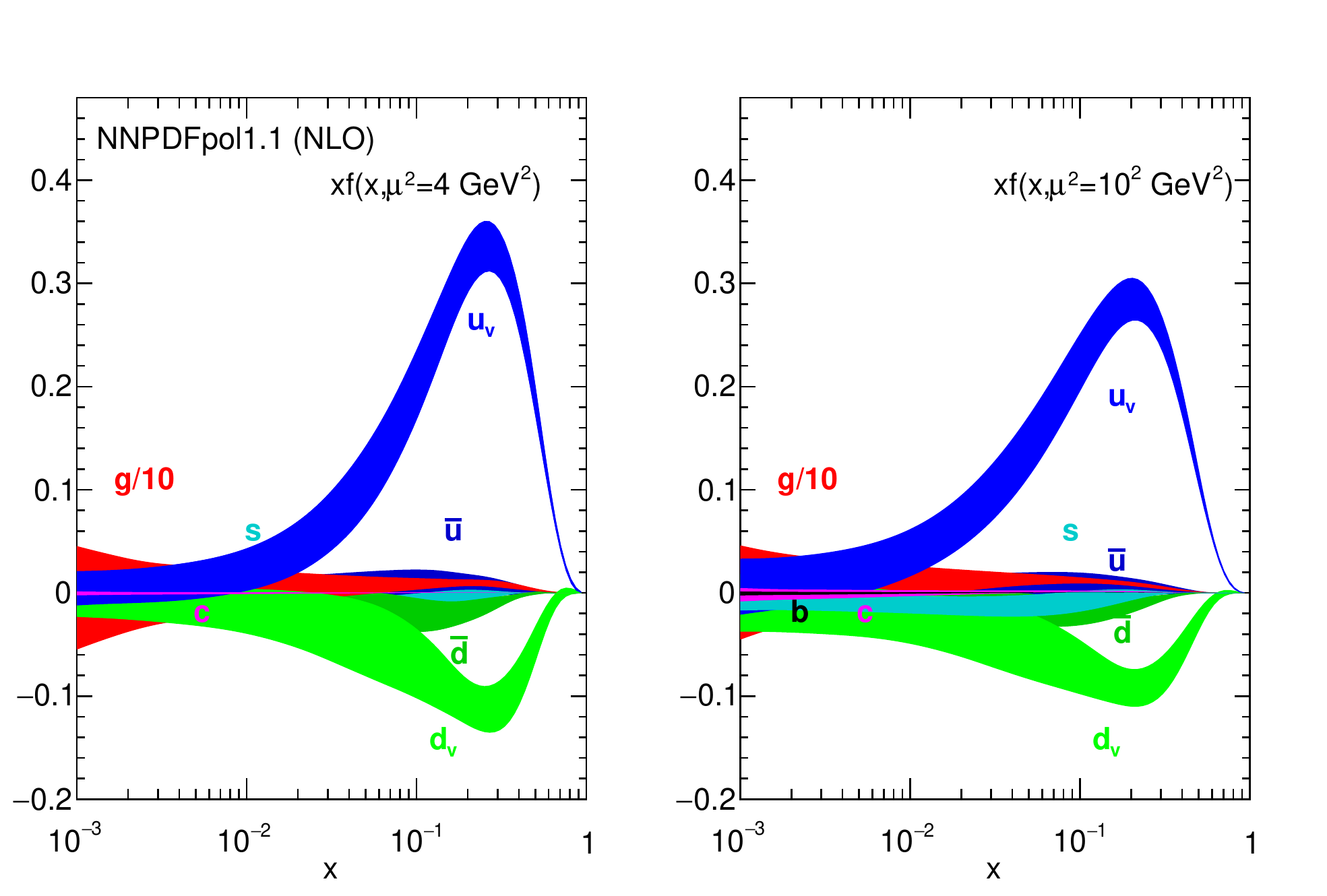}\\
\caption{\small Same as Fig.~\ref{fig:nnlopdfs}, 
but for the polarized NNPDFpol1.1 NLO PDFs~\cite{Nocera:2014gqa}.}
\label{fig:qPDFpol}
\end{figure}

A representative illustration of polarized PDFs obtained from a global
QCD analysis, namely NNDPFpol1.1~\cite{Nocera:2014gqa}, is provided in Fig.~\ref{fig:qPDFpol}.
The format is the same as for the unpolarized case, Fig.~\ref{fig:nnlopdfs},
in order to ease any comparison between the two.
In particular, note the suppression of all polarized PDFs at small values of 
$x$, including polarized sea quark PDFs, with respect to their unpolarized 
counterparts.

\paragraph{State-of-the-art global PDF fits.}

Several modern determinations of polarized PDFs of the proton (up to 
NLO\footnote{A NNLO QCD analysis of polarized PDFs based on inclusive DIS
data only was performed in Refs.~\cite{Shahri:2016uzl,Khanpour:2017cha}.
Inclusive DIS is the only polarized process for which coefficient functions
are known up to NNLO (all others are known up to NLO).} 
and mostly in the $\overline{\rm MS}$ factorization scheme) are available in 
the literature~\cite{Nocera:2014gqa,Nocera:2016xhb,deFlorian:2014yva,deFlorian:2008mr,deFlorian:2009vb,Sato:2016tuz,Leader:2010rb,Blumlein:2010rn,Bourrely:2014uha,Hirai:2008aj}. 
A key goal of these is to unveil the size (and uncertainty) of
$\Delta\Sigma$ and  $\Delta G$ in Eq.~\eqref{eq:moments}. 
The various determinations differ among each other in the data sets included 
in the analysis, in some details of the QCD analysis (like the treatment of 
higher-twist corrections) and in the procedure used to determine PDFs from the 
data (for details, see {\it e.g.} Chap.~3 in Refs.~\cite{Nocera:2014vla} 
and~\cite{Nocera:2016xhb,Jimenez-Delgado:2013sma}). 
The NNDPF procedure and the standard (adopted by DSSV) have 
already been outlined in Sec.~\ref{sec:genframework}. 
We note that DSSV has developed a method based on Mellin moments of the PDFs 
in order to efficiently incorporate NLO computations
of proton-proton cross-sections in the fitting procedure. 
The JAM collaboration has implemented a new approach called 
iterative Monte Carlo procedure~\cite{Sato:2016tuz,Ethier:2017zbq}
in their analyses.

The most recent analyses of polarized PDFs are DSSV14~\cite{deFlorian:2014yva}
and NNPDFpol1.1~\cite{Nocera:2014gqa}.
Motivated by the interest in assessing the impact of RHIC proton-proton 
data, they upgrade the corresponding previous analyses, 
DSSV08~\cite{deFlorian:2008mr,deFlorian:2009vb} and 
NNPDFpol1.0~\cite{Ball:2013lla}, with data respectively on double-spin 
asymmetries for inclusive jet production~\cite{Adamczyk:2014ozi} 
and $\pi^0$ production~\cite{Adare:2014hsq} (DSSV14\footnote{Preliminary 
RHIC results included in Ref.~\cite{deFlorian:2008mr} were replaced in
Ref.~\cite{deFlorian:2014yva} with final results.}), 
and on double-spin asymmetries for high-$p_T$ inclusive jet 
production~\cite{Adamczyk:2014ozi,Adamczyk:2012qj,Adare:2010cc} and single-spin
asymmetries for $W^\pm$ production~\cite{Adamczyk:2014xyw} (NNPDFpol1.1).
The new data have been included in NNPDFpol1.1 
by means of Bayesian reweighting~\cite{Ball:2010gb},
and in DSSV14 by means of a full refit.  

Overall, both the DSSV14 and NNPDFpol1.1 PDF determinations are 
state-of-the-art in the inclusion of the available experimental information. 
The data sets in the two analyses differ between each other only in
fixed-target SIDIS and RHIC $\pi^0$ production measurements, included in 
DSSV14, but not in NNPDFpol1.1. 
The information brought in by these data is complementary to that provided by 
RHIC $W^\pm$ production and inclusive jet production data respectively,
although fraught with larger theoretical uncertainties related to fragmentation.

\begin{figure}[!t]
\centering
\includegraphics[scale=0.33,clip=true,trim= 0 -1.3cm 0 0]{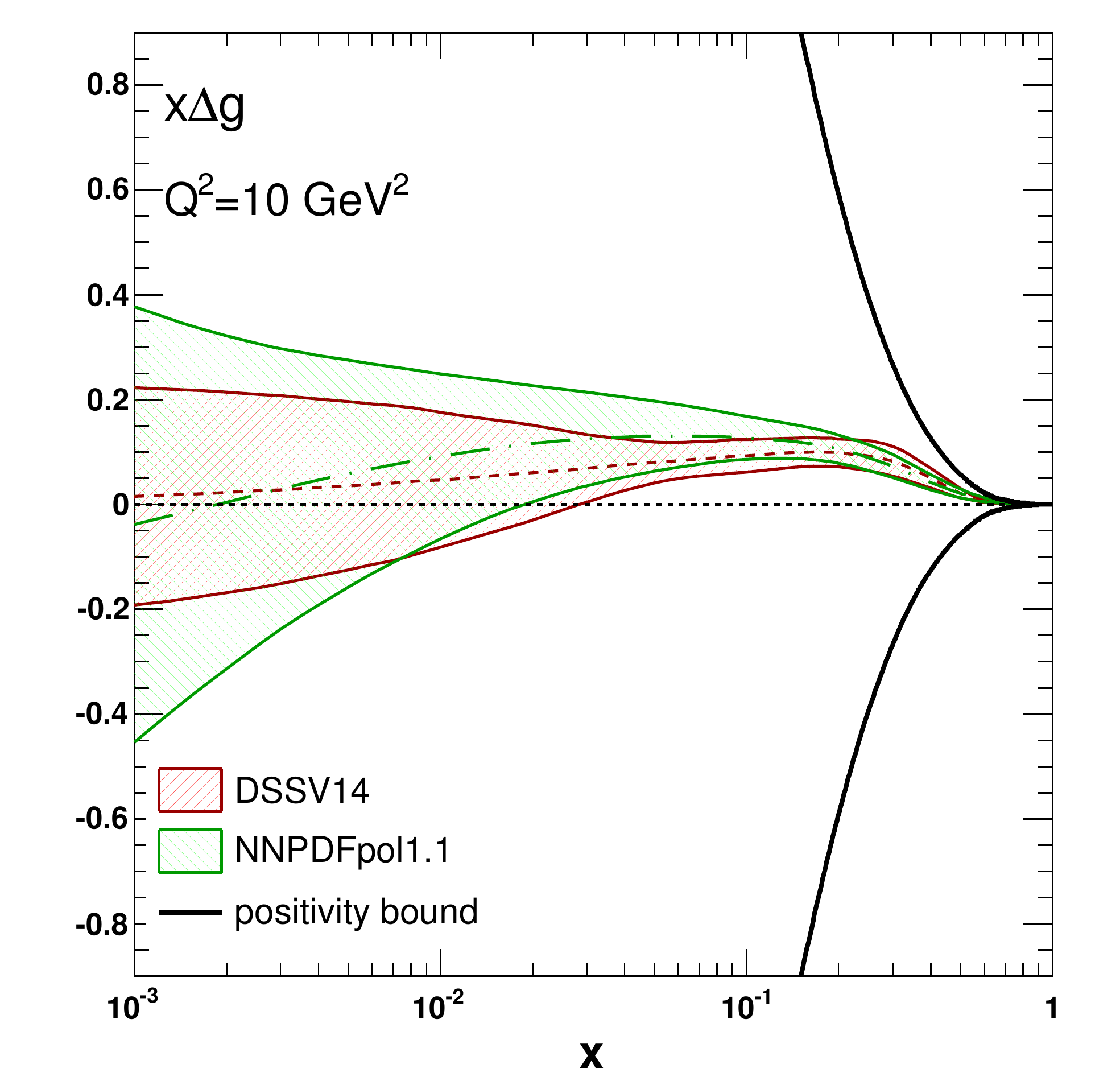}
\includegraphics[scale=0.555,clip=true,trim=0 0 7cm 15cm]{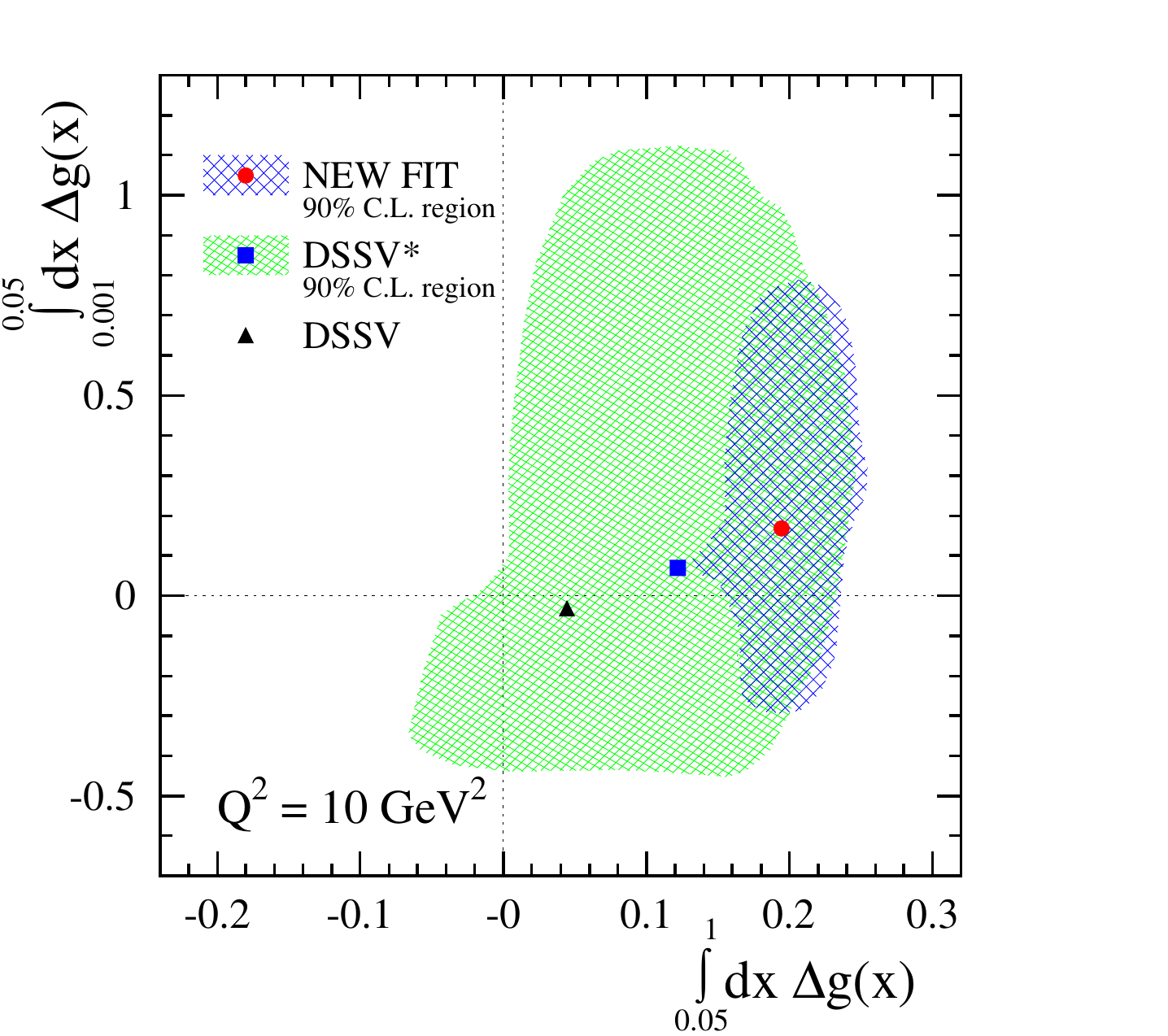}\\
\caption{\small (Left) The polarized gluon momentum distribution  
$x\Delta g$ from the DSSV14 (with $90\%$ C.L. uncertainty band)
and NNPDFpol1.1 PDF sets at $Q^2=10$~GeV$^2$. The NNPDF3.1 positivity
bound is also shown.
(Right) $90\%$ C.L.\ areas in the plane spanned by the truncated moments of
$\Delta g$ computed for $0.05\leq x\leq 1$ and $0.001\leq x\leq 0.05$ at $Q^2=10\,\mathrm{GeV}^2$~\cite{deFlorian:2014yva}.}
\label{fig:RHICpdfs}
\end{figure}

The effect of RHIC data on the polarized PDFs of the proton is twofold:
\begin{itemize}

\item The 2009 STAR and PHENIX data sets on jet and $\pi^0$ 
production~\cite{Adamczyk:2014ozi,Adare:2014hsq}, included in DSSV14
and NNPDFpol1.1, provide the first evidence
of a sizable positive gluon polarization in the proton. 
A comparison of the gluon PDF in the two PDF sets is displayed in 
Fig.~\ref{fig:RHICpdfs} (left panel). 
Comparable results, both central values and uncertainties, are found in the 
$x$ region covered by RHIC data. 
The agreement between the two analyses is optimal in the
range $0.08\leq x \leq 0.2$, where the dominant experimental information comes
from jet data; a slightly smaller central value is found in the DSSV14 
analysis, in comparison to NNPDFpol1.1, in the range 
$0.05\leq x \leq 0.08$, where the dominant experimental information comes from 
$\pi^0$ production data. 
Indeed, these are included in DSSV14 but are not in NNPDFpol1.1. 
Nevertheless, best fits lie well within each other's error
bands, though NNPDFpol1.1 uncertainties tend to be larger than DSSV14
uncertainties outside the region covered by RHIC data.
Very consistent values of the zeroth moment of $\Delta g$, 
Eq.~\eqref{eq:moments}, truncated over the interval $0.05\leq x \leq 1$, are 
found: at $Q^2=10$~GeV$^2$, this is $0.20^{+0.06}_{-0.07}$ for 
DSSV14~\cite{deFlorian:2014yva}, and $0.23\pm 0.06$ for 
NNPDFpol1.1~\cite{Nocera:2014gqa}. The right plot in Fig.~\ref{fig:RHICpdfs} 
shows the corresponding DSSV14 result as an example; the impact of the RHIC
data is clearly visible. 

\item The 2012 STAR data sets on $W$ production~\cite{Adamczyk:2014xyw}, 
included in NNPDFpol1.1, provide evidence of a positive 
$\Delta\bar{u}$ distribution 
and a negative $\Delta\bar{d}$ distribution, with 
$|\Delta\bar{d}|>|\Delta\bar{u}|$~\cite{Nocera:2014gqa}.
The size of the flavor symmetry breaking for polarized sea quarks is 
quantified by the asymmetry $\Delta\bar{u}-\Delta\bar{d}$, which,
in the NNPDFpol1.1 analysis, turn out to be roughly as large as its 
unpolarized counterpart (in absolute value)~\cite{Ball:2017nwa}, 
though much more uncertain~\cite{Nocera:2014rea}. 
Even within this uncertainty, polarized and unpolarized light sea quark 
asymmetries show opposite signs, with the polarized one being clearly positive. 
This trend is also found from analysis of the polarized SIDIS data, 
as revealed by the DSSV08 parton set. 
This result may discriminate among various models of nucleon structure, 
see~\cite{Nocera:2014rea} and references therein. 

\end{itemize}


\paragraph{Open issues.}

Despite the achievements described above, the polarized PDFs presently cannot 
be determined in a global QCD analysis with the same accuracy as their 
unpolarized counterparts.
The experimental data are confined to a relatively narrow range of 
$x$ and $Q^2$.
As a consequence, the size of the contributions of quarks, antiquarks and 
gluons to the nucleon spin, as quantified by their zeroth moments, 
Eq.~\eqref{eq:moments}, are still affected by large uncertainties. 
These come predominantly from the extrapolation into the small-$x$ region 
($x\lesssim 10^{-3}$). 
Here potential modifications in the PDF shape induced by small-$x$ 
evolution~\cite{Bartels:1995iu,Bartels:1996wc,Kovchegov:2015pbl,
Kovchegov:2016weo,Kovchegov:2016zex,Kovchegov:2017jxc,Kovchegov:2017lsr} 
could arise, which presently cannot be tested.
Significant uncertainties also affect the PDFs in the large-$x$ 
{\it valence} region ($x\gtrsim 0.7$). 
This regime is less relevant for the determination of the PDF moments, but it 
is important for comparisons to nonperturbative models of nucleon structure, 
especially in terms of ratios of light-quark polarized to unpolarized PDFs 
(for a comparison between large-$x$ PDFs 
and model predictions, see Ref.~\cite{Nocera:2014uea}).
Finally, the small lever arm of the data in $Q^2$ is a serious limiting factor 
in the determination of $\Delta g$ via evolution, unless the data at low $Q^2$
and large $x$ are included in the fit and carefully analyzed.
This requires an appropriate treatment of power-suppressed corrections and 
possibly a minimization methodology which can iteratively focus on a region 
in parameter space where constraints are not too strong, as done in the 
JAM15 analysis~\cite{Sato:2016tuz}. 

The determination of the total polarized strange distribution $\Delta s^+$ is 
also particularly delicate.
Inclusive DIS data, together with nonsinglet axial couplings, 
Eq.~\eqref{eq:decayconst}, and kaon SIDIS data provide the sole available 
constraint on $\Delta s^+$.
A sizable negative $\Delta s^+$ is found 
consistently in all analyses based on inclusive DIS data only, as a result 
of the constraint from hyperon decays that is usually adopted. 
However, the shape of $\Delta s^+$ may change significantly in analyses that also include
SIDIS data. Typically SIDIS data lead to a trend for $\Delta s^+$ to be
small or even slightly positive in the medium $x$-range, although this depends 
also on the set of kaon FFs used to compute
the corresponding observables~\cite{Leader:2011tm}.  
The recent study in Ref.~\cite{Ethier:2017zbq} sheds some light on this issue
by performing a simultaneous determination of polarized PDFs and unpolarized 
FFs using DIS, SIDIS and single-inclusive annihilation data.
In order to avoid biasing the determination of $\Delta s^+$ by 
assumptions on SU(3) symmetry, the octet axial charge in 
Eq.~\eqref{eq:decayconst} has been allowed to be determined by the data alone.
As a consequence, a slightly positive $\Delta s^+$ distribution, but
compatible with the negative result found from inclusive DIS within its 
large uncertainties, has been obtained.
An octet axial charge about $20\%$ smaller than its quoted experimental value, 
Eq.~\eqref{eq:decayconst}, appears to be preferred by the data.
This implies a zeroth moment $\langle 1\rangle_{\Delta s^+}=-0.03 \pm 0.1$ at 
$\mu^2=1$~GeV$^2$, and hence a larger $\Delta\Sigma$, Eq.~\eqref{eq:singletmom},
than in most other present analyses.
However, we stress that the determination of $\Delta s^+$ from SIDIS data 
also relies on good knowledge of the {\it un}polarized strange distribution. 
Furthermore, unpolarized SIDIS data themselves set constraints on 
FFs and ultimately need to be included as well
to obtain a reliable picture~\cite{Borsa:2017vwy}. 
In any case, further higher precision kaon SIDIS data will be needed 
to reduce the uncertainty on $\Delta s^+$ and further test the degree of 
SU(3) breaking. 

Ongoing and future experimental campaigns at current facilities are
expected to provide additional experimental information
useful to clarify some of the issues outlined above (for an 
assessment of the impact of very recent/forthcoming data, see {\it e.g.}
Refs.~\cite{Aschenauer:2015eha,Aschenauer:2015ata,Nocera:2015vva,
Nocera:2017wep}).
However, a future high-energy, polarized EIC~\cite{Accardi:2012qut} will 
likely be the only facility to be able to address all of the above issues 
with the highest precision. 
The extension of the kinematic reach down to $x\sim 10^{-4}$ and up to
$Q^2=10^4$~GeV$^2$ will allow for an accurate determination of $\Delta g$
via evolution in DIS/SIDIS, of $\Delta\bar{u}$ and 
$\Delta\bar{d}$ via inclusive DIS at high $Q^2$ mediated by electroweak bosons,
and of $\Delta s$ via kaon-tagged SIDIS. 
The potential impact of the longitudinally polarized program at an EIC
has been quantitatively assessed in several dedicated 
studies~\cite{Aschenauer:2012ve,Ball:2013tyh,Aschenauer:2013iia,
Aschenauer:2015ata}.

\section{Benchmarking PDF moments}
\label{sec:benchmarking}

In this section we provide a quantitative comparison between 
current lattice-QCD and global-fit results of the lowest
moments of unpolarized and polarized PDFs.
To this purpose, we identify benchmark quantities
and define the criteria to appraise the determinations
available in the literature.
For each benchmark quantity, we specify a prescription to 
select and combine lattice-QCD calculations and global-fit determinations.
We present our benchmark numbers from each side and compare them.

\subsection{Benchmark criteria}
\label{subsec:BC}

We start by describing our benchmark criteria, which include the definition
of the benchmark quantities and the determination of their reference values,
based on a careful assessment of the lattice-QCD and global-fit results 
available in the literature.

\subsubsection{Benchmark quantities}
\label{subsubsec:BQ}

We identify our benchmark quantities with the following moments of unpolarized 
and polarized PDFs, or of PDF quark flavor combinations.
\begin{itemize}
  \item
$\langle x\rangle_{u^+-d^+}$, $\langle x \rangle_{u^+}$, $\langle x \rangle_{d^+}$, 
$\langle x \rangle_{s^+}$ and $\langle x \rangle_{g}$ in the unpolarized case; 
\item $g_A\equiv\langle 1 \rangle_{\Delta u^+ - \Delta d ^+}$, 
$\langle 1 \rangle_{\Delta u^+}$, $\langle 1 \rangle_{\Delta d^+}$,  
$\langle 1 \rangle_{\Delta s^+}$ and $\langle x \rangle_{\Delta u^- - \Delta d^-}$ 
  in the polarized case.
  \end{itemize}
We adopt the conventional notation described in Appendix~\ref{app:notation}.
We focus on the above quantities because current lattice 
calculations of higher moments and moments of other PDF 
combinations are not sufficiently controlled to allow for a meaningful 
comparison between lattice-QCD and global-fit results. 

\subsubsection{Appraising lattice-QCD calculations}
\label{subsubsec:BClQCD}

To accurately assess current lattice-QCD calculations
available in the literature, we follow a procedure inspired by the review of 
low-energy mesons undertaken by the Flavor Lattice Averaging Group 
(FLAG)~\cite{Aoki:2016frl}. 
For each lattice calculation, we characterize each source of 
uncertainty outlined in Sec.~\ref{Sec:IntroLQCD}. 
We use a rating system inspired by FLAG, awarding a blue star (\bstar) for 
sources of uncertainty that are well controlled or very conservatively 
estimated, a blue circle (\bcirc) for sources of uncertainty that have been 
controlled or estimated to some extent, and a red square (\rsquare) for 
uncertainties that have not met our criteria or for which no estimate is given.
Specifically, the rating system works as follows.

\begin{itemize}
\item {\bfseries Discretization effects and the continuum limit.}
We assume that the lattice actions are ${\cal O}(a)$-improved, {\it i.e.}, 
that the discretization errors vanish quadratically with the lattice spacing. 
For unimproved actions, an additional lattice spacing is required. 
These criteria must be satisfied in each case for at 
least one pion mass below 300~MeV.
\begin{itemize}
\item[\bstar] At least three lattice spacings with at least two lattice 
spacings below 0.1~fm and a range of lattice spacings that satisfies 
$[a_{\mathrm{max}}/a_{\mathrm{min}}]^2 \geq 2$.
\item[\bcirc] At least two lattice spacings with at least one point below 
0.1~fm and a range of lattice spacings that satisfy
$[a_{\mathrm{max}}/a_{\mathrm{min}}]^2 \geq 1.4$.
\end{itemize}
To receive a \bstar~or \bcirc~either a continuum extrapolation must be 
performed, or the results must demonstrate no significant discretization 
effects over the appropriate range of lattice spacings.

\item {\bfseries Unphysical pion masses.}
We define a physical pion mass ensemble to be one with $M_\pi=135\pm 10$~MeV
for the following criteria.
\begin{itemize}
\item[\bstar] One ensemble with a physical pion mass \emph{or} a chiral 
extrapolation with three or more pion masses, with at least two pion masses 
below 250~MeV and at least one below 200~MeV.
\item[\bcirc] A chiral extrapolation with three or more pion masses, two of 
which are below 300~MeV.
\end{itemize}

\item {\bfseries Finite-volume effects.}
\begin{itemize}
\item[\bstar] Ensembles with $M_{\pi,\mathrm{min}}L\geq 4$, \emph{or} at least
three volumes with spatial extent $L>2.5$~fm.
\item[\bcirc] Ensembles with $M_{\pi,\mathrm{min}}L \geq 3.4$, \emph{or} at least
two volumes with spatial extent $L>2.5$~fm.
\end{itemize}
For calculations that use a mixed-action approach, {\it i.e.},
with different lattice actions for the valence and sea quarks,
we apply these criteria to the valence quarks. $M_{\pi,\mathrm{min}}$ is 
the lightest pion mass employed in the calculation. 

\item {\bfseries Excited-state contamination.}
\begin{itemize}
\item[\bstar] At least three source-sink separations or a variational method 
to optimize the operator derived from at least a $3\times 3$ correlator matrix, 
at every pion mass and lattice spacing.
\item[\bcirc] Two source-sink separations at every pion mass and lattice 
spacing, or three or more source-sink separations at one pion mass below 
300~MeV.
For the variational method, an optimized operator derived from a $2\times 2$ 
correlator matrix at every pion mass and lattice spacing, or a $3\times 3$ 
correlator matrix for one pion mass below 300~MeV.
\end{itemize}

\item {\bfseries Renormalization.}
\begin{itemize}
\item[\bstar] Nonperturbative renormalization.
\item[\bcirc] Perturbative renormalization.
\end{itemize}
For $g_A$ we also award a \bstar~for calculations that use fermion actions 
for which $Z_A/Z_V=1$ or employ combinations of quantities for which the 
renormalization is unity by construction.

\item {\bfseries Lattice-spacing determination.}
For lattice-QCD calculations of nucleons, the lattice-spacing determination is 
generally sufficiently precise that it is a very small or negligible source
of systematic uncertainty. 
Therefore we do not include an assessment of the lattice-spacing
determination in our criteria.

\end{itemize}

Another important parameter in lattice-QCD calculations is the number of sea 
quark flavors, $N_f$. 
Following the approach used by FLAG, we prefer to avoid combining calculations 
with differing $N_f$; for more discussion of this issue, see the FLAG 
review~\cite{Aoki:2016frl}.

We now summarize the current status of lattice-QCD calculations of
our benchmark moments of unpolarized and polarized PDFs respectively.
Following FLAG, we consider only those results that are published in 
peer-reviewed journals or that have appeared as preprints. 
Where recent results are a clear update of previously published work, we do 
not include earlier results.
A bibliographical compilation of the results available in the literature 
is given for completeness in Appendix~\ref{sec:LQCDtables},
Tables~\ref{tab:latticebibfirst}--\ref{tab:latticebiblast}.
We characterize the results according to the criteria 
described above, and provide a prescription to combine those results that 
satisfy the criteria into a single benchmark value.

Our criteria and the corresponding ratings are chosen to provide as fair an assessment of the relative merits of various calculations as possible.
Where lattice-QCD results do not meet these standards, we hope that the lattice community will work towards improved calculations and greater precision.
Modifications to this rating system will occur as the lattice-QCD results evolve.

\paragraph{Unpolarized parton distributions.}
We summarize the current status of lattice-QCD calculations of the benchmark 
moments of unpolarized PDFs listed in Sec.~\ref{subsubsec:BQ} in 
Table~\ref{tab:unpolLQCDstatus1}. 
We indicate: the computed moment in the first column; the collaboration who
performed the computation in the second column; the corresponding reference
in the third column; the number of sea quark flavors, $N_f$, in the fourth 
column.
We show whether the calculation has been published~(P) 
or has appeared as a preprint~(PreP) in the fifth column.
In the following five columns, we assess each source of systematic uncertainty
according to the criteria listed above. 
In the last column, we report the computed value at $\mu^2=4\mbox{ GeV}^2$
in the $\overline{{\rm MS}}$ scheme.
We refer the reader to the corresponding references for details on the 
meaning of the errors reported in parentheses.
We do not list results that have not been extrapolated to the physical pion 
mass, nor do we include quenched results in Table~\ref{tab:unpolLQCDstatus1}. 
For completeness, we report these results in  Appendix~\ref{sec:LQCDtables},
Table~\ref{tab:unpolLQCDstatus1B}.

\begin{table}[!t] 
\renewcommand{\arraystretch}{1.2} 
\centering 
\begin{threeparttable}
\begin{tabular}{llcllccccccl}
\toprule
Mom. & Collab. & Ref. & $N_f$ & Status & 
Disc &
QM &
FV &
Ren &
ES &
& Value\\
\midrule
$\langle x\rangle_{u^+-d^+}$ 
& LHPC\,14  
  & \cite{Green:2012ud} 
  & 2+1 
  & P  
  & \rsquare 
  & \bstar   
  & \bstar   
  & \bstar 
  & \bstar 
  & 
  & 0.140(21)\\
& ETMC 17  
  & \cite{Alexandrou:2017oeh} 
  & 2   
  & P
  & \rsquare 
  & \bstar   
  & \rsquare 
  & \bstar 
  & \bstar 
  & $^*$ 
  & 0.194(9)(11)\\
& RQCD 14  
  & \cite{Bali:2014gha} 
  & 2   
  & P  
  & \rsquare 
  & \rsquare 
  & \bcirc   
  & \bstar 
  & \bstar 
  & $^{**}$ 
  & 0.217(9)\\
\midrule
$\langle x\rangle_{u^+}$
&  ETMC 17  
  & \cite{Alexandrou:2017oeh} 
  & 2 
  & P
  & \rsquare 
  & \bstar   
  & \rsquare 
  & \bstar 
  & \bstar 
  & $^{*\triangleright}$ 
  & $0.453(57)(48)$\\
\midrule
$\langle x\rangle_{d^+}$
& ETMC 17  
  & \cite{Alexandrou:2017oeh} 
  & 2 
  & P
  & \rsquare 
  & \bstar   
  & \rsquare 
  & \bstar 
  & \bstar 
  & $^{*\triangleright}$ 
  & $0.259(57)(47)$\\
\midrule
$\langle x\rangle_{s^+}$
& ETMC 17  
  & \cite{Alexandrou:2017oeh} 
  & 2 
  & P
  & \rsquare  
  & \bstar   
  & \rsquare 
  & \bstar 
  & \bstar 
  & $^{*\triangleright}$ & $0.092(41)(0)$\\
\midrule
$\langle x\rangle_{g}$
& ETMC 17  
  & \cite{Alexandrou:2017oeh} 
  & 2 
  & P 
  & \rsquare 
  & \bstar   
  & \rsquare 
  & \bcirc 
  & \bstar 
  & $^*$ 
  & 0.267(22)(27)\\
\bottomrule
\end{tabular}
\begin{tablenotes}
\footnotesize
\item[$\ \,*$] Study employing a single physical pion mass ensemble.
\item[$**$] Study employing a single ensemble with $m_\pi=150$~MeV.
\item[$\ \,\triangleright$] Nonsinglet renormalization is applied.
\end{tablenotes}
\end{threeparttable}
\caption{\small Status of current lattice-QCD calculations of the benchmark 
first moments of unpolarized PDFs listed in Sec.~\ref{subsubsec:BQ}.
A detailed description of each entry, including the symbols used to 
characterize the various sources of systematics, is provided in the text.
Values are shown at $\mu^2=4\mbox{ GeV}^2$.
We refer the reader to the corresponding references for details on the 
errors reported in parentheses.
To denote the various sources of systematic uncertainty, 
we use the abbreviations Disc (discretization),
QM (quark mass), FV (finite volume),
Ren (renormalization) and ES (excited states).
}
\label{tab:unpolLQCDstatus1}
\end{table}

As is apparent from Table~\ref{tab:unpolLQCDstatus1}, there are no lattice 
calculations of the considered first moments for which all systematics 
have been fully explored and controlled.  
In the case of $\langle x\rangle_{u^+-d^+}$ three different results are available 
in the literature.
We present the lattice-QCD benchmark value for this quantity 
as a best-estimate band.
This band extends from the mean of the smallest result minus its error 
to the mean of the largest result plus its error, and includes all results 
listed in Table~\ref{tab:unpolLQCDstatus1} with two or more sea 
quark flavors.
Current studies are not sufficiently precise to distinguish between 
results with different numbers of sea quark flavors.
In the case of $\langle x \rangle_{u^+}$, $\langle x \rangle_{d^+}$, 
$\langle x \rangle_{s^+}$ and $\langle x \rangle_g$, there is only one
lattice result available in the literature:
for these quantities, our lattice-QCD benchmark value is the single result; 
however, it should be noted that these results may underestimate some sources 
of uncertainty. 

The lattice-QCD benchmark numbers for $\langle x\rangle_{u^+-d^+}$,
$\langle x \rangle_{u^+}$, $\langle x \rangle_{d^+}$, 
$\langle x \rangle_{s^+}$ and $\langle x \rangle_g$ will be further
commented below, where they will be collected together with their 
global-fit counterparts in Table~\ref{tab:BMunp}.

Finally, we summarize the current status of lattice-QCD calculations of the 
second moment of the unpolarized valence-quark PDFs, 
$\langle x^2 \rangle_{u^-}$, $\langle x^2 \rangle_{d^-}$ and 
$\langle x^2\rangle_{u^--d^-}$ in Appendix~\ref{sec:LQCDtables},
Table~\ref{tab:unpolLQCDstatus2B}.
The study of these moments is not sufficiently mature to provide benchmark 
values and we only list the results for completeness.

\paragraph{Polarized parton distributions.}
The zeroth moment of the isotriplet polarized PDF combination is related to the 
axial charge of the nucleon, $g_A\equiv \langle 1\rangle_{\Delta u^+-\Delta d^+}$.
This quantity is of central importance to nucleon physics and has long been 
considered an important benchmark for lattice calculations. 
Historically, lattice-QCD calculations of the axial charge have underestimated 
the experimental value $g_A^{\mathrm{exp}} = 1.2723(23)$~\cite{Olive:2016xmw}
(see also Eq.~\eqref{eq:a3}), 
which is most precisely determined from neutron weak decays. 
Thus, the axial charge has been the single most-studied moment in lattice QCD.
We summarize the current status of these calculations in 
Table~\ref{tab:gAstatus} using the same format as in 
Table~\ref{tab:unpolLQCDstatus1}.
All results are quoted at $\mu^2=4\mbox{ GeV}^2$.

\begin{table}[!t]
\renewcommand{\arraystretch}{1.2} 
\centering
\begin{threeparttable}
\begin{tabular}{llcllccccccl}
\toprule
Mom. & Collab. & Ref. & $N_f$ & Status &  
Disc &
QM &
FV &
Ren &
ES &
& Value \\
\midrule
$g_A$
& CalLat\,17 
  & \cite{Berkowitz:2017gql} 
  & 2+1+1 
  & PreP 
  & \rsquare 
  & \bstar  
  & \rsquare 
  & \bstar 
  & \bstar 
  & 
  & 1.278(21)(26) \\
& PNDME\,16  
  & \cite{Bhattacharya:2016zcn} 
  & 2+1+1 
  & P    
  & \bcirc   
  & \bstar  
  & \bcirc   
  & \bstar 
  & \bstar 
  & 
  & 1.195(33)(20)\\
& LHPC\,14    
  & \cite{Green:2012ud} 
  & 2+1 
  & P 
  & \rsquare 
  & \bstar 
  & \bstar 
  & \bstar  
  & \bstar & & 0.97(8)\\
& Mainz\,17   
  & \cite{Capitani:2017qpc} 
  & 2 
  & PreP 
  & \bstar 
  & \bcirc 
  & \bstar 
  & \bstar  
  & \bstar 
  & 
  & $1.278(68)({}^{+0}_{-0.087})$\\
& ETMC\,17    
  & \cite{Alexandrou:2017hac} 
  & 2 
  & P
  & \rsquare  
  & \bstar 
  & \rsquare  
  & \bstar  
  & \bstar 
  & $^*$ 
  & 1.212(33)(22)\\
& RQCD\,15    
  & \cite{Bali:2014nma} 
  & 2 
  & P 
  & \bcirc 
  & \bcirc  
  & \bcirc  
  & \bstar   
  & \bcirc 
  & $^\ddag$
  & 1.280(44)(46) \\
  & QCDSF\,14   
  & \cite{Horsley:2013ayv} 
  & 2 
  & P 
  & \bcirc 
  & \bcirc  
  & \bcirc  
  & \bstar  
  & \rsquare 
  & $^\ddag$
  & 1.29(5)(3) \\
\bottomrule
\end{tabular}
\begin{tablenotes}
\footnotesize
\item[$*$] Study employing a single physical pion mass ensemble.
\item[$^\ddag$] $g_A$ is determined via the ratio $g_A/f_\pi$, employing the 
physical value for $f_\pi$.
\end{tablenotes}
\end{threeparttable}
\caption{\small Same as Table~\ref{tab:unpolLQCDstatus1}, but for the axial 
coupling, $g_A\equiv \langle 1\rangle_{\Delta u^+-\Delta d^+}$. 
Studies with three or more red squares are omitted from this table.
Values are shown at $\mu^2=4\mbox{ GeV}^2$.
}
\label{tab:gAstatus}
\end{table}

As is apparent from Table~\ref{tab:gAstatus}, we consider 
only three calculations of $g_A$ to have all systematics
sufficiently controlled to obtain a blue circle or star.
One of them~\cite{Bhattacharya:2016zcn} is for $N_f=2+1+1$, while two of 
them~\cite{Capitani:2017qpc,Bali:2014nma} are for $N_f=2$.
In the former case, our benchmark value corresponds to the single calculation;
in the latter case, our benchmark value corresponds to a weighted average 
of \cite{Capitani:2017qpc} and \cite{Bali:2014nma}, assuming correlations
between the results, and applying the procedure of \cite{Schmelling:1994pz}.
In summary, our benchmark values are
\begin{equation}\label{eq:gAcriteria}
g_A^{N_f=2+1+1} = 1.195(33)(20)
\,,\qquad \mathrm{and}\qquad 
g_A^{N_f=2} = 1.279(50)\,.
\end{equation}

We observe that the result of~\cite{Berkowitz:2017gql}, although it does
not fulfill all our requirements on systematic uncertainties, uses the same 
gauge configurations as those of \cite{Bhattacharya:2016zcn}.
Therefore, we also carry out a simultaneous fit to the two results for
completeness.
We use a fit function of the form
\begin{eqnarray}
g_A^{\mathrm{fit}}
&=&
c_0 +
f(a) +
c_3M_\pi^2 +
c_4M_\pi^2 \exp(-M_\pi L) +
c_5M_\pi^2 \log\left(\frac{M_\pi^2}{\Lambda_{\chi \mathrm{PT}}^2}\right)\,,
\\
f(a) &=&
\begin{cases}
  c_1a   & \qquad \text{Ref.~\cite{Bhattacharya:2016zcn}} \\
  c_2a^2 & \qquad \text{Ref.~\cite{Berkowitz:2017gql}}\\
\end{cases}
\qquad .
\end{eqnarray}
The coefficient $c_1$ captures ${\cal O}(a)$ effects present in the
valence-quark action of~\cite{Bhattacharya:2016zcn}, while~\cite{Berkowitz:2017gql} 
has discretization effects starting at ${\cal O}(a^2)$. 
The term proportional to $c_4$ captures the leading finite-volume effects, and 
$c_3$ and $c_5$ represent chiral-extrapolation terms. 
Modifications to this fit form, including setting $c_5=0$, have a negligible 
effect on the fit results within extrapolation uncertainties, and the final 
result is in very good agreement with a weighted average of the two 
calculations, assuming 100\% correlations, which is 
$g_A^{N_f=2+1+1,\mathrm{avg}} = 1.243(36)$. 
Based on this fit, we find the best-estimate band of
\begin{equation}\label{eq:gAfit}
g_A^{N_f=2+1+1,\mathrm{fit}} = \numrange{1.22}{1.28}\,.
\end{equation}

\begin{figure}[!t]
\centering
\includegraphics[scale=0.7]{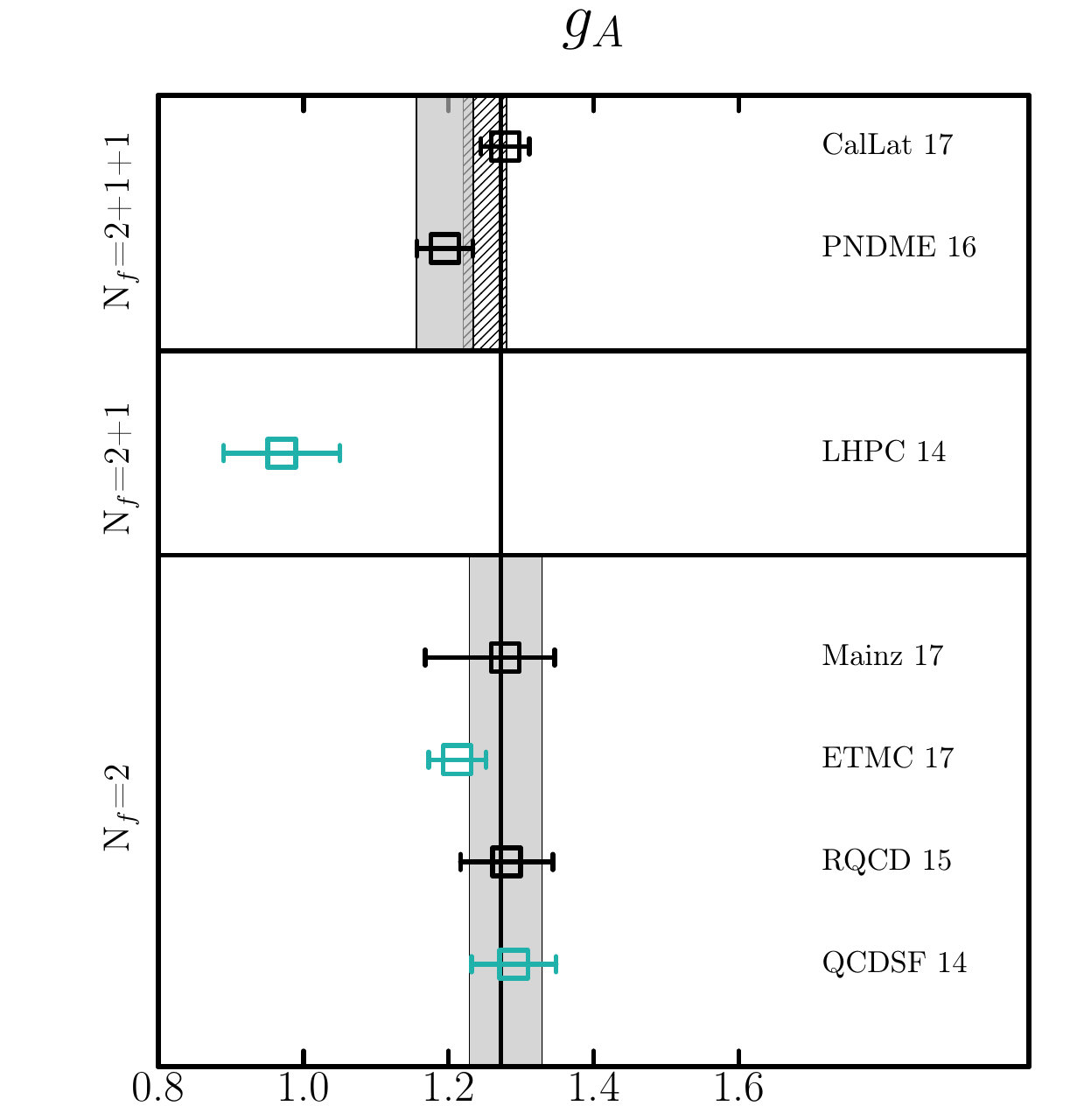}\\
\caption{\small Summary of the current status of lattice-QCD calculations of 
the axial charge, $g_A\equiv \langle 1\rangle_{\Delta u^+-\Delta d^+}$.
The vertical black line represents the current experimental world average 
$g_A^{\mathrm{exp}} = 1.2723(23)$~\cite{Olive:2016xmw}. 
The light gray bands for $N_f=2+1+1$ and $N_f=2$ represent the benchmark 
results of Eq.~\eqref{eq:gAcriteria}, and the dashed gray band for
$N_f=2+1+1$ is the fit band of Eq.~\eqref{eq:gAfit}.}    
\label{fig:gaLQCDstatus}
\end{figure}

We plot all lattice results for the axial coupling, listed in 
Table~\ref{tab:gAstatus}, in Fig.~\ref{fig:gaLQCDstatus}. 
We show the world-average experimental value as a vertical black line. 
The light gray bands for $N_f=2+1+1$ and $N_f=2$ represent the benchmark 
results of Eq.~\eqref{eq:gAcriteria}, and the dashed gray band for
$N_f=2+1+1$ is the combined fit band given in Eq.~\eqref{eq:gAfit}. 

In addition to the axial charge, we summarize the zeroth moments of the 
individual light-quark total polarized distributions in 
Table~\ref{tab:polLQCDstatus0}. 
We summarize the status of lattice-QCD calculations of the
first moments of the polarized PDF combination 
$\langle x \rangle_{\Delta u^- - \Delta d^-}$ in Table~\ref{tab:polLQCDstatus1}. 
We use the same format as in Table~\ref{tab:unpolLQCDstatus1}.
All values are at $\mu^2=4\mbox{ GeV}^2$.
Available results that have not been extrapolated to the physical pion mass
or quenched results are not reported here, but in Appendix~\ref{sec:LQCDtables},
Tables~\ref{tab:polLQCDstatus1B}--\ref{tab:polLQCDstatus2B}, for completeness.

In the case of $\langle 1 \rangle_{\Delta u^+}$ and $\langle 1 \rangle_{\Delta d^+}$,
there is only one result available in the literature for each quantity.
Therefore, although the corresponding systematic uncertainties are not 
completely under control and possibly underestimated, we take the individual 
results as our benchmark values.
In the case of $\langle 1 \rangle_{\Delta s^+}$ and 
$\langle x \rangle_{\Delta u^- - \Delta d^-}$, however, several results are available
in the literature, although without a full characterization of
their systematic uncertainties.
We present our lattice-QCD benchmark value for these quantities as
a best-estimate band extending from the mean minus the error of the 
smallest result to the mean plus the error of the largest. 
We include all results with two or more flavors of sea quarks listed in 
Tables~\ref{tab:polLQCDstatus0} and \ref{tab:polLQCDstatus1}, respectively.

The lattice-QCD benchmark numbers for $g_A$,
$\langle 1 \rangle_{\Delta u^+}$, $\langle 1 \rangle_{\Delta d^+}$,
$\langle 1 \rangle_{\Delta s^+}$ and $\langle x \rangle_{\Delta u^- - \Delta d^-}$
will be further commented below, where they will be collected together 
with their global-fit counterparts in Table~\ref{tab:BMpol}.

\begin{table}[!t]
\renewcommand{\arraystretch}{1.2} 
\centering
\begin{threeparttable}
\begin{tabular}{llcllccccccl}
\toprule
Mom. & Collab. & Ref. & $N_f$ & Status &
Disc &
QM &
FV &
Ren &
ES &
& Value \\
\midrule
$\langle 1\rangle_{\Delta u^+}$
& ETMC\,17 
  & \cite{Alexandrou:2017oeh} 
  & 2 
  & P
  & \rsquare 
  & \bstar 
  & \rsquare 
  & \bstar 
  & \bstar 
  & $^*$ 
  & $0.830(26)(4)$\\
\midrule
$\langle 1\rangle_{\Delta d^+}$
& ETMC\,17  
  & \cite{Alexandrou:2017oeh} 
  & 2 
  & P
  & \rsquare 
  & \bstar 
  & \rsquare  
  & \bstar 
  & \bstar 
  & $^*$ 
  & $-0.386(16)(6)$\\
\midrule
$\langle 1\rangle_{\Delta s^+}$
& $\chi$QCD\,17 
  & \cite{Gong:2015iir} 
  & 2+1 
  & P 
  & \rsquare  
  & \bcirc 
  & \bcirc  
  & \bstar 
  & \bstar
  & $^{\dagger,\triangleleft}$ 
  & $-0.0403(44)(78)$\\
& Engelhardt\,12 
  & \cite{Engelhardt:2012gd} 
  & 2+1 
  & P 
  & \rsquare  
  & \rsquare 
  & \bcirc  
  & \bstar  
  & \bstar  
  & $^\triangleleft$ 
  & $-0.031(17)$\\
& ETMC\,17 
  & \cite{Alexandrou:2017oeh} 
  & 2 
  & P
  & \rsquare  
  & \bstar 
  & \rsquare  
  & \bstar  
  & \bstar 
  & $^*$ 
  & $-0.042(10)(2)$\\
\bottomrule
\end{tabular}
\begin{tablenotes}
\footnotesize
\item[$*$] Study employing a single physical pion mass ensemble.
\item[$\dagger$] Partially quenched simulation with $m_\pi=330$~MeV. 
Criteria applied to the valence quarks. 
\item[$\triangleleft$] Some parts of the renormalization are estimated, 
see references for details.
\end{tablenotes}
\end{threeparttable}
\caption{\small Same as Table~\ref{tab:unpolLQCDstatus1}, but for the 
zeroth moments of the polarized total quark distributions.
Values are shown at $\mu^2=4\mbox{ GeV}^2$.
}
\label{tab:polLQCDstatus0}
\end{table}

\begin{table}[!t] 
\renewcommand{\arraystretch}{1.2}
\centering
\begin{threeparttable}
\begin{tabular}{llcllccccccl}
\toprule
Mom. & Collab. & Ref. & $N_f$ & Status &
Disc &
QM &
FV &
Ren &
ES &
& Value \\
\midrule
$\langle x\rangle_{\Delta u^--\Delta d^-}$
& RBC/ 
  & \multirow{2}{*}{\cite{Aoki:2010xg}} 
  & \multirow{2}{*}{2+1} 
  & \multirow{2}{*}{P} 
  & \multirow{2}{*}{\rsquare}  
  & \multirow{2}{*}{\rsquare} 
  & \multirow{2}{*}{\bstar}  
  & \multirow{2}{*}{\bstar}  
  & \multirow{2}{*}{\rsquare} 
  &  
  & 0.256(23)/\\
& UKQCD\,10 
  &  
  &  
  &  
  &   
  &  
  &   
  &   
  &  
  &  
  & 0.205(59)\\
& LHPC\,10 
  & \cite{Bratt:2010jn} 
  & 2+1 
  & P 
  & \rsquare  
  & \rsquare 
  & \bcirc  
  & \bcirc  
  & \rsquare 
  &  
  & 0.1972(55)\\
& ETMC\,15 
  & \cite{Abdel-Rehim:2015owa} 
  & 2 
  & P 
  & \rsquare  
  & \bstar 
  & \rsquare  
  & \bstar  
  & \bstar 
  & $^*$ 
  & 0.229(33)\\
\bottomrule
\end{tabular}
\begin{tablenotes}
\footnotesize
\item[$*$] Study employing a single physical pion mass ensemble.
\end{tablenotes}
\end{threeparttable}
\caption{\small Same as Table~\ref{tab:unpolLQCDstatus1}, but for the 
first moment of the polarized valence-quark distribution.
Values are shown at $\mu^2=4\mbox{ GeV}^2$.
}
\label{tab:polLQCDstatus1}
\end{table}

\subsubsection{Appraising global-fit results}
\label{subsubsec:GPDFfits}

The current status of global PDF fit determinations and their 
uncertainties has been carefully assessed in dedicated reviews
recently~\cite{Forte:2013wc,Jimenez-Delgado:2013sma}, and further 
summarized in Sec.~\ref{sec:unpPDFs}. 
It is now recognized that PDF uncertainties receive various contributions: 
the measurement uncertainty propagated from the data, uncertainties associated 
with incompatible data sets, procedural uncertainties such as those related to 
the choice of the PDF parametrization, 
and the handling of systematic errors, among others.
As outlined in Sec.~\ref{sec:unpPDFs}, in principle all of these uncertainties 
can be accounted for with suitable methods, both in the Hessian and the 
MC frameworks.
In practice, there is a significant spread in the sophistication 
of these methods between unpolarized and polarized PDF fits.

In Sec.~\ref{sec:unpPDFs}, we also emphasized that there are additional 
theoretical uncertainties on PDFs associated with uncertainty in
the input values of the physical parameters used in the fit (such as the 
reference value of the strong coupling) and with missing higher-order
uncertainties (given that fits are usually performed with fixed-order
perturbation theory).
The size of the former can be accounted for by studying the stability of the 
results upon variation of the input parameters; the size of the latter is
currently unknown, although it is supposed to be sub-dominant.
Therefore, theoretical uncertainties will not be considered in the following.

As far as full moments of PDFs are concerned, global-fit results involve
some degree of extrapolation to the region not covered by experimental data, 
that is not necessarily well accounted for in the PDF error estimates.
Extrapolation is particularly delicate to small $x$ values in the case of 
polarized PDFs: opposite to unpolarized PDFs, the kinematic coverage is 
fairly limited (see Sec.~\ref{sec:polPDFs} and in particular 
Fig.~\ref{fig:kinEIC}) and there is no analog of the momentum sum rule,
Eq.~\eqref{eq:mom}, to further constrain the PDFs.
Extrapolation uncertainties are difficult to quantify, unless
one naively extrapolates uncertainty bands from the measured region.

We now summarize the results for our benchmark moments listed in 
Sec.~\ref{subsubsec:BQ}, based on current global-fit determinations of
unpolarized and polarized PDFs.
We specify how the
available results are combined into a single benchmark value.

\paragraph{Unpolarized parton distributions.}

We summarize the current status of global-fit results of the benchmark
moments of unpolarized PDFs listed in Sec.~\ref{subsubsec:BQ} 
in Table~\ref{tab:unpPDFmoms}.
In the first column we indicate the computed moment, and in the subsequent 
six columns the moment's value, obtained from the most recent available PDF 
determinations: NNPDF3.1~\cite{Ball:2017nwa},
CT14~\cite{Dulat:2015mca}, MMHT2014~\cite{Harland-Lang:2014zoa},
ABMP16~\cite{Alekhin:2017kpj} (with $N_f=4$ flavors), 
CJ15~\cite{Accardi:2016qay} and 
HERAPDF2.0~\cite{Abramowicz:2015mha} respectively.
The most relevant features of these PDF sets have been presented in 
Sec.~\ref{sec:unpPDFs}.
All values in Table~\ref{tab:unpPDFmoms} are displayed
at $\mu^2=4\mbox{ GeV}^2$. 
They have been obtained from the default PDF sets at the highest available 
perturbative order, which is NNLO for all of them except CJ15
for which it is NLO.
The uncertainties for the CT14 PDF set have been rescaled by a factor $1/1.65$ 
to convert from  90\%-CL bands to  68\%-CL bands.
Note that tolerance of $\Delta \chi^2=1$ at 68\% CL is used in the CJ15 PDF 
set; hence, the smaller uncertainties of this set compared to all the other 
PDF sets.
Also, the CJ15 set does  not fit $\langle x \rangle_{s^+}$, therefore the 
corresponding number is not displayed in Table~\ref{tab:unpPDFmoms}. 
In the case of the HERAPDF2.0 set, the error band is the sum in quadrature 
of the statistical, model and parametrization uncertainties.
Taking the results of Table~\ref{tab:unpPDFmoms} at face value,
there are clear discrepancies arising from a variety of 
factors~\cite{Butterworth:2015oua,Accardi:2016ndt};
we examine some of these in the following. 

\begin{table}[!t]
\centering
\renewcommand{\arraystretch}{1.2}
\begin{tabular}{lcccccc}
\toprule
Mom. 
& NNPDF3.1 & CT14 & MMHT2014 & ABMP2016 & CJ15 & HERAPDF2.0 \\
\midrule
$\langle x \rangle_{u^+-d^+}$ 
& 0.152(3) & 0.158(4) & 0.151(4) & 0.167(4) & 0.152(2) & 0.188(3)\ \,\\
$\langle x \rangle_{u^+}$    
& 0.348(4) & 0.348(3) & 0.348(5) & 0.353(3) & 0.348(1) & 0.372(4)\ \,\\
$\langle x \rangle_{d^+}$    
& 0.196(3) & 0.190(3) & 0.197(5) & 0.186(3) & 0.196(1) & 0.185(7)\ \,\\
$\langle x \rangle_{s^+}$    
& 0.039(3) & 0.035(5) & 0.035(9) & 0.041(2) & ---   & 0.035(11)\\
$\langle x \rangle_{g}$     
& 0.410(4) & 0.416(5) & 0.411(9) & 0.412(4) & 0.416(1) & 0.401(10)\\
\bottomrule
\end{tabular}
\caption{\small Status of current global PDF fit determinations of the 
benchmark moments of unpolarized PDFs listed in Sec.~\ref{subsubsec:BQ}.
All values are shown at $\mu^2=4\mbox{ GeV}^2$.
See text for details about the calculation of PDF uncertainties in each case.
}
\label{tab:unpPDFmoms}
\end{table}

In order to provide a benchmark value for the first moments of unpolarized PDFs
listed in Table~\ref{tab:unpPDFmoms}, we follow the latest PDF4LHC 2015 
recommendations~\cite{Butterworth:2015oua}.
Even though the recommendations were primarily formulated for the usage of PDFs
in LHC-related physics, and alternative recommendations have been 
suggested~\cite{Accardi:2016ndt}, we find it useful to apply them here as well.
The reason is twofold.
First, this benchmark exercise aims at accuracy and precision,  
two of the guiding principles underlying the recommendations.
Second, they led to the release of a specific PDF set
that can be easily used to compute all the needed benchmark values.

While we refer the reader to \cite{Butterworth:2015oua} for details,
here we only mention that the PDF4LHC15 PDF set was constructed by means of
a statistical combination~\cite{Carrazza:2015hva,Gao:2013bia,Watt:2012tq,
Carrazza:2015aoa} (an unweighted average) of the 
NNPDF3.0~\cite{Ball:2014uwa}, CT14 and MMHT2014 PDF sets.\footnote{The 
NNPDF3.1 PDF set was not available when the recommendations were formulated.}
The three PDF sets were selected among all the publicly available PDF sets
based on four criteria~\cite{Butterworth:2015oua}.
\begin{itemize}
\item A global data set from a wide variety of observables and processes
should be included in the fit analysis.
\item Theoretical hard cross-sections should be evaluated up to NNLO in a
general-mass variable-flavor number scheme with up to $N_f^\text{max}=5$ 
active quark flavors.
\item The central value of the strong coupling at the $Z$-boson mass,
$\alpha_s(M_Z^2)$ should be fixed at an agreed common value, consistent 
with the PDG world-average~\cite{Olive:2016xmw} ($\alpha_s(M_Z)=0.118$).
\item All known experimental and procedural sources of uncertainty should be 
properly accounted for.
\end{itemize}
The ABMP2016 set (as well as its previous versions) does not meet the second 
and third criteria; the CJ15 set does not meet the first, second and fourth
criteria, while the HERAPDF2.0 set does not meet the first criterion.
Hence, these sets were not included in the PDF4LHC2015 PDF set, although the 
possibility of including them in future versions of the recommendation 
remains open.

In order not to lose important information contained in the PDF sets excluded 
from the PDF4LHC recommendations, we also provide alternative benchmark numbers.
Specifically, we combined all the numbers quoted in Table~\ref{tab:unpPDFmoms}
so that the mean value is an unweighted average of the mean 
values and the error is half of the difference between the smallest and the 
largest result.
The rationale for this choice is that PDF sets entering the PDF4LHC 
recommendations are not benchmarked in the $x\gtrsim 0.1$ region, which can be 
relevant for the moment analysis.
The combination of all results in Table~\ref{tab:unpPDFmoms}, although 
sometimes less precise than the PDF4LHC combination, maximizes the amount of 
experimental information included in the benchmark numbers.
Specifically, it includes the information taken into account 
at large $x$ and small $Q^2$ in the CJ15 and ABMP16 PDF sets, 
which is otherwise excluded from the PDF4LHC set.

The global-fit benchmark numbers for $\langle x\rangle_{u^+-d^+}$,
$\langle x \rangle_{u^+}$, $\langle x \rangle_{d^+}$, 
$\langle x \rangle_{s^+}$ and $\langle x \rangle_g$ will be further
commented below, where they will be collected together with their 
lattice-QCD counterparts in Table~\ref{tab:BMunp}.

\paragraph{Polarized parton distributions.}
We summarize the current status of global-fit results of the benchmark
moments of polarized PDFs listed in Sec.~\ref{subsubsec:BQ} in 
Table~\ref{tab:polPDFmoms}.
In the first column, we indicate the computed moment, and in the subsequent 
three columns, its value as obtained from the most recent available PDF 
determinations: NNPDFpol1.1~\cite{Nocera:2014gqa}, 
DSSV08~\cite{deFlorian:2009vb}~\footnote{The DSSV08 analysis has been updated
by the DSSV14 analysis~\cite{deFlorian:2014yva} essentially 
only in the determination of the gluon PDF. 
The moments in Table~\ref{tab:polPDFmoms} therefore hardly differ
in the two analyses.}, JAM15~\cite{Sato:2016tuz} and 
JAM17~\cite{Ethier:2017zbq}.
The most relevant features of these PDF sets have been presented in
Sec.~\ref{sec:polPDFs}.
All values in Table~\ref{tab:unpPDFmoms} are displayed
at $\mu^2=4\mbox{ GeV}^2$ at NLO.
The uncertainties correspond to 68\%-CL bands with tolerance of 
$\Delta \chi^2=1$ for the DSSV08 PDF set.
In the case of the JAM15 set, we do not provide a value for 
$\langle x \rangle _{\Delta u^--\Delta d^-}$:
the fit is based on inclusive DIS data only, which are not sensitive to 
the valence distribution $\Delta u^- - \Delta d^-$.
We emphasize that, because of extrapolation uncertainties difficult to quantify,
the error estimates in Table~\ref{tab:polPDFmoms} should be interpreted
as a lower bound, especially for the DSSV08 and JAM sets based on 
conventional parametrizations.
In these cases, uncertainty bands are naively extrapolated from the measured 
kinematic region, therefore they are likely to underestimate the contribution 
coming from the small-$x$ region.

\begin{table}[!t]
\centering
\renewcommand{\arraystretch}{1.2}
\begin{tabular}{lcccc}
\toprule
Mom. 
& NNPDFpol1.1 & DSSV08 & JAM15 & JAM17 \\
\midrule
$\langle 1 \rangle_{\Delta u^+-\Delta d^+}$ &
$1.250(16)$ & $1.260(18)$ & $1.314(6)$  & $1.240(41)$\\
$\langle 1 \rangle_{\Delta u^+}$ &
$0.794(46)$ & $0.814(12)$ & $0.831(21)$ & $0.812(22)$\\
$\langle 1 \rangle_{\Delta d^+}$ &  
$-0.453(52)$  &  $-0.456(11)$ &  $-0.476(22)$ &  $-0.428(31)$\\
$\langle 1 \rangle_{\Delta s^+}$ &  
$-0.120(81)$  &  $-0.112(23)$ &  $-0.109(20)$ &  $-0.038(96)$\\
$\langle x \rangle_{\Delta u^- - \Delta d^-}$ &     
$0.195(14)$ &  $0.203(9)$ &  ---        & $0.241(26)$ \\
\bottomrule
\end{tabular}
\caption{\small Status of current global-fit determinations of the 
benchmark moments of polarized PDFs listed in Sec.~\ref{subsubsec:BQ}.
All values are shown at $\mu^2=4\mbox{ GeV}^2$.}
\label{tab:polPDFmoms}
\end{table}

As outlined in Sec.~\ref{sec:polPDFs}, polarized PDFs cannot be determined in 
a global QCD analysis with the same accuracy as their unpolarized counterparts.
Also, because polarized PDFs do not enter precision physics studies at the LHC, 
the selection and combination of different PDF sets has received much less
attention.
No recommendations analogous to those from the PDF4LHC working group
exist for polarized PDFs.

To provide a benchmark value for the relevant moments of 
polarized PDFs listed in Table~\ref{tab:polPDFmoms}, we apply an unweighted 
average of the NNPDFpol1.1, DSSV08 and JAM15 PDF sets.
The rationale for this choice is twofold.
On the one hand, we maximize the amount of experimental information 
that can determine the central value of our benchmark moments.
As explained in Sec.~\ref{sec:polPDFs}, the NNPDFpol1.1 and the DSSV08 PDF 
sets are based on a very similar set of inclusive DIS data, while the JAM15 
PDF set is based on a much wider inclusive DIS data set.
This wider set can help constrain the moments of the total quark 
distributions.
The NNPDFpol1.1 and the DSSV08 PDF sets are based respectively on $pp$ and 
SIDIS data to disentangle the quark and antiquark distributions.
This can help constrain the moments of the valence distributions.
On the other hand, we balance the rather different uncertainties among the 
three PDF sets, specifically the larger NNPDFpol1.1 estimate
against the smaller DSSV08 and JAM15 values.
This way, we avoid a possible underestimation of the procedural uncertainties 
induced for example by the choice of a simple PDF parametrization 
in the DSSV08 and JAM15 fits, or by the extrapolation to the small-$x$ region.
Because the JAM17 set is unique in fitting simultaneously polarized PDFs and 
FFs, we do not include it in our benchmark average, but quote it as a useful 
comparison.

The global-fit benchmark numbers for $g_A$,
$\langle 1 \rangle_{\Delta u^+}$, $\langle 1 \rangle_{\Delta d^+}$,
$\langle 1 \rangle_{\Delta s^+}$ and $\langle x \rangle_{\Delta u^- - \Delta d^-}$
will be further commented below, where they will be collected
with their lattice-QCD counterparts in Table~\ref{tab:BMpol}.

\subsection{Comparing lattice-QCD and global-fit benchmark moments}
\label{subsec:BN}

We can now compare the lattice-QCD and global PDF fit results presented in 
Secs.~\ref{subsubsec:BClQCD}--\ref{subsubsec:GPDFfits} for the unpolarized
and polarized PDF moments respectively.

\begin{figure}[!t]
\centering
\includegraphics[scale=0.44,angle=270]{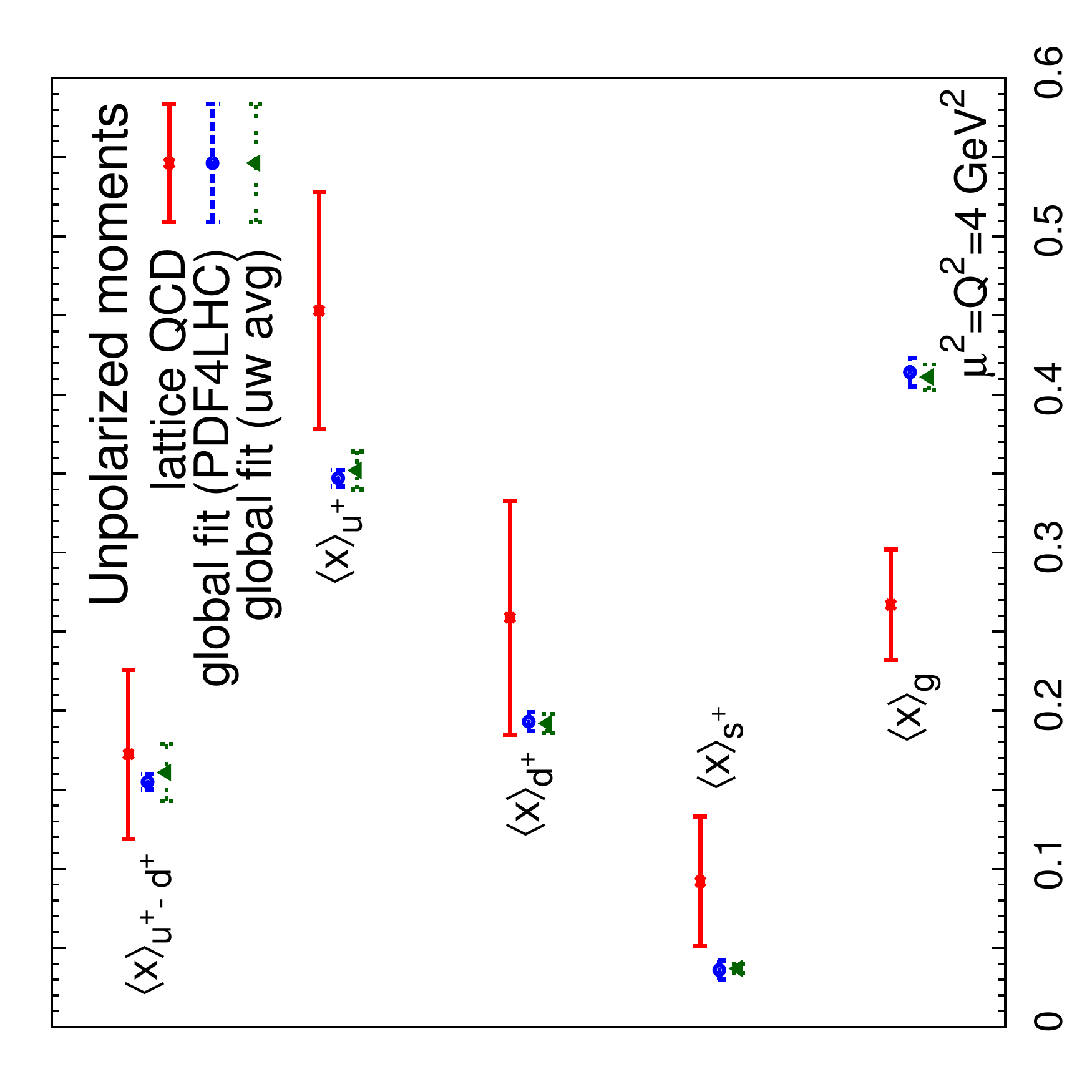}
\includegraphics[scale=0.44,angle=270]{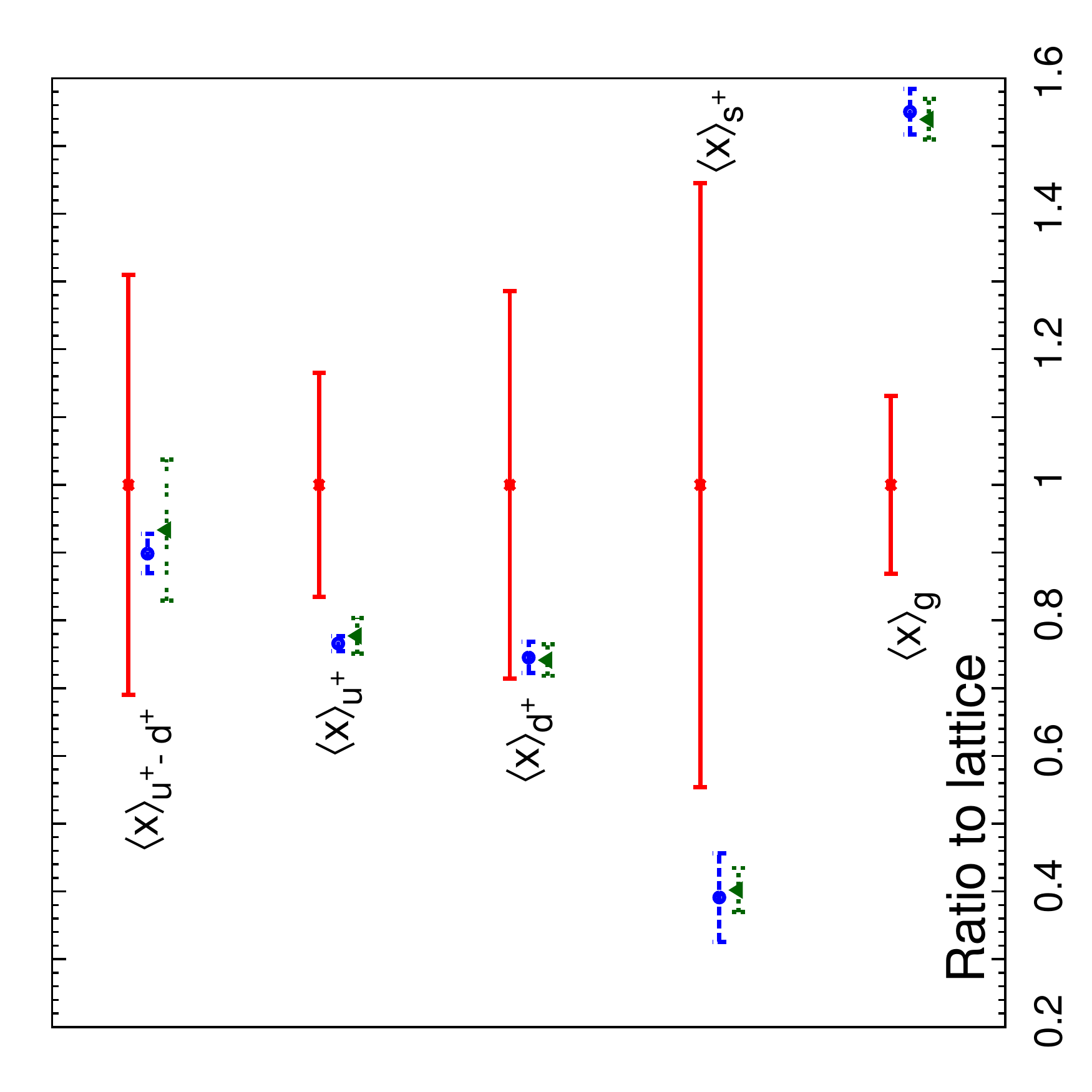}\\
\caption{\small A comparison of the unpolarized PDF benchmark moments 
between the lattice QCD computations and global fit determinations.
Results are displayed both in terms of absolute values (left) and ratios to
the lattice values (right) at $\mu^2=4$~GeV$^2$.}
\label{fig:Bmomsunp}
\end{figure}

\paragraph{Unpolarized parton distributions.}
The benchmark values of the first moments of the unpolarized PDFs, obtained
as described in Secs.~\ref{subsubsec:BClQCD}--\ref{subsubsec:GPDFfits}, 
are summarized in Table~\ref{tab:BMunp}.
Both the PDF4LHC and the unweighted average (uw avg) are displayed in the case 
of global fits.
The results from a single lattice calculation, which might underestimate some 
sources of uncertainty, are denoted with a superscript~$\dagger$.
All values shown here are at $\mu^2=4\mbox{ GeV}^2$.
For ease of comparison, these benchmark results are also graphically
compared in Fig.~\ref{fig:Bmomsunp}, both in terms of absolute values 
(left panel) and of uncorrelated ratios to the lattice central values 
(right panel).

\begin{table}[!t]
\centering
\renewcommand{\arraystretch}{1.2}
\begin{tabular}{lccc}
\toprule
Moment & Lattice QCD & Global Fit (PDF4LHC) & Global fit (uw avg)\\
\midrule
$\langle x \rangle_{u^+ -d^+}$ 
& \numrange{0.119}{0.226} 
& 0.155(5)
& \, 0.161(18)\\
$\langle x \rangle_{u^+}$     
& 0.453(75)$^\dagger$ 
& 0.347(5)
& \, 0.352(12)\\
$\langle x \rangle_{d^+}$     
& 0.259(74)$^\dagger$ 
& 0.193(6)
& 0.192(6)\\
$\langle x \rangle_{s^+}$     
& 0.092(41)$^\dagger$ 
& 0.036(6)
& 0.037(3)\\
$\langle x\rangle_{g}$       
& 0.267(35)$^\dagger$ 
& 0.414(9)
& 0.411(8)\\
\bottomrule
\end{tabular}
\caption{\small Benchmark values for lattice-QCD calculations and global-fit 
determinations of the benchmark moments of unpolarized PDFs.
All values are shown at $\mu^2=4\mbox{ GeV}^2$.
Results with a superscript~$\dagger$ are from a single lattice 
calculation; they may underestimate some sources of uncertainty.}
\label{tab:BMunp}
\end{table}

As is apparent from Table~\ref{tab:BMunp} and Fig.~\ref{fig:Bmomsunp}, there is 
a significant difference in the uncertainties between the lattice QCD and 
global fit results, with the latter always about one order of magnitude 
smaller than the former.
Moreover, even within their large uncertainties, the lattice-QCD results 
for the first moments of the total up and strange quark and the gluon PDFs
are not compatible with their global-fit counterparts.
In the case of quarks, the discrepancy is below $2\sigma$ (in units of the 
lattice-QCD uncertainty), while in the case of the gluon the discrepancy is
slightly larger than $3\sigma$.

On the lattice-QCD side, we note that in the flavor-singlet sector calculations
neglected part of the renormalization and computed some other parts only 
perturbatively.
Most of the discrepancies between lattice-QCD and global-fit results are 
observed in the flavor-singlet sector.
Progress in taking into account the renormalization properly
could shift lattice-QCD results significantly, and reconcile them 
with the global fits in the future.
We also note that the momentum sum rule, Eq.~\eqref{eq:mom}, usually is not 
imposed in lattice-QCD calculations.
In the ETMC\,17 analysis~\cite{Alexandrou:2017oeh}, it turns
out to be $1.071(93)(72)$, see Table~\ref{tab:unpolLQCDstatus1}, if 
uncertainties are assumed to be uncorrelated.
Although there is no evidence for a violation of the momentum sum rule 
based on this result, one must be careful combining results from different 
calculations to account for correlations and other sources of error. 
Also, note that the ETMC\,17 analysis is performed with $N_f=2$ flavors,
hence the strange quark should not participate in the sum rule.

On the global-fit side, we note that the amount of experimental information 
that constrains the total up-quark distribution is the largest among all 
distributions.
Therefore, it seems unlikely that its global-fit central value could vary 
significantly in the future, and become compatible with the current
lattice result.
Conversely, the amount of experimental information that constrains the
total strange distribution in a global fit is less abundant and less accurate.
A slight shift in its central value, towards the current lattice-QCD results,
might be observed in the future, as soon as new data sensitive to the strange 
quark becomes available.
Finally, in an attempt to reconcile the lattice-QCD and the global-fit results
of the first moment of the gluon PDF, one could assume a completely
different behavior of the gluon PDF below the HERA kinematic
coverage, $x\sim 10^{-5}$ (see Fig.~\ref{fig:kinplot-report}).
While such a kinematic region remains completely unexplored,
in general the contribution of this region to the moments is negligible
and thus unlikely to resolve the situation. 

All these remarks apply irrespective of the benchmark value used for 
global fits, either the PDF4LHC or the unweighted average.
They also still hold if individual lattice-QCD and/or global-fit
results in Tables~\ref{tab:unpolLQCDstatus1}--\ref{tab:unpPDFmoms} are 
compared instead of their benchmark values in Table~\ref{tab:BMunp}. 
These results suggest that both greater accuracy and greater precision are
required in lattice-QCD calculations to match the accuracy and 
precision of the first moments of unpolarized PDFs determined from a global
fit.

\begin{figure}[!t]
\centering
\includegraphics[scale=0.44,angle=270]{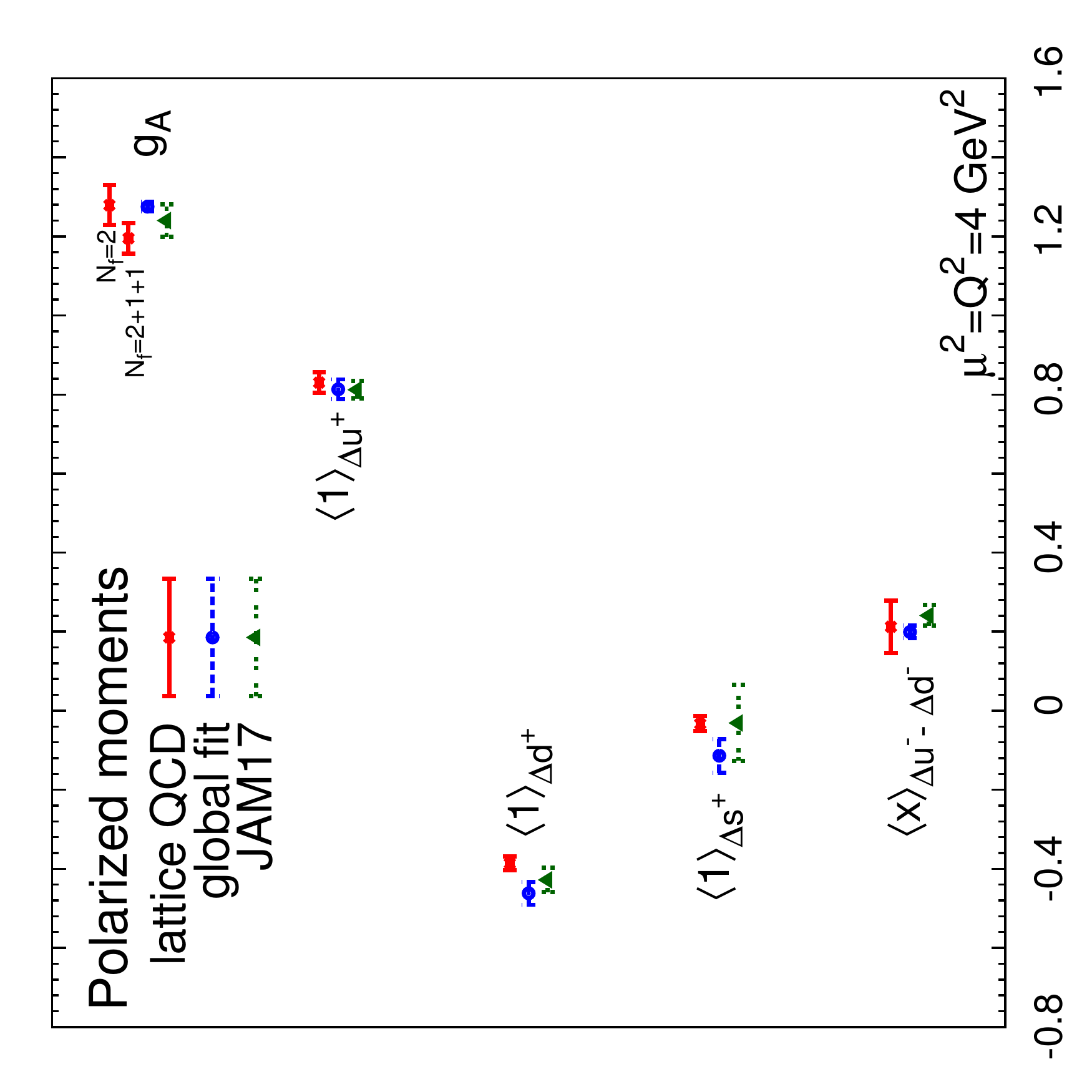}
\includegraphics[scale=0.44,angle=270]{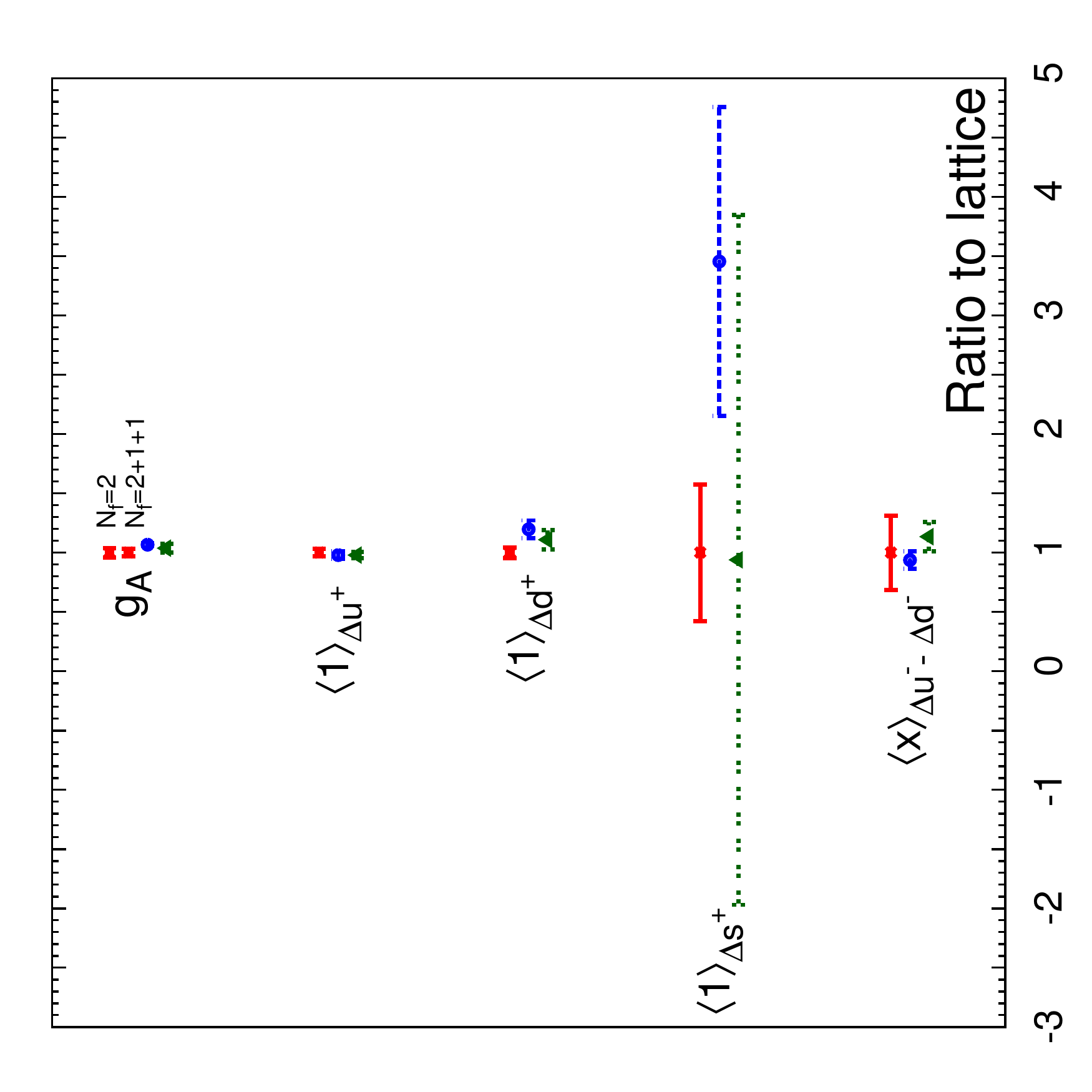}\\
\caption{\small Same as Fig.~\ref{fig:Bmomsunp}, but for the polarized
benchmark moments.}
\label{fig:Bmomspol}
\end{figure}

\paragraph{Polarized parton distributions.}
The benchmark values of the first moments of the unpolarized PDFs, obtained
as described in Secs.~\ref{subsubsec:BClQCD}--\ref{subsubsec:GPDFfits}, 
are summarized in Table~\ref{tab:BMpol}.
Results from a single lattice calculation, which might underestimate some 
sources of uncertainty, are denoted with a superscript~$\dagger$.
In the case of $g_A$, we report the two values with $N_f=2+1+1$ and
$N_f=2$ sea quarks from lattice QCD.
The value of $g_A$ is scale-independent, and we quote all other results at 
$\mu^2=4\mbox{ GeV}^2$.
For ease of comparison, these values are also displayed in 
Fig.~\ref{fig:Bmomspol} in the same format as in Fig.~\ref{fig:Bmomsunp}.
In the case of $g_A$, the result with $N_f=2+1+1$ is used as normalization
factor in the right panel of Fig.~\ref{fig:Bmomspol}.
Results from the JAM17 analysis~\cite{Ethier:2017zbq}, see 
Table~\ref{tab:polPDFmoms}, are displayed separately.
The reason for this is that, 
in contrast with the NNPDFpol1.1, DSSV08 and JAM15 fits, in the JAM17 fit 
the experimental value of $g_A$, Eq.~\eqref{eq:a3}, 
is not an input of the fit, but it is fitted alongside the PDFs.
Furthermore, in JAM17 PDFs are fitted alongside FFs.
\begin{table}[!t]
\centering
\renewcommand{\arraystretch}{1.2}
\begin{tabular}{lcc}
\toprule
Moment & Lattice QCD & Global Fit\\
\midrule
\multirow{2}{*}{$g_A\equiv\langle 1\rangle_{\Delta u^+ - \Delta d^+}$} 
& $1.195(39)$ ($N_f=2+1+1$) 
& \multirow{2}{*}{$1.275(12)$} \\
& $1.279(50)$ ($N_f=2$) & \\
$\langle 1 \rangle_{\Delta u^+}$     
& $0.830(26)^\dagger$ 
& $0.813(25)$\\
$\langle 1 \rangle_{\Delta d^+}$     
& $-0.386(17)^\dagger$ 
& $-0.462(29)$\\
$\langle 1 \rangle_{\Delta s^+}$     
& $-0.052$ -- $-0.014$
& $-0.114(43)$\\
$\langle x\rangle_{\Delta u^- - \Delta d^-}$       
& \numrange{0.146}{0.279} 
& $0.199(16)$\\
\bottomrule
\end{tabular}
\caption{\small Same as Table~\ref{tab:BMunp}, but for the polarized benchmark 
moments.}
\label{tab:BMpol}
\end{table}

As is apparent from Table~\ref{tab:BMpol} and Fig.~\ref{fig:Bmomspol}, the 
size of the uncertainties on the moments is in general comparable between the 
lattice-QCD and the global-fit results, opposite to the unpolarized 
case\footnote{Note that the uncertainty of 
$\langle 1 \rangle_{\Delta s^+}$ in the right panel of Fig.~\ref{fig:Bmomspol}
appears much larger than the uncertainty of other quark moments because
of the normalization value, which is very close to zero.}.
The corresponding central values are also in reasonable agreement within their
mutual uncertainties.

In the case of $g_A$, the global-fit result obtained from the unweighted 
average of the NNPDFpol1.1, DSSV08 and JAM15 fits shows a preference for the
lattice-QCD result obtained with $N_f=2$ sea quarks (compared to that with 
$N_f=2+1+1$ sea quarks).
Its uncertainty is, however, four times smaller than that of both lattice 
results.
This is not unexpected, since, in all the three fits that are combined, the 
experimental value of $g_A$ is imposed in the fits themselves.
The final uncertainty on the global-fit value of $g_A$ is thus reduced by 
the uncertainty of its experimental value $g_A^\text{exp}$, which is almost
one order of magnitude smaller than the uncertainty on the lattice-QCD results
(see Fig.~\ref{fig:gaLQCDstatus}).
If the experimental value of $g_A$ is not imposed as a boundary condition in 
the fit, as in the JAM17 analysis, the size of the uncertainty on $g_A$ is 
comparable to that of the lattice results, although it is not able to 
discriminate between the $N_f=2$ or the $N_f=2=1+1$ results.
Overall, this is a noteworthy confirmation of SU(2) symmetry in QCD to
almost 2\%.

In the case of the zeroth moments of the total polarized quark distributions,
the uncertainty on the lattice-QCD result is comparable to (in the case
of $\langle 1 \rangle_{\Delta u^+}$) or smaller than (in the case
of $\langle 1 \rangle_{\Delta d^+}$ and $\langle 1 \rangle_{\Delta s^+}$)
the uncertainty on the global-fit result.
However, in the case of the zeroth moments of the total down- and strange-quark 
distributions, the lattice-QCD and the global-fit results are discrepant
by about two $\sigma$ (in units of the
lattice QCD uncertainty).
On the one hand, we observe that the uncertainty on the lattice-QCD results 
might have been underestimated because of the lack of full control over
all systematics (see Sec.~\ref{subsubsec:BClQCD}).
On the other hand, we observe that the global-fit result has been obtained
by requiring SU(3) symmetry, {\it i.e.}, by imposing in the individual fits 
the experimental value (with a possibly inflated uncertainty) of the octet PDF 
combination, as explained in Sec.~\ref{sec:polPDFs}.
Relaxing this constraint can reconcile the discrepancy observed between 
the lattice-QCD and the global-fit result for the zeroth moments of the 
total down and strange PDFs.
This is demonstrated by comparison with the JAM17 result, whose uncertainty 
band nicely includes both the lattice-QCD and the global-fit benchmark values.

In the case of the first moment of the valence distribution 
$\Delta u^--\Delta d^-$, the lattice-QCD and the global-fit results are 
in excellent agreement, although the uncertainty of the former is five times
larger than that of the latter.

All these remarks still hold if individual lattice-QCD and/or global-fit
results in Tables~\ref{tab:gAstatus}--
\ref{tab:polPDFmoms}
are compared instead of their benchmark values in Table~\ref{tab:BMpol}.
These results suggest that lattice-QCD calculations could provide a useful
input to global fits of polarized PDFs, especially in limiting the
extrapolation uncertainty into the completely unknown small-$x$ region.
This will become more and more useful as full control over all sources of
systematic uncertainties is achieved.

\section{Improving PDF fits with lattice-QCD calculations}
\label{sec:projections}

In this section, we provide an estimate of the potential
impact of future lattice-QCD calculations
in global unpolarized and polarized PDF fits.
This study is carried out with two publicly available
tools: the Bayesian reweighting
method~\cite{Ball:2011gg,Ball:2010gb} applied to the
NNPDF3.1~\cite{Ball:2017nwa} and NNPDFpol1.1~\cite{Nocera:2014gqa} sets; 
and the Hessian profiling method~\cite{Camarda:2015zba} applied to
HERAPDF2.0 set~\cite{Abramowicz:2015mha}.
Both methods allow us to quantify the impact of new measurements
(or of future measurements, if pseudo-data are used) on PDFs without
repeating the global analysis.
The main limitation of these methods is that they are maximally reliable
if the amount of information carried in by the new (pseudo-)data is moderate
in comparison to that already included in the fit.

For simplicity, we limit our study to the impact of a subset of the moments 
that can be computed using lattice QCD, focusing on those that can be 
currently calculated with the highest precision.
Therefore, we restrict ourselves to the benchmark moments discussed in 
Sec.~\ref{sec:benchmarking}.
We also consider pseudo-data based on $x$-space
lattice-QCD calculations from the quasi-PDF approach
discussed in Sec.~\ref{sec:xdependence}.
As we show, particularly in the unpolarized case, the
constraining power of direct $x$-space calculations is
superior to that of PDF moments.

\subsection{Impact of lattice calculations of PDF moments}

We start by quantifying the constraining power of projected lattice-QCD 
calculations of PDF moments on both unpolarized and polarized global fits.
We define the settings for our study and present our results following
Bayesian reweighting and Hessian profiling, respectively.

\subsubsection{Analysis settings}
\label{sec:projections:settings}

In the unpolarized case,  we consider
the first moments (momentum fractions) of $q^+=q+\bar{q}$ (with $q=u,d,s$),
of the gluon, and of the isovector combination $u^+-d^+$.
In the polarized case, we consider the zeroth moments (spin fractions) 
of $\Delta q^+$ (with $q=u,d,s$) and of the isovector combination 
$g_A=\Delta u^+-\Delta d^+$, and the first moment of the 
$\Delta u^- - \Delta d^-$ combination.
Using the notation outlined in Appendix~\ref{app:notation}, we have
\begin{align}
\la x\ra_{u^+}\, , \
\la x\ra_{d^+}\, , \
\la x\ra_{s^+}\, , \
\la x\ra_{g}\, ,  \
\la x\ra_{u^+-d^+} & \qquad\text{for the unpolarized case} \, ,
\label{eq:UM}\\
\la 1\ra_{\Delta u^+}\, , \
\la 1\ra_{\Delta d^+}\, , \
\la 1\ra_{\Delta s^+}\, , \
\la x\ra_{\Delta u^--\Delta d^-}\, ,\
\la 1\ra_{\Delta u^+ - \Delta d^+} & \qquad\text{for the polarized case}\, .
\label{eq:PM}
\end{align}

We look at three different scenarios, which we denote
as Scenario~A, B, and C, for the projected total systematic
uncertainty associated with lattice-QCD calculations.
Our choice for this uncertainty is denoted by $\delta_L^{(i)}$.
It is summarized in Table~\ref{tab:scenarios} for each PDF moment $i$ in 
Eqs.~\eqref{eq:UM}--\eqref{eq:PM} and for each scenario.
Current uncertainties on lattice-QCD results 
(see Secs.~\ref{subsubsec:BClQCD}--\ref{subsec:BN})
are also quoted for comparison.
We emphasize that, while trying to be reasonably
realistic, we do not associate a given scenario
with a specific time scale for the calculation.
Our results provide a guide to the potential
constraining power of future lattice-QCD calculations
of PDF moments once included in global analyses. 
 
\begin{table}[!t]
\centering
\footnotesize
\renewcommand{\arraystretch}{1.3} 
\begin{tabular}{cccccc}
\toprule
Scenario &  \multicolumn{5}{c}{$\delta_L^{(i)}$ for unpolarized moments} \\
& $\la x\ra_{u^+}$ 
& $\la x\ra_{d^+}$ 
& $\la x\ra_{s^+}$  
& $\la x\ra_{g}$  
&   $\la x\ra_{u^+-d^+}$  \\
\midrule
Current  
& $\sim 16\%$  
& $\sim 30\%$ 
& $\sim 45\%$  
& $\sim 13\%$  
& $\sim 60\%$ \\
A   & 3\%  & 3\% &  5\% &  3\% &  5\% \\
B   & 2\%  & 2\% &  4\% &  2\% &  4\%  \\
C   & 1\%  & 1\% &  3\% &  1\% &  3\%  \\
\bottomrule
\\
\toprule
Scenario & \multicolumn{5}{c}{$\delta_L^{(i)}$ for polarized moments} \\ 
& $\la 1\ra_{\Delta u^+}$  
& $\la 1\ra_{\Delta d^+}$  
& $\la 1\ra_{\Delta s^+}$
& $\la x\ra_{\Delta u^--\Delta d^-}$  
& $\la 1\ra_{\Delta u^+ - \Delta d^+}$\\
\midrule
Current  
& $\sim 3\%$  
& $\sim 5\%$ 
& $\sim 70\%$ 
& $\sim 65\%$ 
& $\sim 3\%$ \\
A   & 5\% & 10\%  & 100\% & 70\%  & 5\% \\
B   & 3\% &  5\%  &  50\% & 30\%  & 3\% \\
C   & 1\% &  2\%  &  20\% & 15\%  & 1\% \\
\bottomrule
\end{tabular}
\caption{\small The three scenarios assumed for the total percentage
systematic uncertainty $\delta_L^{(i)}$ in future lattice-QCD calculations.
The unpolarized (upper table) and polarized (lower table) PDF moments
included in the analysis are shown.
Current systematic uncertainties in state-of-the-art lattice-QCD calculations
are also displayed according to the benchmark exercise performed in
Sec.~\ref{sec:benchmarking} (see also Tables~\ref{tab:BMunp} 
and \ref{tab:BMpol} for the unpolarized and polarized cases, respectively).
\label{tab:scenarios}
}
\end{table}

Our choice of uncertainties in Table~\ref{tab:scenarios}
is rather different for the unpolarized and polarized cases.
For the unpolarized case, Scenario~A is based on values
of $\delta_L^{(i)}$ rather smaller than the typical uncertainties that affect 
state-of-the-art lattice-QCD calculations, see Table~\ref{tab:BMunp}.
As expected from Fig.~\ref{fig:Bmomsunp}, and as we have explicitly verified,
including lattice-QCD pseudo-data with uncertainties of similar size as those 
of Table~\ref{tab:BMunp} leaves unpolarized PDFs essentially unchanged.
Significantly reduced uncertainties $\delta_L^{(i)}$ must be assumed to 
demonstrate any impact on global fits.
We assume that $\delta_L^{(i)}$ is typically larger for $\la x\ra_{s^+}$
and $\la x\ra_{u^+-d^+}$, compared to the other moments, in line with 
what is observed from Table~\ref{tab:BMunp}.
Scenarios~B and C are rather optimistic, in that they require systematic 
uncertainties to decrease by roughly a factor of two and a factor of four 
with respect to Scenario~A.
For the polarized case, Scenario~A assumes that the uncertainties 
$\delta_L^{(i)}$ are similar to current uncertainties in
state-of-the-art lattice-QCD calculations, see Sec.~\ref{sec:benchmarking},
and Table~\ref{tab:BMpol}.

We note that a total systematic error of $\delta_L^{(i)}\sim 1\%$
is probably the best that one can achieve within a lattice-QCD calculation 
in the near future, since at that level several other effects, such as QED 
corrections, become relevant. These are much more difficult to deal with.
For both the polarized and the unpolarized case, the generalization of these 
projections to other conceivable scenarios
is straightforward and can be obtained from the authors upon request.

\subsubsection{Bayesian reweighting analysis}
\label{sec:projections:rw}

To quantify the impact of future lattice-QCD calculations on global fits 
in each of the three scenarios in Table~\ref{tab:scenarios},
we use a procedure based on Bayesian reweighting analysis.
We briefly describe this procedure here, and refer 
to~\cite{Ball:2011gg,Ball:2010gb} for additional details.

\begin{itemize}

\item We first generate pseudo-data for the lattice-QCD calculations
of $\la x\ra_{u^+}$, $\la x\ra_{d^+}$, $\la x\ra_{s^+}$,
$\la x\ra_{g}$, and $\la x\ra_{u^+-d^+}$ (for the unpolarized case), and
$\la 1\ra_{\Delta u^+}$, $\la 1\ra_{\Delta d^+}$,
$\la 1\ra_{\Delta s^+}$, $\la x\ra_{\Delta u^--\Delta d^-}$, and
$\la 1\ra_{\Delta u^+ - \Delta d^+}$ (for the polarized case).
We denote generically these moments by $\mathcal{F}_i$.
  
\item We construct the associated pseudo-data, denoted by 
$\mathcal{F}_i^\text{(exp)}$, by taking the central values from
the corresponding NNPDF fits, NNPDF3.1 NNLO for the unpolarized case and 
NNPDFpol1.1 NLO for the polarized case.
That is, we {\it assume} for simplicity that the central value
of such future lattice calculations would coincide with the current ones
from the global fit.\footnote{ The exercise can be repeated
 with the actual lattice-QCD central values. However, this 
 requires some choices, such as how to impose 
 the momentum sum rule.
 This is beyond the scope of the present studies.}
As discussed in Sec.~\ref{sec:unpPDFs}, this corresponds to computing
the mean over the Monte Carlo replica sample,
\be
\label{eq:pseudodatadef}
\mathcal{F}_i^\text{(exp)} \equiv \frac{1}{N_\text{rep}}\sum_{k=1}^{N_\text{rep}}
\mathcal{F}_i^{(k)} \, , \quad i=1,\ldots,N_\text{mom} \, ,
\ee
where $N_\text{mom}$ is the number of PDF moments that will be included
in the reweighting; here $N_\text{mom}=5$ both for the unpolarized and 
polarized cases.
To be consistent with the calculations in Sec.~\ref{sec:benchmarking},
the central values of the pseudo-data, Eq.~\eqref{eq:pseudodatadef},
are also evaluated at $Q^2=4\text{ GeV}^2$ 
(see Tables~\ref{tab:unpPDFmoms} and \ref{tab:polPDFmoms}).

\item The uncertainty in the pseudo-data, denoted by 
$\delta\mathcal{F}_i^\text{(exp)} $, is taken to be the value indicated in
Table~\ref{tab:scenarios} for each of the three scenarios.
Thus, the absolute uncertainty on the $i$-th moment
is given by 
$\delta\mathcal{F}_i^\text{(exp)}=\delta_L^{(i)}\mathcal{F}_i^\text{(exp)} $.

\item Using the pseudo-data (central values and total uncertainties)
as defined above, we compute the Bayesian weights $\omega_k$.
These weights quantify the agreement between each $k$-th replica in 
the input PDF set and the corresponding lattice pseudo-data.
We compute the $\chi^2$ for each replica $k$ as
\be
\chi^{2(k)}= \sum_{i=1}^{N_\text{mom}} \frac{\lp
\mathcal{F}_i^{(k)} -\mathcal{F}_i^\text{(exp)} \rp^2}{
\lp \delta\mathcal{F}_i^\text{(exp)}\rp^2} \, , \quad k=1,\ldots,N_\text{rep} \, ,
\ee
assuming that there are no correlations between different $N_\text{mom}$ moments.
This assumption in general might not be a good approximation, since most 
lattice-QCD systematic errors are correlated among different moments, 
and can be avoided, provided the full breakdown of systematic errors 
for each quantity is available.
  
Once the values of the $\chi^2$ have been evaluated,
we compute the corresponding weights for each replica.
The relation between the weights $w_k$  and the values of
the $\chi^{2(k)}$ of each replica is~\cite{Ball:2011gg,Ball:2010gb}
\be
\omega_k =\frac{\lp \chi^{2(k)} \rp^{(N_\text{mom}-1)/2}\exp(-\chi^{2(k)}/2)}{
\sum_{k=1}^{N_\text{rep}} \lc \lp \chi^{2(k)} \rp^{(N_\text{mom}-1)/2}\exp(-\chi^{2(k)}/2)\rc} \, ,
\ee
where the denominator ensures that the weight admits
a probabilistic interpretation, that is, $\sum_k w_k=1$.
These weights represent a measure of the agreement of the individual replicas 
with the new pseudo-data.
For instance, replicas which have associated values of the moments far from 
the pseudo-data (within uncertainties) will have a large $\chi^2$ and a 
very small weight, being thus effectively discarded.

\item These weights are used to recompute the PDFs, their moments,
and generic cross-sections.
This procedure emulates the
impact of adding lattice-QCD pseudo-data to a complete PDF fit.
For instance, after reweighting, the mean value of
the PDF moments is
\be
\label{eq:pseudodatadef1}
\mathcal{F}_i^\text{(rw)} = \sum_{k=1}^{N_\text{rep}}\omega_k
\mathcal{F}_i^{(k)} \, , \quad i=1,\ldots,N_\text{mom} \, ,
\ee
with a similar relation for the associated uncertainties.
\end{itemize}

One limitation of the reweighting procedure is that it is maximally 
reliable if the effective number of replicas $N_\text{eff}$ that survive the 
reweighting procedure (which is a measure of the amount
of information left) is not too small.
This effective number of replicas is quantified in terms of the Shannon 
entropy~\cite{Ball:2011gg,Ball:2010gb}
\be
\label{eq:effnrep}
N_\text{eff}\equiv \exp\lc \frac{1}{N_\text{rep}}\sum_{k=1}^{N_\text{rep}}\omega_k
\log \lp N_\text{rep}/\omega_k\rp\rc \, .
\ee
Finding $N_\text{eff}\ll N_\text{rep}$ means that the pseudo-data
have a large impact on the fit, potentially leading to a large
reduction of the PDF uncertainties.
If either the effective number of replicas becomes too small 
(say $N_\text{eff}\lsim 25$), or the relative fraction is small 
(say, $N_\text{eff}/N_\text{rep}\lsim 0.10$), then the results become unreliable, 
since they are affected by large statistical fluctuations.

Therefore, before considering the effects
of the lattice-QCD pseudo-data at the PDF
level, we need to ensure that the
three scenarios defined
in Table~\ref{tab:scenarios} still lead
to values of $N_\text{eff}$ large enough for
the reweighting procedure to be reliable.
In Table~\ref{tab:neff} we indicate the effective number of replicas
$N_\text{eff}$, Eq.~\eqref{eq:effnrep}, remaining when the pseudo-data
are included in the global
fit according to the scenarios in Table~\ref{tab:scenarios}.
For completeness, we also quote the original number
of replicas $N_\text{rep}$ for the prior
PDF sets, NNPDF3.1 and NNPDFpol1.1, respectively.
As we can see, there is a marked decrease of $N_\text{rep}$
for the three scenarios, indicating that adding the
PDF moments leads to non-trivial constraints on the global fit.
For instance, in the most optimistic scenario, Scenario~C, the effective 
number of replicas is around two (five) times smaller than the starting 
number of replicas in the unpolarized (polarized) case.

\begin{table}[!t]
\centering
\footnotesize
\renewcommand{\arraystretch}{1.3} 
\begin{tabular}{lcc}
\toprule
&  NNPDF3.1  &  NNPDFpol1.1 \\
\midrule
$N_\text{rep}$ original   &   1000 &  100   \\
$N_\text{eff}$ Scenario A    &   740  &  72   \\
$N_\text{eff}$ Scenario B    &   750   &   59  \\
$N_\text{eff}$ Scenario C   &   510  &   20  \\
\bottomrule
\end{tabular}
\caption{\small The effective number of replicas
$N_\text{eff}$, Eq.~\eqref{eq:effnrep}, remaining after pseudo-data
on the PDF moments are included in the global
fit according to the scenarios outlined
in Table~\ref{tab:scenarios}.
For completeness, we also indicate the original number
of replicas $N_\text{rep}$ for the prior
PDF sets, NNPDF3.1 and NNPDFpol1.1.
\label{tab:neff}}
\end{table}

\paragraph{Impact on unpolarized global fits.}
\label{subsec:upolfits}
We now discuss the results of applying the reweighting procedure to NNPDF3.1.
In Table~\ref{tab:unpolmomentsrw} we summarize
the values of the unpolarized PDF moments
used as pseudo-data $\mathcal{F}_i^{(\rm exp)}$,
and the corresponding results
after the reweighting has been performed, for the
three scenarios summarized 
in Table~\ref{tab:scenarios};
PDF uncertainties correspond to 68\%-CL intervals.
We recall that, as explained above, the three scenarios exhibit
uncertainties $\delta_L^{(i)}$ for the lattice-QCD pseudo-data rather smaller
than those of current state-of-the-art
calculations (see Table~\ref{tab:BMunp}).

\begin{table}[!t]
\centering
\footnotesize
\renewcommand{\arraystretch}{1.4} 
\begin{tabular}{lcccc}
\toprule 
&  Original  & Scenario A  &  Scenario B  &  Scenario C  \\
\midrule
$\la x\ra_{u^+}$     
& $0.348 \pm 0.005$ 
& $0.349 \pm 0.004$ 
& $0.349 \pm 0.004$ 
& $0.349 \pm 0.003$ \\
$\la x\ra_{d^+}$     
& $0.196 \pm 0.004$     
& $0.196 \pm 0.004$       
& $0.196 \pm 0.003$ 
& $0.196 \pm 0.002$ \\
$\la x\ra_{s^+}$     
& $0.0393 \pm 0.0036$   
& $0.0389 \pm 0.0030$   
& $0.0389 \pm 0.0024$   
& $0.0389 \pm 0.0014$  \\
$\la x\ra_{g}$       
& $0.4097 \pm 0.0042$    
& $0.4097 \pm 0.0043$    
& $0.4097 \pm 0.0040$ 
& $0.4097 \pm 0.0029$  \\
$\la x\ra_{u^+-d^+}$  
& $0.1522 \pm 0.0033$   
& $0.1521 \pm 0.0037$   
& $0.1521 \pm 0.0035$ 
& $0.1521 \pm 0.0029$ \\
\bottomrule
\end{tabular}
\caption{\small Values of the unpolarized PDF moments
  used as pseudo-data, as well as the corresponding results
  after the reweighting has been performed, for the
  three scenarios summarized 
  in Table~\ref{tab:scenarios}.
  The PDF uncertainties quoted correspond in all cases to 68\% CL intervals.
\label{tab:unpolmomentsrw}}
\end{table}

From Table~\ref{tab:unpolmomentsrw} we see that a significant
reduction in the uncertainties in the unpolarized PDF moments is challenging 
to achieve unless we assume the most aggressive scenarios.
For instance, in Scenario~C, which is about the best precision that
can be achieved from lattice-QCD in the near future, the PDF uncertainties 
of the first moments (that is, the momentum fractions) for $u^+,d^+,s^+$ and 
$g$ decrease by around 30\%--60\%.
The most marked decrease is for the strange momentum fraction, since this is 
affected by the largest PDF error in the prior fit.
In contrast, the nonsinglet combination $\la x\ra_{u^+-d^+}$ is essentially
unchanged in all three scenarios.
Note that, in Table~\ref{tab:unpolmomentsrw}, the central values of the PDF 
moments are stable, since we assume that the central values of the 
pseudo-data correspond to those of the input PDFs. 
In a realistic situation, this is not necessarily the case and 
central values of the PDFs could also vary.

\begin{figure}[!t]
\centering
\includegraphics[angle=270,scale=0.35]{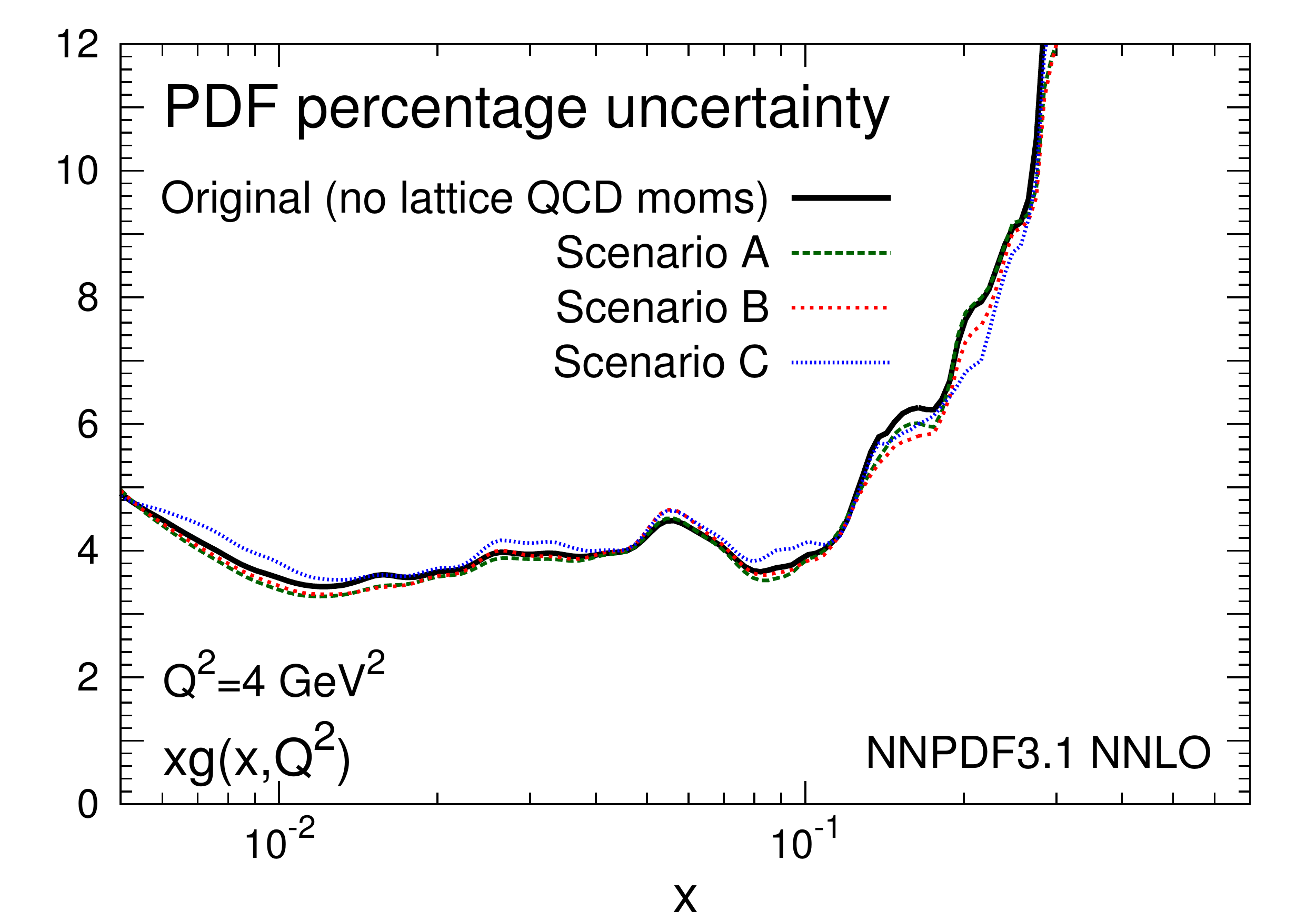}
\includegraphics[angle=270,scale=0.35]{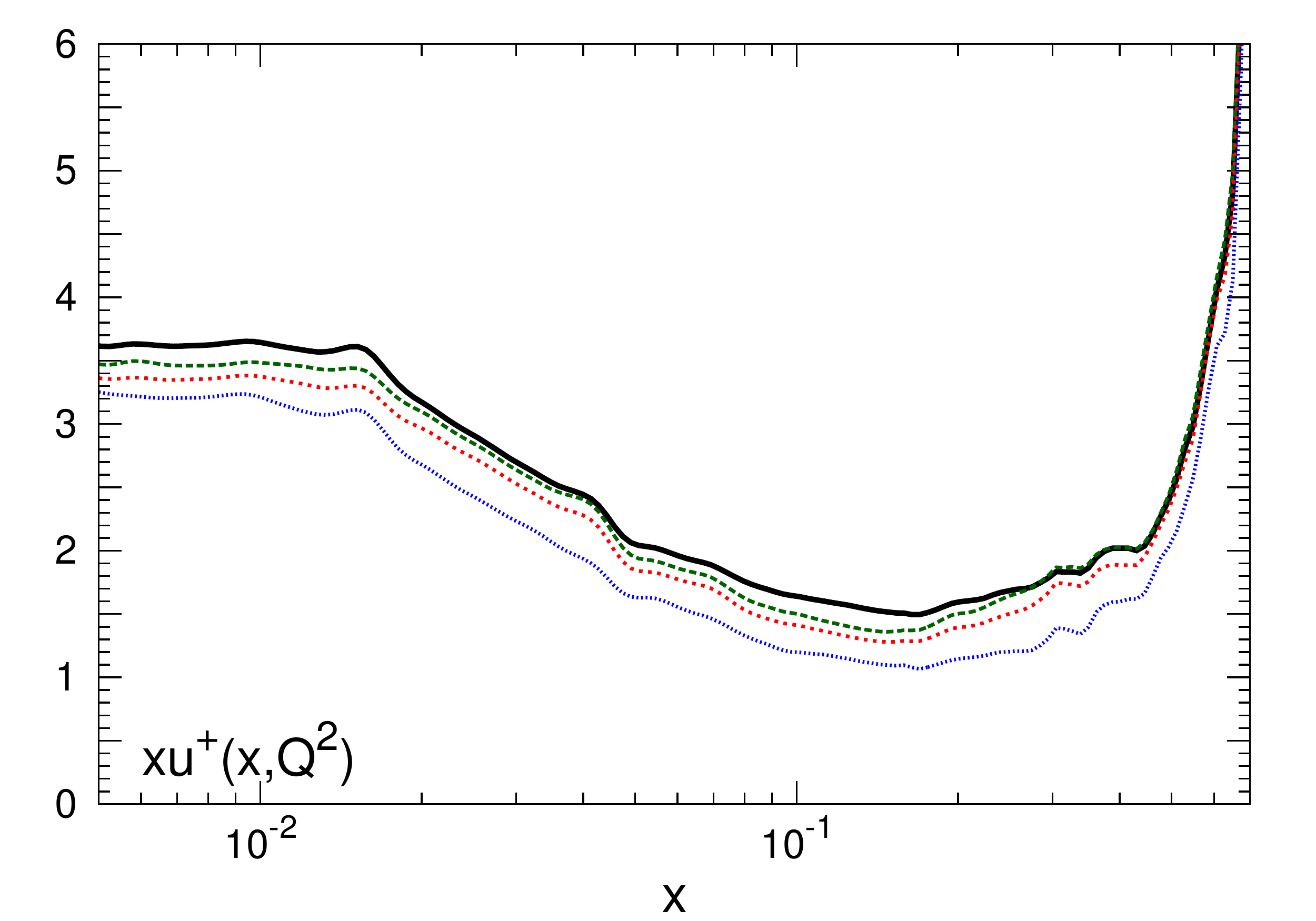}\\
\includegraphics[angle=270,scale=0.35]{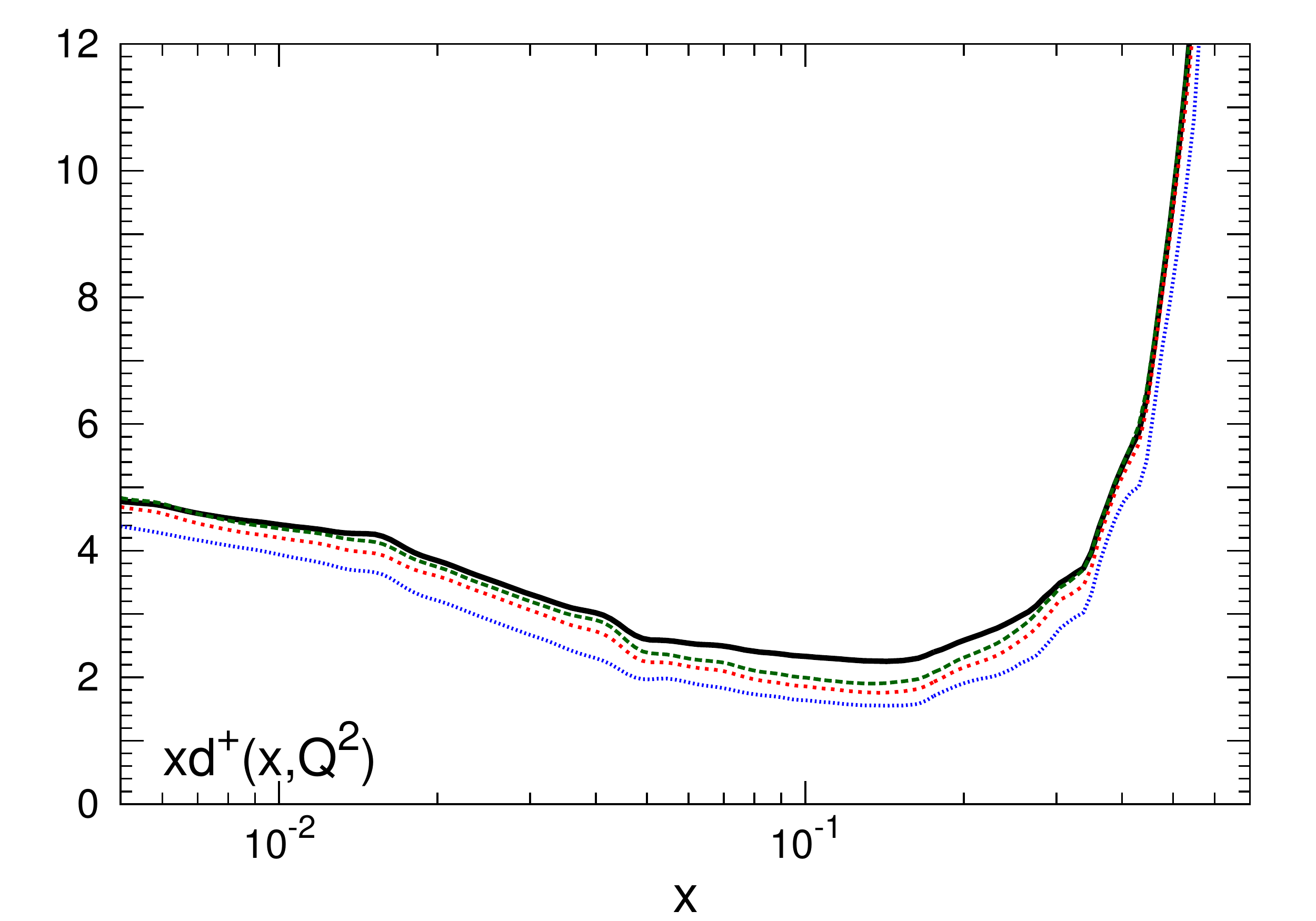}
\includegraphics[angle=270,scale=0.35]{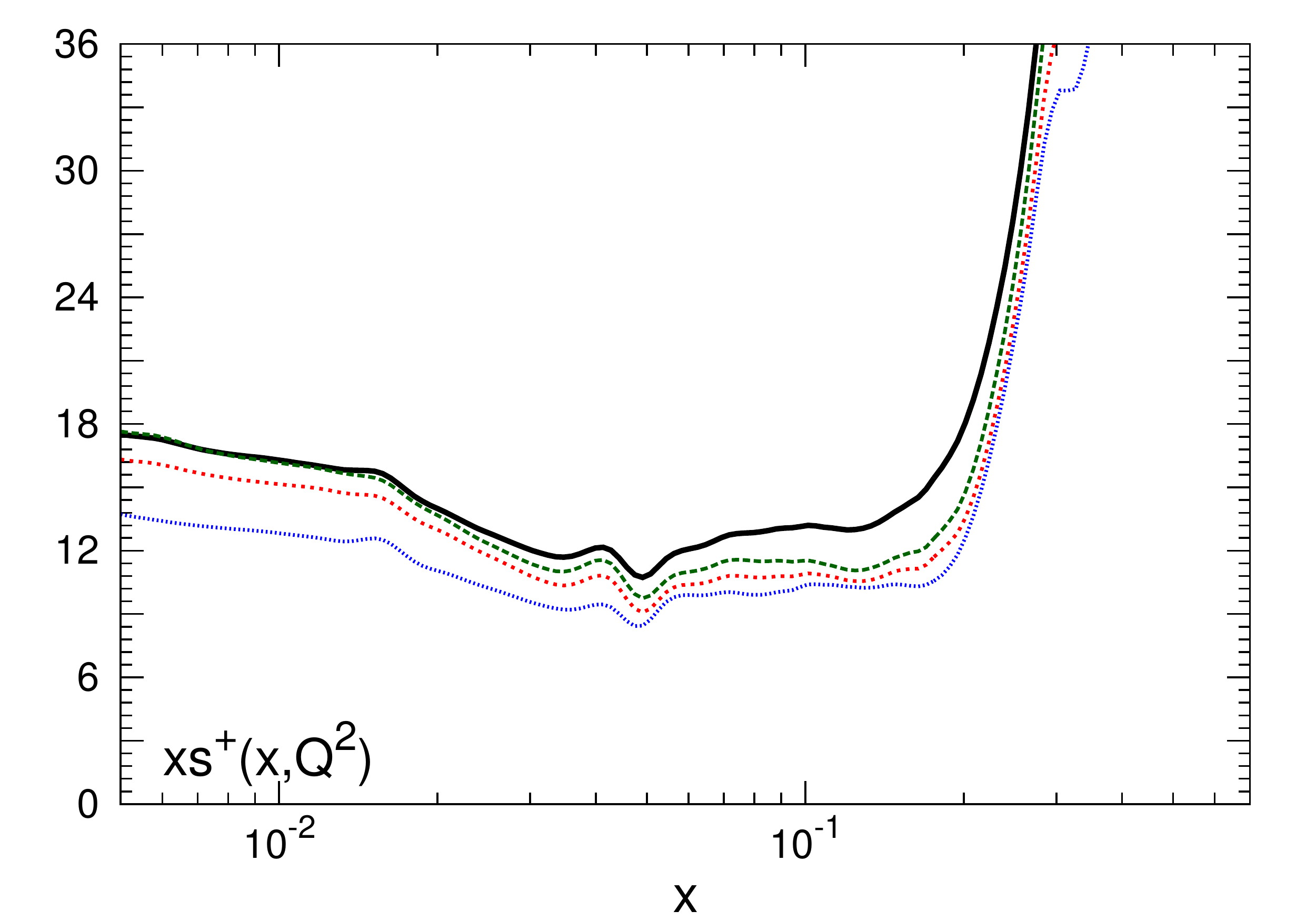}\\
\caption{\small The percentage PDF uncertainty in NNPDF3.1  
  for the gluon and the $u^+$, $d^+$ and $s^+$ quark PDFs at
  $Q^2=4\text{ GeV}^2$,
  compared to the results of including lattice-QCD pseudo-data for
  moments of PDFs in the fit, according to the three
  scenarios in Table~\ref{tab:scenarios}.
  See text for details.
}    
\label{fig:impactUnpol}
\end{figure}

Further evidence that reducing uncertainties in unpolarized PDFs will be 
challenging is shown in Fig.~\ref{fig:impactUnpol}, which displays the 
percentage PDF uncertainties in NNPDF3.1 for the gluon and the
$u^+$, $d^+$ and $s^+$ quark PDFs at $Q^2=4\text{ GeV}^2$, compared to the 
corresponding results including lattice-QCD pseudo-data.
In the case of the $u^+,d^+$ and $s^+$, we observe a slight reduction
of the PDF uncertainties, which is more marked as we move
from Scenarios~A to C.
For instance, in the latter case the percentage PDF
uncertainty on $u^+$ ($d^+$ and $s^+$) at $x\simeq 0.1$
decreases from 1.8\% to 1.2\% (from 2.2\% to 1.7\% and from 13\% to 10\%, 
respectively).
The PDF uncertainties of the gluon PDF, however,
are essentially unchanged even in the most optimistic scenario.

We also observe the trend that the reduction of the uncertainty 
of the PDF moments (see Table~\ref{tab:unpolmomentsrw})
is more significant than the PDF uncertainty as a function of $x$  
(Fig.~\ref{fig:impactUnpol}).
We will see that this pattern also persists for the polarized PDF case.
As the PDF moments integrate across all $x$ values (with emphasis on 
smaller $x$ values), this suggests that there are correlations which could be 
driving this result.
In particular we note that in Scenario~C the uncertainty on the moment for 
$s^+$ is less than $4\%$ while for the PDF at any $x$ it is always greater 
than  $8\%$, a result which can only be achieved due to strong anticorrelation
between different $x$ regions. 
Additional studies examining the PDF correlations before and after inclusion 
of the lattice-QCD input could prove enlightening. 

\begin{figure}[!t]
\centering
\includegraphics[scale=0.45]{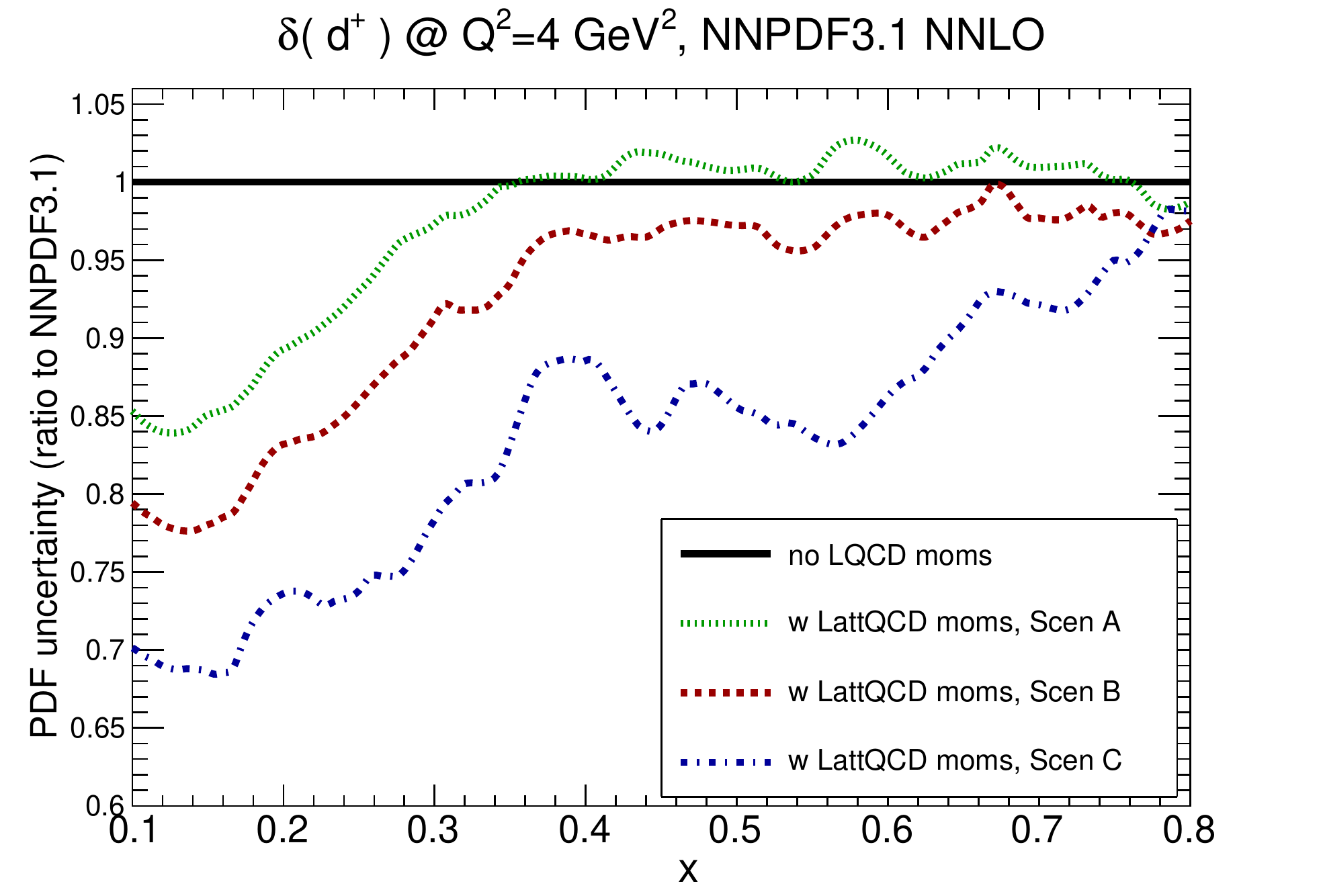}
\includegraphics[scale=0.45]{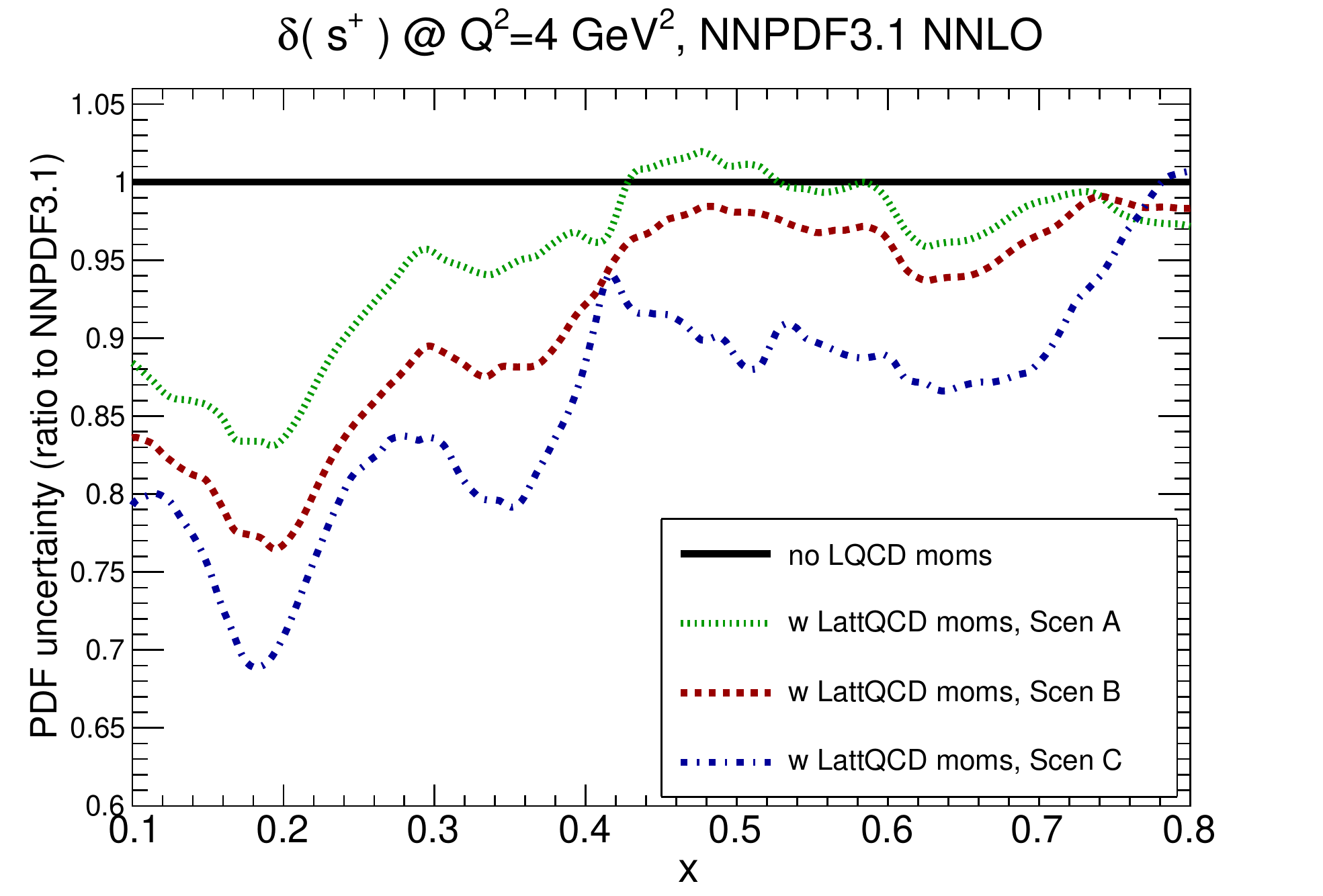}
\caption{\small Same as Fig.~\ref{fig:impactUnpol}, now focusing
  on the large-$x$ region, and showing the ratio of the
  PDF uncertainty in the fits based on the three scenarios
  to the original
  PDF uncertainty in the NNPDF3.1 set, for the $d^+$ (left)
  and $s^+$ (right) total quark PDFs.
}    
\label{fig:impactUnpollargex}
\end{figure}

Focusing on the large-$x$ region, where the
impact of the PDF moments considered here is expected to be more marked, in
Fig.~\ref{fig:impactUnpollargex} we show the ratio 
of the uncertainty in each scenario to the prior PDF 
uncertainty in the NNPDF3.1 set, for the $d^+$
and $s^+$ total quark PDFs.
This comparison clearly illustrates that the relative reduction
of the PDF uncertainties upon addition of lattice-QCD
pseudo-data is not completely flat, and that it exhibits some structure.
The constraints from lattice-QCD calculations of these 
PDF moments decrease for larger values of $x$.

\paragraph{Impact on polarized global fits.}
Now we turn to apply the reweighting procedure to NNPDFpol1.1.
In Table~\ref{tab:polmomentsrw}
we list the values of the polarized PDF moments
used as pseudo-data, and the corresponding results
after the reweighting has been performed for the
three scenarios summarized in Table~\ref{tab:scenarios}.
As in the unpolarized case, the PDF uncertainties quoted correspond in 
all cases to 68\%-CL intervals.
As we can see from this comparison, in Scenario~A
(which assumes lattice-QCD pseudo-data with uncertainties similar
to existing calculations) there is a marked impact on the
polarized PDF moments.
For both $\la 1\ra_{\Delta u^+}$ and $\la 1\ra_{\Delta d^+}$
the PDF uncertainties are roughly halved, with a similar, but less marked,
trend for $\la 1\ra_{\Delta s^+}$.
At this level, there is no impact on the nonsinglet
combinations $\la 1\ra_{\Delta u^+ - \Delta d^+}$
and $\la x\ra_{\Delta u^--\Delta d^-}$.

\begin{table}[!t]
\centering
\footnotesize
\renewcommand{\arraystretch}{1.4} 
\begin{tabular}{lcccc}
\toprule
& Original & Scenario A &  Scenario B & Scenario C \\
\midrule
$\la 1\ra_{\Delta u^+}$    
& $+0.788 \pm 0.079$   
& $+0.798 \pm 0.039$     
& $+0.797 \pm 0.023$ 
& $+0.790 \pm 0.009$ \\
$\la 1\ra_{\Delta d^+}$   
& $-0.450 \pm 0.083$  
& $-0.450 \pm 0.042$  
& $-0.456 \pm 0.026$    
& $-0.465 \pm 0.012$ \\
$\la 1\ra_{\Delta s^+}$    
& $-0.124 \pm 0.108$  
& $-0.120 \pm 0.070$  
& $-0.121 \pm 0.076$    
& $-0.111 \pm 0.029$ \\
$\la 1\ra_{\Delta u^+ - \Delta d^+}$  
& $+1.250 \pm 0.024$   
& $+1.250 \pm 0.022$  
& $+1.253 \pm 0.016$ 
& $+1.256 \pm 0.012$ \\
$\la x\ra_{\Delta u^--\Delta d^-}$     
& $+0.196 \pm 0.014$    
& $+0.195 \pm 0.014$
& $+0.196 \pm 0.016$     
& $+0.198 \pm 0.012$ \\
\bottomrule
\end{tabular}
\caption{\small Same as Table~\ref{tab:unpolmomentsrw}, now for
  the polarized PDF moments computed with NNPDFpol1.1.
  The corresponding impact at the PDF level is shown in
  Fig.~\ref{fig:impactPol}.
\label{tab:polmomentsrw}
}
\end{table}

As we further decrease the assumed uncertainties in the lattice-QCD
pseudo-data, we observe a corresponding reduction of the uncertainties
in the global fit.
In Scenario~C, the most optimistic, we find that for both
$\la 1\ra_{\Delta u^+}$ and $\la 1\ra_{\Delta d^+}$ there is an uncertainty
reduction by about an order of magnitude compared to the current values,
and by about a factor of five for $\la 1\ra_{\Delta s^+}$.
Therefore, future lattice-QCD calculations of
polarized PDF moments can potentially lead to a much more
precise understanding of the spin structure of the proton.
The other quark combinations exhibit less sensitivity to the inclusion
of the PDF moments in the global fit, because
they are already quite well constrained by available experimental
data.
The PDF uncertainties for  $\la 1\ra_{\Delta u^+ - \Delta d^+}$
are reduced by a factor of two in this quite optimistic scenario, while
those of $\la x\ra_{\Delta u^--\Delta d^-}$ are essentially unaffected even
in the most optimistic scenario.

\begin{figure}[!t]
\centering
\includegraphics[angle=270,scale=0.35]{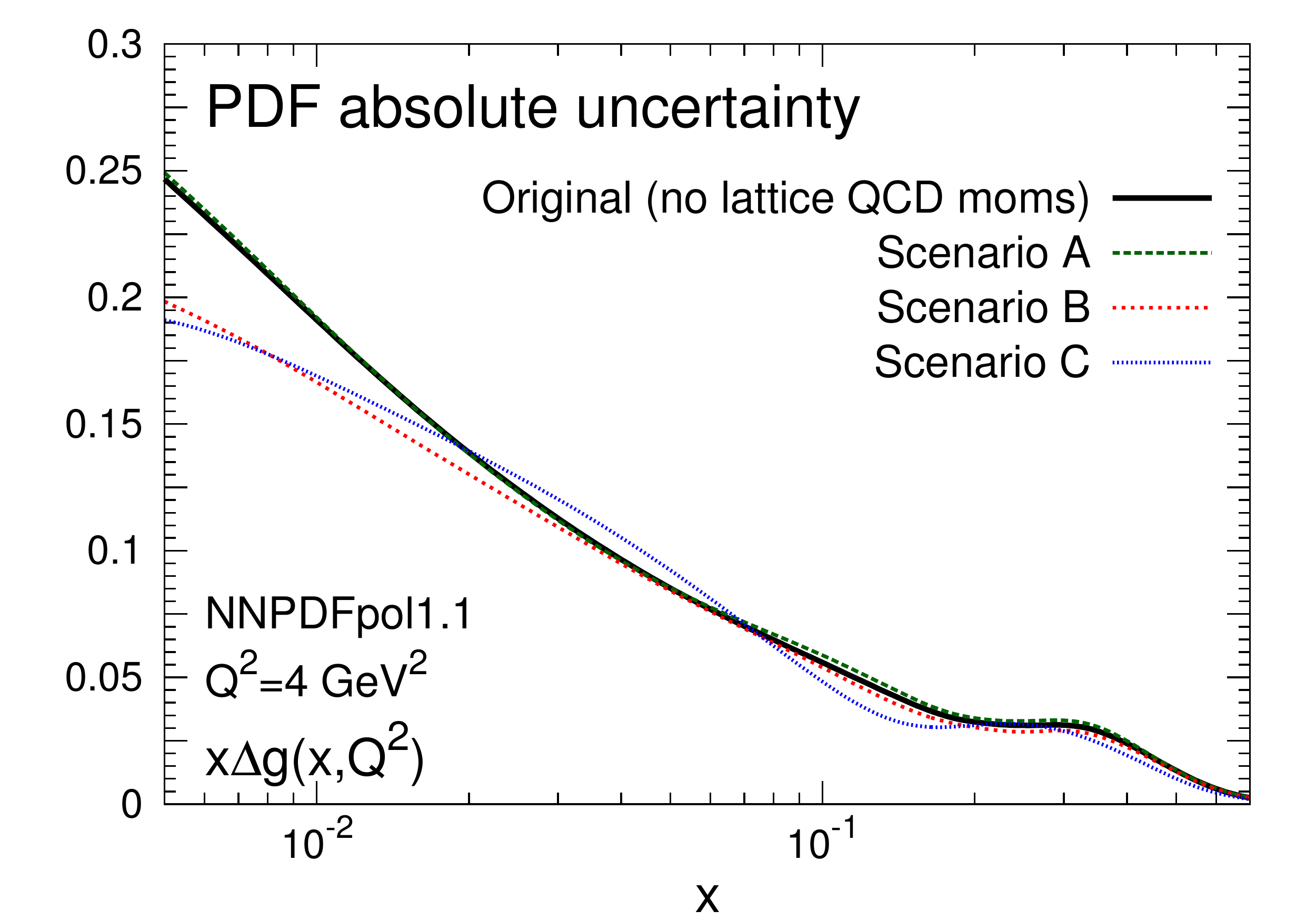}
\includegraphics[angle=270,scale=0.35]{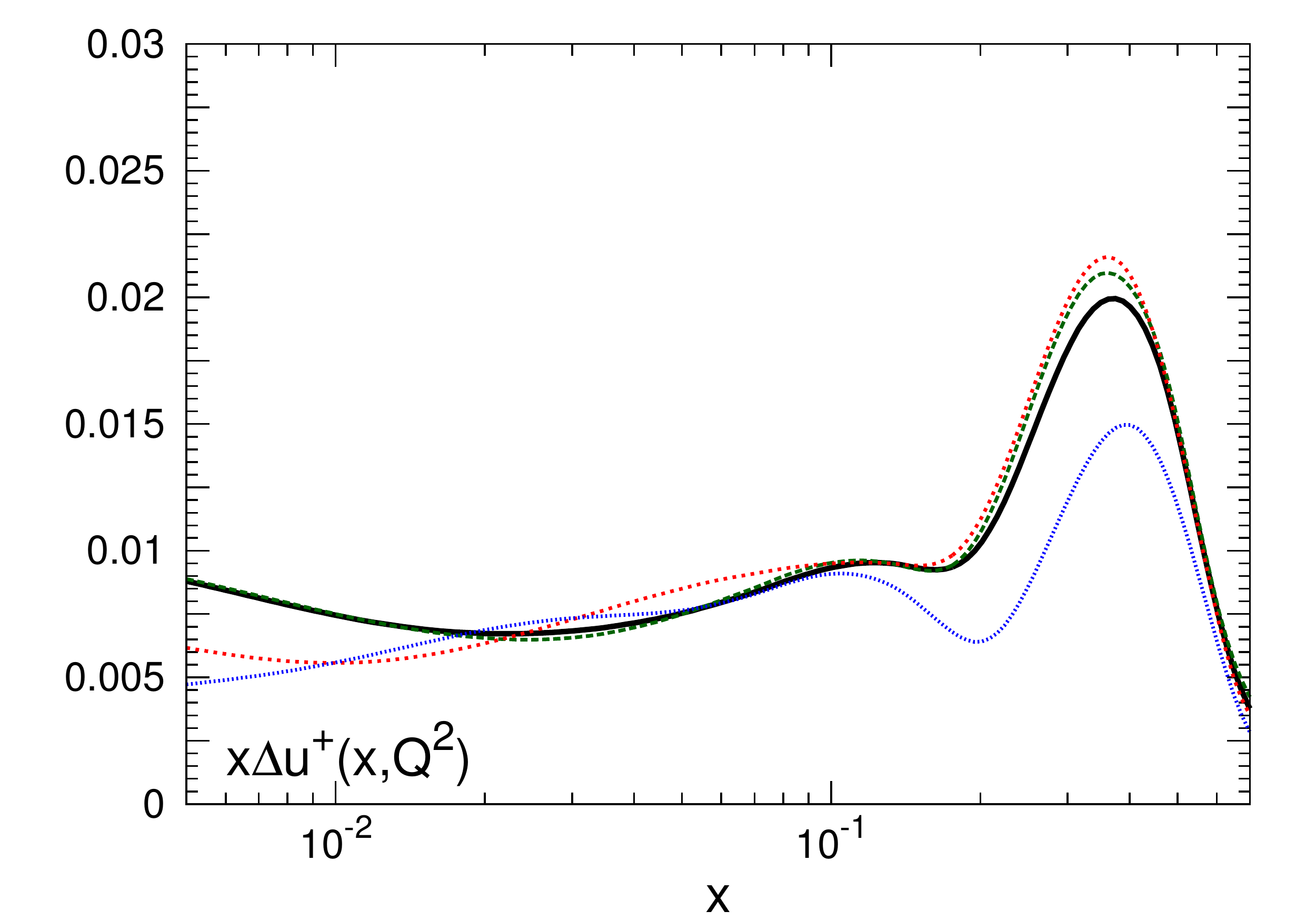}
\includegraphics[angle=270,scale=0.35]{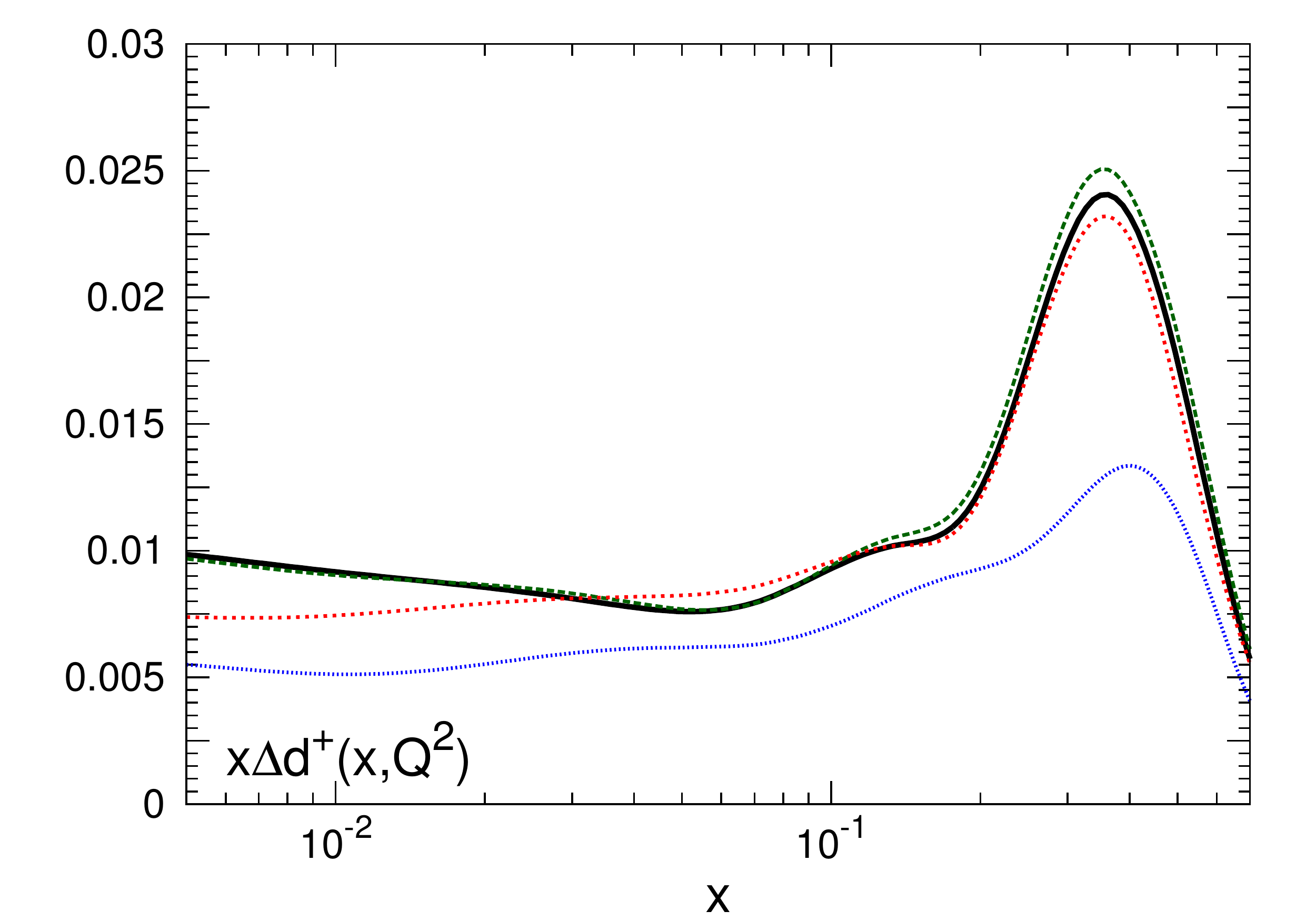}
\includegraphics[angle=270,scale=0.35]{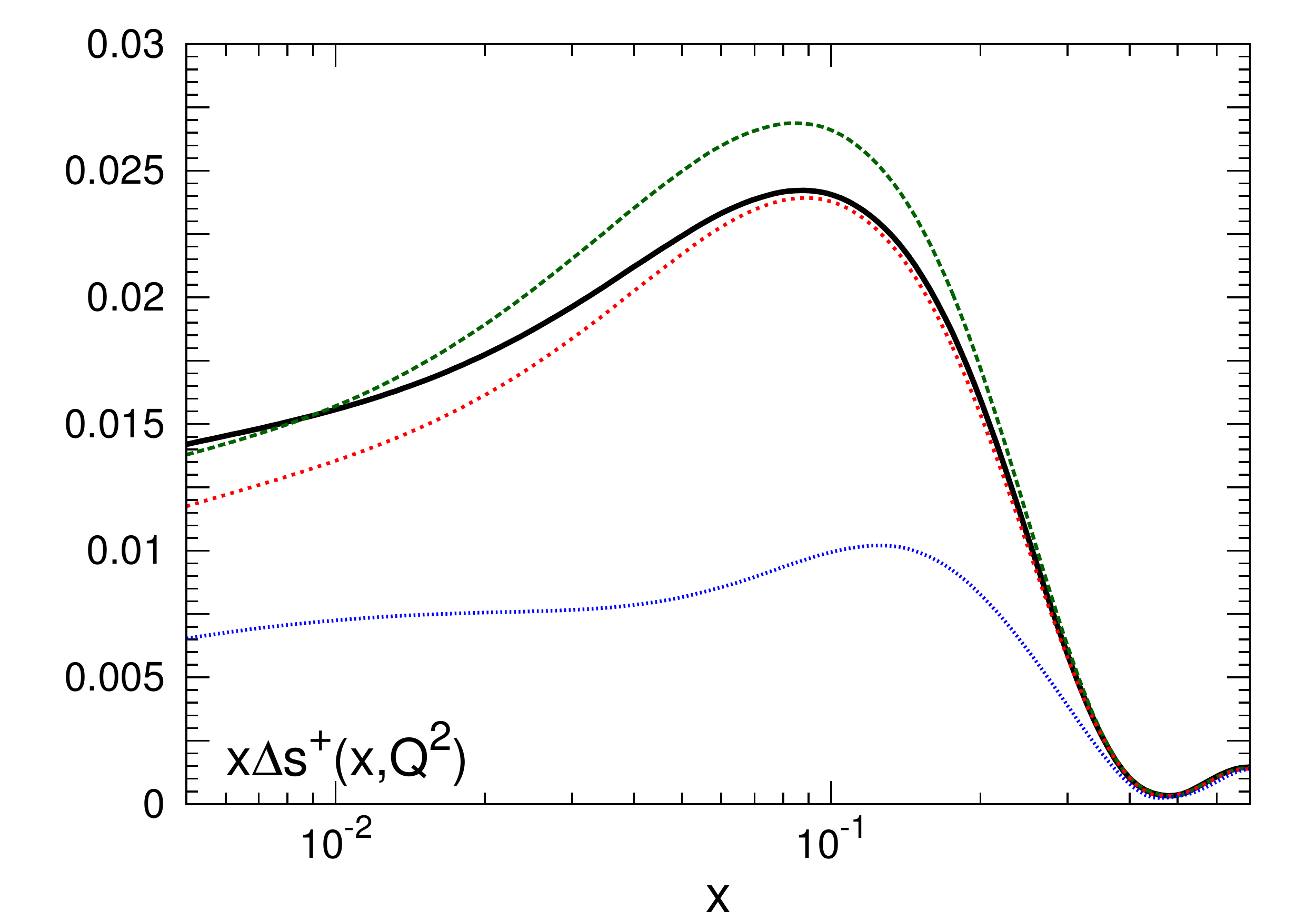}
\caption{\small Same as Fig.~\ref{fig:impactUnpol}, now
  showing the absolute PDF uncertainties of the NNPDFpol1.1 fit
  $Q^2=4\text{ GeV}^2$, compared to the corresponding results once the lattice 
  pseudo-data on polarized moments are included in the analysis via reweighting.
}    
\label{fig:impactPol}
\end{figure}

In Fig.~\ref{fig:impactPol} we compare the absolute PDF uncertainties
of the NNPDFpol1.1 fit to the corresponding results once the lattice 
pseudo-data on polarized moments are included in the analysis by means of the
reweighting.
We show absolute rather than relative uncertainties
because, unlike unpolarized PDFs, polarized PDFs often exhibit nodes
(in particular for strangeness and the gluon) and in the nearby regions
the concept of relative uncertainty becomes ill-defined.
  
From Fig.~\ref{fig:impactPol} we see that for scenarios
A and B there is only a very moderate reduction (or even a slight increase)
of the PDF uncertainties, seemingly at odds with the results
for their moments in Table~\ref{tab:polmomentsrw}.
The reason is that the first PDF moments alone provide only limited
information on the shape of the PDFs themselves, and therefore in some
cases one finds a larger error reduction on the moments (since these
are the fitted quantities) than on the PDFs themselves (which are
only indirectly constrained).
Once, however, the lattice-QCD pseudo-data uncertainties
decrease beyond a certain level, these uncertainties start to influence the 
PDF shape, as we can see from the results of Scenario~C.
In that case we find that the PDF uncertainties can decrease by up to a factor
of two (three) for $\Delta d^+(x,Q)$ ($\Delta s^+(x,Q)$).
We also see the apparently simple feature that relative reduction of PDF 
uncertainties is more or less constant along the whole range of $x$. 
For the strange quark this is perhaps roughly consistent with a simple 
reduction in the normalization uncertainty independent of $x$.
However, similarly to the unpolarized case, for the down quark this 
decrease is a much smaller factor than the decrease in the uncertainty 
of the moments, meaning that there must be some anticorrelation between 
PDFs at different $x$ values.

\subsubsection{Hessian profiling analysis}
\label{sec:hessianprofiling}

To complement the results obtained
with the Bayesian reweighting approach,
we use a profiling method, suitable
for Hessian PDF sets, to estimate the effect of including
lattice-QCD pseudo-data into the fit~\cite{Paukkunen:2014zia,Camarda:2015zba}.
We choose HERAPDF2.0~\cite{Abramowicz:2015mha}
as a representative set of Hessian PDFs.
As in the case of the Bayesian reweighting
exercise presented in the previous section
we consistently use the same lattice-QCD
pseudo-data on PDF moments to estimate the impact on HERAPDF2.0.
An additional advantage of the HERAPDF2.0 set is
the use
of a standard tolerance
$\Delta\chi^2=1$ for defining the 68\%-CL PDF
uncertainties,
which enables a robust framework for applying the profiling method.

For Hessian PDF sets, the Hessian profiling method
can be used to both check the compatibility of new data with a given PDF set,
and also  estimate the impact these data will have on the PDFs. 
In the following we describe the essential components of the profiling method, 
and assume  that the  Hessian PDF set uses a tolerance of $\Delta\chi^2=1$, 
which corresponds to 68\%~CL uncertainties,
as is the case with the HERAPDF2.0 set.\footnote{In this exercise
we consider only the {\it experimental} HERAPDF2.0
uncertainties, but not the {\it model} and {\it parametrization}
variations, which are not suited for profiling.}
The central values of the considered moments are obtained using the central PDFs and the corresponding
errors are calculated according to:
\begin{equation}
\delta\mathcal{F}_i = \frac{1}{2} \sqrt{\sum_{k}\left(\mathcal{F}_i(f_k^+)-\mathcal{F}_i(f_k^-)\right)^2}\, ,
\quad i=1,\ldots,N_\text{mom} \, ,
\end{equation}
where $k$ labels the number of error PDFs (Hessian eigenvectors)
which have both a positive and negative direction.
In the profiling method, one considers a $\chi^2$ function in which the $\chi^2$ of the new
data has been added to the initial $\chi^2_0$, namely
\begin{equation}
\label{eq:newchi2}
\chi^2_{\text{new}} = \chi_0^2 + \sum_{k}^{N_{\text{eig}}} z_k^2
                    + \sum_{i=1}^{N_{\text{data}}}
                      \frac{\lp \mathcal{F}_i - \mathcal{F}_i^{\rm(exp)}\rp^2}
                           {\lp\delta\mathcal{F}_i^{\rm(exp)}\rp^2}\,,
\end{equation}
where $\chi^2_0$ is the value of the $\chi^2$ function in the minimum of the initial PDF set,
$z_k$ are the parameters diagonalizing the Hessian matrix of the initial PDF set,
$N_{\text{eig}}$ is the dimension of the eigenvector space in which initial Hessian errors are defined
(half of the number of error PDFs), $\mathcal{F}_i^{\rm(exp)}$ is the new
\hbox{(pseudo-)data},
and $\mathcal{F}_i$ the corresponding theory prediction.

In the spirit of the Hessian method, the new theory predictions $\mathcal{F}_i$ can be expanded
using a linear approximation:
\begin{equation}
\mathcal{F}_i \simeq \mathcal{F}_i[S_0] + \sum_k \frac{\partial\mathcal{F}_i[S]}{\partial z_k}\bigg|_{S=S_0} z_k \quad
              \simeq \mathcal{F}_i[S_0] + \sum_k D_{ik} w_k \ ,
\end{equation}
where $S_0$ represents the central PDF and we have defined
\be
D_{ik}=\frac{1}{2}(\mathcal{F}_i[S_k^+]-\mathcal{F}_i[S_k^-]) \, ;
\ee
here the  derivative has been approximated by a finite difference of the 
Hessian PDF error sets $S_k^{\pm}$.
The new $\chi^2$ of Eq.~\eqref{eq:newchi2} can now be minimized with respect to the parameters $w_k$,
which results in:
\begin{equation}
%
w_k^{min}  = \sum_n \ -B_{kn}^{-1} \, a_n \quad ,
\end{equation}
where we have introduced
\begin{equation}
B_{kn} = \sum_i \frac{D_{ik}D_{in}}{\lp\delta\mathcal{F}_i^{\rm(exp)}\rp^2} + \delta_{kn},
\qquad
\qquad
a_k = \sum_i \frac{D_{ik}(\mathcal{F}_i[S_0] - \mathcal{F}_i^{\rm(exp)})}{\lp\delta\mathcal{F}_i^{\rm(exp)}\rp^2} \, . 
\end{equation}

The key result of the Hessian profiling method
is that now the components of the solution 
$w_k^{min}$
define a new set
of PDFs representing a global minimum after including the new data:
\begin{equation}
f_{\text{new}} = f_{S_0} + \sum_{k=1}^{N_{\text{eig}}} \frac{f_{S_k^+}-f_{S_k^-}}{2} w_k^{\text{min}} \ .
\end{equation}
A set of new error PDFs can be also defined; in this case the matrix $B_{kn}$ plays the role of
the Hessian matrix from which the PDF uncertainties
can be obtained. 

We performed this study using the xFitter program~\cite{Alekhin:2014irh}
assuming the same three scenarios for the lattice-QCD pseudo-data as 
in Table~\ref{tab:scenarios}. 
The results are shown in Table~\ref{tab:unpolmomentsProf}, where we tabulate 
the uncertainties of the input HERAPDF2.0 PDF in column two and the 
corresponding uncertainties for each scenario in columns three to five. 
The analogous results from the reweighting method, applied to the 
NNPDF3.1 data set, were listed in Table~\ref{tab:unpolmomentsrw}.

\begin{table}[!t]
\centering
\footnotesize
\renewcommand{\arraystretch}{1.3} 
\begin{tabular}{lcccc}
\toprule 
&  Original  & Scenario A  &  Scenario B  &  Scenario C  \\
\midrule
  $\la x\ra_{u^+}$     
&  $0.3720\pm 0.0036$  
&  $0.3720\pm 0.0030$  
&  $0.3720\pm 0.0027$  
&  $0.3720\pm 0.0020$ \\
  $\la x\ra_{d^+}$     
&  $0.1845\pm 0.0053$  
&  $0.1845\pm 0.0028$  
&  $0.1845\pm 0.0023$  
&  $0.1845\pm 0.0015$ \\
  $\la x\ra_{s^+}$     
&  $0.0346\pm 0.0037$  
&  $0.0346\pm 0.0015$  
&  $0.0346\pm 0.0012$  
&  $0.0346\pm 0.0009$ \\
  $\la x\ra_{g}$       
&  $0.4006\pm 0.0078$  
&  $0.4006\pm 0.0042$  
&  $0.4006\pm 0.0035$  
&  $0.4006\pm 0.0024$ \\
  $\la x\ra_{u^+-d^+}$ 
&  $0.1875\pm 0.0074$  
&  $0.1875\pm 0.0045$  
&  $0.1875\pm 0.0039$  
&  $0.1875\pm 0.0027$ \\
\bottomrule
\end{tabular}
\caption{\small Values of the unpolarized PDF moments
  used as lattice-QCD pseudo-data, as well as the corresponding results
  after the profiling  for the
three scenarios summarized in Table~\ref{tab:scenarios}.
The HERAPDF2.0 PDFs were used, and the PDF uncertainties quoted correspond in all cases to 68\%~CL intervals.
The corresponding results of applying the reweighting method
to NNPDF3.1 were listed in Table~\ref{tab:unpolmomentsrw}.
\label{tab:unpolmomentsProf}
}
\end{table}

From a comparison of the constraining power of the lattice-QCD pseudo-data  
displayed in Table~\ref{tab:unpolmomentsProf} to Table~\ref{tab:unpolmomentsrw},
we observe a consistent trend between Bayesian reweighting of NNDPF3.1 and 
Hessian profiling of HERAPDF2.0.
The PDF uncertainties for $\la x\ra_{d^+}$ ($\la x\ra_{s^+}$
and  $\la x\ra_g$) reduce by a factor of roughly
four (four and three, respectively) compared to the original
HERAPDF2.0 uncertainties.
When comparing with Sec.~\ref{sec:projections:rw},
the initial uncertainties of the HERAPDF2.0  analysis 
are affected by the choice of data (DIS data only), and 
the number and form of the parametrization (14 parameter HERAPDF form);
the final uncertainties are determined by the profiling procedure. 
In particular the profiling for the HERAPDF2.0 study assigns an effective 
uncertainty on the pseudodata corresponding to $\Delta\chi^2=1$, whereas the 
constraint in the NNPDF study is weaker, as it would be for a PDF set with 
eigenvectors, but which applied a tolerance criterion. 
While these initial studies are instructive, 
further comparisons of these analyses would be valuable. 

\begin{figure}[!t]
\centering
\includegraphics[width=0.45\textwidth]{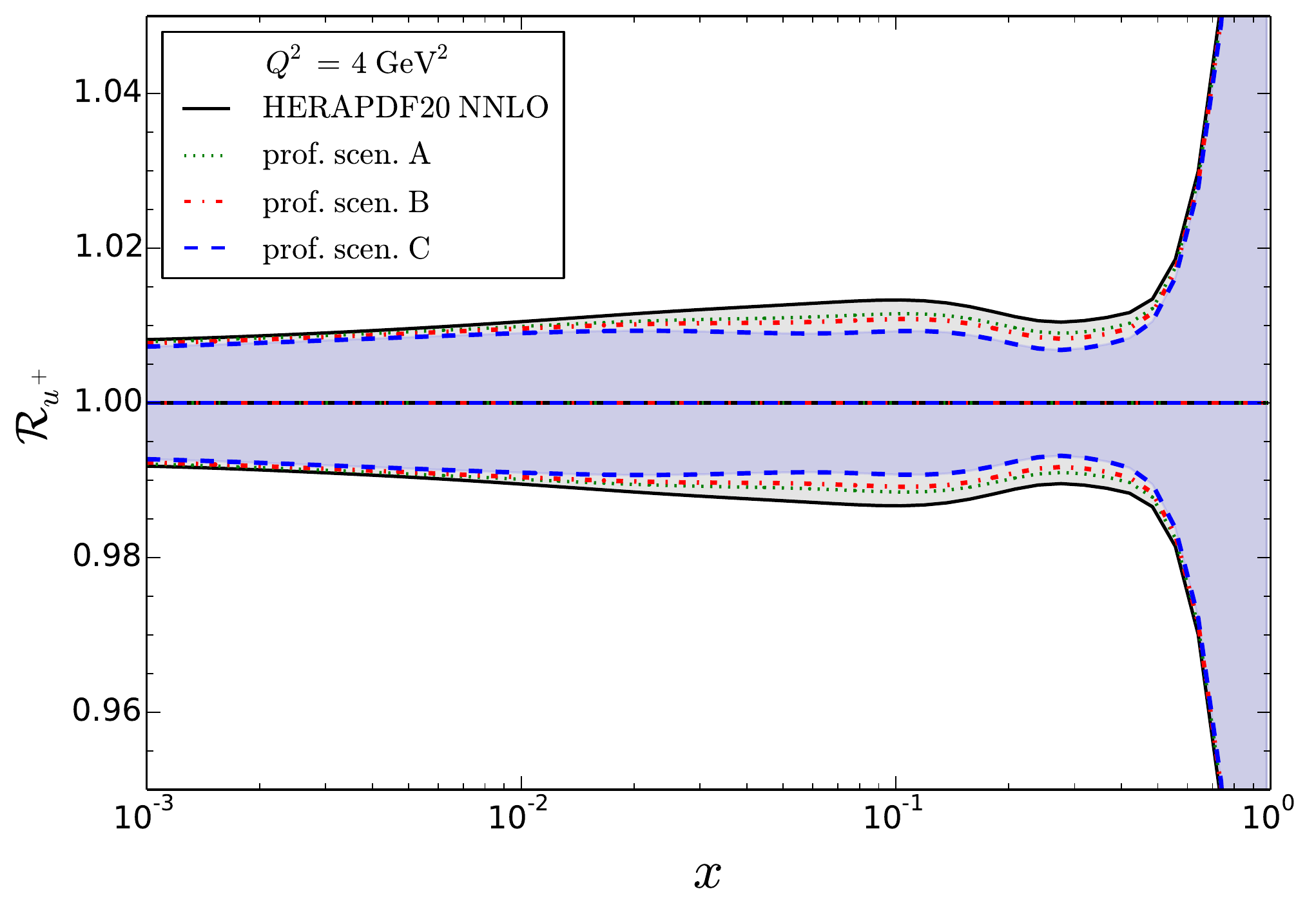}
\includegraphics[width=0.45\textwidth]{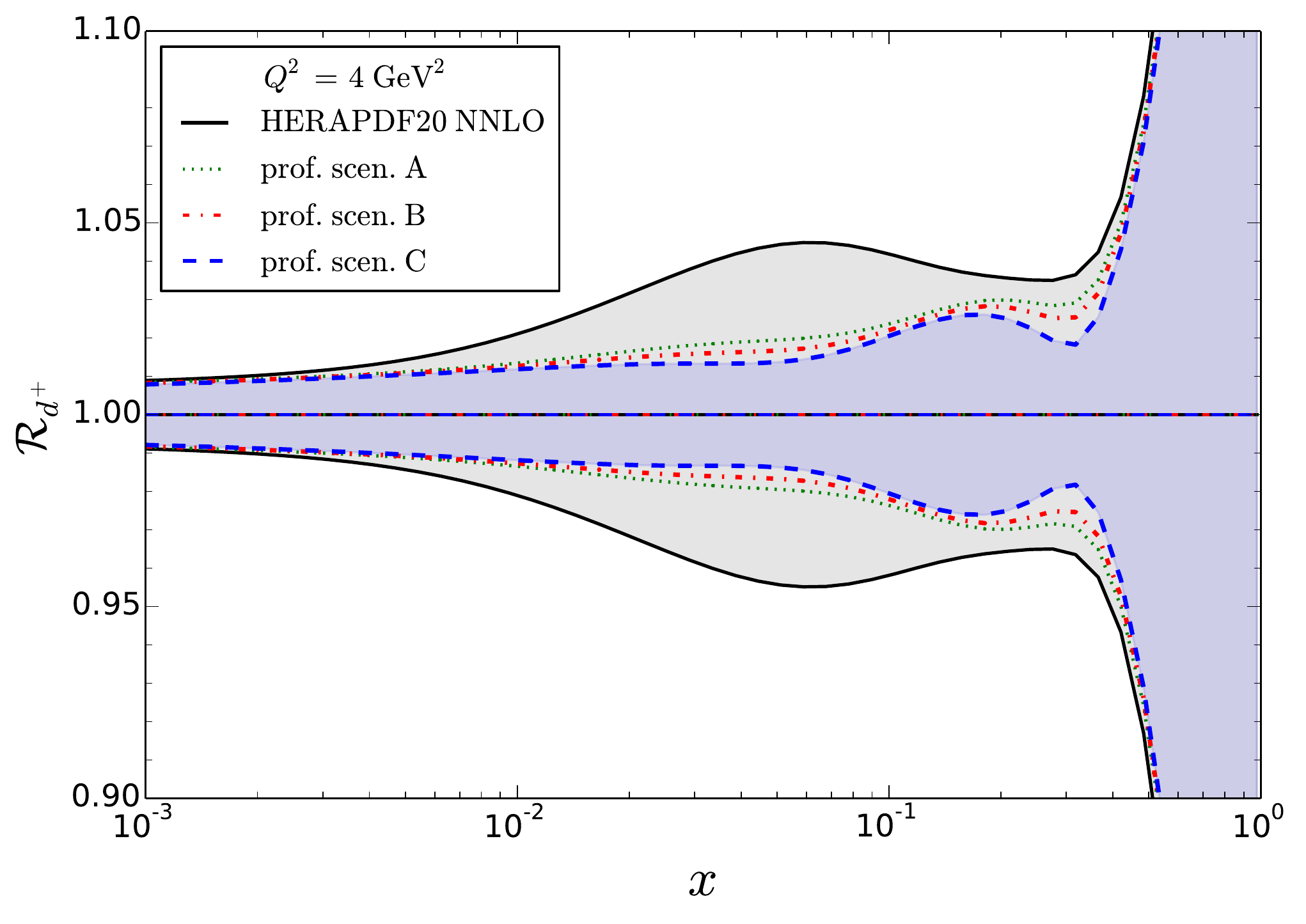}\\
\includegraphics[width=0.45\textwidth]{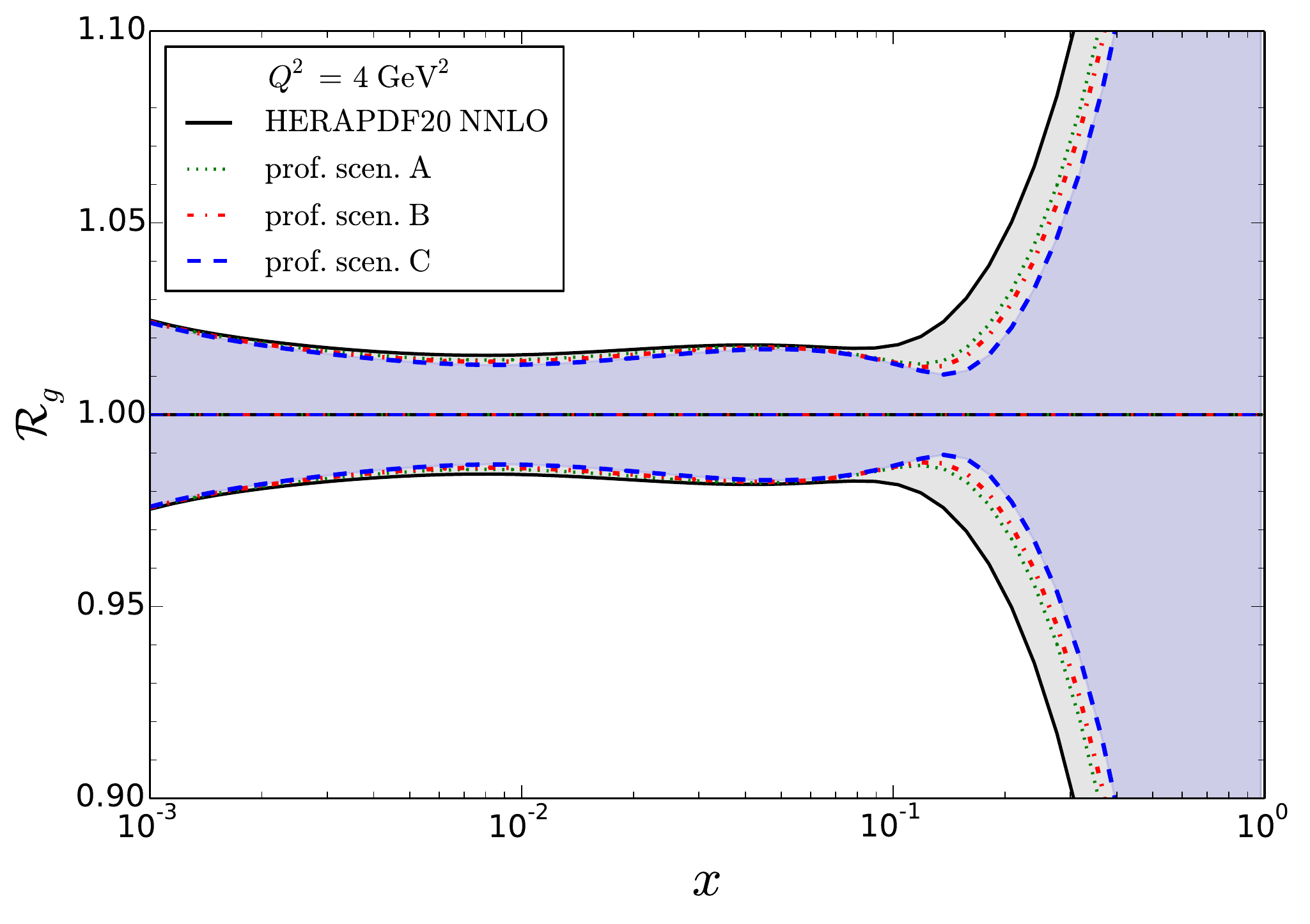}
\includegraphics[width=0.45\textwidth]{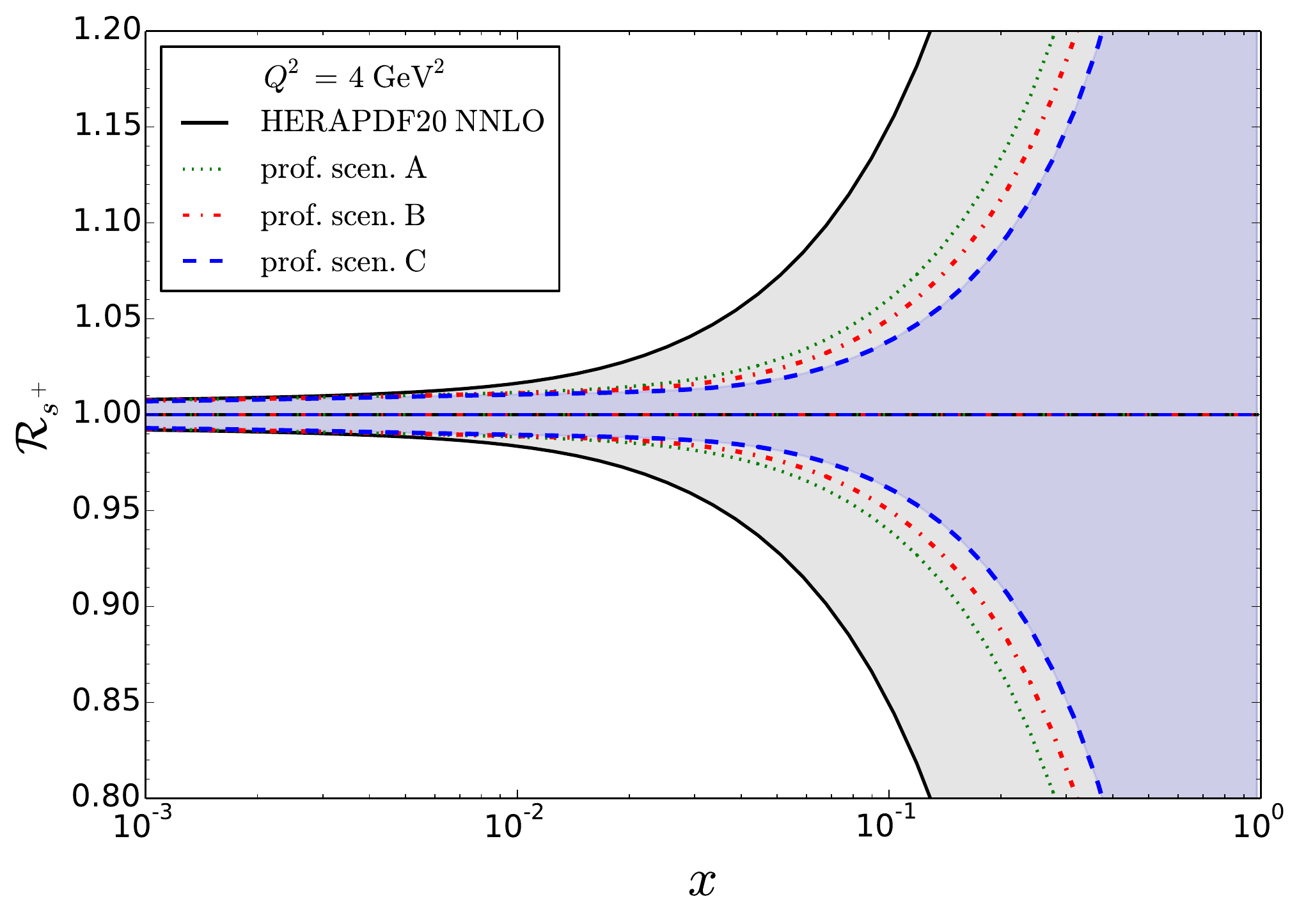}
\caption{\small Comparison
of the $u^+$, $d^+$, $g$, and $s^+$ PDFs at the scale of $Q^2=4\text{ GeV}^2$
between the original  HERAPDF2.0 set and the results of the profiling
exercise accounting for the constraints of
the lattice-QCD moments
pseudo-data in Scenarios~A, B and C.
Only the {\it experimental} PDF uncertainties are shown,
but not the {\it model} and {\it parametrization} variations.
}
\label{fig:pdfsProf}
\end{figure}

In Fig.~\ref{fig:pdfsProf} we present a comparison of the
$u^+$, $d^+$, $g$, and $s^+$ PDFs at the scale of $Q^2=4\text{ GeV}^2$
between the original  HERAPDF2.0 set and the results of the profiling
exercise for Scenarios~A, B and C.
Only the {\it experimental} PDF uncertainties are shown in this comparison,
but not the {\it model} and {\it parametrization} variations.
The corresponding results based on the reweighting
of NNPDF3.1 were shown in Figs.~\ref{fig:impactUnpol}
and~\ref{fig:impactUnpollargex}.

From Fig.~\ref{fig:pdfsProf} we see that, as expected, the impact of the 
lattice pseudo-data is greatest in the medium and large-$x$ regions.
The precise impact on the PDFs is rather
similar for the three scenarios, with the most optimistic
Scenario~C leading to the largest reduction in uncertainties.
The quark flavor combinations that are most affected by the
lattice-QCD pseudo-data are the $d^{+}$ and $s^{+}$ PDFs,
and, to a lesser extent, the gluon PDF.
The improvement in the PDF uncertainties for $d^{+}$ and $s^{+}$
occurs because the DIS data
used in HERAPDF2.0 include only limited constraints
on quark flavor separation, and, for these PDFs, the lattice-QCD 
pseudo-data add important new information.

\subsection{Impact of lattice calculations of  $x$-space PDFs}
\label{sec:projectionsxspace}

In the previous section, we studied the impact of
lattice-QCD calculations of PDF moments. 
We now perform an exploration of the
potential constraints that future lattice QCD calculations
of $x$-space PDFs can provide on global analyses.
We focus on the isotriplet
combination $x u-x d$ (and $x\Delta u - x\Delta d$
in the polarized case), the quark combination 
on which initial studies have been focused, as
it is the simplest to calculate, owing to the lack of disconnected
diagrams and the absence of mixing with other quark flavors or with gluons.

Following the same Bayesian reweighting procedure employed for PDF moments
in Sec.~\ref{sec:projections:rw},
we have generated pseudo-data for the isotriplet
combinations
\be
\label{eq:isotriplet_unpol}
u(x_i,Q^2)-d(x_i,Q^2) \, \quad\text{and} \, \quad
\bar{u}(x_i,Q^2)-\bar{d}(x_i,Q^2) \, , \quad i=1,\ldots,N_x \, ,
\ee
for the unpolarized case, and for
\be
\label{eq:isotriplet_pol}
\Delta u(x_i,Q^2)-\Delta d(x_i,Q^2) \, \quad\text{and} \, \quad
\Delta\bar{u}(x_i,Q^2)-\Delta\bar{d}(x_i,Q^2) \, , \quad i=1,\ldots,N_x \, ,
\ee
for the polarized case, with $N_x$ being the number of points
in $x$-space that are being sampled.
We take $Q^2=4\text{ GeV}^2$, consistent with our choices for the exercise 
performed in Secs.~\ref{sec:projections:rw}--\ref{sec:hessianprofiling}.

We consider three scenarios, denoted by Scenarios D, E, and F,
for the total uncertainty $\delta_L^{(i)}$
that will be assigned to
the lattice-QCD calculations of the specific quark
combinations listed in Eqs.~\eqref{eq:isotriplet_unpol}
and \eqref{eq:isotriplet_pol}.
Lattice-QCD computations are expected to have the smallest systematic 
uncertainties at large $x$, so we choose the $N_x=5$ points to be
\be
x_i = 0.70\, ,0.75,\, 0.80,\, 0.85, \, 0.90 \, .
\ee
For each scenario, we assume the same relative error for each value of 
$\{x_i\}$, and we neglect possible correlations between 
neighboring $x$-points.
We assume uncertainties of $\delta_{L}^{(i)}=12\%, 6\%$ and 3\% for scenarios
D, E, and F, respectively.
Note that we assume the same values of $\delta_{L}^{(i)}$ for the polarized
and unpolarized cases, as well as for both the quark
and antiquark isotriplet combinations Eqs.~\eqref{eq:isotriplet_unpol}
and \eqref{eq:isotriplet_pol}.

\begin{figure}[!t]
\centering
\includegraphics[scale=0.45]{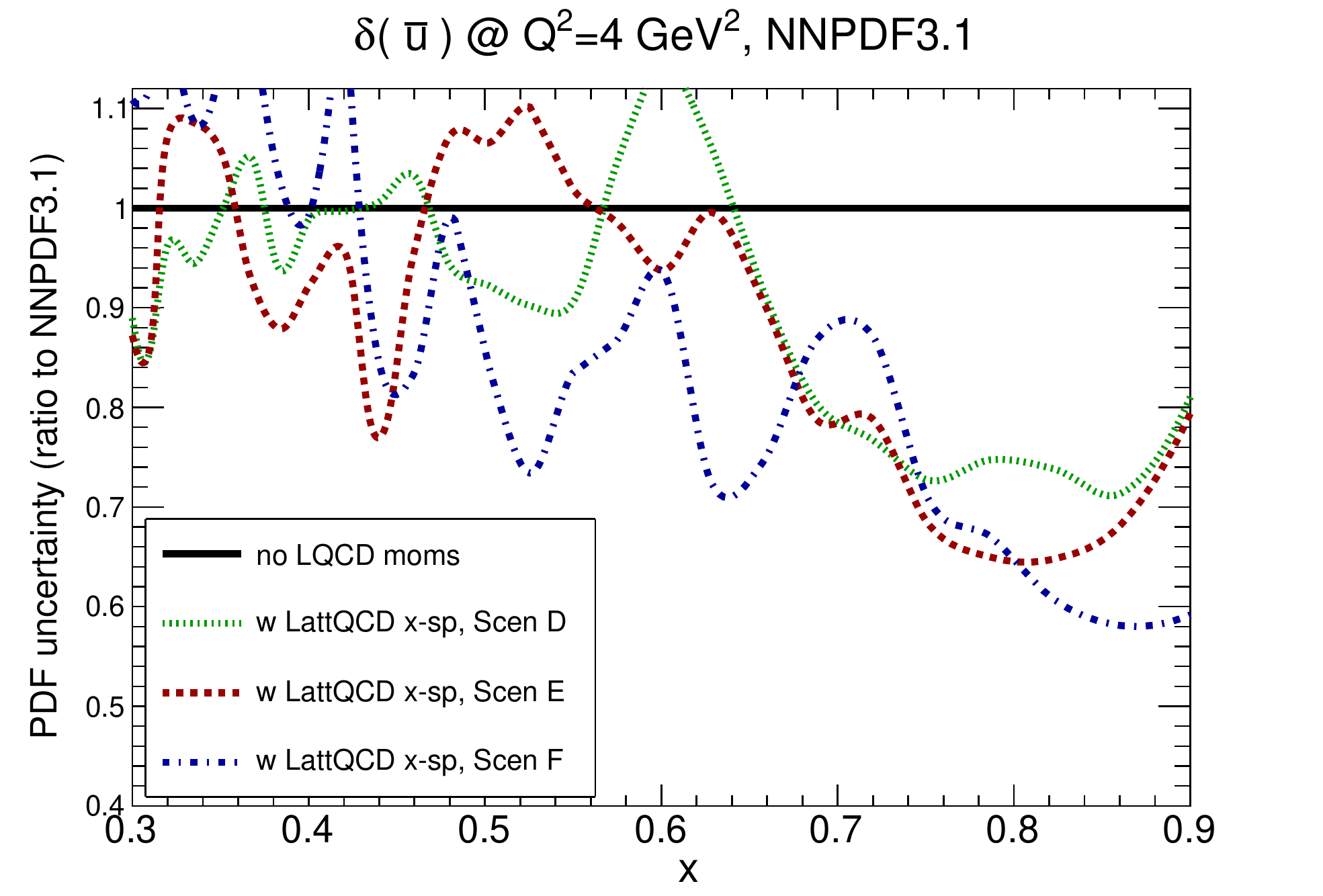}
\includegraphics[scale=0.45]{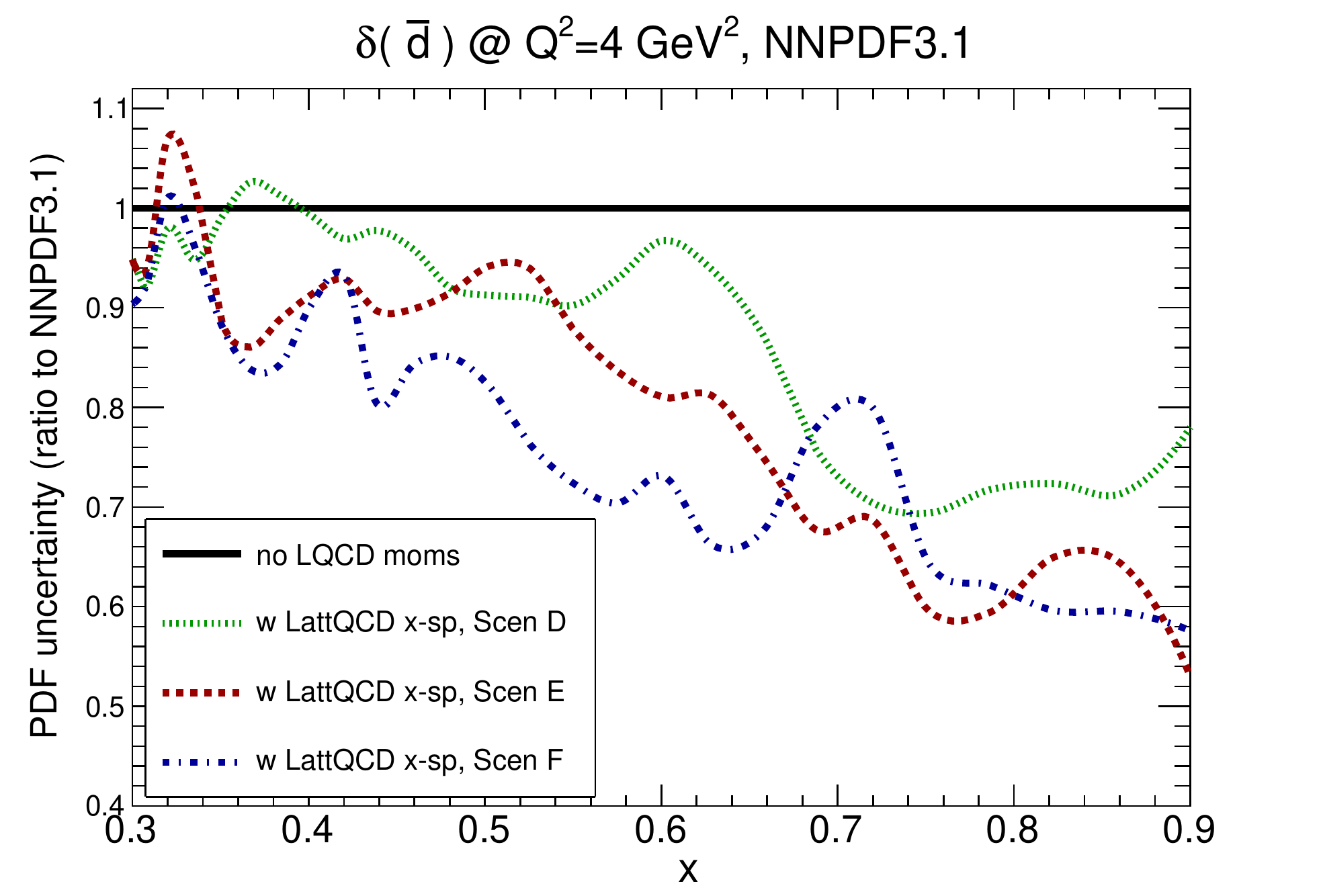}
\includegraphics[scale=0.45]{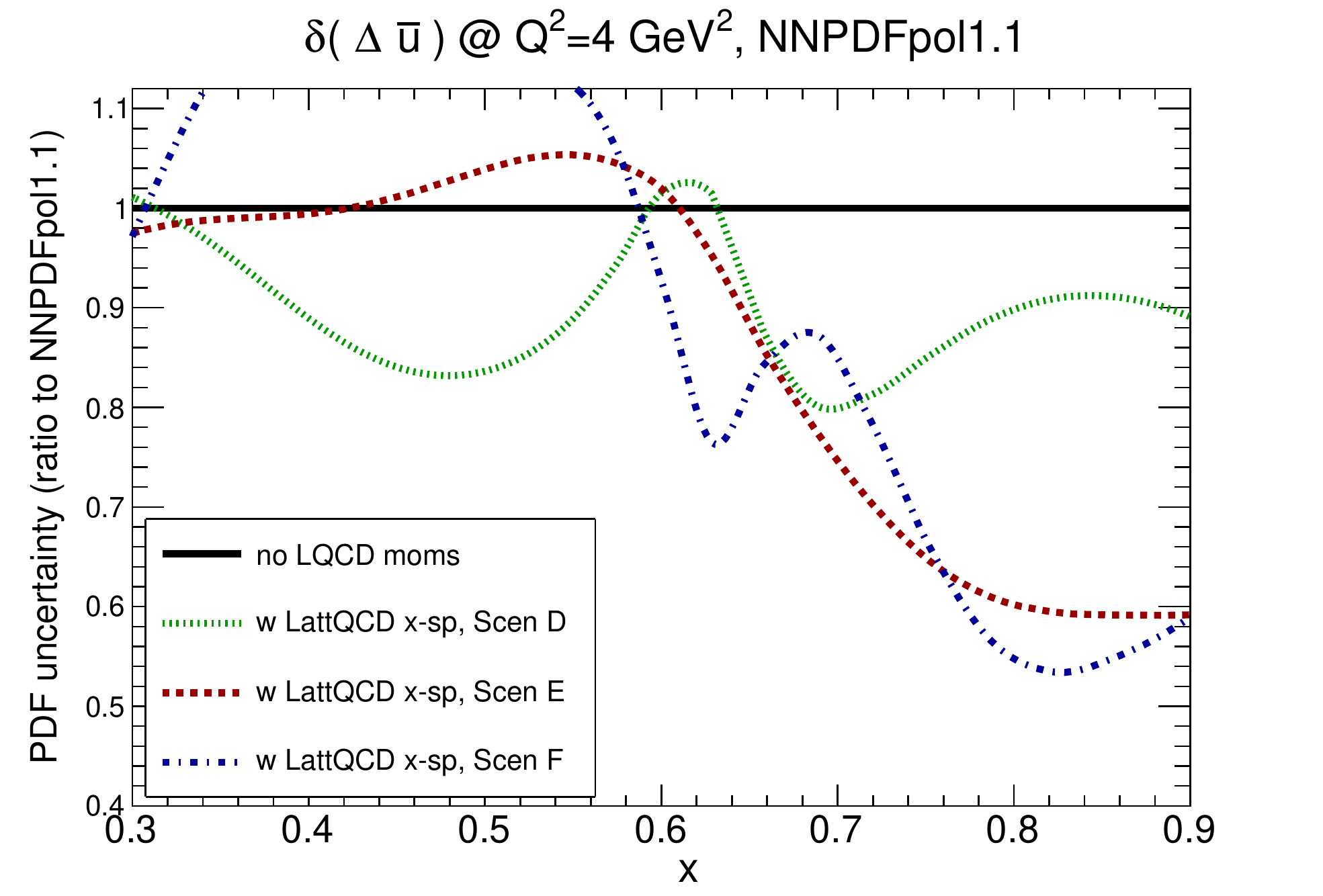}
\includegraphics[scale=0.45]{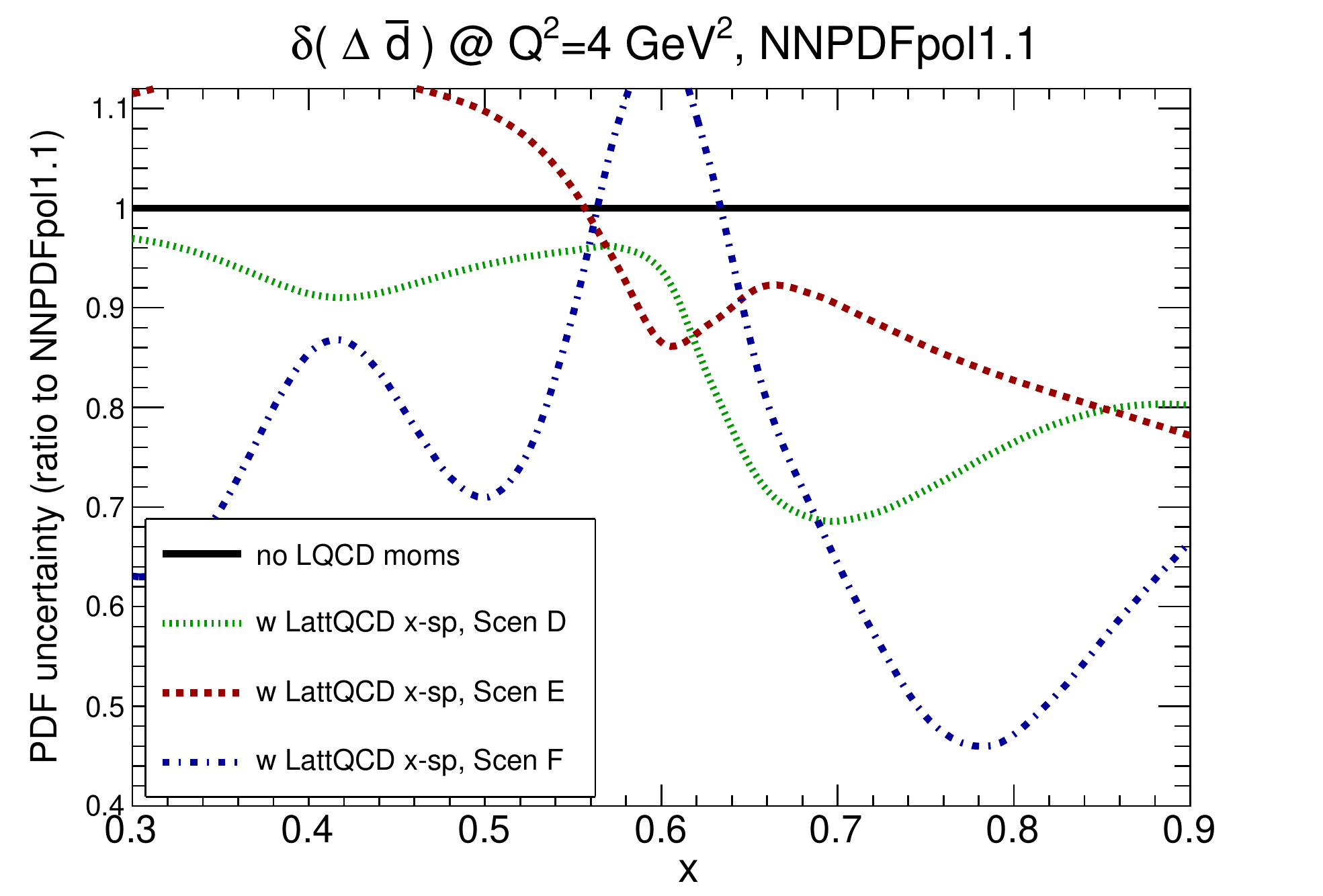}
\caption{\small The ratio of PDF uncertainties to the original
  NNPDF3.1 (NNPDFpol1.1) in the fits where lattice-QCD pseudo-data
  on $x$-space PDFs have been added to the global unpolarized
  (polarized) analysis.
  Specifically, we show the impact on the PDF uncertainties
  in $\bar{u}$ and $\bar{d}$ at large-$x$ in the upper
  plots, with the corresponding comparison for $\Delta\bar{u}$
  and $\Delta\bar{d}$ in the lower plots.
}    
\label{fig:impactxspace}
\end{figure}

We summarize the results of this exercise in Fig.~\ref{fig:impactxspace}, 
where we plot the ratio of the PDF uncertainties in each Scenarios A, B and C 
(D, E and F) to the uncertainty of the original
NNPDF3.1 (NNPDFpol1.1) set.
We show the impact on the PDF uncertainties
in $\bar{u}$ and $\bar{d}$ at large-$x$ in the upper
plots, with the corresponding comparison for $\Delta\bar{u}$
and $\Delta\bar{d}$ in the lower plots.
We concentrate on the results for the individual quark flavors, even though 
the constraints are imposed on differences between flavors, as the former are 
of the more direct interest for phenomenology. 
From this comparison, we find that lattice-QCD calculations of the 
$x$-dependence of PDFs can significantly reduce the uncertainties for both 
unpolarized and polarized antiquarks in the large-$x$ region.
Taking into account that the PDF uncertainties on the large-$x$
antiquarks are rather large, and that they
enter a number of important Beyond the Standard Model (BSM) search channels
(such as for instance for production of new heavy gauge bosons $W'$ and $Z'$),
our analysis demonstrates that such calculations would have direct
phenomenological implications.
We note however that the curves in Fig.~\ref{fig:impactxspace}
fluctuate by a rather large amount.
This might be due to the fact that the uncertainties of the original PDFs
fluctuate, particularly at low scales.

Fig.~\ref{fig:impactxspace} shows that
in the unpolarized case the large-$x$ PDF uncertainties could be reduced
to $60\%$ of their original value.
We also find that there are no large
differences between the three
scenarios,
probably because the constraint is on quark differences not on individual 
flavors, so there is freedom for $\bar u$ and $\bar d$ to vary in a correlated 
fashion while still satisfying the constraint. 
However, it does suggest 
that a direct lattice-QCD calculation
of $x \bar{u}-x \bar{d}$ does not need to reach uncertainties
at the few-percent level to influence global fits.
For the polarized PDFs, Fig.~\ref{fig:impactxspace} demonstrates that the
reduction in PDF uncertainties could be significantly more marked.
For instance, in the case of $\Delta \bar{d}$, at $x\simeq 0.8$
the resulting PDF uncertainty from Scenario~F is less than 50\%
of the original uncertainty.

In Table~\ref{tab:neffxspace} we indicate the effective number of replicas
$N_\text{eff}$, Eq.~\eqref{eq:effnrep}, remaining when
the lattice-QCD pseudo-data for Eqns.~\eqref{eq:isotriplet_unpol}
and~\eqref{eq:isotriplet_pol} are included in the global
unpolarized and polarized fits.
Here we find a marked decrease in $N_\text{rep}$
for the three scenarios,
in particular for the unpolarized case.
For example, in the most optimistic Scenario~F, only
64 effective replicas remain out of the
original sample of $N_\text{rep}=1000$ replicas.
See Table~\ref{tab:neff} for the corresponding
information at the level of PDF moments.
   
\begin{table}[!t]
\centering
\footnotesize
\renewcommand{\arraystretch}{1.3} 
\begin{tabular}{lcc}
\toprule
&  NNPDF3.1  &  NNPDFpol1.1 \\
\midrule
$N_\text{rep}$ original   & 1000 & 100 \\
$N_\text{eff}$ Scenario D &  376 &  41 \\
$N_\text{eff}$ Scenario E &  173 &  35 \\
$N_\text{eff}$ Scenario F &  64  &  22 \\
\bottomrule
\end{tabular}
\caption{\small The effective number of replicas
  $N_\text{eff}$, Eq.~\eqref{eq:effnrep}, remaining when the pseudo-data
  on the lattice-QCD calculations
  of Eqns.~\eqref{eq:isotriplet_unpol}
  and~\eqref{eq:isotriplet_pol} 
  are included in the global
  unpolarized and polarized fits. 
  \label{tab:neffxspace}
  }
\end{table}

We emphasize that the results of this exercise must be interpreted
with some care.
First of all, the results depend sensitively on the specific values of
$\left\{ x_i \right\}$
that we have assumed for the lattice-QCD calculation,
and on the values
of the associated uncertainties $\delta_L^{(i)}$.
The quantitative results depend on the choice of input PDF set and would 
vary if, for example, the input set were the HERAPDF2.0 set used for the 
Hessian profiling exercise of Sec.~\ref{sec:hessianprofiling}.
Even with these caveats, our analysis makes clear that a direct
computation of the isotriplet combination $x u-x d$ on the lattice
has the potential to constrain the large-$x$ PDFs in
a more significant way than corresponding PDF moment calculations,
particularly in the unpolarized case.
Given the importance of antiquark PDFs in the large-$x$ region for LHC 
phenomenology (especially for a high-luminosity run), 
pursuing these calculations should be high on the list
of priorities for the lattice-QCD community.

\subsection{Discussion}

We conclude this section with a brief discussion of the main lessons that
can be learned from this exercise, which provides the first quantitative 
estimate of the impact of present and future lattice-QCD calculations of PDF 
moments and $x$-space PDFs, for both polarized and unpolarized PDFs.

First, we have demonstrated that in the polarized case,
even with current uncertainties, lattice-QCD calculations of
selected PDF moments can impose sizable constraints on several
important polarized quark combinations.
This suggests that global polarized PDF analyses should consider
including existing lattice-QCD calculations in their fits to constrain some
of the least known quark combinations, such as the total strangeness.
The situation is rather different in the unpolarized case,
where a reduction of the current lattice-QCD uncertainties by a factor of 
between five and ten seems to be required to influence global fits.
This difference arises because unpolarized PDFs are known with much higher 
precision than polarized PDFs, thanks to the much wider amount of experimental 
data sensitive to unpolarized PDFs,
including the constraints from recent high-precision measurements at the LHC.
Thus, in addition to the differences highlighted  in Fig.~\ref{fig:Bmomsunp},
much more precise lattice-QCD calculations than in the polarized case 
need to be used to be competitive with current PDF fits.

Second, lattice-QCD calculations of the quark isotriplet combinations
$xu-xd$ and $x\bar{u}-x\bar{d}$ would be instrumental in constraining
quark PDFs at large $x$.
Even a calculation with $\delta_L\simeq 10\%$ uncertainties at large-$x$ would
start to provide useful constraints on global fits.
Moreover, we find that, in the unpolarized case, the information on the
PDFs that could be derived from a direct $x$-space calculation
from lattice-QCD is clearly superior to the information that can be obtained
from PDF moments alone, at least for the subset of PDFs and moments used in 
the present exercise.

The profiling studies presented in this section could be extended in
a number of directions.
In the polarized case, one could include the current lattice-QCD
values of the moments listed in Table~\ref{tab:BMpol} in global analyses: 
indeed, we have demonstrated that at the current level of uncertainties one 
expects to find some non-trivial constraints.
In this respect, a crucial topic to investigate is the compatibility 
(or lack thereof) of the existing lattice-QCD numbers compared to constraints 
from experimental data.
For both unpolarized and polarized PDFs, it would be interesting to include the 
effects of other moments and flavor combinations.
Higher moments, in particular, typically probe regions of higher $x$, compared
to lower moments, and in the large-$x$ regions uncertainties in the global-fit 
PDFs are more marked.
One could also consider the effects of the quark combinations for which 
$x$-space calculations might be available, for example those related to the 
proton strangeness.
Finally, a more refined analysis should include the theoretical correlations
expected in lattice-QCD calculations, for instance, in the case of $x$-space 
calculations, one expects neighboring points in $x$ to be highly correlated.

\section{Outlook}
\label{sec:outlook}

The study of the PDFs of the proton is an active interdisciplinary research 
field lying at the crossroads of high-energy, hadronic and nuclear physics, 
with important applications in astroparticle physics.
In this white paper, we have reviewed our current knowledge of PDFs as
determined from both the global analysis framework
and from lattice-QCD computations. 
We have established a common language between the two communities, 
to facilitate interactions between them.
We have presented a first systematic comparison between state-of-the-art 
lattice-QCD calculations of PDF moments and the corresponding results from
global analyses both in the unpolarized and polarized cases.
Our results suggest that the improvement in accuracy and precision required 
in lattice-QCD calculations to match the first moments of PDFs 
determined from a global fit is larger in the unpolarized case than in the 
polarized.
We have provided additional benchmark numbers from the global fits 
for the higher moments not used in this benchmark comparison, which can be
used to validate future lattice-QCD calculations.

The main outcome of this white paper is the first
quantitative study of the impact of lattice-QCD calculations
in the global fits, based on both PDF moments and on Bjorken-$x$ dependence 
pseudo-data, assuming a number of different scenarios for
the associated uncertainties.
In the case of unpolarized PDFs, we have demonstrated that a reduction 
of the uncertainties of current lattice-QCD calculations is needed 
to provide any impact on global PDF fits.
In the case of polarized PDFs, we have shown that current lattice-QCD
calculations can provide useful input into global-PDF analyses.
Although the studies presented here are still in an initial exploratory phase, 
they provide strong motivation for global fitters to begin consider 
incorporating lattice-QCD constraints into their global analyses.

The studies presented in this white paper can be extended in a number of 
directions.

First, we have restricted our benchmark comparison only to the lowest moments 
of polarized and unpolarized PDFs, whose various sources of systematic 
uncertainties have been computed with the greatest control.
Future work should extend this comparison to higher PDF moments,
which will have some impact on PDF fits, provided the precision and accuracy
of lattice-QCD calculations keep improving.
More elaborate benchmarks could be performed, for instance on PDF moments
truncated from below to better take into account the fiducial $x$ region.
Appropriate benchmarks could also be devised to test lattice-QCD methods
that aim at the determination of the $x$ dependence of the PDFs.
We emphasize that modifications to the rating system adopted to
characterize lattice-QCD calculations of PDF moments should occur 
as lattice-QCD results will evolve.

Second, a similar benchmark exercise between global fit results and 
lattice-QCD calculations should be performed at the level of
$x$-space calculations.
It will be important to compare in detail the available lattice-QCD results 
with state-of-the-art global fits, to validate the former and
thereby demonstrate to what extent lattice-QCD calculations of $x$-space PDFs 
can contribute to global fits.

Third, it should be possible to assess the impact of lattice-QCD 
calculations on other nonperturbative objects
determined from global analyses of experimental data.
Examples of these include the transversity (see Ref.~\cite{Lin:2017stx}
for a recent study), transverse-momentum dependent 
PDFs (TMDs), generalized PDFs (GPDs) (see, {\it e.g.},~\cite{Angeles-Martinez:2015sea,Musch:2011er,Engelhardt:2015xja,Yoon:2017qzo} and references therein), 
or collinear PDFs for hadrons other than 
protons~\cite{Sutton:1991ay,Burkardt:2001jg}.
All these quantities are known with much less precision than unpolarized
and polarized PDFs, given that the corresponding experimental information
is rather scarce. 
In this case, lattice-QCD calculations could have the potential
to provide new information, without the need of high precision.

In summary, the aim of this study has been to build a bridge between the 
lattice-QCD and global-fit communities.
Our final goal is for lattice-QCD calculations to provide novel inputs into 
polarized and unpolarized PDF fits. 
Precise lattice-QCD results could reduce the uncertainties of
global PDF fits and/or discriminate between different sets.
We hope this white paper motivates the lattice-QCD and global-fit
communities to continue fruitful interactions to improve our knowledge of PDFs.

\subsection*{Acknowledgments}

We are grateful to Jacqueline Gills and Michelle Bosher for their help in the
organization of the workshop.
We thank Rodolfo Sassot and Jacob Ethier for providing us with the 
DSSV14 predictions in Table~\ref{tab:polgmom} and with the JAM17 predictions in
Tables~\ref{tab:polPDFmoms},~\ref{tab:polHmoms} and~\ref{tab:polgmom}.
The workshop was partly supported by the European Research Council (ERC) via 
the Starting Grant {\it PDF4BSM - Parton Distributions in the Higgs Boson Era}.
We also thank the Department of Energy's (DoE) Institute of Nuclear Theory 
(INT) at the University of Washington in Seattle for partial support during 
the completion of this work.
This work was also partially funded by the U.S. DoE contract 
No.~DE-AC05-06OR23177, under which Jefferson Science Associates, 
LLC operates Jefferson Lab. 
H.-W.L. is supported by the U.S. National Science Foundation (NSF) under Grant 
PHY 1653405; E.R.N. by the U.K. Science and Technology Facilities Council 
(STFC) via the Rutherford Grant ST/M003787/1 and the Consolidated Grant 
ST/P000630/1; K.O. by Jefferson Science Associates, LLC under U.S. 
DoE contract No.~DE-AC05-06OR23177, 
by U.S. DoE under Grant No.~DE-FG02-04ER41302, 
and by STFC via the Consolidated Grant ST/P000681/1; 
F.I.O. and P.M.N. by the U.S. 
DoE under Grant No.~DE-SC0010129; J.R. by the ERC via the Starting 
Grant {\it PDF4BSM - Parton Distributions in the Higgs Boson Era} and by the 
Dutch Organization for Scientific Research (NWO) under Grant N0.~680-91-105;
A.A. acknowledges support by the U.S. DoE under contract No.~DE-SC008791;
A.B. and G.B. by the ERC via the Consolidator Grant {\it 3DSPIN - Mapping the
proton in 3D};
J.-W.C. partly by the Ministry of Science and Technology, Taiwan,
under Grant No. 105-2112-M-002-017-MY3 and the Kenda Foundation;
S.C. by the Deutsche Forschungsgemeinschaft via the SFB/TRR 55 project;
M.C. by the U.S. DoE, Office of Science, Office of Nuclear Physics, within the 
framework of the TMD Topical Collaboration, as well as by the NSF under Grant 
No.~PHY-1714407;
L.D.D. by the Royal Society, Wolfson Research Merit Award,
Grant No. WM140078:
M.E. by the U.S. DoE, Office of Science, Office of Nuclear Physics through 
Grant DE-FG02-96ER40965 as well as through the TMD Topical Collaboration;
R.G. by the U.S. DoE, Office of Science of High Energy Physics,
under Contract No.~DE-KA-1401020;
L.A.H.-L. and R.S.T. by the STFC via Grant awards ST/L000377/1 and ST/P000274/1;
S.L. by the U.S. DoE through Grant No.~DE-SC0016286;
C.J.M. by the U.S. DoE through Grant No.~DE-FG02-00ER41132;
T.I. by Science and Technology Commission of Shanghai Municipality 
(Grants No.~16DZ2260200) and in part by the DoE, Laboratory Directed Research 
and Development (LDRD) funding of BNL, under contract No.~DE-EC0012704;
C.-P.Y. by the NSF under Grant No.~PHY-1417326.

\appendix

\section{Definition of the PDF moments}
\label{app:notation}

In this appendix, we summarize the conventions adopted in this paper to denote 
the moments of relevant unpolarized and polarized PDF combinations.
We focus on the quantities which can be presently computed in lattice QCD,
although those used for benchmarks in Sec.~\ref{sec:benchmarking} are only
a subset of them.
In the equations below, we use the shorthand notation
\begin{equation}
q^\pm \equiv q\pm\bar{q}\, 
\quad\text{ and }\quad
\Delta q^\pm \equiv \Delta q\pm\Delta\bar{q}\, 
,\qquad q=u,d,s,c \,,
\end{equation}
for unpolarized and polarized PDFs respectively.
We identify $\mu$ with the QCD factorization scale and $Q$ with the 
characteristic scale of a given hard-scattering process.
The use of the following notation is strongly recommended for any comparison 
between lattice-QCD computations and global-fit determinations of 
PDF moments.

\begin{itemize}

\item Unpolarized moments.

\begin{enumerate}

\item The first moment of the total $u^+-d^+$ PDF combination
\begin{equation}
\left.\langle x\rangle_{u^+-d^+}(\mu^2)\right|_{\mu^2=Q^2}
=
\int_0^1 dx\, x\left\{u(x,Q^2)+\bar{u}(x,Q^2)-d(x,Q^2)-\bar{d}(x,Q^2)\right\} \, .
\label{eq:unpfmumdtot}
\end{equation}

\item The second moment of the valence $u^--d^-$ PDF combination
\begin{equation}
\left.\langle x^2\rangle_{u^--d^-}(\mu^2)\right|_{\mu^2=Q^2}
=
\int_0^1 dx\, x^2\left\{u(x,Q^2)-\bar{u}(x,Q^2)-d(x,Q^2)+\bar{d}(x,Q^2)\right\} \, .
\label{eq:unpsmumdval}  
\end{equation}

\item The first moment of the individual quark $q^+$ total PDF combination
\begin{equation}
\left.\langle x\rangle_{q^+=u^+,d^+,s^+,c^+}(\mu^2)\right|_{\mu^2=Q^2}
=
\int_0^1 dx\, x\left\{q(x,Q^2)+\bar{q}(x,Q^2)\right\} \, .
\label{eq:unpfmiqtot}
\end{equation}

\item The second moment of the individual quark $q^-$ valence PDF combination
\begin{equation}
\left.\langle x^2\rangle_{q^-=u^-,d^-,s^-,c^-}(\mu^2)\right|_{\mu^2=Q^2}
=
\int_0^1 dx\, x^2\left\{q(x,Q^2)-\bar{q}(x,Q^2)\right\} \, .
\label{eq:unpsmiqval}
\end{equation}

\item The first moment of the gluon PDF
\begin{equation}
\left.\langle x \rangle_g(\mu^2)\right|_{\mu^2=Q^2}
=
\int_0^1 dx\, x\, g(x,Q^2) \, .
\label{eq:unpfmg}
\end{equation}

\end{enumerate}

\item Polarized moments.

\begin{enumerate}

\item The zeroth moment of the total $u^+-d^+$ PDF combination
\begin{equation}
\left.\langle 1 \rangle_{\Delta u^+-\Delta d^+}(\mu^2)\right|_{\mu^2=Q^2}
=
\int_0^1 dx \left\{\Delta u(x,Q^2)+\Delta\bar{u}(x,Q^2)
-\Delta d(x,Q^2)-\Delta\bar{d}(x,Q^2)\right\} \, .
\label{eq:polzmumdtot}
\end{equation}

\item The first moment of the valence $u^--d^-$ PDF combination
\begin{equation}
\left.\langle x\rangle_{\Delta u^--\Delta d^-}(\mu^2)\right|_{\mu^2=Q^2}
=
\int_0^1 dx\, x\left\{\Delta u(x,Q^2)-\Delta\bar{u}(x,Q^2)-\Delta d(x,Q^2)+\Delta \bar{d}(x,Q^2)\right\}
\label{eq:polfmumdval}  
\end{equation}

\item The zeroth moment of the individual quark $q^+$ total PDF combination
\begin{equation}
\left.\langle 1\rangle_{q^+=\Delta u^+,\Delta d^+,\Delta s^+,\Delta c^+}(\mu^2)\right|_{\mu^2=Q^2}
=
\int_0^1 dx \left\{\Delta q(x,Q^2)+\Delta\bar{q}(x,Q^2)\right\} \, .
\label{eq:polzmiqtot}
\end{equation}

\item The first moment of the individual quark $q^-$ valence PDF combination
\begin{equation}
\left.\langle x\rangle_{\Delta q^-=\Delta u^-,\Delta d^-,\Delta s^-,\Delta c^-}(\mu^2)\right|_{\mu^2=Q^2}
=
\int_0^1 dx\, x\left\{\Delta q(x,Q^2)-\Delta\bar{q}(x,Q^2)\right\} \, .
\label{eq:polfmiqval}
\end{equation}

\end{enumerate}

\end{itemize}

Some of these moments have a direct physical interpretation, see 
Sec.~\ref{Sec:IntroPDFs}.
For instance, Eq.~\eqref{eq:unpfmiqtot} and Eq.~\eqref{eq:polzmiqtot}
correspond respectively to the proton's momentum and spin fractions carried
by a given quark flavor (and its corresponding antiquark) at the scale 
$\mu^2=Q^2$.
Higher moments and/or moments of other flavor combinations are readily
computable from any phenomenological PDF set.
We do not consider them though, as the corresponding lattice-QCD
computations are outside the current reach.

\section{PDF moments from lattice QCD}
\label{sec:LQCDtables}

In this appendix, we summarize additional results for the moments of 
unpolarized and polarized PDFs from lattice QCD that were not discussed in 
Sec.~\ref{subsubsec:BClQCD}, either because the calculations were 
performed in the quenched approximation, or because they were not extrapolated
to the physical pion mass.

\begin{itemize}

\item In Table~\ref{tab:unpolLQCDstatus1B}, we show the first moments of 
unpolarized PDFs $\la x\ra_{u^+-d^+}$, $\la x\ra_{q^+}$ and $\la x\ra_g$
that were not included in Table~\ref{tab:unpolLQCDstatus1}.

\item In Table~\ref{tab:unpolLQCDstatus2B}, we show the second moments of 
unpolarized PDFs $\la x^2\ra_{u^--d^-}$, $\la x^2\ra_{u^-}$ and $\la x^2\ra_{d^-}$.

\item In Table~\ref{tab:polLQCDstatus1B}, we show the zeroth moments of 
polarized PDFs $\la 1\ra_{\Delta u^+}$, $\la 1\ra_{\Delta d^+}$ and 
$\la 1\ra_{\Delta s^+}$ that were not included in Table~\ref{tab:polLQCDstatus0}.

\item In Table~\ref{tab:polLQCDstatus2B}, we show the first moments of 
polarized PDFs $\la x\ra_{\Delta u^--\Delta d^-}$, $\la x\ra_{\Delta u^-}$ and  
$\la x\ra_{\Delta d^-}$ that were not included in Table~\ref{tab:polLQCDstatus1}.

\end{itemize}
All results are displayed at $\mu^2=4$~GeV$^2$.
The characterization of each source of systematic uncertainty follows the
conventions delineated in Sec.~\ref{subsubsec:BClQCD}, to which the reader 
is referred for the meaning of each symbol in 
Tables~\ref{tab:unpolLQCDstatus1B}--\ref{tab:polLQCDstatus2B}.
Moments are denoted according to the notation introduced in 
Appendix~\ref{app:notation}.

\begin{table}[!t]
\renewcommand{\arraystretch}{1.2} 
\centering 
\footnotesize
\begin{threeparttable}
\begin{tabular}{llcllccccccl}
\toprule
Mom. & Collab. & Ref. & $N_f$ & Status & Disc~[fm] & QM & FV & Ren & ES & & \\
\midrule
$\langle x\rangle_{u^+-d^+}$
& ETMC\,15
  & \cite{Abdel-Rehim:2015owa} 
  & 2+1+1 
  & P 
  & 0.06,0.08
  & ---  
  & \rsquare,\bstar 
  & \bstar,\bstar 
  & \rsquare,\bstar  
  &  
  & Fig.~\ref{fig:latt_res}~(a) \\
& ETMC\,15 
  & \cite{Abdel-Rehim:2015owa} 
  & 2 
  & P 
  & 0.06--0.09 
  & ---  
  & \bcirc 
  & \bstar 
  & \rsquare  
  &  
  & Fig.~\ref{fig:latt_res}~(a) \\
& RQCD\,14 
  & \cite{Bali:2014gha} 
  & 2 
  & P 
  & 0.06--0.08
  & --- 
  & \bcirc 
  & \bstar  
  & \bcirc  
  &  
  & Fig.~\ref{fig:latt_res}~(a) \\
\midrule
$\langle x\rangle_{q^+}$
& ETMC\,13 
  & \cite{Abdel-Rehim:2013wlz} 
  & 2+1+1 
  & P 
  & 0.08
  & --- 
  & \bstar  
  & \bstar  
  & \bstar  
  & $\&$ 
  & Fig.~\ref{fig:latt_res}~(b) \\
& $\chi$QCD\,13 
  & \cite{Deka:2013zha} 
  & 0 
  & P 
  & \rsquare  
  & \rsquare 
  & \rsquare  
  & \bcirc  
  & \rsquare
  & $\dagger\ddag$ 
  & $\langle x\rangle_{u^+}=0.451(37)$,\\
& $\chi$QCD\,13 
  & \cite{Deka:2013zha} 
  & 0 
  & P 
  & \rsquare  
  & \rsquare 
  & \rsquare  
  & \bcirc  
  & \rsquare
  & $\dagger\ddag$ 
  & $\langle x\rangle_{d^+}=0.188(20)$,\\
& $\chi$QCD\,13 
  & \cite{Deka:2013zha} 
  & 0 
  & P 
  & \rsquare  
  & \rsquare 
  & \rsquare  
  & \bcirc  
  & \rsquare 
  & $\dagger\ddag$ 
  & $\langle x\rangle_{s^+}=0.024(6)$\\
\midrule
$\langle x\rangle_{g}$
& ETMC\,13 
  & \cite{Alexandrou:2016ekb} 
  & 2+1+1 
  & P 
  & 0.08
  & --- 
  & \bstar  
  & \bcirc  
  & \bstar  
  &
  & Fig.~\ref{fig:latt_res}~(c) \\
& $\chi$QCD\,13 
  & \cite{Deka:2013zha} 
  & 0 
  & P 
  & \rsquare  
  & \rsquare 
  & \rsquare  
  & \bcirc   
  & \bstar 
  & $\ddag$ & 0.334(55) \\
& QCDSF\,12 
  & \cite{Horsley:2012pz} 
  & 0 
  & P 
  & \rsquare  
  & \rsquare 
  & \bstar  
  & \bstar  
  & --- 
  & $\dagger$ & 0.43(7)(5) \\
\bottomrule
\end{tabular}
\begin{tablenotes}
\scriptsize
\item[$\&$] Nonsinglet renormalization is applied.
\item[$\dagger$] The lightest $m_\pi$ has $Lm_\pi\ge 4.0$, however, 
$L\sim 1.6$~fm.
\item[$\ddag$] The connected contribution is only evaluated at one $t_{sep}$.
\end{tablenotes}
\end{threeparttable}
\caption{\small Status of current lattice-QCD calculations of the first 
moments of unpolarized PDFs.
All results are quoted at $\mu^2=4\mbox{ GeV}^2$.
We use the abbreviations Disc (discretization), QM (quark mass),
FV (finite volume), Ren (renormalization) and ES (excited states)
to denote the corresponding sources of uncertainty.}
\label{tab:unpolLQCDstatus1B}
\end{table}

\begin{table}[!t]
\renewcommand{\arraystretch}{1.2} 
\centering
\footnotesize
\begin{threeparttable}
\begin{tabular}{llcllccccccl}
\toprule
Mom. & Collab. & Ref. & $N_f$ & Status & Disc & QM & FV & Ren & ES & & \\
\midrule
$\langle x^2\rangle_{u^--d^-}$
& LHPC and SESAM\,02 
  & \cite{Dolgov:2002zm} 
  & 2 
  & P 
  & \rsquare 
  & \rsquare 
  & \rsquare 
  & \bcirc 
  & \rsquare 
  &  
  & 0.145(69)\\
& QCDSF\,05 
  &\cite{Gockeler:2004wp} 
  & 0 
  & P 
  & \rsquare  
  & \rsquare 
  & \rsquare  
  & \bstar  
  & \rsquare 
  &  
  & 0.083(17)\\
& LHPC and SESAM\,02 
  &\cite{Dolgov:2002zm} 
  & 0 
  & P 
  & \rsquare 
  & \rsquare 
  & \rsquare 
  & \bcirc 
  & \rsquare 
  &  
  & 0.090(68)\\
\midrule
$\langle x^2\rangle_{u^-}$
  & $\chi$QCD\,09 
  & \cite{Deka:2008xr} 
  & 0 
  & P 
  & \rsquare  
  & \rsquare 
  & \rsquare  
  & \bcirc  
  & \rsquare 
  & $\ast$  
  & $0.117(18)$ \\
\midrule
$\langle x^2\rangle_{d^-}$
  & $\chi$QCD\,09 
  & \cite{Deka:2008xr} 
  & 0 
  & P 
  & \rsquare  
  & \rsquare 
  & \rsquare  
  & \bcirc  
  & \rsquare 
  & $\ast$  
  & $0.052(9)$\\
\bottomrule
\end{tabular}
\begin{tablenotes}
\scriptsize
\item[$\ast$] Only the connected contribution is included.
\end{tablenotes}
\end{threeparttable}
\caption{\small Same as Table~\ref{tab:unpolLQCDstatus1B}, but for 
second moments of unpolarized PDFs.}
\label{tab:unpolLQCDstatus2B} 
\end{table}

\begin{table}[!t]
\renewcommand{\arraystretch}{1.2} 
\centering
\footnotesize
\begin{threeparttable}
\begin{tabular}{llcllccccccl}
\toprule
Mom. & Collab. & Ref. & $N_f$ & Status & Disc~[fm] & QM & FV & Ren & ES & & \\
\midrule
$\langle 1\rangle_{\Delta u^+, \Delta d^+}$
& ETMC\,13 
  &\cite{Abdel-Rehim:2013wlz} 
  & 2+1+1 
  & P 
  & 0.08  
  & --- 
  & \bstar  
  & \bstar  
  & \bstar  
  & $\&$ 
  & Fig.~\ref{fig:latt_res}~(e)\\
& LHPC\,17 
  & \cite{Green:2017keo} 
  & 2+1 
  & P 
  & 0.11 
  & --- 
  & \bstar  
  & \bstar  
  & \bstar 
  &  
  & Fig.~\ref{fig:latt_res}~(e)\\
& QCDSF/CSSM\,15 
  & \cite{Chambers:2015bka}  
  & 2+1 
  & P 
  & 0.07  
  & --- 
  & \bstar 
  & \bstar  
  & \bstar   
  & 
  & Fig.~\ref{fig:latt_res}~(e) \\
& QCDSF\,11 
  & \cite{QCDSF:2011aa}  
  & 2 
  & P 
  & 0.07  
  & --- 
  & \bstar 
  & \bstar  
  & \rsquare   
  &   
  & Fig.~\ref{fig:latt_res}~(e)\\
\midrule
$\langle 1\rangle_{\Delta s^+}$
  & ETMC\,13 
  & \cite{Abdel-Rehim:2013wlz} 
  & 2+1+1 
  & P 
  & 0.08  
  & --- 
  & \bstar  
  & \bstar  
  & \bstar  
  & $\&$ 
  & Fig.~\ref{fig:latt_res}~(d)\\
& LHPC\,17 
  & \cite{Green:2017keo} 
  & 2+1 
  & P 
  & 0.11 
  & --- 
  & \bstar  
  & \bstar  
  & \bstar 
  &  
  & Fig.~\ref{fig:latt_res}~(d) \\
& QCDSF/CSSM\,15 
  &\cite{Chambers:2015bka}  
  & 2+1 
  & P 
  & 0.07  
  & --- 
  & \bstar 
  & \bstar  
  & \bstar   
  & 
  & Fig.~\ref{fig:latt_res}~(d) \\
& QCDSF\,11 
  & \cite{QCDSF:2011aa}  
  & 2 
  & P 
  & 0.07  
  & --- 
  & \bstar 
  & \bstar  
  & \bstar   
  &   
  & Fig.~\ref{fig:latt_res}~(d) \\
\bottomrule
\end{tabular}
\begin{tablenotes}
\scriptsize
\item[$\&$] Nonsinglet renormalization is applied.
excited state analysis for $\langle x\rangle_g$ is considered.
\end{tablenotes}
\end{threeparttable}
\caption{\small Same as Table~\ref{tab:unpolLQCDstatus1B}, but for 
zeroth moments of polarized PDFs.}
\label{tab:polLQCDstatus1B}
\end{table}

\begin{table}[!t]
\renewcommand{\arraystretch}{1.2} 
\centering
\footnotesize
\begin{threeparttable}
\begin{tabular}{llcllccccccl}
\toprule
Mom. & Collab. & Ref. & $N_f$ & Status &  
Disc~[fm] & QM & FV & Ren & ES & &  \\
\midrule
$\langle x\rangle_{\Delta u^--\Delta d^-}$
& ETMC\,15 
  & \cite{Abdel-Rehim:2015owa} 
  & 2+1+1 
  & P 
  & 0.06,0.08  
  & --- 
  & \rsquare,\bstar 
  & \bstar,\bstar 
  & \rsquare,\bstar  
  &   
  & Fig.~\ref{fig:latt_res}~(f) \\
& ETMC\,15 
  & \cite{Abdel-Rehim:2015owa} 
  & 2 
  & P 
  & 0.06--0.09  
  & --- 
  & \bcirc 
  & \bstar 
  & \rsquare 
  &  
  & Fig.~\ref{fig:latt_res}~(f) \\
\midrule
$\langle x\rangle_{\Delta u^-}$
& ETMC\,13 
  &\cite{Abdel-Rehim:2013wlz} 
  & 2+1+1 
  & P 
  & 0.08  
  & $373$~MeV 
  & \bstar  
  & \bstar  
  & \bstar 
  & $\&$ 
  &  $0.214(11)$\\
\midrule
$\langle x\rangle_{\Delta d^-}$
& ETMC\,13 
  & \cite{Abdel-Rehim:2013wlz} 
  & 2+1+1 
  & P 
  & 0.08  
  & $373$~MeV 
  & \bstar  
  & \bstar  
  & \bstar 
  & $\&$ 
  & $0.083(11)$\\
\bottomrule
\end{tabular}
\begin{tablenotes}
\scriptsize
\item[$\&$] Nonsinglet renormalization is applied.
\end{tablenotes}
\end{threeparttable}
\caption{\small Same as Table~\ref{tab:unpolLQCDstatus1B}, but for
first moments of polarized PDFs.}
\label{tab:polLQCDstatus2B}
\end{table}

We also provide tables with full bibliographic details.
\begin{itemize}
\item In Table~\ref{tab:latticebibfirst}, for the axial coupling 
$g_A\equiv\langle 1\rangle_{\Delta u^+-\Delta d^+}$.
We do not include quenched results, perturbatively renormalized results, 
and conference proceedings results.

\item In Table~\ref{tablenonisovectorquarkspins}, for the non-isovector quark 
spins.

\item In Table~\ref{tab:unpolarizedisotriplet}, for $\langle x\rangle_{u^+-d^+}$.
We omit quenched and non-renormalized results.

\item In Table~\ref{tab:nonisovectormomfrac}, for the non-isovector momentum 
fractions.

\item In Table~\ref{tab:nonisopolcase}, for 
$\langle x\rangle_{\Delta u^--\Delta d^-}$.

\item In Table~\ref{tab:latticebiblast} for  higher moments of unpolarized
  and polarized PDFs.

\end{itemize}

\begin{table}[!t]
\renewcommand{\arraystretch}{1.2} 
\centering
\footnotesize
\begin{threeparttable}
\begin{tabular}{llllll}
\toprule
Ref. & Sea quarks & Valence quarks & Renormalization & 
$N_{\Delta t}$ & $m_\pi$ (MeV)\\
\midrule
  Mainz '17b* \cite{Capitani:2017qpc} &
  2 clover & clover & Schr\"odinger functional & 4--6 & 193--473\\

  ETMC '17b \cite{Alexandrou:2017hac} &
  2 clover-TM & clover-TM & Rome-Southampton & 3 & 131\\

  CalLat '17b \cite{Berkowitz:2017gql} &
  2+1+1 staggered & domain wall & Rome-Southampton & all & 131--313 \\

  LHPC '17 \cite{Green:2017keo} &
  2+1 clover & clover & Rome-Southampton & 5 & 317 \\

  NME '17 \cite{Yoon:2016jzj} &
  2+1 clover & clover & Rome-Southampton & 1**,4--5 & 172--285 \\

  Mainz '17a \cite{vonHippel:2016wid} &
  2 clover & clover & Schr\"odinger functional & 4--6 & 193--456\\

  Dragos et al.\ '16 \cite{Dragos:2016rtx} &
  3 clover & clover & Rome-Southampton & 1,2**,5 & 460 \\

  PNDME '16 \cite{Bhattacharya:2016zcn} &
  2+1+1 staggered & clover & Rome-Southampton & 3--5 & 128--319\\

  $\chi$QCD '16 \cite{Yang:2015zja} &
  2+1 domain wall & overlap & $Z_A/Z_V=1$ & 3 & 330 \\

  ETMC '15b \cite{Abdel-Rehim:2015owa} &
    2 clover-TM & \multicolumn{4}{l}{superseded by ETMC '17} \\
  & 2 twisted mass & twisted mass & Rome-Southampton & 1 & 262--470\\
  & 2+1+1 twisted mass & twisted mass & & 1, 4 & 213, 373\\

  RQCD '15 \cite{Bali:2014nma} &
  2 clover & clover & Rome-Southampton & 1--5 & 150--490\\

  PNDME '14 \cite{Bhattacharya:2013ehc} &
  \multicolumn{5}{l}{superseded by PNDME '16} \\

  QCDSF '14 \cite{Horsley:2013ayv} &
  2 clover & clover & $g_A/f_\pi \times f_\pi^\text{phys}$ & 1,5 & 157--1591 \\

  LHPC '14 \cite{Green:2012ud} &
  2+1 clover & clover & Rome-Southampton & 3 & 149--356\\

  ETMC '13 \cite{Alexandrou:2013joa} &
  \multicolumn{5}{l}{superseded by ETMC '15b} \\

  CSSM '13 \cite{Owen:2012ts} &
  2+1 clover & clover & Schr\"odinger functional & 1**$^\dagger$ & 290 \\

  Mainz '12 \cite{Capitani:2012gj} &
  \multicolumn{5}{l}{superseded by Mainz '17b} \\

  ETMC '11 \cite{Alexandrou:2011nr} &
  \multicolumn{5}{l}{superseded by ETMC '15b} \\

  LHPC '10 \cite{Bratt:2010jn} &
  2+1 staggered & domain wall & $A_\mu/\mathcal{A}_\mu$ ratio & 1--2 & 293--758 \\

  RBC-UKQCD '09 \cite{Yamazaki:2009zq} &
  2+1 domain wall & domain wall & $Z_A/Z_V=1$ & 1 & 329--668 \\

  RBC-UKQCD '08 \cite{Yamazaki:2008py} &
  \multicolumn{5}{l}{superseded by RBC-UKQCD '09} \\

  RBC '08 \cite{Lin:2008uz} &
  2 domain wall & domain wall & $Z_A/Z_V=1$ & 1--2 & 493--695 \\

  LHPC '08 \cite{Hagler:2007xi} &
  \multicolumn{5}{l}{superseded by LHPC '10} \\

  Alexandrou et al.\ '07 \cite{Alexandrou:2007xj} &
  2 Wilson & Wilson & Rome-Southampton & 1 & 384--691 \\

  LHPC '06 \cite{Edwards:2005ym} &
  \multicolumn{5}{l}{superseded by LHPC '10} \\

  QCDSF '06 \cite{Khan:2006de} &
  \multicolumn{5}{l}{superseded by QCDSF '14} \\
\bottomrule
\end{tabular}
\begin{tablenotes}
\scriptsize
\item[$*$] Preprint.
\item[$**$] A variationally optimized interpolating operator is employed.
\item[$\dagger$] Carried out with a single fixed source-operator separation 
and all source-sink separations.
\end{tablenotes}
\end{threeparttable}
\caption{\small Full details of lattice-QCD calculations of the axial 
coupling $g_A\equiv\langle 1\rangle_{\Delta u^+-\Delta d^+}$.
We omit quenched results, perturbatively renormalized results, and conference 
proceedings.}
\label{tab:latticebibfirst}
\end{table}

\begin{table}[!t]
\renewcommand{\arraystretch}{1.2} 
\centering
\footnotesize
\begin{threeparttable}
\begin{tabular}{lllll}
\toprule
Ref. & Flavors & Sea quarks & Valence quarks & Renormalization \\
\midrule

  ETMC '17b \cite{Alexandrou:2017hac} &
  $u,d,s,c$ & 2 clover-TM & clover-TM & Rome-Southampton \\

  ETMC '17c \cite{Alexandrou:2017oeh} &
  $u,d,s$ & 2 clover-TM & clover-TM & Rome-Southampton \\

  $\chi$QCD '17b \cite{Gong:2015iir} &
  $s,c$ & 2+1 domain wall & overlap & single-flavor anomalous WI \\

  LHPC '17 \cite{Green:2017keo} &
  $u,d,s$ & 2+1 clover & clover & Rome-Southampton \\

  CSSM and &
  $u+d+s$ &
  2+1, 3 clover & clover & Rome-Southampton \\
  QCDSF/UKQCD '15 \cite{Chambers:2015bka} & conn.\ / disc. & & & \\

  ETMC '14 \cite{Abdel-Rehim:2013wlz} &
  $u+d,s$ & 2+1+1 twisted mass & twisted mass & nonsinglet Rome-Southampton\\

  Engelhardt '12 \cite{Engelhardt:2012gd} &
  $s$ & 2+1 staggered & domain wall & nonsinglet $A_\mu/\mathcal{A}_\mu$ ratio \\

  QCDSF '12 \cite{QCDSF:2011aa} &
  $u,d,s$ & 2 clover & clover & nonsinglet Rome-Southampton \\
  & & & &+ two-loop singlet-nonsinglet\\

  Babich et al.\ '10 \cite{Babich:2010at} &
  $s$ & 2 aniso-clover & aniso-clover & none \\

  SESAM '99 \cite{Gusken:1999as} &
  $u,d,s$ & 2 Wilson & Wilson & one loop \\

  $\chi$QCD '95 \cite{Dong:1995rx} &
  $u,d,s$ & quenched & Wilson & one loop \\

  Fukugita et al.\ '95 \cite{Fukugita:1994fh} &
  $u,d,s$ & quenched & Wilson & one loop \\

  Gupta and Mandula '94 \cite{Gupta:1994qw} &
  singlet* & quenched & Wilson & anomalous Ward identity \\

  Allés et al.\ '94 \cite{Alles:1994ss} &
  singlet* & quenched & Wilson & anomalous Ward identity \\

  Altmeyer et al.\ '94 \cite{Altmeyer:1992nt} &
  singlet & 4 staggered & staggered & anomalous Ward identity \\

  Mandula and Ogilvie '93 \cite{Mandula:1992bc} &
  $s$* & quenched & Wilson & none \\
  \bottomrule
\end{tabular}
\begin{tablenotes}
\scriptsize
\item[$*$] No signal available.
\end{tablenotes}
\end{threeparttable}
\caption{\small Full details of lattice-QCD calculations of the non-isovector 
quark spins.
The earliest results are summarized in Ref.~\cite{Liu:1995kb}.}
\label{tablenonisovectorquarkspins}
\end{table}

\begin{table}[!t]
\renewcommand{\arraystretch}{1.2} 
\centering
\footnotesize
\begin{threeparttable}
\begin{tabular}{llllll}
\toprule
Ref. & Sea quarks & Valence quarks & Renormalization 
& $N_{\Delta t}$ & $m_\pi$ (MeV)\\
\midrule

  $\chi$QCD '16 \cite{Yang:2015zja} &
  2+1 domain wall & overlap & one loop & 3 & 330 \\

  ETMC '15b \cite{Abdel-Rehim:2015owa} &
    2 clover-TM & clover-TM & Rome-Southampton & 3 & 131 \\
  & 2 twisted mass & twisted mass & & 1 & 262--470\\
  & 2+1+1 twisted mass & twisted mass & & 1, 5 & 213, 373\\

  ETMC '15a \cite{Alexandrou:2015qia} &
  2+1+1 twisted mass & twisted mass & Rome-Southampton & 1 & 302--466 \\

  RQCD '14 \cite{Bali:2014gha} &
  2 clover & clover & Rome-Southampton & 1--6 & 149--490 \\

  LHPC '14 \cite{Green:2012ud} &
  2+1 clover & clover & Rome-Southampton & 3 & 149--356\\

  ETMC '13 \cite{Alexandrou:2013joa} &
  \multicolumn{5}{l}{superseded by ETMC '15b} \\

  RQCD '12 \cite{Bali:2012av} &
  \multicolumn{5}{l}{superseded by RQCD '14} \\

  ETMC '11 \cite{Alexandrou:2011nr} &
  \multicolumn{5}{l}{superseded by ETMC '15b} \\

  QCDSF/UKQCD '11* \cite{Pleiter:2011gw} &
  2 clover & clover & Rome-Southampton & 1 & 170--670 \\

  LHPC '11* \cite{Syritsyn:2011vk} &
  2+1 domain wall & domain wall & Rome-Southampton & 1 & 297--403 \\

  LHPC '10 \cite{Bratt:2010jn} &
  2+1 staggered & domain wall & one-loop $Z_\mathcal{O}/Z_A$ & 1--2 & 293--758 \\

  RBC-UKQCD '10 \cite{Aoki:2010xg} &
  2+1 domain wall & domain wall & Rome-Southampton & 1 & 329--668 \\

  RBC '08 \cite{Lin:2008uz} &
  2 domain wall & domain wall & Rome-Southampton & 1--2 & 493--695 \\

  LHPC '08 \cite{Hagler:2007xi} &
  \multicolumn{5}{l}{superseded by LHPC '10} \\

  LHPC and &
  2 Wilson & Wilson & one loop & 1--2 & 490\\
  SESAM '02 \cite{Dolgov:2002zm} &
  and quenched & & & \\
\bottomrule
\end{tabular}
\begin{tablenotes}
\scriptsize
\item[$*$] Conference proceedings.
\end{tablenotes}
\end{threeparttable}
\caption{\small Full details of lattice-QCD calculations of 
$\langle x\rangle_{u^+-d^+}$. We omit quenched and non-renormalized results.}
\label{tab:unpolarizedisotriplet}
\end{table}

\begin{table}[!t]
\renewcommand{\arraystretch}{1.2} 
\centering
\footnotesize
\begin{threeparttable}
\begin{tabular}{lllll}
\toprule
Ref. & Flavors & Sea quarks & Valence quarks & Renormalization \\
\midrule

  ETMC '17a \cite{Alexandrou:2016ekb} & $g$
    & 2+1+1 twisted mass & twisted mass & one loop \\
  & & 2 clover-TM & clover-TM & \\

  ETMC '17c \cite{Alexandrou:2017oeh} & $u,d,s,g$
    & 2 clover-TM & clover-TM & Rome-Southampton ($q$)\\
    & & & & one-loop ($g$)\\   
 
  ETMC '15a \cite{Alexandrou:2015qia} &
  $u+d-2s$ &  2+1+1 twisted mass & twisted mass & Rome-Southampton \\

  ETMC '14 \cite{Abdel-Rehim:2013wlz} &
  $u+d$ & 2+1+1 twisted mass & twisted mass & nonsinglet\\ 
  & & & & Rome-Southampton\\

  $\chi$QCD '15 \cite{Deka:2013zha} &
  $u,d,s,g$ & quenched & Wilson & sum rule + one-loop \\

  QCDSF-UKQCD '12 \cite{Horsley:2012pz} &
  $g$ & quenched & clover & nonperturbative \\
\bottomrule
\end{tabular}
\end{threeparttable}
\caption{\small Full details of lattice-QCD calculations of the non-isovector 
momentum fractions.}
\label{tab:nonisovectormomfrac}
\end{table}

\begin{table}[!t]
\renewcommand{\arraystretch}{1.2} 
\centering
\footnotesize
\begin{threeparttable}
\begin{tabular}{lllll}
\toprule
Ref. & Sea quarks & Valence quarks & Renormalization & $N_{\Delta t}$ \\
\midrule

  ETMC '15b \cite{Abdel-Rehim:2015owa} &
    2 clover-TM & clover-TM & Rome-Southampton & 3 \\
  & 2 twisted mass & twisted mass & & 1 \\
  & 2+1+1 twisted mass & twisted mass & & 1 or 4 \\

  ETMC '13 \cite{Alexandrou:2013joa} &
  \multicolumn{4}{l}{superseded by ETMC '15b} \\

  ETMC '11 \cite{Alexandrou:2011nr} &
  \multicolumn{4}{l}{superseded by ETMC '15b} \\

  QCDSF/UKQCD '11* \cite{Pleiter:2011gw} &
  2 clover & clover & Rome-Southampton & 1 \\

  LHPC '10 \cite{Bratt:2010jn} &
  2+1 staggered & domain wall & one-loop $Z_\mathcal{O}/Z_A$ & 1--2 \\

  RBC-UKQCD '10 \cite{Aoki:2010xg} &
  2+1 domain wall & domain wall & Rome-Southampton & 1 \\

  RBC '08 \cite{Lin:2008uz} &
  2 domain wall & domain wall & Rome-Southampton & 1--2 \\

  LHPC '08 \cite{Hagler:2007xi} &
  \multicolumn{4}{l}{superseded by LHPC '10} \\

  LHPC and &
  2 Wilson & Wilson & one loop & 1--2 \\
  SESAM '02 \cite{Dolgov:2002zm} &
  and quenched & & & \\

  QCDSF '97 \cite{Gockeler:1997zr} &
  quenched & Wilson & one loop & 1 \\
\bottomrule
\end{tabular}
\begin{tablenotes}
\scriptsize
\item[$*$] Conference proceedings.
\end{tablenotes}
\end{threeparttable}
\caption{\small Full details of lattice-QCD calculations of 
$\langle x\rangle_{\Delta u^--\Delta d^-}$.}
\label{tab:nonisopolcase}
\end{table}

\begin{table}[!t]
\renewcommand{\arraystretch}{1.2} 
\centering
\footnotesize
\begin{threeparttable}
\begin{tabular}{lllll}
\toprule
Ref. & Observables & Sea quarks & Valence quarks & Renormalization \\
\midrule

  LHPC '10$^{\dagger}$
  \cite{Bratt:2010jn} &
  $\langle x\rangle_{u^+-d^+}$,
  $\langle x^2\rangle_{u^--d^-}$, &
  2+1 staggered &
  domain wall &
  one-loop $Z_\mathcal{O}/Z_A$ \\
  &   $g_A$,
  $\langle x\rangle_{\Delta u^--\Delta d^-}$,
  $\langle x^2\rangle_{\Delta u^+-\Delta d^+}$ & & & \\

  $\chi$QCD '09 \cite{Deka:2008xr} &
  $\langle x\rangle_{u^+,d^+,s^+}$ (superseded by $\chi$QCD '15), &
  quenched &
  Wilson &
  one loop \\
  & $\langle x^2 \rangle_{u^-,d^-,s^-}$ & & &\\

  LHPC '08 \cite{Hagler:2007xi} &
  superseded by LHPC '10 & & &\\

  QCDSF '05c \cite{Gockeler:2005vw} &
  $\langle x^2\rangle_{\Delta u^+-\Delta d^+}$ &
  2 clover & clover & Rome-Southampton \\

  QCDSF '05b \cite{Gockeler:2004wp} &
  $\langle x\rangle_{u^+-d^+}$,
  $\langle x^2\rangle_{u^--d^-}$,
  $\langle x^3\rangle_{u^+-d^+}$ &
  quenched &
  clover &
  Rome-Southampton \\

  QCDSF '05a* \cite{Gockeler:2004vx} &
  $\langle x\rangle_{u^+-d^+}$,
  $\langle x^2\rangle_{u^--d^-}$,
  $\langle x^3\rangle_{u^+-d^+}$ &
  2 clover & clover & one loop \\

  LHPC and &
  $\langle x\rangle_{u^+-d^+}$,
  $\langle x^2\rangle_{u^--d^-}$,
  $\langle x^3\rangle_{u^+-d^+}$, &
  2 Wilson & Wilson & one loop \\
  SESAM '02 \cite{Dolgov:2002zm} &
  $g_A$,
  $\langle x\rangle_{\Delta u^--\Delta d^-}$,
  $\langle x^2\rangle_{\Delta u^+-\Delta d^+}$ &
  and quenched & & \\

  QCDSF '01 \cite{Gockeler:2000ja} &
  $\langle x^2\rangle_{\Delta u^+-\Delta d^+}$ &
  quenched & clover & Rome-Southampton \\

  QCDSF '96 \cite{Gockeler:1995wg} &
  $\langle x\rangle_{u^+-d^+}$,
  $\langle x^2\rangle_{u^--d^-}$,
  $\langle x^3\rangle_{u^+-d^+}$, &
  quenched & Wilson & one loop \\
  & $g_A$,
  $\langle x^2\rangle_{\Delta u^+-\Delta d^+}$ & & &\\
\bottomrule
\end{tabular}
\begin{tablenotes}
\scriptsize
\item[$*$] Conference proceedings.
\item[$\dagger$] The moment $\langle x^2\rangle_{u-d}=A_{30}^{u-d}(0)$ is plotted 
in the ratio of form factors $A_{30}(t)/A_{10}(t)$, where we can use 
$A_{10}^{u-d}(0)=1$. 
The moment $\langle x^2\rangle_{\Delta u-\Delta d}=\tilde A_{30}^{u-d}(0)$ is plotted 
in the ratio of form factors $\tilde A_{30}(t)/\tilde A_{10}(t)$ and we can use 
$\tilde A_{10}^{u-d}(0)=g_A$.
\end{tablenotes}
\end{threeparttable}
\caption{\small Full details of lattice-QCD calculations of higher moments of 
unpolarized and polarized PDFs.}
\label{tab:latticebiblast}
\end{table}

\begin{figure}[!p]
\begin{center}
\centerline{
\subfloat[]{\includegraphics[width=0.49\textwidth]{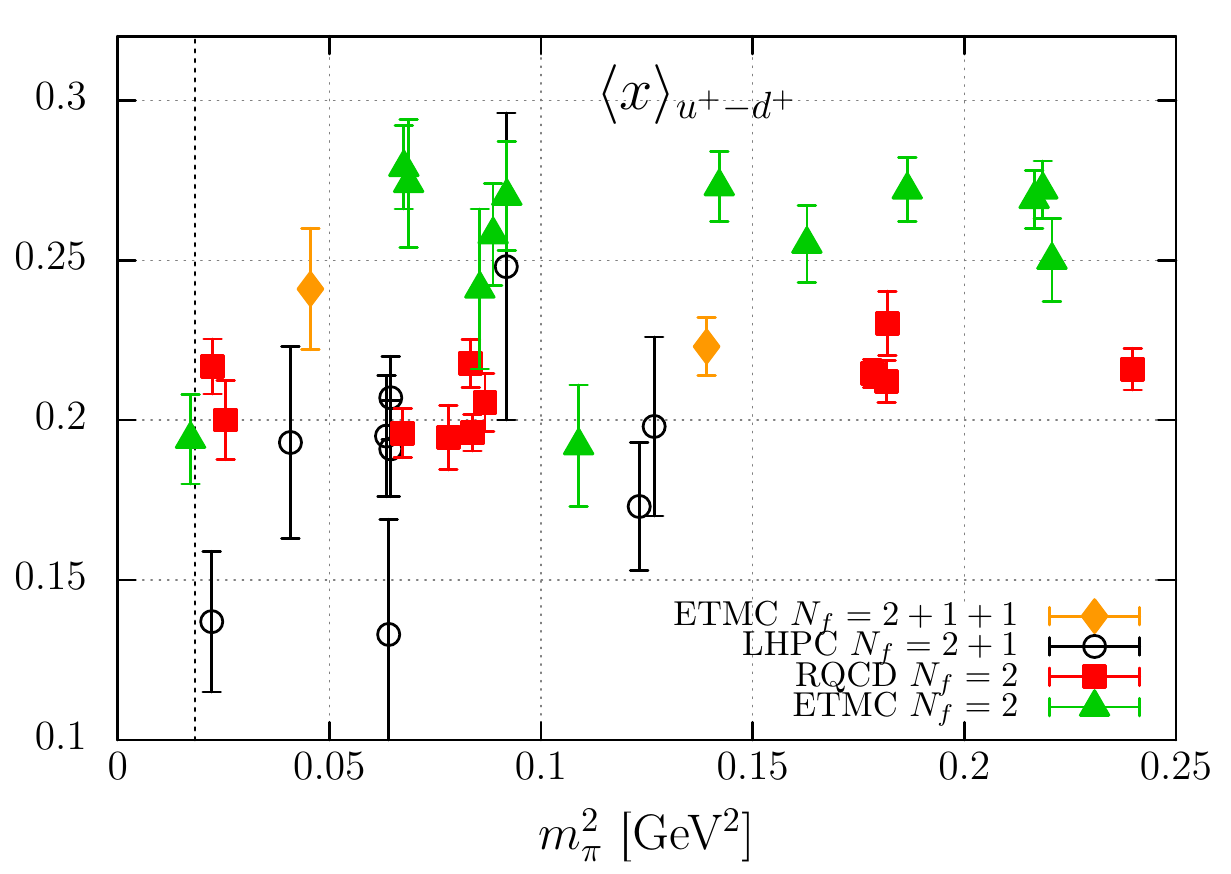}}
\subfloat[]{\includegraphics[width=0.47\textwidth]{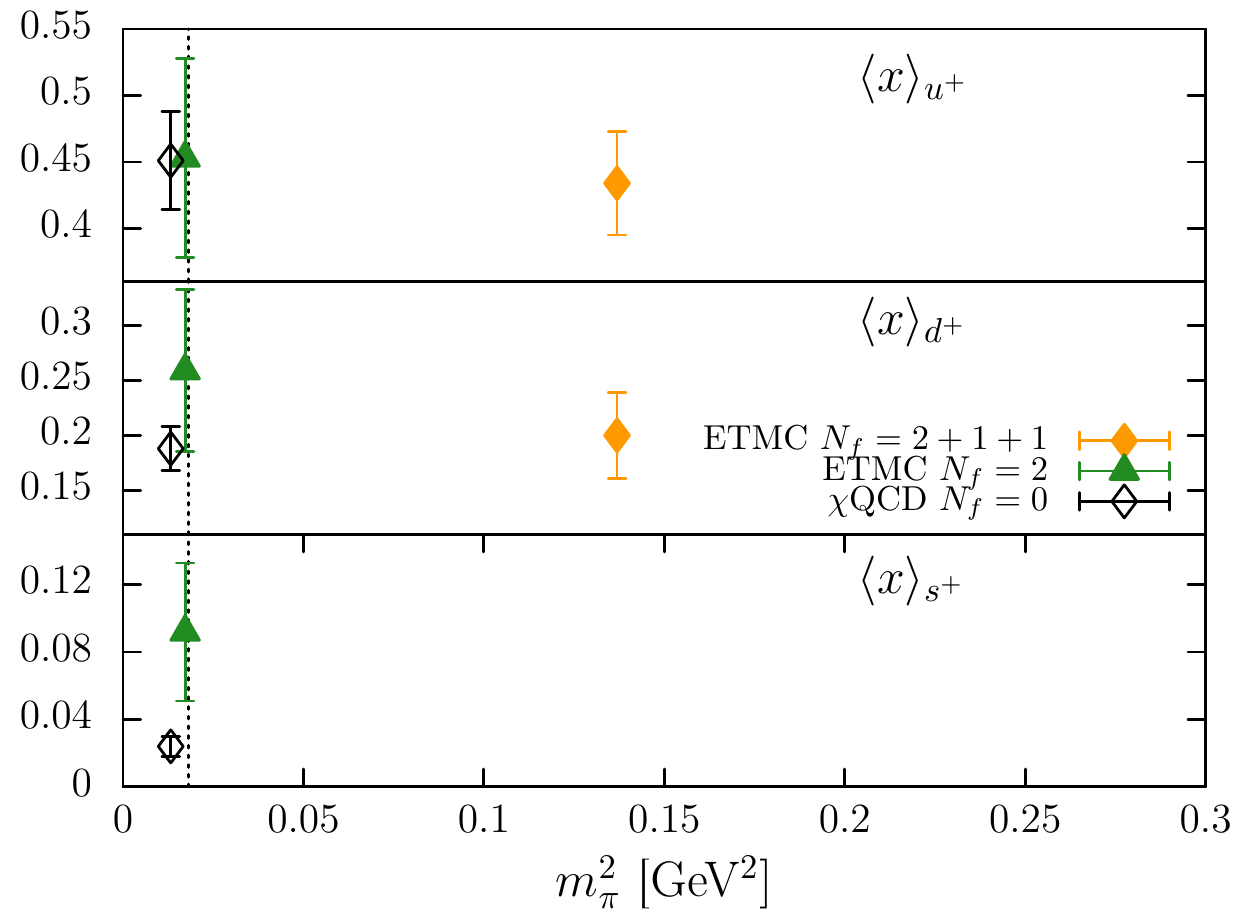}}
}
\centerline{
\subfloat[]{\includegraphics[width=0.49\textwidth]{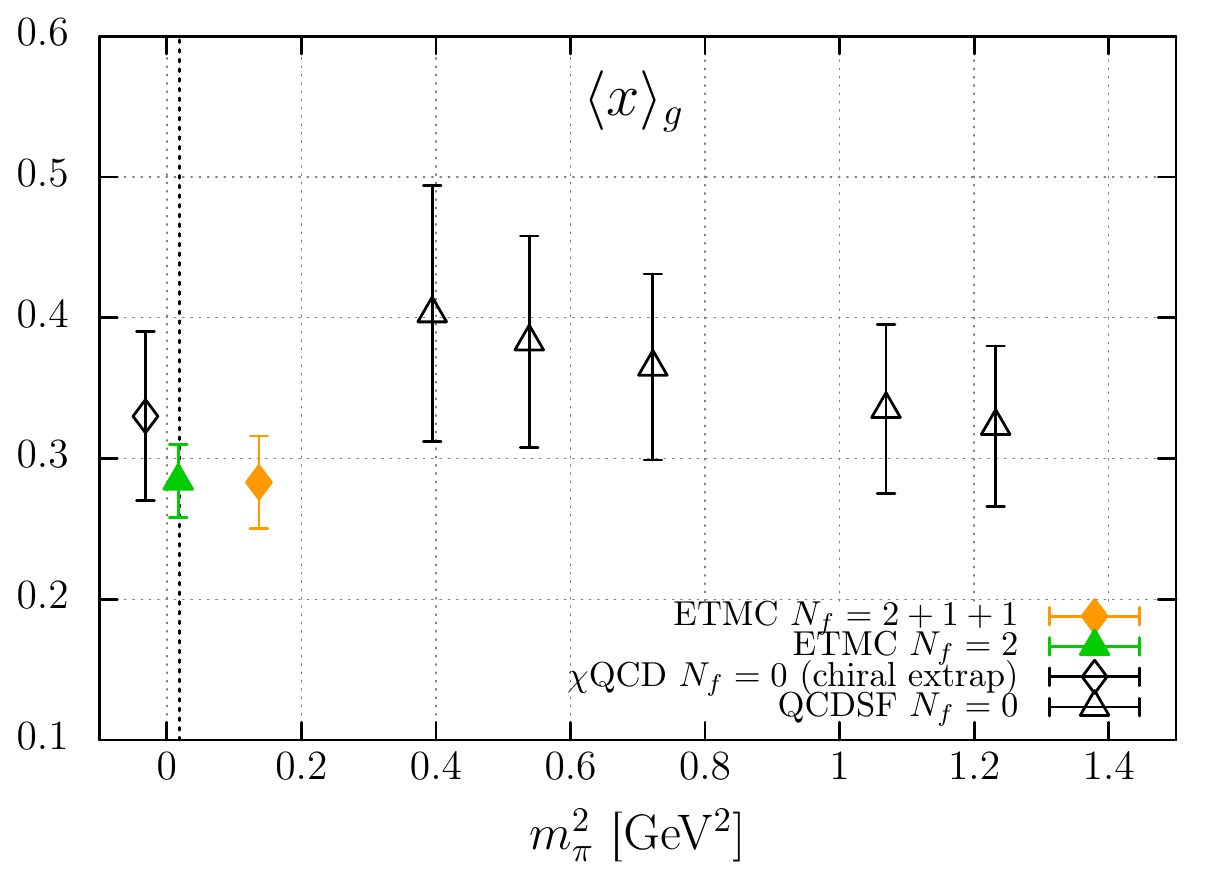}}
\subfloat[]{\includegraphics[width=0.49\textwidth]{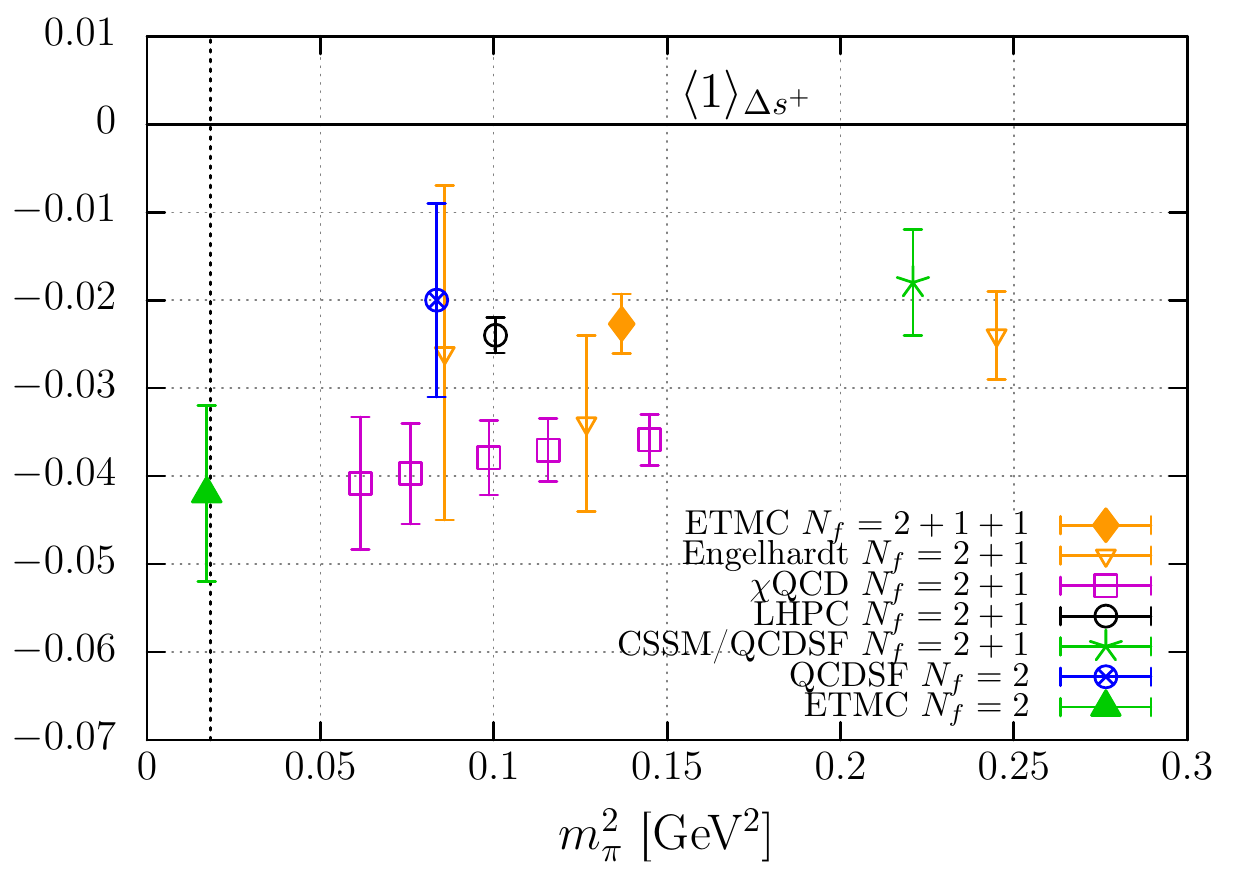}}
}
\centerline{
\subfloat[]{\includegraphics[width=0.49\textwidth]{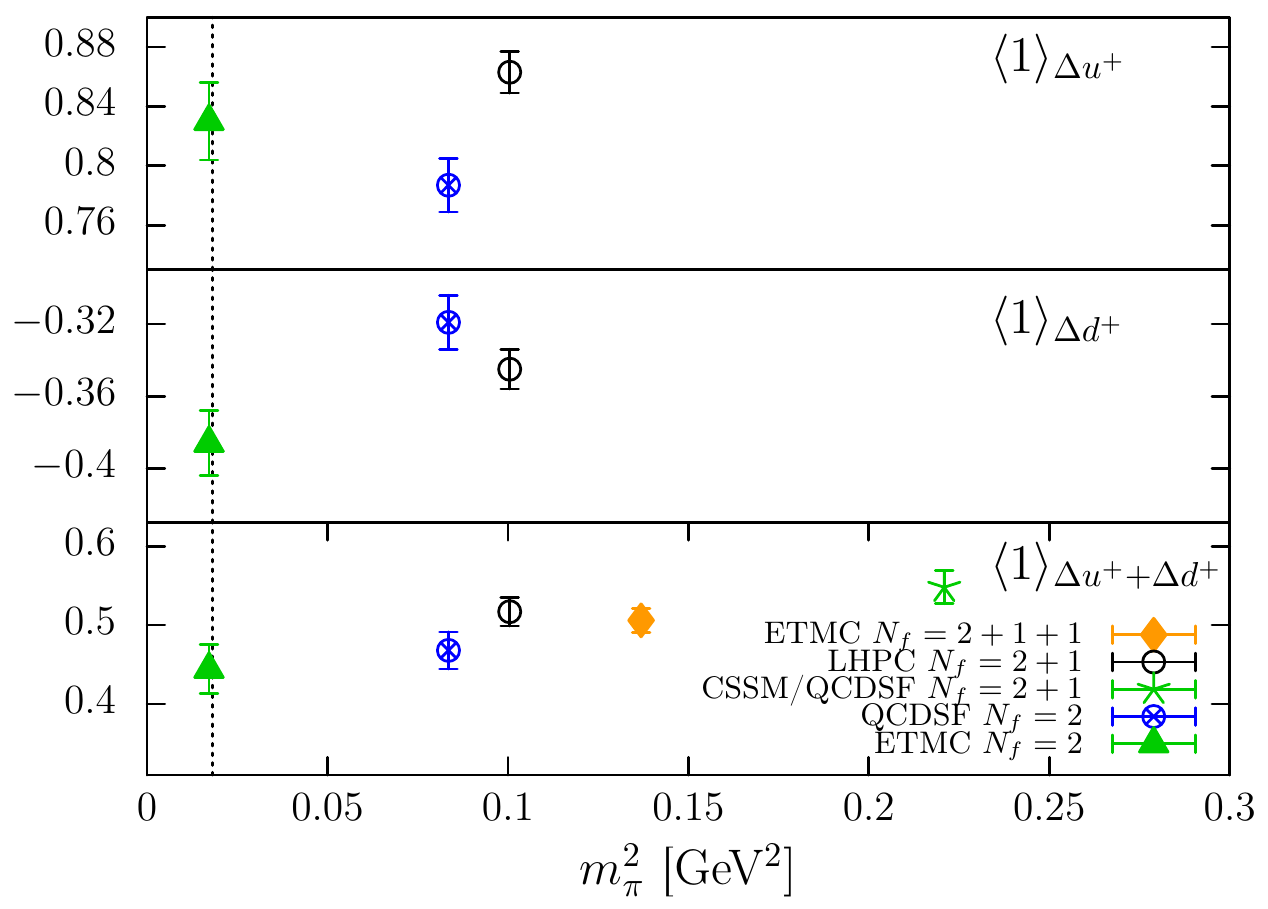}}
\ \ 
\subfloat[]{\includegraphics[width=0.49\textwidth]{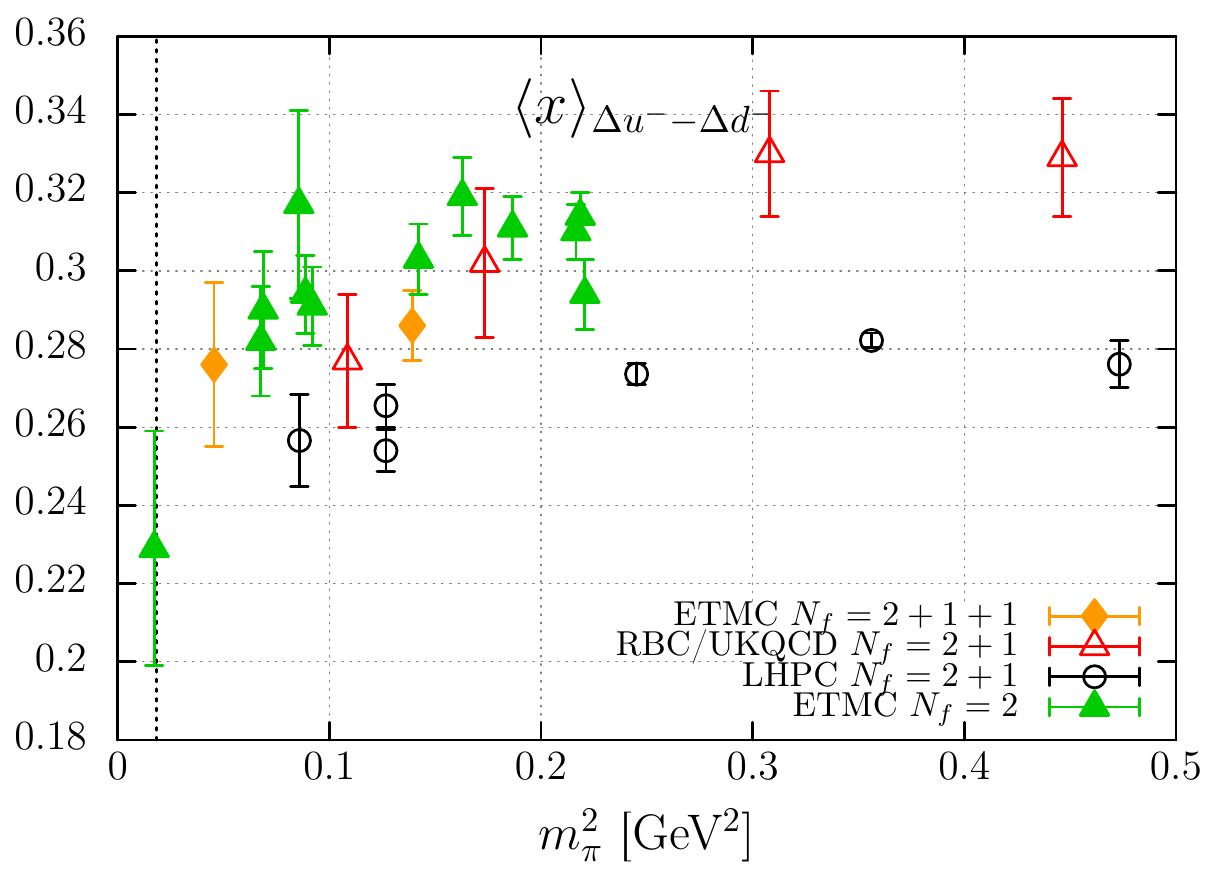}}
}
\end{center}
\caption{\small Comparison of lattice-QCD results for
  the zeroth and first moments
  of unpolarized and polarized PDFs.
  From left to right and from top to bottom, we show
  results for $\la x\ra_{u^+-d^+}$, $\la x\ra_{q^+}$, $\la x\ra_{g}$,
  $\la 1 \ra_{\Delta s^+}$, $\la 1\ra_{\Delta q^+}$ and
  $\la x\ra_{\Delta u^- - \Delta d^-}$;
  see Tables~\ref{tab:unpolLQCDstatus1}, \ref{tab:polLQCDstatus0},
  \ref{tab:polLQCDstatus1}, \ref{tab:unpolLQCDstatus1B},
  \ref{tab:polLQCDstatus1B} and \ref{tab:polLQCDstatus2B}
  for details.
}
\label{fig:latt_res}
\end{figure}

A representative subset of the results contained in
Tables~\ref{tab:unpolLQCDstatus1}, \ref{tab:polLQCDstatus0},
\ref{tab:polLQCDstatus1}, \ref{tab:unpolLQCDstatus1B},
\ref{tab:polLQCDstatus1B} and \ref{tab:polLQCDstatus2B}
is displayed in Fig.~\ref{fig:latt_res}.
Specifically, we show from left to right and from top to bottom
results for $\la x\ra_{u^+-d^+}$, $\la x\ra_{q^+}$, $\la x\ra_{g}$,
$\la 1 \ra_{\Delta s^+}$, $\la 1\ra_{\Delta q^+}$ and
$\la x\ra_{\Delta u^- - \Delta d^-}$;
see the corresponding entries of each table for details.

\section{PDF fit results for higher moments }
\label{app:Hmoms}

In this appendix, we summarize the current values of the higher moments of 
unpolarized and polarized PDFs from global fits, not previously 
listed in Sec.~\ref{subsubsec:GPDFfits}.
Even though these moments were not selected for the benchmark exercise 
performed in Sec.~\ref{subsec:BN}, we find it useful to collect them 
here for future reference.

\begin{itemize}

\item In Table~\ref{tab:unpHmoms} we display the values of the second moments 
of the unpolarized quark valence distributions $\langle x^2\rangle_{u^-}$, 
$\langle x^2\rangle_{d^-}$, $\langle x^2\rangle_{s^-}$ and 
$\langle x^2\rangle_{u^--d^-}$.

\item In Table~\ref{tab:polHmoms} we display the values of the first moments 
of the polarized quark valence distributions $\langle x\rangle_{\Delta u^-}$, 
$\langle x\rangle_{\Delta d^-}$ and $\langle x\rangle_{\Delta s^-}$.

\item In Table~\ref{tab:polgmom} we display the value of the zeroth moment
of the polarized gluon PDF truncated in the region $[x_{\rm min}, x_{\rm max}]$,
$\langle 1\rangle_{\Delta g}^{[x_{\rm min}, x_{\rm max}]}=\int_{x_{\rm min}}^{x_{\rm max}}dx\,\Delta g(x,Q^2)$.
The truncated moment is shown instead of the full moment
because the latter is potentially affected by a large extrapolation 
uncertainty difficult to quantify (see Sec.~\ref{sec:polPDFs}). 
Three different truncated ranges are considered: $[10^{-5},1]$,  $[10^{-3},1]$
and  $[10^{-2},1]$.

\end{itemize}
All values in Tables~\ref{tab:unpHmoms}--\ref{tab:polgmom} are computed at 
$\mu^2=Q^2=4$~GeV$^2$.
For the description of the corresponding PDF sets and their uncertainties, see
Secs.~\ref{sec:unpPDFs}, \ref{sec:polPDFs} and \ref{subsubsec:GPDFfits}.

\begin{table}[!t]
\centering
\small
\begin{tabular}{lccccccc}
\toprule
Mom. & NNPDF3.1 & CT14 & MMHT14 & ABMP16 & CJ15 & HERAPDF2.0 & PDF4LHC15 \\
\midrule
$\langle x^2\rangle_{u^-}$ 
& 0.0851(27) & 0.0841(13) & 0.0831(14)    
& 0.0845(8) & 0.0853(3) & 0.0886(29) & 0.0833(15) \\
$\langle x^2\rangle_{d^-}$
& 0.0284(27) & 0.0295(10) & 0.0305(11)    
& 0.0267(7) & 0.0305(3) & 0.0334(18) & 0.0305(17) \\ 
$\langle x^2\rangle_{s^-}$
& 0.0010(31) & ---        & 0.0006(8)\ \, 
& ---       & ---       & ---        & 0.0011(11) \\
$\langle x^2\rangle_{u^--d^-}$
& 0.0571(27) & 0.0546(19) & 0.0526(19)    
& 0.0578(9) & 0.0548(3) & 0.0553(17) & 0.0530(24) \\
\bottomrule
\end{tabular}
\caption{\small Second moments of unpolarized valence PDFs from 
global PDF fits at $\mu^2=Q^2=4$~GeV$^2$.}
\label{tab:unpHmoms}
\end{table}

\begin{table}[!t]
\centering
\footnotesize
\begin{tabular}{lccc}
\toprule
Mom. & NNPDFpol1.1 & DSSV08 & JAM17\\
\midrule
$\langle x\rangle_{\Delta u^-}$ 
& \ 0.1493(85) & \ 0.1624(56) & \ 0.181(14)\\
$\langle x\rangle_{\Delta d^-}$ 
&  $-0.0468(79)$ &  $-0.0410(55)$ &  $-0.060(18)$\\
\bottomrule
\end{tabular}
\caption{\small First moments of polarized valence PDFs from global 
PDF fits at $\mu^2=Q^2=4$~GeV$^2$.}
\label{tab:polHmoms}
\end{table}

\begin{table}[!t]
\centering
\footnotesize
\begin{tabular}{lcccc}
\toprule
Mom. & NNPDFpol1.1 & DSSV14 & JAM15 & JAM17\\
\midrule
$\langle 1\rangle_{\Delta g}^{[10^{-5},1]}$  
& $-0.1(1.7)$ & $0.27(^{+1.72}_{-1.44})$  & 1.08(87) & 0.18(28)\\
$\langle 1\rangle_{\Delta g}^{[10^{-3},1]}$  
& 0.14(78)  & $0.27(^{+0.63}_{-0.54})$  & 0.74(40) & 0.20(24)\\
$\langle 1\rangle_{\Delta g}^{[10^{-2},1]}$  
& 0.23(24)  & $0.24(^{+0.23}_{-0.15})$  & 0.52(19) & 0.18(20)\\
\bottomrule
\end{tabular}
\caption{\small Truncated zeroth moments of the polarized gluon PDF from 
global fits at $\mu^2=Q^2=4$~GeV$^2$.}
\label{tab:polgmom}
\end{table}

\clearpage 

\newpage 
\bibliography{PDFLattice2017}

\providecommand{\href}[2]{#2}\begingroup\raggedright\begin{thebibliography}{100}

\bibitem{Ball:2016spl}
R.~D. Ball, E.~R. Nocera, and J.~Rojo {\em Eur. Phys. J.} {\bf C76} (2016),
  no.~7 383, [\href{http://arxiv.org/abs/1604.00024}{{\tt arXiv:1604.00024}}].

\bibitem{Nocera:2014uea}
E.~R. Nocera {\em Phys.Lett.} {\bf B742} (2015) 117--125,
  [\href{http://arxiv.org/abs/1410.7290}{{\tt arXiv:1410.7290}}].

\bibitem{Perez:2012um}
E.~Perez and E.~Rizvi {\em Rep.Prog.Phys.} {\bf 76} (2013) 046201,
  [\href{http://arxiv.org/abs/1208.1178}{{\tt arXiv:1208.1178}}].

\bibitem{DeRoeck:2011na}
A.~De~Roeck and R.~S. Thorne {\em Prog.Part.Nucl.Phys.} {\bf 66} (2011) 727,
  [\href{http://arxiv.org/abs/1103.0555}{{\tt arXiv:1103.0555}}].

\bibitem{Alekhin:2011sk}
S.~Alekhin et~al. \href{http://arxiv.org/abs/1101.0536}{{\tt arXiv:1101.0536}}.

\bibitem{Ball:2012wy}
R.~D. Ball et~al. {\em JHEP} {\bf 04} (2013) 125,
  [\href{http://arxiv.org/abs/1211.5142}{{\tt arXiv:1211.5142}}].

\bibitem{Forte:2013wc}
S.~Forte and G.~Watt {\em Ann. Rev. Nucl. Part. Sci.} {\bf 63} (2013) 291--328,
  [\href{http://arxiv.org/abs/1301.6754}{{\tt arXiv:1301.6754}}].

\bibitem{Jimenez-Delgado:2013sma}
P.~Jimenez-Delgado, W.~Melnitchouk, and J.~F. Owens {\em J. Phys.} {\bf G40}
  (2013) 093102, [\href{http://arxiv.org/abs/1306.6515}{{\tt
  arXiv:1306.6515}}].

\bibitem{Rojo:2015acz}
J.~Rojo et~al. {\em J. Phys.} {\bf G42} (2015) 103103,
  [\href{http://arxiv.org/abs/1507.00556}{{\tt arXiv:1507.00556}}].

\bibitem{Butterworth:2015oua}
J.~Butterworth et~al. {\em J. Phys.} {\bf G43} (2016) 023001,
  [\href{http://arxiv.org/abs/1510.03865}{{\tt arXiv:1510.03865}}].

\bibitem{Accardi:2016ndt}
A.~Accardi et~al. {\em Eur. Phys. J.} {\bf C76} (2016), no.~8 471,
  [\href{http://arxiv.org/abs/1603.08906}{{\tt arXiv:1603.08906}}].

\bibitem{Gao:2017yyd}
J.~Gao, L.~Harland-Lang, and J.~Rojo
  \href{http://arxiv.org/abs/1709.04922}{{\tt arXiv:1709.04922}}.

\bibitem{Ball:2017nwa}
{\bf NNPDF} Collaboration, R.~D. Ball et~al. {\em Eur. Phys. J.} {\bf C77}
  (2017), no.~10 663, [\href{http://arxiv.org/abs/1706.00428}{{\tt
  arXiv:1706.00428}}].

\bibitem{Harland-Lang:2014zoa}
L.~A. Harland-Lang, A.~D. Martin, P.~Motylinski, and R.~S. Thorne {\em Eur.
  Phys. J.} {\bf C75} (2015), no.~5 204,
  [\href{http://arxiv.org/abs/1412.3989}{{\tt arXiv:1412.3989}}].

\bibitem{Dulat:2015mca}
S.~Dulat, T.-J. Hou, J.~Gao, M.~Guzzi, J.~Huston, P.~Nadolsky, J.~Pumplin,
  C.~Schmidt, D.~Stump, and C.~P. Yuan {\em Phys. Rev.} {\bf D93} (2016), no.~3
  033006, [\href{http://arxiv.org/abs/1506.07443}{{\tt arXiv:1506.07443}}].

\bibitem{Alekhin:2017kpj}
S.~Alekhin, J.~Blümlein, S.~Moch, and R.~Placakyte {\em Phys. Rev.} {\bf D96}
  (2017), no.~1 014011, [\href{http://arxiv.org/abs/1701.05838}{{\tt
  arXiv:1701.05838}}].

\bibitem{Accardi:2016qay}
A.~Accardi, L.~T. Brady, W.~Melnitchouk, J.~F. Owens, and N.~Sato {\em Phys.
  Rev.} {\bf D93} (2016), no.~11 114017,
  [\href{http://arxiv.org/abs/1602.03154}{{\tt arXiv:1602.03154}}].

\bibitem{Nocera:2014gqa}
{\bf NNPDF} Collaboration, E.~R. Nocera, R.~D. Ball, S.~Forte, G.~Ridolfi, and
  J.~Rojo {\em Nucl. Phys.} {\bf B887} (2014) 276--308,
  [\href{http://arxiv.org/abs/1406.5539}{{\tt arXiv:1406.5539}}].

\bibitem{deFlorian:2009vb}
D.~de~Florian, R.~Sassot, M.~Stratmann, and W.~Vogelsang {\em Phys. Rev.} {\bf
  D80} (2009) 034030, [\href{http://arxiv.org/abs/0904.3821}{{\tt
  arXiv:0904.3821}}].

\bibitem{Sato:2016tuz}
{\bf Jefferson Lab Angular Momentum} Collaboration, N.~Sato, W.~Melnitchouk,
  S.~E. Kuhn, J.~J. Ethier, and A.~Accardi {\em Phys. Rev.} {\bf D93} (2016),
  no.~7 074005, [\href{http://arxiv.org/abs/1601.07782}{{\tt
  arXiv:1601.07782}}].

\bibitem{Hirai:2008aj}
{\bf Asymmetry Analysis} Collaboration, M.~Hirai and S.~Kumano {\em Nucl.
  Phys.} {\bf B813} (2009) 106--122,
  [\href{http://arxiv.org/abs/0808.0413}{{\tt arXiv:0808.0413}}].

\bibitem{Abramowicz:2015mha}
{\bf ZEUS, H1} Collaboration, H.~Abramowicz et~al. {\em Eur. Phys. J.} {\bf
  C75} (2015), no.~12 580, [\href{http://arxiv.org/abs/1506.06042}{{\tt
  arXiv:1506.06042}}].

\bibitem{Zenaiev:2015rfa}
{\bf PROSA} Collaboration, O.~Zenaiev et~al. {\em Eur. Phys. J.} {\bf C75}
  (2015), no.~8 396, [\href{http://arxiv.org/abs/1503.04581}{{\tt
  arXiv:1503.04581}}].

\bibitem{Gauld:2016kpd}
R.~Gauld and J.~Rojo {\em Phys. Rev. Lett.} {\bf 118} (2017), no.~7 072001,
  [\href{http://arxiv.org/abs/1610.09373}{{\tt arXiv:1610.09373}}].

\bibitem{Boughezal:2017nla}
R.~Boughezal, A.~Guffanti, F.~Petriello, and M.~Ubiali {\em JHEP} {\bf 07}
  (2017) 130, [\href{http://arxiv.org/abs/1705.00343}{{\tt arXiv:1705.00343}}].

\bibitem{Currie:2016bfm}
J.~Currie, E.~W.~N. Glover, and J.~Pires {\em Phys. Rev. Lett.} {\bf 118}
  (2017), no.~7 072002, [\href{http://arxiv.org/abs/1611.01460}{{\tt
  arXiv:1611.01460}}].

\bibitem{Czakon:2016olj}
M.~Czakon, N.~P. Hartland, A.~Mitov, E.~R. Nocera, and J.~Rojo {\em JHEP} {\bf
  04} (2017) 044, [\href{http://arxiv.org/abs/1611.08609}{{\tt
  arXiv:1611.08609}}].

\bibitem{Guzzi:2014wia}
M.~Guzzi, K.~Lipka, and S.-O. Moch {\em JHEP} {\bf 01} (2015) 082,
  [\href{http://arxiv.org/abs/1406.0386}{{\tt arXiv:1406.0386}}].

\bibitem{deFlorian:2014yva}
D.~de~Florian, R.~Sassot, M.~Stratmann, and W.~Vogelsang {\em Phys. Rev. Lett.}
  {\bf 113} (2014), no.~1 012001, [\href{http://arxiv.org/abs/1404.4293}{{\tt
  arXiv:1404.4293}}].

\bibitem{Olive:2016xmw}
{\bf Particle Data Group} Collaboration, C.~Patrignani et~al. {\em Chin. Phys.}
  {\bf C40} (2016), no.~10 100001.

\bibitem{Gupta:1997nd}
R.~Gupta, {\it {Introduction to lattice QCD: Course}},  in {\em {Probing the
  standard model of particle interactions. Proceedings, Summer School in
  Theoretical Physics, NATO Advanced Study Institute, 68th session, Les
  Houches, France, July 28-September 5, 1997. Pt. 1, 2}}, pp.~83--219, 1997.
\newblock \href{http://arxiv.org/abs/hep-lat/9807028}{{\tt hep-lat/9807028}}.

\bibitem{Constantinou:2014tga}
M.~Constantinou {\em PoS} {\bf LATTICE2014} (2015) 001,
  [\href{http://arxiv.org/abs/1411.0078}{{\tt arXiv:1411.0078}}].

\bibitem{Syritsyn:2014saa}
S.~Syritsyn {\em PoS} {\bf LATTICE2013} (2014) 009,
  [\href{http://arxiv.org/abs/1403.4686}{{\tt arXiv:1403.4686}}].

\bibitem{Lin:2012ev}
H.-W. Lin {\em PoS} {\bf LATTICE2012} (2012) 013,
  [\href{http://arxiv.org/abs/1212.6849}{{\tt arXiv:1212.6849}}].

\bibitem{Lin:2014zya}
H.-W. Lin, J.-W. Chen, S.~D. Cohen, and X.~Ji {\em Phys. Rev.} {\bf D91} (2015)
  054510, [\href{http://arxiv.org/abs/1402.1462}{{\tt arXiv:1402.1462}}].

\bibitem{Alexandrou:2015rja}
C.~Alexandrou, K.~Cichy, V.~Drach, E.~Garcia-Ramos, K.~Hadjiyiannakou,
  K.~Jansen, F.~Steffens, and C.~Wiese {\em Phys. Rev.} {\bf D92} (2015)
  014502, [\href{http://arxiv.org/abs/1504.07455}{{\tt arXiv:1504.07455}}].

\bibitem{Chen:2016utp}
J.-W. Chen, S.~D. Cohen, X.~Ji, H.-W. Lin, and J.-H. Zhang {\em Nucl. Phys.}
  {\bf B911} (2016) 246--273, [\href{http://arxiv.org/abs/1603.06664}{{\tt
  arXiv:1603.06664}}].

\bibitem{Alexandrou:2016jqi}
C.~Alexandrou, K.~Cichy, M.~Constantinou, K.~Hadjiyiannakou, K.~Jansen,
  F.~Steffens, and C.~Wiese {\em Phys. Rev.} {\bf D96} (2017), no.~1 014513,
  [\href{http://arxiv.org/abs/1610.03689}{{\tt arXiv:1610.03689}}].

\bibitem{Ellis:1991qj}
R.~K. Ellis, W.~J. Stirling, and B.~R. Webber, {\em {QCD and collider
  physics}}.
\newblock Cambridge University Press, 1996.

\bibitem{Leader:2001gr}
E.~Leader, {\em {Spin in particle physics}}.
\newblock Cambridge University Press, 2001.

\bibitem{Collins:2011zzd}
J.~Collins, {\em {Foundations of perturbative QCD}}.
\newblock Cambridge University Press, 2013.

\bibitem{DeGrand:2006zz}
T.~DeGrand and C.~E. De~Tar, {\em {Lattice methods for quantum
  chromodynamics}}.
\newblock 2006.

\bibitem{Gattringer:2010zz}
C.~Gattringer and C.~B. Lang, {\em {Quantum chromodynamics on the lattice}}.
\newblock Springer, 2010.

\bibitem{Gross:1973ju}
D.~J. Gross and F.~Wilczek {\em Phys. Rev.} {\bf D8} (1973) 3633--3652.

\bibitem{Gross:1973id}
D.~J. Gross and F.~Wilczek {\em Phys. Rev. Lett.} {\bf 30} (1973) 1343--1346.

\bibitem{Gross:1974cs}
D.~J. Gross and F.~Wilczek {\em Phys. Rev.} {\bf D9} (1974) 980--993.

\bibitem{Politzer:1974fr}
H.~D. Politzer {\em Phys. Rept.} {\bf 14} (1974) 129--180.

\bibitem{Collins:1987pm}
J.~C. Collins and D.~E. Soper {\em Ann. Rev. Nucl. Part. Sci.} {\bf 37} (1987)
  383--409.

\bibitem{Collins:1989gx}
J.~C. Collins, D.~E. Soper, and G.~F. Sterman {\em Adv. Ser. Direct. High
  Energy Phys.} {\bf 5} (1989) 1--91,
  [\href{http://arxiv.org/abs/hep-ph/0409313}{{\tt hep-ph/0409313}}].

\bibitem{Campbell:2006wx}
J.~M. Campbell, J.~W. Huston, and W.~J. Stirling {\em Rept. Prog. Phys.} {\bf
  70} (2007) 89, [\href{http://arxiv.org/abs/hep-ph/0611148}{{\tt
  hep-ph/0611148}}].

\bibitem{Gasser:1983yg}
J.~Gasser and H.~Leutwyler {\em Annals Phys.} {\bf 158} (1984) 142.

\bibitem{Leader:2013jra}
E.~Leader and C.~Lorcé {\em Phys. Rept.} {\bf 541} (2014), no.~3 163--248,
  [\href{http://arxiv.org/abs/1309.4235}{{\tt arXiv:1309.4235}}].

\bibitem{tHooft:1973mfk}
G.~'t~Hooft {\em Nucl. Phys.} {\bf B61} (1973) 455--468.

\bibitem{Weinberg:1951ss}
S.~Weinberg {\em Phys. Rev.} {\bf D8} (1973) 3497--3509.

\bibitem{Manohar:1992tz}
A.~V. Manohar, {\it {An Introduction to spin dependent deep inelastic
  scattering}},  in {\em {Lake Louise Winter Institute: Symmetry and Spin in
  the Standard Model Lake Louise, Alberta, Canada, February 23-29, 1992}},
  pp.~1--46, 1992.
\newblock \href{http://arxiv.org/abs/hep-ph/9204208}{{\tt hep-ph/9204208}}.

\bibitem{Anselmino:1993tc}
M.~Anselmino, P.~Gambino, and J.~Kalinowski {\em Z. Phys.} {\bf C64} (1994)
  267--274, [\href{http://arxiv.org/abs/hep-ph/9401264}{{\tt hep-ph/9401264}}].

\bibitem{Anselmino:1992rn}
M.~Anselmino and E.~Leader {\em Phys. Lett.} {\bf B293} (1992) 216--218.

\bibitem{Collins:1981uw}
J.~C. Collins and D.~E. Soper {\em Nucl. Phys.} {\bf B194} (1982) 445--492.

\bibitem{Curci:1980uw}
G.~Curci, W.~Furmanski, and R.~Petronzio {\em Nucl. Phys.} {\bf B175} (1980)
  27--92.

\bibitem{Baulieu:1979mr}
L.~Baulieu, E.~G. Floratos, and C.~Kounnas {\em Nucl. Phys.} {\bf B166} (1980)
  321--339.

\bibitem{Dokshitzer:1977sg}
Y.~L. Dokshitzer {\em Sov. Phys. JETP} {\bf 46} (1977) 641--653. [Zh. Eksp.
  Teor. Fiz.73,1216(1977)].

\bibitem{Gribov:1972ri}
V.~N. Gribov and L.~N. Lipatov {\em Sov. J. Nucl. Phys.} {\bf 15} (1972)
  438--450. [Yad. Fiz.15,781(1972)].

\bibitem{Altarelli:1977zs}
G.~Altarelli and G.~Parisi {\em Nucl. Phys.} {\bf B126} (1977) 298--318.

\bibitem{Moch:2004pa}
S.~Moch, J.~A.~M. Vermaseren, and A.~Vogt {\em Nucl. Phys.} {\bf B688} (2004)
  101--134, [\href{http://arxiv.org/abs/hep-ph/0403192}{{\tt hep-ph/0403192}}].

\bibitem{Vogt:2004mw}
A.~Vogt, S.~Moch, and J.~A.~M. Vermaseren {\em Nucl. Phys.} {\bf B691} (2004)
  129--181, [\href{http://arxiv.org/abs/hep-ph/0404111}{{\tt hep-ph/0404111}}].

\bibitem{Davies:2016jie}
J.~Davies, A.~Vogt, B.~Ruijl, T.~Ueda, and J.~A.~M. Vermaseren {\em Nucl.
  Phys.} {\bf B915} (2017) 335--362,
  [\href{http://arxiv.org/abs/1610.07477}{{\tt arXiv:1610.07477}}].

\bibitem{Moch:2017uml}
S.~Moch, B.~Ruijl, T.~Ueda, J.~A.~M. Vermaseren, and A.~Vogt {\em JHEP} {\bf
  10} (2017) 041, [\href{http://arxiv.org/abs/1707.08315}{{\tt
  arXiv:1707.08315}}].

\bibitem{Moch:2014sna}
S.~Moch, J.~A.~M. Vermaseren, and A.~Vogt {\em Nucl. Phys.} {\bf B889} (2014)
  351--400, [\href{http://arxiv.org/abs/1409.5131}{{\tt arXiv:1409.5131}}].

\bibitem{Vogt:2004ns}
A.~Vogt {\em Comput. Phys. Commun.} {\bf 170} (2005) 65--92,
  [\href{http://arxiv.org/abs/hep-ph/0408244}{{\tt hep-ph/0408244}}].

\bibitem{Salam:2008qg}
G.~P. Salam and J.~Rojo {\em Comput. Phys. Commun.} {\bf 180} (2009) 120--156,
  [\href{http://arxiv.org/abs/0804.3755}{{\tt arXiv:0804.3755}}].

\bibitem{Botje:2010ay}
M.~Botje {\em Comput. Phys. Commun.} {\bf 182} (2011) 490--532,
  [\href{http://arxiv.org/abs/1005.1481}{{\tt arXiv:1005.1481}}].

\bibitem{Bertone:2013vaa}
V.~Bertone, S.~Carrazza, and J.~Rojo {\em Comput. Phys. Commun.} {\bf 185}
  (2014) 1647--1668, [\href{http://arxiv.org/abs/1310.1394}{{\tt
  arXiv:1310.1394}}].

\bibitem{Bertone:2015cwa}
V.~Bertone, S.~Carrazza, and E.~R. Nocera {\em JHEP} {\bf 03} (2015) 046,
  [\href{http://arxiv.org/abs/1501.00494}{{\tt arXiv:1501.00494}}].

\bibitem{Giele:2002hx}
W.~Giele et~al., {\it {The QCD / SM working group: Summary report}},  in {\em
  {Physics at TeV colliders. Proceedings, Euro Summer School, Les Houches,
  France, May 21-June 1, 2001}}, pp.~275--426, 2002.
\newblock \href{http://arxiv.org/abs/hep-ph/0204316}{{\tt hep-ph/0204316}}.

\bibitem{Dittmar:2005ed}
M.~Dittmar et~al. \href{http://arxiv.org/abs/hep-ph/0511119}{{\tt
  hep-ph/0511119}}.

\bibitem{DiPierro:2000nt}
M.~Di~Pierro, {\it {From Monte Carlo integration to lattice quantum
  chromodynamics: An Introduction}},  in {\em {GSA Summer School on Physics on
  the Frontier and in the Future Batavia, Illinois, July 31-August 7, 2000}},
  2000.
\newblock \href{http://arxiv.org/abs/hep-lat/0009001}{{\tt hep-lat/0009001}}.

\bibitem{Lepage:1998dt}
G.~P. Lepage, {\it {Lattice QCD for novices}},  in {\em {Strong interactions at
  low and intermediate energies. Proceedings, 13th Annual Hampton University
  Graduate Studies, HUGS'98, Newport News, USA, May 26-June 12, 1998}},
  pp.~49--90, 1998.
\newblock \href{http://arxiv.org/abs/hep-lat/0506036}{{\tt hep-lat/0506036}}.

\bibitem{Luscher:1998pe}
M.~Lüscher, {\it {Advanced lattice QCD}},  in {\em {Probing the standard model
  of particle interactions. Proceedings, Summer School in Theoretical Physics,
  NATO Advanced Study Institute, 68th session, Les Houches, France, July
  28-September 5, 1997. Pt. 1, 2}}, pp.~229--280, 1998.
\newblock \href{http://arxiv.org/abs/hep-lat/9802029}{{\tt hep-lat/9802029}}.

\bibitem{Binder:2015klx}
D.~Landau and K.~Binder, {\em {A Guide to Monte Carlo Simulations in
  Statistical Physics}}.
\newblock Cambridge University Press, 2015.

\bibitem{Newman:1999mng}
M.~Newman and G.~Barkema, {\em {Monte Carlo Methods in Statistical Physics}}.
\newblock Oxford University Press, 1999.

\bibitem{Brodsky:2015aia}
S.~J. Brodsky, A.~L. Deshpande, H.~Gao, R.~D. McKeown, C.~A. Meyer, Z.-E.
  Meziani, R.~G. Milner, J.~Qiu, D.~G. Richards, and C.~D. Roberts
  \href{http://arxiv.org/abs/1502.05728}{{\tt arXiv:1502.05728}}.

\bibitem{Aschenauer:2014twa}
E.-C. Aschenauer et~al. {\em Eur. Phys. J.} {\bf A53} (2017), no.~4 71,
  [\href{http://arxiv.org/abs/1410.8831}{{\tt arXiv:1410.8831}}].

\bibitem{Aoki:2016frl}
S.~Aoki et~al. {\em Eur. Phys. J.} {\bf C77} (2017), no.~2 112,
  [\href{http://arxiv.org/abs/1607.00299}{{\tt arXiv:1607.00299}}].

\bibitem{Creutz:2011hy}
M.~Creutz {\em Acta Phys. Slov.} {\bf 61} (2011) 1--127,
  [\href{http://arxiv.org/abs/1103.3304}{{\tt arXiv:1103.3304}}].

\bibitem{Vladikas:2011bp}
A.~Vladikas, {\it {Three Topics in Renormalization and Improvement}},  in {\em
  {Modern perspectives in lattice QCD: Quantum field theory and high
  performance computing. Proceedings, International School, 93rd Session, Les
  Houches, France, August 3-28, 2009}}, pp.~161--222, 2011.
\newblock \href{http://arxiv.org/abs/1103.1323}{{\tt arXiv:1103.1323}}.

\bibitem{Chandrasekharan:2004cn}
S.~Chandrasekharan and U.~J. Wiese {\em Prog. Part. Nucl. Phys.} {\bf 53}
  (2004) 373--418, [\href{http://arxiv.org/abs/hep-lat/0405024}{{\tt
  hep-lat/0405024}}].

\bibitem{Golterman:2009kw}
M.~Golterman, {\it {Applications of chiral perturbation theory to lattice
  QCD}},  in {\em {Modern perspectives in lattice QCD: Quantum field theory and
  high performance computing. Proceedings, International School, 93rd Session,
  Les Houches, France, August 3-28, 2009}}, pp.~423--515, 2009.
\newblock \href{http://arxiv.org/abs/0912.4042}{{\tt arXiv:0912.4042}}.

\bibitem{Luscher:1985dn}
M.~Lüscher {\em Commun. Math. Phys.} {\bf 104} (1986) 177.

\bibitem{Luscher:1986pf}
M.~Lüscher {\em Commun. Math. Phys.} {\bf 105} (1986) 153--188.

\bibitem{Lepage:1989hd}
G.~P. Lepage, {\it {The Analysis of Algorithms for Lattice Field Theory}},  in
  {\em {Boulder ASI 1989:97-120}}, pp.~97--120, 1989.

\bibitem{Martinelli:1994ty}
G.~Martinelli, C.~Pittori, C.~T. Sachrajda, M.~Testa, and A.~Vladikas {\em
  Nucl.Phys.} {\bf B445} (1995) 81--108,
  [\href{http://arxiv.org/abs/hep-lat/9411010}{{\tt hep-lat/9411010}}].

\bibitem{Gockeler:1996mu}
M.~Göckeler, R.~Horsley, E.-M. Ilgenfritz, H.~Perlt, P.~E.~L. Rakow,
  G.~Schierholz, and A.~Schiller {\em Phys. Rev.} {\bf D54} (1996) 5705--5714,
  [\href{http://arxiv.org/abs/hep-lat/9602029}{{\tt hep-lat/9602029}}].

\bibitem{Gockeler:2004wp}
{\bf QCDSF} Collaboration, M.~Göckeler, R.~Horsley, D.~Pleiter, P.~E.~L.
  Rakow, and G.~Schierholz {\em Phys. Rev.} {\bf D71} (2005) 114511,
  [\href{http://arxiv.org/abs/hep-ph/0410187}{{\tt hep-ph/0410187}}].

\bibitem{Luscher:2010iy}
M.~Lüscher {\em JHEP} {\bf 08} (2010) 071,
  [\href{http://arxiv.org/abs/1006.4518}{{\tt arXiv:1006.4518}}]. [Erratum:
  JHEP03,092(2014)].

\bibitem{Borsanyi:2012zs}
S.~Borsanyi et~al. {\em JHEP} {\bf 09} (2012) 010,
  [\href{http://arxiv.org/abs/1203.4469}{{\tt arXiv:1203.4469}}].

\bibitem{Durr:2008zz}
S.~Dürr et~al. {\em Science} {\bf 322} (2008) 1224--1227,
  [\href{http://arxiv.org/abs/0906.3599}{{\tt arXiv:0906.3599}}].

\bibitem{Kendall:2008zz}
{\bf HPQCD} Collaboration, I.~D. Kendall, C.~T.~H. Davies, C.~McNeile, G.~P.
  Lepage, and J.~Shigemitsu {\em PoS} {\bf LATTICE2008} (2008) 223.

\bibitem{Detmold:2003rq}
W.~Detmold, W.~Melnitchouk, and A.~W. Thomas {\em Mod.Phys.Lett.} {\bf A18}
  (2003) 2681--2698, [\href{http://arxiv.org/abs/hep-lat/0310003}{{\tt
  hep-lat/0310003}}].

\bibitem{Zimmermann:1972tv}
W.~Zimmermann {\em Annals Phys.} {\bf 77} (1973) 570--601. [Lect. Notes
  Phys.558,278(2000)].

\bibitem{Gockeler:1995wg}
M.~Göckeler, R.~Horsley, E.-M. Ilgenfritz, H.~Perlt, P.~E.~L. Rakow,
  G.~Schierholz, and A.~Schiller {\em Phys. Rev.} {\bf D53} (1996) 2317--2325,
  [\href{http://arxiv.org/abs/hep-lat/9508004}{{\tt hep-lat/9508004}}].

\bibitem{Blumlein:2008kz}
J.~Blümlein and H.~Böttcher {\em Phys. Lett.} {\bf B662} (2008) 336--340,
  [\href{http://arxiv.org/abs/0802.0408}{{\tt arXiv:0802.0408}}].

\bibitem{Martinelli:1996pk}
G.~Martinelli and C.~T. Sachrajda {\em Nucl. Phys.} {\bf B478} (1996) 660--686,
  [\href{http://arxiv.org/abs/hep-ph/9605336}{{\tt hep-ph/9605336}}].

\bibitem{Davoudi:2012ya}
Z.~Davoudi and M.~J. Savage {\em Phys. Rev.} {\bf D86} (2012) 054505,
  [\href{http://arxiv.org/abs/1204.4146}{{\tt arXiv:1204.4146}}].

\bibitem{Monahan:2015lha}
C.~Monahan and K.~Orginos {\em Phys. Rev.} {\bf D91} (2015), no.~7 074513,
  [\href{http://arxiv.org/abs/1501.05348}{{\tt arXiv:1501.05348}}].

\bibitem{Detmold:2005gg}
W.~Detmold and C.~J.~D. Lin {\em Phys. Rev.} {\bf D73} (2006) 014501,
  [\href{http://arxiv.org/abs/hep-lat/0507007}{{\tt hep-lat/0507007}}].

\bibitem{Braun:2007wv}
V.~Braun and D.~Mueller {\em Eur. Phys. J.} {\bf C55} (2008) 349--361,
  [\href{http://arxiv.org/abs/0709.1348}{{\tt arXiv:0709.1348}}].

\bibitem{Liu:1993cv}
K.-F. Liu and S.-J. Dong {\em Phys. Rev. Lett.} {\bf 72} (1994) 1790--1793,
  [\href{http://arxiv.org/abs/hep-ph/9306299}{{\tt hep-ph/9306299}}].

\bibitem{Liu:1999ak}
K.-F. Liu {\em Phys. Rev.} {\bf D62} (2000) 074501,
  [\href{http://arxiv.org/abs/hep-ph/9910306}{{\tt hep-ph/9910306}}].

\bibitem{Hansen:2017mnd}
M.~T. Hansen, H.~B. Meyer, and D.~Robaina {\em Phys. Rev.} {\bf D96} (2017),
  no.~9 094513, [\href{http://arxiv.org/abs/1704.08993}{{\tt
  arXiv:1704.08993}}].

\bibitem{Liu:2016djw}
K.-F. Liu {\em PoS} {\bf LATTICE2015} (2016) 115,
  [\href{http://arxiv.org/abs/1603.07352}{{\tt arXiv:1603.07352}}].

\bibitem{Liu:2017lpe}
K.-F. Liu {\em Phys. Rev.} {\bf D96} (2017), no.~3 033001,
  [\href{http://arxiv.org/abs/1703.04690}{{\tt arXiv:1703.04690}}].

\bibitem{Chambers:2017dov}
A.~J. Chambers, R.~Horsley, Y.~Nakamura, H.~Perlt, P.~E.~L. Rakow,
  G.~Schierholz, A.~Schiller, K.~Somfleth, R.~D. Young, and J.~M. Zanotti {\em
  Phys. Rev. Lett.} {\bf 118} (2017), no.~24 242001,
  [\href{http://arxiv.org/abs/1703.01153}{{\tt arXiv:1703.01153}}].

\bibitem{Horsley:2012pz}
{\bf UKQCD, QCDSF} Collaboration, R.~Horsley, R.~Millo, Y.~Nakamura, H.~Perlt,
  D.~Pleiter, P.~E.~L. Rakow, G.~Schierholz, A.~Schiller, F.~Winter, and J.~M.
  Zanotti {\em Phys. Lett.} {\bf B714} (2012) 312--316,
  [\href{http://arxiv.org/abs/1205.6410}{{\tt arXiv:1205.6410}}].

\bibitem{Chambers:2014qaa}
{\bf QCDSF/UKQCD, CSSM} Collaboration, A.~J. Chambers et~al. {\em Phys. Rev.}
  {\bf D90} (2014), no.~1 014510, [\href{http://arxiv.org/abs/1405.3019}{{\tt
  arXiv:1405.3019}}].

\bibitem{Chambers:2015bka}
A.~J. Chambers et~al. {\em Phys. Rev.} {\bf D92} (2015), no.~11 114517,
  [\href{http://arxiv.org/abs/1508.06856}{{\tt arXiv:1508.06856}}].

\bibitem{Ji:2001wha}
X.~Ji and C.~Jung {\em Phys. Rev. Lett.} {\bf 86} (2001) 208,
  [\href{http://arxiv.org/abs/hep-lat/0101014}{{\tt hep-lat/0101014}}].

\bibitem{Ji:2013dva}
X.~Ji {\em Phys.Rev.Lett.} {\bf 110} (2013), no.~26 262002,
  [\href{http://arxiv.org/abs/1305.1539}{{\tt arXiv:1305.1539}}].

\bibitem{Ji:2014gla}
X.~Ji {\em Sci.China Phys.Mech.Astron.} {\bf 57} (2014) 1407--1412,
  [\href{http://arxiv.org/abs/1404.6680}{{\tt arXiv:1404.6680}}].

\bibitem{Xiong:2013bka}
X.~Xiong, X.~Ji, J.-H. Zhang, and Y.~Zhao {\em Phys. Rev.} {\bf D90} (2014),
  no.~1 014051, [\href{http://arxiv.org/abs/1310.7471}{{\tt arXiv:1310.7471}}].

\bibitem{Radyushkin:2016hsy}
A.~Radyushkin {\em Phys. Lett.} {\bf B767} (2017) 314--320,
  [\href{http://arxiv.org/abs/1612.05170}{{\tt arXiv:1612.05170}}].

\bibitem{Radyushkin:2017cyf}
A.~V. Radyushkin {\em Phys. Rev.} {\bf D96} (2017), no.~3 034025,
  [\href{http://arxiv.org/abs/1705.01488}{{\tt arXiv:1705.01488}}].

\bibitem{Orginos:2017kos}
K.~Orginos, A.~Radyushkin, J.~Karpie, and S.~Zafeiropoulos {\em Phys. Rev.}
  {\bf D96} (2017), no.~9 094503, [\href{http://arxiv.org/abs/1706.05373}{{\tt
  arXiv:1706.05373}}].

\bibitem{Gamberg:2014zwa}
L.~Gamberg, Z.-B. Kang, I.~Vitev, and H.~Xing {\em Phys. Lett.} {\bf B743}
  (2015) 112--120, [\href{http://arxiv.org/abs/1412.3401}{{\tt
  arXiv:1412.3401}}].

\bibitem{Bali:2016lva}
G.~S. Bali, B.~Lang, B.~U. Musch, and A.~Schäfer {\em Phys. Rev.} {\bf D93}
  (2016), no.~9 094515, [\href{http://arxiv.org/abs/1602.05525}{{\tt
  arXiv:1602.05525}}].

\bibitem{Green:2017xeu}
J.~Green, K.~Jansen, and F.~Steffens
  \href{http://arxiv.org/abs/1707.07152}{{\tt arXiv:1707.07152}}.

\bibitem{Wang:2017qyg}
W.~Wang, S.~Zhao, and R.~Zhu \href{http://arxiv.org/abs/1708.02458}{{\tt
  arXiv:1708.02458}}.

\bibitem{Ma:2014jla}
Y.-Q. Ma and J.-W. Qiu \href{http://arxiv.org/abs/1404.6860}{{\tt
  arXiv:1404.6860}}.

\bibitem{Ma:2014jga}
Y.-Q. Ma and J.-W. Qiu {\em Int. J. Mod. Phys. Conf. Ser.} {\bf 37} (2015)
  1560041, [\href{http://arxiv.org/abs/1412.2688}{{\tt arXiv:1412.2688}}].

\bibitem{Ma:2017pxb}
Y.-Q. Ma and J.-W. Qiu {\em Phys. Rev. Lett.} {\bf 120} (2018), no.~2 022003,
  [\href{http://arxiv.org/abs/1709.03018}{{\tt arXiv:1709.03018}}].

\bibitem{Li:2016amo}
H.-n. Li {\em Phys. Rev.} {\bf D94} (2016), no.~7 074036,
  [\href{http://arxiv.org/abs/1602.07575}{{\tt arXiv:1602.07575}}].

\bibitem{Carlson:2017gpk}
C.~E. Carlson and M.~Freid {\em Phys. Rev.} {\bf D95} (2017), no.~9 094504,
  [\href{http://arxiv.org/abs/1702.05775}{{\tt arXiv:1702.05775}}].

\bibitem{Briceno:2017cpo}
R.~A. Briceño, M.~T. Hansen, and C.~J. Monahan {\em Phys. Rev.} {\bf D96}
  (2017), no.~1 014502, [\href{http://arxiv.org/abs/1703.06072}{{\tt
  arXiv:1703.06072}}].

\bibitem{Dotsenko:1979wb}
V.~S. Dotsenko and S.~N. Vergeles {\em Nucl. Phys.} {\bf B169} (1980) 527--546.

\bibitem{Arefeva:1980zd}
I.~{\relax Ya}. Arefeva {\em Phys. Lett.} {\bf B93} (1980) 347--353.

\bibitem{Craigie:1980qs}
N.~S. Craigie and H.~Dorn {\em Nucl. Phys.} {\bf B185} (1981) 204--220.

\bibitem{Stefanis:1983ke}
N.~G. Stefanis {\em Nuovo Cim.} {\bf A83} (1984) 205.

\bibitem{Dorn:1986dt}
H.~Dorn {\em Fortsch. Phys.} {\bf 34} (1986) 11--56.

\bibitem{Ji:2017oey}
X.~Ji, J.-H. Zhang, and Y.~Zhao \href{http://arxiv.org/abs/1706.08962}{{\tt
  arXiv:1706.08962}}.

\bibitem{Ishikawa:2017faj}
T.~Ishikawa, Y.-Q. Ma, J.-W. Qiu, and S.~Yoshida {\em Phys. Rev.} {\bf D96}
  (2017), no.~9 094019, [\href{http://arxiv.org/abs/1707.03107}{{\tt
  arXiv:1707.03107}}].

\bibitem{Musch:2010ka}
B.~U. Musch, P.~Hagler, J.~W. Negele, and A.~Schafer {\em Phys. Rev.} {\bf D83}
  (2011) 094507, [\href{http://arxiv.org/abs/1011.1213}{{\tt
  arXiv:1011.1213}}].

\bibitem{Ishikawa:2016znu}
T.~Ishikawa, Y.-Q. Ma, J.-W. Qiu, and S.~Yoshida
  \href{http://arxiv.org/abs/1609.02018}{{\tt arXiv:1609.02018}}.

\bibitem{Chen:2016fxx}
J.-W. Chen, X.~Ji, and J.-H. Zhang {\em Nucl. Phys.} {\bf B915} (2017) 1--9,
  [\href{http://arxiv.org/abs/1609.08102}{{\tt arXiv:1609.08102}}].

\bibitem{Monahan:2016bvm}
C.~Monahan and K.~Orginos {\em JHEP} {\bf 03} (2017) 116,
  [\href{http://arxiv.org/abs/1612.01584}{{\tt arXiv:1612.01584}}].

\bibitem{Xiong:2017jtn}
X.~Xiong, T.~Luu, and U.-G. Meissner
  \href{http://arxiv.org/abs/1705.00246}{{\tt arXiv:1705.00246}}.

\bibitem{Constantinou:2017sej}
M.~Constantinou and H.~Panagopoulos {\em Phys. Rev.} {\bf D96} (2017), no.~5
  054506, [\href{http://arxiv.org/abs/1705.11193}{{\tt arXiv:1705.11193}}].

\bibitem{Chen:2017mzz}
J.-W. Chen, T.~Ishikawa, L.~Jin, H.-W. Lin, Y.-B. Yang, J.-H. Zhang, and
  Y.~Zhao {\em Phys. Rev.} {\bf D97} (2018), no.~1 014505,
  [\href{http://arxiv.org/abs/1706.01295}{{\tt arXiv:1706.01295}}].

\bibitem{Chen:2017mie}
J.-W. Chen, T.~Ishikawa, L.~Jin, H.-W. Lin, Y.-B. Yang, J.-H. Zhang, and
  Y.~Zhao \href{http://arxiv.org/abs/1710.01089}{{\tt arXiv:1710.01089}}.

\bibitem{Luscher:1991wu}
M.~Lüscher, P.~Weisz, and U.~Wolff {\em Nucl. Phys.} {\bf B359} (1991)
  221--243.

\bibitem{Alexandrou:2017huk}
C.~Alexandrou, K.~Cichy, M.~Constantinou, K.~Hadjiyiannakou, K.~Jansen,
  H.~Panagopoulos, and F.~Steffens {\em Nucl. Phys.} {\bf B923} (2017)
  394--415, [\href{http://arxiv.org/abs/1706.00265}{{\tt arXiv:1706.00265}}].

\bibitem{Rossi:2017muf}
G.~C. Rossi and M.~Testa {\em Phys. Rev.} {\bf D96} (2017), no.~1 014507,
  [\href{http://arxiv.org/abs/1706.04428}{{\tt arXiv:1706.04428}}].

\bibitem{Alexandrou:2016eyt}
C.~Alexandrou, K.~Cichy, M.~Constantinou, K.~Hadjiyiannakou, K.~Jansen,
  F.~Steffens, and C.~Wiese {\em PoS} {\bf LATTICE2016} (2016) 151,
  [\href{http://arxiv.org/abs/1612.08728}{{\tt arXiv:1612.08728}}].

\bibitem{Lin:2017ani}
H.-W. Lin, J.-W. Chen, T.~Ishikawa, and J.-H. Zhang
  \href{http://arxiv.org/abs/1708.05301}{{\tt arXiv:1708.05301}}.

\bibitem{Alexandrou:2017dzj}
C.~Alexandrou, S.~Bacchio, K.~Cichy, M.~Constantinou, K.~Hadjiyiannakou,
  K.~Jansen, G.~Koutsou, A.~Scapellato, and F.~Steffens, {\it {Computation of
  parton distributions from the quasi-PDF approach at the physical point}},  in
  {\em {35th International Symposium on Lattice Field Theory (Lattice 2017)
  Granada, Spain, June 18-24, 2017}}, 2017.
\newblock \href{http://arxiv.org/abs/1710.06408}{{\tt arXiv:1710.06408}}.

\bibitem{Ioffe:1969kf}
B.~L. Ioffe {\em Phys. Lett.} {\bf 30B} (1969) 123--125.

\bibitem{Braun:1994jq}
V.~Braun, P.~Gornicki, and L.~Mankiewicz {\em Phys. Rev.} {\bf D51} (1995)
  6036--6051, [\href{http://arxiv.org/abs/hep-ph/9410318}{{\tt
  hep-ph/9410318}}].

\bibitem{Ji:2017rah}
X.~Ji, J.-H. Zhang, and Y.~Zhao {\em Nucl. Phys.} {\bf B924} (2017) 366--376,
  [\href{http://arxiv.org/abs/1706.07416}{{\tt arXiv:1706.07416}}].

\bibitem{Radyushkin:2017lvu}
A.~V. Radyushkin \href{http://arxiv.org/abs/1710.08813}{{\tt
  arXiv:1710.08813}}.

\bibitem{Karpie:2017bzm}
J.~Karpie, K.~Orginos, A.~Radyushkin, and S.~Zafeiropoulos
  \href{http://arxiv.org/abs/1710.08288}{{\tt arXiv:1710.08288}}.

\bibitem{Broniowski:2017gfp}
W.~Broniowski and E.~Ruiz~Arriola \href{http://arxiv.org/abs/1711.03377}{{\tt
  arXiv:1711.03377}}.

\bibitem{Chen:2017lnm}
J.-W. Chen, T.~Ishikawa, L.~Jin, H.-W. Lin, A.~Schäfer, Y.-B. Yang, J.-H.
  Zhang, and Y.~Zhao \href{http://arxiv.org/abs/1711.07858}{{\tt
  arXiv:1711.07858}}.

\bibitem{Stratmann:2001pb}
M.~Stratmann and W.~Vogelsang {\em Phys. Rev.} {\bf D64} (2001) 114007,
  [\href{http://arxiv.org/abs/hep-ph/0107064}{{\tt hep-ph/0107064}}].

\bibitem{Nadolsky:2003fz}
P.~M. Nadolsky and C.~P. Yuan {\em Nucl. Phys.} {\bf B666} (2003) 3--30,
  [\href{http://arxiv.org/abs/hep-ph/0304001}{{\tt hep-ph/0304001}}].

\bibitem{Collins:1978wz}
J.~C. Collins, F.~Wilczek, and A.~Zee {\em Phys. Rev.} {\bf D18} (1978) 242.

\bibitem{Brodsky:1973kr}
S.~J. Brodsky and G.~R. Farrar {\em Phys. Rev. Lett.} {\bf 31} (1973)
  1153--1156.

\bibitem{Forte:2002fg}
S.~Forte, L.~Garrido, J.~I. Latorre, and A.~Piccione {\em JHEP} {\bf 05} (2002)
  062, [\href{http://arxiv.org/abs/hep-ph/0204232}{{\tt hep-ph/0204232}}].

\bibitem{DelDebbio:2004xtd}
{\bf NNPDF} Collaboration, L.~Del~Debbio, S.~Forte, J.~I. Latorre, A.~Piccione,
  and J.~Rojo {\em JHEP} {\bf 03} (2005) 080,
  [\href{http://arxiv.org/abs/hep-ph/0501067}{{\tt hep-ph/0501067}}].

\bibitem{Ethier:2017zbq}
J.~J. Ethier, N.~Sato, and W.~Melnitchouk {\em Phys. Rev. Lett.} {\bf 119}
  (2017), no.~13 132001, [\href{http://arxiv.org/abs/1705.05889}{{\tt
  arXiv:1705.05889}}].

\bibitem{Brodsky:2015fna}
S.~J. Brodsky, A.~Kusina, F.~Lyonnet, I.~Schienbein, H.~Spiesberger, and
  R.~Vogt {\em Adv. High Energy Phys.} {\bf 2015} (2015) 231547,
  [\href{http://arxiv.org/abs/1504.06287}{{\tt arXiv:1504.06287}}].

\bibitem{Ball:2016neh}
{\bf NNPDF} Collaboration, R.~D. Ball, V.~Bertone, M.~Bonvini, S.~Carrazza,
  S.~Forte, A.~Guffanti, N.~P. Hartland, J.~Rojo, and L.~Rottoli {\em Eur.
  Phys. J.} {\bf C76} (2016), no.~11 647,
  [\href{http://arxiv.org/abs/1605.06515}{{\tt arXiv:1605.06515}}].

\bibitem{Hou:2017khm}
T.-J. Hou, S.~Dulat, J.~Gao, M.~Guzzi, J.~Huston, P.~Nadolsky, C.~Schmidt,
  J.~Winter, K.~Xie, and C.~P. Yuan \href{http://arxiv.org/abs/1707.00657}{{\tt
  arXiv:1707.00657}}.

\bibitem{Gao:2013xoa}
J.~Gao, M.~Guzzi, J.~Huston, H.-L. Lai, Z.~Li, P.~Nadolsky, J.~Pumplin,
  D.~Stump, and C.~P. Yuan {\em Phys. Rev.} {\bf D89} (2014), no.~3 033009,
  [\href{http://arxiv.org/abs/1302.6246}{{\tt arXiv:1302.6246}}].

\bibitem{Ball:2009qv}
{\bf NNPDF} Collaboration, R.~D. Ball, L.~Del~Debbio, S.~Forte, A.~Guffanti,
  J.~I. Latorre, J.~Rojo, and M.~Ubiali {\em JHEP} {\bf 05} (2010) 075,
  [\href{http://arxiv.org/abs/0912.2276}{{\tt arXiv:0912.2276}}].

\bibitem{DAgostini:2003syq}
G.~D'Agostini, {\em {Bayesian reasoning in data analysis: A critical
  introduction}}.
\newblock 2003.

\bibitem{DAgostini:1993arp}
G.~D'Agostini {\em Nucl. Instrum. Meth.} {\bf A346} (1994) 306--311.

\bibitem{Stump:2001gu}
D.~Stump, J.~Pumplin, R.~Brock, D.~Casey, J.~Huston, J.~Kalk, H.~L. Lai, and
  W.~K. Tung {\em Phys. Rev.} {\bf D65} (2001) 014012,
  [\href{http://arxiv.org/abs/hep-ph/0101051}{{\tt hep-ph/0101051}}].

\bibitem{Pumplin:2001ct}
J.~Pumplin, D.~Stump, R.~Brock, D.~Casey, J.~Huston, J.~Kalk, H.~L. Lai, and
  W.~K. Tung {\em Phys. Rev.} {\bf D65} (2001) 014013,
  [\href{http://arxiv.org/abs/hep-ph/0101032}{{\tt hep-ph/0101032}}].

\bibitem{Pumplin:2002vw}
J.~Pumplin, D.~R. Stump, J.~Huston, H.~L. Lai, P.~M. Nadolsky, and W.~K. Tung
  {\em JHEP} {\bf 07} (2002) 012,
  [\href{http://arxiv.org/abs/hep-ph/0201195}{{\tt hep-ph/0201195}}].

\bibitem{Giele:1998gw}
W.~T. Giele and S.~Keller {\em Phys. Rev.} {\bf D58} (1998) 094023,
  [\href{http://arxiv.org/abs/hep-ph/9803393}{{\tt hep-ph/9803393}}].

\bibitem{Giele:2001mr}
W.~T. Giele, S.~A. Keller, and D.~A. Kosower
  \href{http://arxiv.org/abs/hep-ph/0104052}{{\tt hep-ph/0104052}}.

\bibitem{Watt:2012tq}
G.~Watt and R.~S. Thorne {\em JHEP} {\bf 08} (2012) 052,
  [\href{http://arxiv.org/abs/1205.4024}{{\tt arXiv:1205.4024}}].

\bibitem{Hou:2016sho}
T.-J. Hou et~al. {\em JHEP} {\bf 03} (2017) 099,
  [\href{http://arxiv.org/abs/1607.06066}{{\tt arXiv:1607.06066}}].

\bibitem{Gao:2013bia}
J.~Gao and P.~Nadolsky {\em JHEP} {\bf 07} (2014) 035,
  [\href{http://arxiv.org/abs/1401.0013}{{\tt arXiv:1401.0013}}].

\bibitem{Carrazza:2015aoa}
S.~Carrazza, S.~Forte, Z.~Kassabov, J.~I. Latorre, and J.~Rojo {\em Eur. Phys.
  J.} {\bf C75} (2015), no.~8 369, [\href{http://arxiv.org/abs/1505.06736}{{\tt
  arXiv:1505.06736}}].

\bibitem{Binoth:2010nha}
{\bf SM and NLO Multileg Working Group} Collaboration, T.~Binoth et~al., {\it
  {The SM and NLO Multileg Working Group: Summary report}},  in {\em {Physics
  at TeV colliders. Proceedings, 6th Workshop, dedicated to Thomas Binoth, Les
  Houches, France, June 8-26, 2009}}, pp.~21--189, 2010.
\newblock \href{http://arxiv.org/abs/1003.1241}{{\tt arXiv:1003.1241}}.

\bibitem{Thorne:2012az}
R.~S. Thorne {\em Phys. Rev.} {\bf D86} (2012) 074017,
  [\href{http://arxiv.org/abs/1201.6180}{{\tt arXiv:1201.6180}}].

\bibitem{Martin:2003sk}
A.~D. Martin, R.~G. Roberts, W.~J. Stirling, and R.~S. Thorne {\em Eur. Phys.
  J.} {\bf C35} (2004) 325--348,
  [\href{http://arxiv.org/abs/hep-ph/0308087}{{\tt hep-ph/0308087}}].

\bibitem{Accardi:2009br}
A.~Accardi, M.~E. Christy, C.~E. Keppel, P.~Monaghan, W.~Melnitchouk, J.~G.
  Morfin, and J.~F. Owens {\em Phys. Rev.} {\bf D81} (2010) 034016,
  [\href{http://arxiv.org/abs/0911.2254}{{\tt arXiv:0911.2254}}].

\bibitem{Martin:2009iq}
A.~D. Martin, W.~J. Stirling, R.~S. Thorne, and G.~Watt {\em Eur. Phys. J.}
  {\bf C63} (2009) 189--285, [\href{http://arxiv.org/abs/0901.0002}{{\tt
  arXiv:0901.0002}}].

\bibitem{Ball:2009mk}
{\bf NNPDF} Collaboration, R.~D. Ball, L.~Del~Debbio, S.~Forte, A.~Guffanti,
  J.~I. Latorre, A.~Piccione, J.~Rojo, and M.~Ubiali {\em Nucl. Phys.} {\bf
  B823} (2009) 195--233, [\href{http://arxiv.org/abs/0906.1958}{{\tt
  arXiv:0906.1958}}].

\bibitem{Martin:2012da}
A.~D. Martin, A.~J. T.~M. Mathijssen, W.~J. Stirling, R.~S. Thorne, B.~J.~A.
  Watt, and G.~Watt {\em Eur. Phys. J.} {\bf C73} (2013), no.~2 2318,
  [\href{http://arxiv.org/abs/1211.1215}{{\tt arXiv:1211.1215}}].

\bibitem{Accardi:2011fa}
A.~Accardi, W.~Melnitchouk, J.~F. Owens, M.~E. Christy, C.~E. Keppel, L.~Zhu,
  and J.~G. Morfin {\em Phys. Rev.} {\bf D84} (2011) 014008,
  [\href{http://arxiv.org/abs/1102.3686}{{\tt arXiv:1102.3686}}].

\bibitem{Manohar:2016nzj}
A.~Manohar, P.~Nason, G.~P. Salam, and G.~Zanderighi {\em Phys. Rev. Lett.}
  {\bf 117} (2016), no.~24 242002, [\href{http://arxiv.org/abs/1607.04266}{{\tt
  arXiv:1607.04266}}].

\bibitem{Manohar:2017eqh}
A.~V. Manohar, P.~Nason, G.~P. Salam, and G.~Zanderighi
  \href{http://arxiv.org/abs/1708.01256}{{\tt arXiv:1708.01256}}.

\bibitem{Campbell:2016lzl}
J.~M. Campbell, R.~K. Ellis, and C.~Williams {\em Phys. Rev. Lett.} {\bf 118}
  (2017), no.~22 222001, [\href{http://arxiv.org/abs/1612.04333}{{\tt
  arXiv:1612.04333}}].

\bibitem{Czakon:2016dgf}
M.~Czakon, D.~Heymes, and A.~Mitov {\em JHEP} {\bf 04} (2017) 071,
  [\href{http://arxiv.org/abs/1606.03350}{{\tt arXiv:1606.03350}}].

\bibitem{Li:2012wna}
Y.~Li and F.~Petriello {\em Phys. Rev.} {\bf D86} (2012) 094034,
  [\href{http://arxiv.org/abs/1208.5967}{{\tt arXiv:1208.5967}}].

\bibitem{Carli:2010rw}
T.~Carli, D.~Clements, A.~Cooper-Sarkar, C.~Gwenlan, G.~P. Salam, F.~Siegert,
  P.~Starovoitov, and M.~Sutton {\em Eur. Phys. J.} {\bf C66} (2010) 503--524,
  [\href{http://arxiv.org/abs/0911.2985}{{\tt arXiv:0911.2985}}].

\bibitem{Wobisch:2011ij}
{\bf fastNLO} Collaboration, M.~Wobisch, D.~Britzger, T.~Kluge, K.~Rabbertz,
  and F.~Stober \href{http://arxiv.org/abs/1109.1310}{{\tt arXiv:1109.1310}}.

\bibitem{Bertone:2014zva}
V.~Bertone, R.~Frederix, S.~Frixione, J.~Rojo, and M.~Sutton {\em JHEP} {\bf
  08} (2014) 166, [\href{http://arxiv.org/abs/1406.7693}{{\tt
  arXiv:1406.7693}}].

\bibitem{Owens:2012bv}
J.~F. Owens, A.~Accardi, and W.~Melnitchouk {\em Phys. Rev.} {\bf D87} (2013),
  no.~9 094012, [\href{http://arxiv.org/abs/1212.1702}{{\tt arXiv:1212.1702}}].

\bibitem{Carrazza:2014gfa}
S.~Carrazza, A.~Ferrara, D.~Palazzo, and J.~Rojo {\em J. Phys.} {\bf G42}
  (2015), no.~5 057001, [\href{http://arxiv.org/abs/1410.5456}{{\tt
  arXiv:1410.5456}}].

\bibitem{Carrazza:2015hva}
S.~Carrazza, J.~I. Latorre, J.~Rojo, and G.~Watt {\em Eur. Phys. J.} {\bf C75}
  (2015) 474, [\href{http://arxiv.org/abs/1504.06469}{{\tt arXiv:1504.06469}}].

\bibitem{Carrazza:2016htc}
S.~Carrazza, S.~Forte, Z.~Kassabov, and J.~Rojo {\em Eur. Phys. J.} {\bf C76}
  (2016), no.~4 205, [\href{http://arxiv.org/abs/1602.00005}{{\tt
  arXiv:1602.00005}}].

\bibitem{Altarelli:1998gn}
G.~Altarelli, S.~Forte, and G.~Ridolfi {\em Nucl. Phys.} {\bf B534} (1998)
  277--296, [\href{http://arxiv.org/abs/hep-ph/9806345}{{\tt hep-ph/9806345}}].

\bibitem{Cabibbo:2003cu}
N.~Cabibbo, E.~C. Swallow, and R.~Winston {\em Ann. Rev. Nucl. Part. Sci.} {\bf
  53} (2003) 39--75, [\href{http://arxiv.org/abs/hep-ph/0307298}{{\tt
  hep-ph/0307298}}].

\bibitem{FloresMendieta:1998ii}
R.~Flores-Mendieta, E.~E. Jenkins, and A.~V. Manohar {\em Phys. Rev.} {\bf D58}
  (1998) 094028, [\href{http://arxiv.org/abs/hep-ph/9805416}{{\tt
  hep-ph/9805416}}].

\bibitem{deFlorian:2014xna}
D.~de~Florian, R.~Sassot, M.~Epele, R.~J. Hernández-Pinto, and M.~Stratmann
  {\em Phys. Rev.} {\bf D91} (2015), no.~1 014035,
  [\href{http://arxiv.org/abs/1410.6027}{{\tt arXiv:1410.6027}}].

\bibitem{deFlorian:2017lwf}
D.~de~Florian, M.~Epele, R.~J. Hernández-Pinto, R.~Sassot, and M.~Stratmann
  {\em Phys. Rev.} {\bf D95} (2017), no.~9 094019,
  [\href{http://arxiv.org/abs/1702.06353}{{\tt arXiv:1702.06353}}].

\bibitem{Hirai:2016loo}
M.~Hirai, H.~Kawamura, S.~Kumano, and K.~Saito {\em PTEP} {\bf 2016} (2016),
  no.~11 113B04, [\href{http://arxiv.org/abs/1608.04067}{{\tt
  arXiv:1608.04067}}].

\bibitem{Sato:2016wqj}
N.~Sato, J.~J. Ethier, W.~Melnitchouk, M.~Hirai, S.~Kumano, and A.~Accardi {\em
  Phys. Rev.} {\bf D94} (2016), no.~11 114004,
  [\href{http://arxiv.org/abs/1609.00899}{{\tt arXiv:1609.00899}}].

\bibitem{Bertone:2017tyb}
{\bf NNPDF} Collaboration, V.~Bertone, S.~Carrazza, N.~P. Hartland, E.~R.
  Nocera, and J.~Rojo {\em Eur. Phys. J.} {\bf C77} (2017), no.~8 516,
  [\href{http://arxiv.org/abs/1706.07049}{{\tt arXiv:1706.07049}}].

\bibitem{Borsa:2017vwy}
I.~Borsa, R.~Sassot, and M.~Stratmann
  \href{http://arxiv.org/abs/1708.01630}{{\tt arXiv:1708.01630}}.

\bibitem{Aschenauer:2015eha}
E.-C. Aschenauer et~al. \href{http://arxiv.org/abs/1501.01220}{{\tt
  arXiv:1501.01220}}.

\bibitem{Bourrely:1993dd}
C.~Bourrely and J.~Soffer {\em Phys. Lett.} {\bf B314} (1993) 132--138.

\bibitem{Bourrely:1990pz}
C.~Bourrely, J.~P. Guillet, and J.~Soffer {\em Nucl. Phys.} {\bf B361} (1991)
  72--92.

\bibitem{Dudek:2012vr}
J.~Dudek et~al. {\em Eur. Phys. J.} {\bf A48} (2012) 187,
  [\href{http://arxiv.org/abs/1208.1244}{{\tt arXiv:1208.1244}}].

\bibitem{Accardi:2012qut}
A.~Accardi et~al. {\em Eur. Phys. J.} {\bf A52} (2016), no.~9 268,
  [\href{http://arxiv.org/abs/1212.1701}{{\tt arXiv:1212.1701}}].

\bibitem{Aschenauer:2014cki}
E.~C. Aschenauer et~al. \href{http://arxiv.org/abs/1409.1633}{{\tt
  arXiv:1409.1633}}.

\bibitem{Shahri:2016uzl}
F.~Taghavi-Shahri, H.~Khanpour, S.~Atashbar~Tehrani, and Z.~Alizadeh~Yazdi {\em
  Phys. Rev.} {\bf D93} (2016), no.~11 114024,
  [\href{http://arxiv.org/abs/1603.03157}{{\tt arXiv:1603.03157}}].

\bibitem{Khanpour:2017cha}
H.~Khanpour, S.~T. Monfared, and S.~Atashbar~Tehrani {\em Phys. Rev.} {\bf D95}
  (2017), no.~7 074006, [\href{http://arxiv.org/abs/1703.09209}{{\tt
  arXiv:1703.09209}}].

\bibitem{Nocera:2016xhb}
E.~Nocera and S.~Pisano {\em PoS} {\bf DIS2016} (2016) 284,
  [\href{http://arxiv.org/abs/1608.08575}{{\tt arXiv:1608.08575}}].

\bibitem{deFlorian:2008mr}
D.~de~Florian, R.~Sassot, M.~Stratmann, and W.~Vogelsang {\em Phys. Rev. Lett.}
  {\bf 101} (2008) 072001, [\href{http://arxiv.org/abs/0804.0422}{{\tt
  arXiv:0804.0422}}].

\bibitem{Leader:2010rb}
E.~Leader, A.~V. Sidorov, and D.~B. Stamenov {\em Phys. Rev.} {\bf D82} (2010)
  114018, [\href{http://arxiv.org/abs/1010.0574}{{\tt arXiv:1010.0574}}].

\bibitem{Blumlein:2010rn}
J.~Blümlein and H.~Böttcher {\em Nucl. Phys.} {\bf B841} (2010) 205--230,
  [\href{http://arxiv.org/abs/1005.3113}{{\tt arXiv:1005.3113}}].

\bibitem{Bourrely:2014uha}
C.~Bourrely and J.~Soffer {\em Phys. Lett.} {\bf B740} (2015) 168--171,
  [\href{http://arxiv.org/abs/1408.7057}{{\tt arXiv:1408.7057}}].

\bibitem{Nocera:2014vla}
E.~R. Nocera, {\em {Unbiased spin-dependent Parton Distribution Functions}}.
\newblock PhD thesis, Milan U., 2014.
\newblock \href{http://arxiv.org/abs/1403.0440}{{\tt arXiv:1403.0440}}.

\bibitem{Ball:2013lla}
{\bf NNPDF} Collaboration, R.~D. Ball, S.~Forte, A.~Guffanti, E.~R. Nocera,
  G.~Ridolfi, and J.~Rojo {\em Nucl. Phys.} {\bf B874} (2013) 36--84,
  [\href{http://arxiv.org/abs/1303.7236}{{\tt arXiv:1303.7236}}].

\bibitem{Adamczyk:2014ozi}
{\bf STAR} Collaboration, L.~Adamczyk et~al. {\em Phys. Rev. Lett.} {\bf 115}
  (2015), no.~9 092002, [\href{http://arxiv.org/abs/1405.5134}{{\tt
  arXiv:1405.5134}}].

\bibitem{Adare:2014hsq}
{\bf PHENIX Collaboration} Collaboration, A.~Adare et~al. {\em Phys.Rev.} {\bf
  D90} (2014), no.~1 012007--, [\href{http://arxiv.org/abs/1402.6296}{{\tt
  arXiv:1402.6296}}].

\bibitem{Adamczyk:2012qj}
{\bf STAR Collaboration} Collaboration, L.~Adamczyk et~al. {\em Phys.Rev.} {\bf
  D86} (2012) 032006--, [\href{http://arxiv.org/abs/1205.2735}{{\tt
  arXiv:1205.2735}}].

\bibitem{Adare:2010cc}
{\bf PHENIX Collaboration} Collaboration, A.~Adare et~al. {\em Phys.Rev.} {\bf
  D84} (2011) 012006--, [\href{http://arxiv.org/abs/1009.4921}{{\tt
  arXiv:1009.4921}}].

\bibitem{Adamczyk:2014xyw}
{\bf STAR Collaboration} Collaboration, L.~Adamczyk et~al. {\em Phys.Rev.Lett.}
  {\bf 113} (2014) 072301--, [\href{http://arxiv.org/abs/1404.6880}{{\tt
  arXiv:1404.6880}}].

\bibitem{Ball:2010gb}
{\bf NNPDF} Collaboration, R.~D. Ball, V.~Bertone, F.~Cerutti, L.~Del~Debbio,
  S.~Forte, A.~Guffanti, J.~I. Latorre, J.~Rojo, and M.~Ubiali {\em Nucl.
  Phys.} {\bf B849} (2011) 112--143,
  [\href{http://arxiv.org/abs/1012.0836}{{\tt arXiv:1012.0836}}]. [Erratum:
  Nucl. Phys.B855,927(2012)].

\bibitem{Nocera:2014rea}
E.~R. Nocera {\em PoS} {\bf DIS2014} (2014) 204.

\bibitem{Bartels:1995iu}
J.~Bartels, B.~I. Ermolaev, and M.~G. Ryskin {\em Z. Phys.} {\bf C70} (1996)
  273--280, [\href{http://arxiv.org/abs/hep-ph/9507271}{{\tt hep-ph/9507271}}].

\bibitem{Bartels:1996wc}
J.~Bartels, B.~I. Ermolaev, and M.~G. Ryskin {\em Z. Phys.} {\bf C72} (1996)
  627--635, [\href{http://arxiv.org/abs/hep-ph/9603204}{{\tt hep-ph/9603204}}].

\bibitem{Kovchegov:2015pbl}
Y.~V. Kovchegov, D.~Pitonyak, and M.~D. Sievert {\em JHEP} {\bf 01} (2016) 072,
  [\href{http://arxiv.org/abs/1511.06737}{{\tt arXiv:1511.06737}}]. [Erratum:
  JHEP10,148(2016)].

\bibitem{Kovchegov:2016weo}
Y.~V. Kovchegov, D.~Pitonyak, and M.~D. Sievert {\em Phys. Rev. Lett.} {\bf
  118} (2017), no.~5 052001, [\href{http://arxiv.org/abs/1610.06188}{{\tt
  arXiv:1610.06188}}].

\bibitem{Kovchegov:2016zex}
Y.~V. Kovchegov, D.~Pitonyak, and M.~D. Sievert {\em Phys. Rev.} {\bf D95}
  (2017), no.~1 014033, [\href{http://arxiv.org/abs/1610.06197}{{\tt
  arXiv:1610.06197}}].

\bibitem{Kovchegov:2017jxc}
Y.~V. Kovchegov, D.~Pitonyak, and M.~D. Sievert {\em Phys. Lett.} {\bf B772}
  (2017) 136--140, [\href{http://arxiv.org/abs/1703.05809}{{\tt
  arXiv:1703.05809}}].

\bibitem{Kovchegov:2017lsr}
Y.~V. Kovchegov, D.~Pitonyak, and M.~D. Sievert
  \href{http://arxiv.org/abs/1706.04236}{{\tt arXiv:1706.04236}}.

\bibitem{Leader:2011tm}
E.~Leader, A.~V. Sidorov, and D.~B. Stamenov {\em Phys. Rev.} {\bf D84} (2011)
  014002, [\href{http://arxiv.org/abs/1103.5979}{{\tt arXiv:1103.5979}}].

\bibitem{Aschenauer:2015ata}
E.~C. Aschenauer, R.~Sassot, and M.~Stratmann {\em Phys. Rev.} {\bf D92}
  (2015), no.~9 094030, [\href{http://arxiv.org/abs/1509.06489}{{\tt
  arXiv:1509.06489}}].

\bibitem{Nocera:2015vva}
E.~R. Nocera {\em J. Phys. Conf. Ser.} {\bf 678} (2016), no.~1 012030,
  [\href{http://arxiv.org/abs/1510.04248}{{\tt arXiv:1510.04248}}].

\bibitem{Nocera:2017wep}
E.~R. Nocera, {\it {Impact of Recent RHIC Data on Helicity-Dependent Parton
  Distribution Functions}},  in {\em {22nd International Symposium on Spin
  Physics (SPIN 2016) Urbana, IL, USA, September 25-30, 2016}}, 2017.
\newblock \href{http://arxiv.org/abs/1702.05077}{{\tt arXiv:1702.05077}}.

\bibitem{Aschenauer:2012ve}
E.~C. Aschenauer, R.~Sassot, and M.~Stratmann {\em Phys. Rev.} {\bf D86} (2012)
  054020, [\href{http://arxiv.org/abs/1206.6014}{{\tt arXiv:1206.6014}}].

\bibitem{Ball:2013tyh}
{\bf NNPDF} Collaboration, R.~D. Ball, S.~Forte, A.~Guffanti, E.~R. Nocera,
  G.~Ridolfi, and J.~Rojo {\em Phys. Lett.} {\bf B728} (2014) 524--531,
  [\href{http://arxiv.org/abs/1310.0461}{{\tt arXiv:1310.0461}}].

\bibitem{Aschenauer:2013iia}
E.~C. Aschenauer, T.~Burton, T.~Martini, H.~Spiesberger, and M.~Stratmann {\em
  Phys. Rev.} {\bf D88} (2013) 114025,
  [\href{http://arxiv.org/abs/1309.5327}{{\tt arXiv:1309.5327}}].

\bibitem{Green:2012ud}
J.~R. Green, M.~Engelhardt, S.~Krieg, J.~W. Negele, A.~V. Pochinsky, and S.~N.
  Syritsyn {\em Phys. Lett.} {\bf B734} (2014) 290--295,
  [\href{http://arxiv.org/abs/1209.1687}{{\tt arXiv:1209.1687}}].

\bibitem{Alexandrou:2017oeh}
C.~Alexandrou, M.~Constantinou, K.~Hadjiyiannakou, K.~Jansen, C.~Kallidonis,
  G.~Koutsou, A.~V. Avilés-Casco, and C.~Wiese {\em Phys. Rev. Lett.} {\bf
  119} (2017) 142002, [\href{http://arxiv.org/abs/1706.02973}{{\tt
  arXiv:1706.02973}}].

\bibitem{Bali:2014gha}
G.~S. Bali, S.~Collins, B.~Gläßle, M.~Göckeler, J.~Najjar, R.~H. Rödl,
  A.~Schäfer, R.~W. Schiel, A.~Sternbeck, and W.~Söldner {\em Phys. Rev.}
  {\bf D90} (2014), no.~7 074510, [\href{http://arxiv.org/abs/1408.6850}{{\tt
  arXiv:1408.6850}}].

\bibitem{Berkowitz:2017gql}
E.~Berkowitz et~al. \href{http://arxiv.org/abs/1704.01114}{{\tt
  arXiv:1704.01114}}.

\bibitem{Bhattacharya:2016zcn}
T.~Bhattacharya, V.~Cirigliano, S.~Cohen, R.~Gupta, H.-W. Lin, and B.~Yoon {\em
  Phys. Rev.} {\bf D94} (2016), no.~5 054508,
  [\href{http://arxiv.org/abs/1606.07049}{{\tt arXiv:1606.07049}}].

\bibitem{Capitani:2017qpc}
S.~Capitani, M.~Della~Morte, D.~Djukanovic, G.~M. von Hippel, J.~Hua,
  B.~Jäger, P.~M. Junnarkar, H.~B. Meyer, T.~D. Rae, and H.~Wittig
  \href{http://arxiv.org/abs/1705.06186}{{\tt arXiv:1705.06186}}.

\bibitem{Alexandrou:2017hac}
C.~Alexandrou, M.~Constantinou, K.~Hadjiyiannakou, K.~Jansen, C.~Kallidonis,
  G.~Koutsou, and A.~Vaquero Avilés-Casco {\em Phys. Rev.} {\bf D96} (2017),
  no.~5 054507, [\href{http://arxiv.org/abs/1705.03399}{{\tt
  arXiv:1705.03399}}].

\bibitem{Bali:2014nma}
G.~S. Bali, S.~Collins, B.~Glässle, M.~Göckeler, J.~Najjar, R.~H. Rödl,
  A.~Schäfer, R.~W. Schiel, W.~Söldner, and A.~Sternbeck {\em Phys. Rev.}
  {\bf D91} (2015), no.~5 054501, [\href{http://arxiv.org/abs/1412.7336}{{\tt
  arXiv:1412.7336}}].

\bibitem{Horsley:2013ayv}
R.~Horsley, Y.~Nakamura, A.~Nobile, P.~E.~L. Rakow, G.~Schierholz, and J.~M.
  Zanotti {\em Phys. Lett.} {\bf B732} (2014) 41--48,
  [\href{http://arxiv.org/abs/1302.2233}{{\tt arXiv:1302.2233}}].

\bibitem{Schmelling:1994pz}
M.~Schmelling {\em Phys. Scripta} {\bf 51} (1995) 676--679.

\bibitem{Gong:2015iir}
{\bf $\chi$QCD} Collaboration, M.~Gong, Y.-B. Yang, J.~Liang, A.~Alexandru,
  T.~Draper, and K.-F. Liu {\em Phys. Rev.} {\bf D95} (2017), no.~11 114509,
  [\href{http://arxiv.org/abs/1511.03671}{{\tt arXiv:1511.03671}}].

\bibitem{Engelhardt:2012gd}
M.~Engelhardt {\em Phys. Rev.} {\bf D86} (2012) 114510,
  [\href{http://arxiv.org/abs/1210.0025}{{\tt arXiv:1210.0025}}].

\bibitem{Aoki:2010xg}
Y.~Aoki, T.~Blum, H.-W. Lin, S.~Ohta, S.~Sasaki, R.~Tweedie, J.~Zanotti, and
  T.~Yamazaki {\em Phys. Rev.} {\bf D82} (2010) 014501,
  [\href{http://arxiv.org/abs/1003.3387}{{\tt arXiv:1003.3387}}].

\bibitem{Bratt:2010jn}
{\bf LHPC} Collaboration, J.~D. Bratt et~al. {\em Phys. Rev. D} {\bf 82} (2010)
  094502, [\href{http://arxiv.org/abs/1001.3620}{{\tt arXiv:1001.3620}}].

\bibitem{Abdel-Rehim:2015owa}
A.~Abdel-Rehim et~al. {\em Phys. Rev.} {\bf D92} (2015), no.~11 114513,
  [\href{http://arxiv.org/abs/1507.04936}{{\tt arXiv:1507.04936}}]. [Erratum:
  Phys. Rev.D93,no.3,039904(2016)].

\bibitem{Ball:2014uwa}
{\bf NNPDF} Collaboration, R.~D. Ball et~al. {\em JHEP} {\bf 04} (2015) 040,
  [\href{http://arxiv.org/abs/1410.8849}{{\tt arXiv:1410.8849}}].

\bibitem{Ball:2011gg}
R.~D. Ball, V.~Bertone, F.~Cerutti, L.~Del~Debbio, S.~Forte, A.~Guffanti, N.~P.
  Hartland, J.~I. Latorre, J.~Rojo, and M.~Ubiali {\em Nucl. Phys.} {\bf B855}
  (2012) 608--638, [\href{http://arxiv.org/abs/1108.1758}{{\tt
  arXiv:1108.1758}}].

\bibitem{Camarda:2015zba}
{\bf HERAFitter developers' Team} Collaboration, S.~Camarda et~al. {\em Eur.
  Phys. J.} {\bf C75} (2015), no.~9 458,
  [\href{http://arxiv.org/abs/1503.05221}{{\tt arXiv:1503.05221}}].

\bibitem{Paukkunen:2014zia}
H.~Paukkunen and P.~Zurita {\em JHEP} {\bf 12} (2014) 100,
  [\href{http://arxiv.org/abs/1402.6623}{{\tt arXiv:1402.6623}}].

\bibitem{Alekhin:2014irh}
S.~Alekhin et~al. {\em Eur. Phys. J.} {\bf C75} (2015), no.~7 304,
  [\href{http://arxiv.org/abs/1410.4412}{{\tt arXiv:1410.4412}}].

\bibitem{Lin:2017stx}
H.-W. Lin, W.~Melnitchouk, A.~Prokudin, N.~Sato, and H.~Shows
  \href{http://arxiv.org/abs/1710.09858}{{\tt arXiv:1710.09858}}.

\bibitem{Angeles-Martinez:2015sea}
R.~Angeles-Martinez et~al. {\em Acta Phys. Polon.} {\bf B46} (2015), no.~12
  2501--2534, [\href{http://arxiv.org/abs/1507.05267}{{\tt arXiv:1507.05267}}].

\bibitem{Musch:2011er}
B.~U. Musch, P.~Hagler, M.~Engelhardt, J.~W. Negele, and A.~Schafer {\em Phys.
  Rev.} {\bf D85} (2012) 094510, [\href{http://arxiv.org/abs/1111.4249}{{\tt
  arXiv:1111.4249}}].

\bibitem{Engelhardt:2015xja}
M.~Engelhardt, P.~Hägler, B.~Musch, J.~Negele, and A.~Schäfer {\em Phys.
  Rev.} {\bf D93} (2016), no.~5 054501,
  [\href{http://arxiv.org/abs/1506.07826}{{\tt arXiv:1506.07826}}].

\bibitem{Yoon:2017qzo}
B.~Yoon, M.~Engelhardt, R.~Gupta, T.~Bhattacharya, J.~R. Green, B.~U. Musch,
  J.~W. Negele, A.~V. Pochinsky, A.~Schäfer, and S.~N. Syritsyn {\em Phys.
  Rev.} {\bf D96} (2017), no.~9 094508,
  [\href{http://arxiv.org/abs/1706.03406}{{\tt arXiv:1706.03406}}].

\bibitem{Sutton:1991ay}
P.~J. Sutton, A.~D. Martin, R.~G. Roberts, and W.~J. Stirling {\em Phys. Rev.}
  {\bf D45} (1992) 2349--2359.

\bibitem{Burkardt:2001jg}
M.~Burkardt and S.~Dalley {\em Prog. Part. Nucl. Phys.} {\bf 48} (2002)
  317--362, [\href{http://arxiv.org/abs/hep-ph/0112007}{{\tt hep-ph/0112007}}].

\bibitem{Abdel-Rehim:2013wlz}
A.~Abdel-Rehim, C.~Alexandrou, M.~Constantinou, V.~Drach, K.~Hadjiyiannakou,
  K.~Jansen, G.~Koutsou, and A.~Vaquero {\em Phys. Rev.} {\bf D89} (2014),
  no.~3 034501, [\href{http://arxiv.org/abs/1310.6339}{{\tt arXiv:1310.6339}}].

\bibitem{Deka:2013zha}
M.~Deka et~al. {\em Phys. Rev.} {\bf D91} (2015), no.~1 014505,
  [\href{http://arxiv.org/abs/1312.4816}{{\tt arXiv:1312.4816}}].

\bibitem{Alexandrou:2016ekb}
C.~Alexandrou, M.~Constantinou, K.~Hadjiyiannakou, K.~Jansen, H.~Panagopoulos,
  and C.~Wiese {\em Phys. Rev.} {\bf D96} (2017), no.~5 054503,
  [\href{http://arxiv.org/abs/1611.06901}{{\tt arXiv:1611.06901}}].

\bibitem{Dolgov:2002zm}
{\bf LHPC, SESAM} Collaboration, D.~Dolgov et~al. {\em Phys. Rev.} {\bf D66}
  (2002) 034506, [\href{http://arxiv.org/abs/hep-lat/0201021}{{\tt
  hep-lat/0201021}}].

\bibitem{Deka:2008xr}
M.~Deka, T.~Streuer, T.~Doi, S.~J. Dong, T.~Draper, K.~F. Liu, N.~Mathur, and
  A.~W. Thomas {\em Phys. Rev.} {\bf D79} (2009) 094502,
  [\href{http://arxiv.org/abs/0811.1779}{{\tt arXiv:0811.1779}}].

\bibitem{Green:2017keo}
J.~Green, N.~Hasan, S.~Meinel, M.~Engelhardt, S.~Krieg, J.~Laeuchli, J.~Negele,
  K.~Orginos, A.~Pochinsky, and S.~Syritsyn {\em Phys. Rev.} {\bf D95} (2017),
  no.~11 114502, [\href{http://arxiv.org/abs/1703.06703}{{\tt
  arXiv:1703.06703}}].

\bibitem{QCDSF:2011aa}
{\bf QCDSF} Collaboration, G.~S. Bali et~al. {\em Phys. Rev. Lett.} {\bf 108}
  (2012) 222001, [\href{http://arxiv.org/abs/1112.3354}{{\tt
  arXiv:1112.3354}}].

\bibitem{Yoon:2016jzj}
B.~Yoon et~al. {\em Phys. Rev.} {\bf D95} (2017), no.~7 074508,
  [\href{http://arxiv.org/abs/1611.07452}{{\tt arXiv:1611.07452}}].

\bibitem{vonHippel:2016wid}
G.~von Hippel, T.~D. Rae, E.~Shintani, and H.~Wittig {\em Nucl. Phys.} {\bf
  B914} (2017) 138--159, [\href{http://arxiv.org/abs/1605.00564}{{\tt
  arXiv:1605.00564}}].

\bibitem{Dragos:2016rtx}
J.~Dragos, R.~Horsley, W.~Kamleh, D.~B. Leinweber, Y.~Nakamura, P.~E.~L. Rakow,
  G.~Schierholz, R.~D. Young, and J.~M. Zanotti {\em Phys. Rev.} {\bf D94}
  (2016), no.~7 074505, [\href{http://arxiv.org/abs/1606.03195}{{\tt
  arXiv:1606.03195}}].

\bibitem{Yang:2015zja}
Y.-B. Yang, A.~Alexandru, T.~Draper, M.~Gong, and K.-F. Liu {\em Phys. Rev.}
  {\bf D93} (2016), no.~3 034503, [\href{http://arxiv.org/abs/1509.04616}{{\tt
  arXiv:1509.04616}}].

\bibitem{Bhattacharya:2013ehc}
T.~Bhattacharya, S.~D. Cohen, R.~Gupta, A.~Joseph, H.-W. Lin, and B.~Yoon {\em
  Phys. Rev.} {\bf D89} (2014), no.~9 094502,
  [\href{http://arxiv.org/abs/1306.5435}{{\tt arXiv:1306.5435}}].

\bibitem{Alexandrou:2013joa}
C.~Alexandrou, M.~Constantinou, S.~Dinter, V.~Drach, K.~Jansen, C.~Kallidonis,
  and G.~Koutsou {\em Phys. Rev.} {\bf D88} (2013), no.~1 014509,
  [\href{http://arxiv.org/abs/1303.5979}{{\tt arXiv:1303.5979}}].

\bibitem{Owen:2012ts}
B.~J. Owen, J.~Dragos, W.~Kamleh, D.~B. Leinweber, M.~S. Mahbub, B.~J. Menadue,
  and J.~M. Zanotti {\em Phys. Lett.} {\bf B723} (2013) 217--223,
  [\href{http://arxiv.org/abs/1212.4668}{{\tt arXiv:1212.4668}}].

\bibitem{Capitani:2012gj}
S.~Capitani, M.~Della~Morte, G.~von Hippel, B.~Jäger, A.~Jüttner,
  B.~Knippschild, H.~B. Meyer, and H.~Wittig {\em Phys. Rev.} {\bf D86} (2012)
  074502, [\href{http://arxiv.org/abs/1205.0180}{{\tt arXiv:1205.0180}}].

\bibitem{Alexandrou:2011nr}
C.~Alexandrou, J.~Carbonell, M.~Constantinou, P.~A. Harraud, P.~Guichon,
  K.~Jansen, C.~Kallidonis, T.~Korzec, and M.~Papinutto {\em Phys. Rev.} {\bf
  D83} (2011) 114513, [\href{http://arxiv.org/abs/1104.1600}{{\tt
  arXiv:1104.1600}}].

\bibitem{Yamazaki:2009zq}
T.~Yamazaki, Y.~Aoki, T.~Blum, H.-W. Lin, S.~Ohta, S.~Sasaki, R.~Tweedie, and
  J.~Zanotti {\em Phys. Rev.} {\bf D79} (2009) 114505,
  [\href{http://arxiv.org/abs/0904.2039}{{\tt arXiv:0904.2039}}].

\bibitem{Yamazaki:2008py}
{\bf RBC+UKQCD} Collaboration, T.~Yamazaki, Y.~Aoki, T.~Blum, H.~W. Lin, M.~F.
  Lin, S.~Ohta, S.~Sasaki, R.~J. Tweedie, and J.~M. Zanotti {\em Phys. Rev.
  Lett.} {\bf 100} (2008) 171602, [\href{http://arxiv.org/abs/0801.4016}{{\tt
  arXiv:0801.4016}}].

\bibitem{Lin:2008uz}
H.-W. Lin, T.~Blum, S.~Ohta, S.~Sasaki, and T.~Yamazaki {\em Phys. Rev.} {\bf
  D78} (2008) 014505, [\href{http://arxiv.org/abs/0802.0863}{{\tt
  arXiv:0802.0863}}].

\bibitem{Hagler:2007xi}
{\bf LHPC} Collaboration, P.~Hägler et~al. {\em Phys. Rev.} {\bf D77} (2008)
  094502, [\href{http://arxiv.org/abs/0705.4295}{{\tt arXiv:0705.4295}}].

\bibitem{Alexandrou:2007xj}
C.~Alexandrou, G.~Koutsou, T.~Leontiou, J.~W. Negele, and A.~Tsapalis {\em
  Phys. Rev.} {\bf D76} (2007) 094511,
  [\href{http://arxiv.org/abs/0706.3011}{{\tt arXiv:0706.3011}}]. [Erratum:
  Phys. Rev.D80,099901(2009)].

\bibitem{Edwards:2005ym}
{\bf LHPC} Collaboration, R.~G. Edwards, G.~T. Fleming, P.~Hägler, J.~W.
  Negele, K.~Orginos, A.~V. Pochinsky, D.~B. Renner, D.~G. Richards, and
  W.~Schroers {\em Phys. Rev. Lett.} {\bf 96} (2006) 052001,
  [\href{http://arxiv.org/abs/hep-lat/0510062}{{\tt hep-lat/0510062}}].

\bibitem{Khan:2006de}
A.~A. Khan et~al. {\em Phys. Rev.} {\bf D74} (2006) 094508,
  [\href{http://arxiv.org/abs/hep-lat/0603028}{{\tt hep-lat/0603028}}].

\bibitem{Babich:2010at}
R.~Babich, R.~C. Brower, M.~A. Clark, G.~T. Fleming, J.~C. Osborn, C.~Rebbi,
  and D.~Schaich {\em Phys. Rev.} {\bf D85} (2012) 054510,
  [\href{http://arxiv.org/abs/1012.0562}{{\tt arXiv:1012.0562}}].

\bibitem{Gusken:1999as}
{\bf SESAM} Collaboration, S.~Güsken, P.~Ueberholz, J.~Viehoff, N.~Eicker,
  T.~Lippert, K.~Schilling, A.~Spitz, and T.~Struckmann {\em Phys. Rev.} {\bf
  D59} (1999) 114502, [\href{http://arxiv.org/abs/hep-lat/9901009}{{\tt
  hep-lat/9901009}}].

\bibitem{Dong:1995rx}
S.~J. Dong, J.~F. Lagae, and K.~F. Liu {\em Phys. Rev. Lett.} {\bf 75} (1995)
  2096--2099, [\href{http://arxiv.org/abs/hep-ph/9502334}{{\tt
  hep-ph/9502334}}].

\bibitem{Fukugita:1994fh}
M.~Fukugita, Y.~Kuramashi, M.~Okawa, and A.~Ukawa {\em Phys. Rev. Lett.} {\bf
  75} (1995) 2092--2095, [\href{http://arxiv.org/abs/hep-lat/9501010}{{\tt
  hep-lat/9501010}}].

\bibitem{Gupta:1994qw}
R.~Gupta and J.~E. Mandula {\em Phys. Rev.} {\bf D50} (1994) 6931--6938,
  [\href{http://arxiv.org/abs/hep-lat/9402018}{{\tt hep-lat/9402018}}].

\bibitem{Alles:1994ss}
B.~Allés, M.~Campostrini, L.~Del~Debbio, A.~Di~Giacomo, H.~Panagopoulos, and
  E.~Vicari {\em Phys. Lett.} {\bf B336} (1994) 248--250,
  [\href{http://arxiv.org/abs/hep-lat/9402019}{{\tt hep-lat/9402019}}].

\bibitem{Altmeyer:1992nt}
R.~Altmeyer, M.~Göckeler, R.~Horsley, E.~Laermann, and G.~Schierholz {\em
  Phys. Rev.} {\bf D49} (1994) 3087--3090.

\bibitem{Mandula:1992bc}
J.~E. Mandula and M.~C. Ogilvie {\em Phys. Lett.} {\bf B312} (1993) 327--332,
  [\href{http://arxiv.org/abs/hep-lat/9208009}{{\tt hep-lat/9208009}}].

\bibitem{Liu:1995kb}
K.-F. Liu, {\it {Comments on lattice calculations of proton spin components}},
  in {\em {Future physics with light and heavy nuclei at HERMES. Contributions
  to the proceedings of the '95 - '96 Workshop on Future Physics at HERA,
  working group on light and heavy nuclei at HERA. Part 2}}, 1995.
\newblock \href{http://arxiv.org/abs/hep-lat/9510046}{{\tt hep-lat/9510046}}.

\bibitem{Alexandrou:2015qia}
C.~Alexandrou, M.~Constantinou, S.~Dinter, V.~Drach, K.~Hadjiyiannakou,
  K.~Jansen, G.~Koutsou, and A.~Vaquero {\em JHEP} {\bf 06} (2015) 068,
  [\href{http://arxiv.org/abs/1501.03734}{{\tt arXiv:1501.03734}}].

\bibitem{Bali:2012av}
G.~S. Bali, S.~Collins, M.~Deka, B.~Gläßle, M.~Göckeler, J.~Najjar,
  A.~Nobile, D.~Pleiter, A.~Schäfer, and A.~Sternbeck {\em Phys. Rev.} {\bf
  D86} (2012) 054504, [\href{http://arxiv.org/abs/1207.1110}{{\tt
  arXiv:1207.1110}}].

\bibitem{Pleiter:2011gw}
{\bf QCDSF/UKQCD} Collaboration, D.~Pleiter et~al. {\em PoS} {\bf LATTICE2010}
  (2010) 153, [\href{http://arxiv.org/abs/1101.2326}{{\tt arXiv:1101.2326}}].

\bibitem{Syritsyn:2011vk}
S.~N. Syritsyn, J.~R. Green, J.~W. Negele, A.~V. Pochinsky, M.~Engelhardt,
  P.~Hagler, B.~Musch, and W.~Schroers {\em PoS} {\bf LATTICE2011} (2011) 178,
  [\href{http://arxiv.org/abs/1111.0718}{{\tt arXiv:1111.0718}}].

\bibitem{Gockeler:1997zr}
M.~Göckeler, R.~Horsley, L.~Mankiewicz, H.~Perlt, P.~E.~L. Rakow,
  G.~Schierholz, and A.~Schiller {\em Phys. Lett.} {\bf B414} (1997) 340--346,
  [\href{http://arxiv.org/abs/hep-ph/9708270}{{\tt hep-ph/9708270}}].

\bibitem{Gockeler:2005vw}
M.~Göckeler, R.~Horsley, D.~Pleiter, P.~E.~L. Rakow, A.~Schäfer,
  G.~Schierholz, H.~Stüben, and J.~M. Zanotti {\em Phys. Rev.} {\bf D72}
  (2005) 054507, [\href{http://arxiv.org/abs/hep-lat/0506017}{{\tt
  hep-lat/0506017}}].

\bibitem{Gockeler:2004vx}
{\bf QCDSF} Collaboration, M.~Göckeler, P.~Hägler, R.~Horsley, D.~Pleiter,
  P.~E.~L. Rakow, A.~Schäfer, G.~Schierholz, and J.~M. Zanotti {\em Nucl.
  Phys. Proc. Suppl.} {\bf 140} (2005) 399--404,
  [\href{http://arxiv.org/abs/hep-lat/0409162}{{\tt hep-lat/0409162}}].
  [,399(2004)].

\bibitem{Gockeler:2000ja}
M.~Göckeler, R.~Horsley, W.~Kürzinger, H.~Oelrich, D.~Pleiter, P.~E.~L.
  Rakow, A.~Schäfer, and G.~Schierholz {\em Phys. Rev.} {\bf D63} (2001)
  074506, [\href{http://arxiv.org/abs/hep-lat/0011091}{{\tt hep-lat/0011091}}].

\end{thebibliography}\endgroup

\end{document}